\documentclass{elsart}
\usepackage{graphics,epsf}
\usepackage{rotating}
\usepackage{bbm}

\newcommand{\G}{\gamma}
\newcommand{\GG}{{\gamma\gamma}}
\newcommand{\EPEM}{e^+e^-}
\newcommand{\MUPM}{\mu^+\mu^-}
\newcommand{\A}{\alpha}
\newcommand{\ZA}{Z\alpha}
\newcommand{\BE}{\begin{equation}}
\newcommand{\EE}{\end{equation}}
\renewcommand{\P}{{\mathbbm P}}

\makeatletter
\def\lesssim{\mathrel{\mathpalette\vereq<}}
\def\vereq#1#2{\lower3pt\vbox{\baselineskip1.5pt \lineskip1.5pt
\ialign{$\m@th#1\hfill##\hfil$\crcr#2\crcr\sim\crcr}}}
\def\gtrsim{\mathrel{\mathpalette\vereq>}}
\makeatother

\begin{document}
\runauthor{Baur, Hencken, Trautmann, Sadovsky, Kharlov}
\begin{frontmatter}
\title{Coherent $\gamma\gamma$ and $\gamma$A interactions in very peripheral
collisions at relativistic ion colliders.}
\author[FZJ]{Gerhard Baur}
\author[UBA]{Kai Hencken}
\author[UBA]{Dirk Trautmann}
\author[IHEP]{Serguei Sadovsky}
\author[IHEP]{Yuri Kharlov}

\address[FZJ]{Forschungszentrum J\"ulich, J\"ulich, Germany}
\address[UBA]{Universit\"at Basel, Basel, Switzerland}
\address[IHEP]{IHEP, Protvino, Russia}
\begin{abstract}
  Due to coherence, there are strong electromagnetic fields of short
  duration in very peripheral collisions. They give rise to
  photon-photon and photon-nucleus collisions with a high flux up to an
  invariant mass region hitherto unexplored experimentally. After a
  general survey of the field equivalent photon numbers and photon-photon 
  luminosities, especially for relativistic heavy ion collisions, are 
  discussed. Special care needs to be taken to include the effects of the
  strong interaction and nuclear size in this case. Photon-photon and
  photon-hadron physics at various invariant mass scales are then
  discussed. The maximum equivalent photon energy in the lab-system
  (collider frame) are typically of the order of 3 GeV for RHIC and
  100 GeV for LHC. Diffractive processes are an important background
  process. Lepton-pair, especially electron-positron pair production is 
  copious. Due to the strong fields there will be new phenomena, like 
  multiple $\EPEM$ pair production. The experimental techniques
  to select $\GG$-processes are finally discussed together with important
  background processes.
\end{abstract}
\begin{keyword}
photon-photon processes, photon-hadron processes, relativistic heavy ion
collisions.
\end{keyword}
\end{frontmatter}

\tableofcontents

\section{Introduction and Purpose}
\label{sec_intro}

With the first collisions at the relativistic heavy ion collider RHIC
at Brookhaven (BNL) in June 2000 heavy ion physics has entered a new stage.
The interaction of nuclei at such high energies has only been 
observed up to now in cosmic ray interactions. Many interesting 
physics topics can now be studied in the laboratory. The 
main aim is to study the violent 
central collisions. One is looking for the formation and the 
signature of a new state of hadronic matter, the Quark Gluon Plasma.
The present status of this field is exposed in the Proceedings of 
Quark Matter QM2001 \cite{QM2001}.

Relativistic heavy ions are also very important tools for other physics 
investigations. We mention here projects (sometimes referred to 
as {\it ``non-QGP'' physics}, a term which we will avoid in the following)
like the search for new physics at very high rapidities in the CASTOR
subproject at ALICE. This project is related to cosmic ray physics and
searches for the so called Centauro events at the LHC \cite{Angelis97}.
Other ``exotic'' physics topics like  the possible occurrence of CP 
violation or ``disoriented chiral condensates'' (DCC), see
\cite{Anselm1989,Bjorken1997}, have also been investigated in the past.

\begin{figure}[tbh]
\begin{center}
\resizebox{4.5cm}{!}{\includegraphics{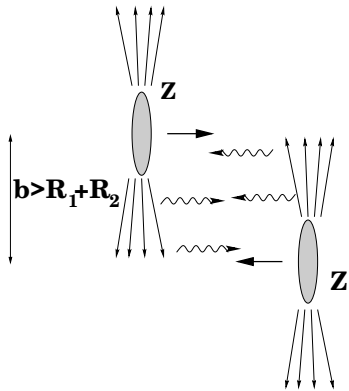}}
\end{center}
\caption{
A fast moving nucleus with charge $Ze$ is surrounded by a strong
electromagnetic field. This can be viewed as a cloud of virtual
photons. These photons can often be considered as real. They are
called ``{\it equivalent}'' or ``{\it quasireal photons}''.
In the collision of two ions these quasireal photons can collide with 
each other and with the other nucleus. For very peripheral collisions with 
impact parameters $b>2R$, this is useful for photon-photon as well as 
photon-nucleus collisions.}
\label{fig_collisions}
\end{figure}

It is the purpose of this Report to review the physics 
of very peripheral\footnote{In the following we will always use 
``very peripheral'' to denote the distant collisions with $b>R_1+R_2$ we are 
interested in. Here $b$ denotes the impact parameter and $R_1$ and $R_2$
are the radii of the two nuclei. The term ``ultraperipheral''  collisions
is also sometimes used for them. This is an attempt to distinguish it clearly 
from what is sometimes called ``peripheral collisions'', where 
$b\approx R_1+R_2$.}
collisions. We put the  main emphasis on the energy regions of the 
colliders RHIC (Relativistic Heavy Ion Collider) at Brookhaven and the 
forthcoming LHC (Large Hadron Collider) at CERN/Geneva.  
Due to the coherent action of all the protons
in the nucleus, the electromagnetic field surrounding the ions 
is very strong. It acts for a very short time. 
According to Fermi \cite{Fermi24}
``{\it this time-dependent electromagnetic field can be 
replaced by the field of radiation with a corresponding frequency 
distribution}'', see also Fig.~\ref{fig_collisions}. He called this  
``\"aquivalente Strahlung''.
The spectrum of these photons can be calculated from the kinematics of 
the process. The "equivalent photon method" (EPA) is often called also the 
"Weizs\"acker-Williams method''. For a more popular 
introduction see, e.g., \cite{BertulaniB94}, and also the corresponding 
remarks in \cite{JacksonED}.

A very useful view on the electromagnetic processes is the parton
picture with the photons seen as the partons. We give the basic
argument here in the introduction, a detailed explanation will then
follow in Chap.~\ref{sec_lum}. In this picture the scattering is 
described as an incoherent superposition of the scattering of the various 
constituents. For example, nuclei consist of nucleons which in turn consist 
of quarks and gluons, photons consist of lepton pairs, electrons consist of
photons, etc.. Relativistic nuclei have photons as an important constituent, 
especially for low virtuality and a low ratio of the photon energy compared 
to the one of the ion (see below for quantitative arguments).
This is due to the coherent action of all the charges in the nucleus: for 
these conditions the wavelength of the photon is larger than the size of the
nucleus, therefore it does not resolve the individual nucleons but sees the 
coherent action of them. This has some similarity to the ``low-x physics''
and the Weizs\"acker-Williams approach of \cite{Krasnitz:2001ph}
to gluons in the initial state of heavy ion collisions.

The coherence condition limits the virtuality $Q^2=-q^2$ of the photon to 
very low values
\BE
Q^2 \lesssim 1/R^2, 
\label{eq:cond_coherent}
\EE
where the radius of a nucleus is
approximately $R=1.2\mbox{ fm}\ A^{1/3}$ with $A$ the nucleon
number. This is due to the rapid decrease of the 
nuclear electromagnetic form factor for high $Q^2$ values.
For most purposes these photons can therefore be considered as 
real (``quasireal'').
From the kinematics of the process one has a photon four-momentum of 
$q^\mu=(\omega,\vec q_\perp,q_3=\omega/v)$, where $\omega$ and $q_\perp$ 
are energy and transverse momentum of the quasireal photon in a given
frame, where the projectile moves with velocity $v$. This leads to an 
invariant four-momentum transfer of 
\BE
Q^2=\frac{\omega^2}{\G^2}+q_\perp^2,
\EE
where the Lorentz factor is $\gamma=E/m=1/\sqrt{1-v^2}$.
The condition Eq.~(\ref{eq:cond_coherent}) limits the maximum energy of the 
quasireal photon to 
\BE
\omega<\omega_{max} \approx \frac{\G}{R},
\label{eq_wmax}
\EE
and the perpendicular component of its momentum to
\BE
q_\perp \lesssim \frac{1}{R}.
\label{pt_max}
\EE
At LHC energies this means a maximum photon energy of about 
100~GeV in the laboratory system, at RHIC this number is 
about 3~GeV. We define the
ratio $x=\omega/E$, where $E$ denotes
the energy of the nucleus $E= M_N \G A$ and $M_N$ is the nucleon mass.
It is therefore smaller than
\BE
x< x_{max}=\frac{1}{R M_N A} = \frac{\lambda_C(A)}{R},
\EE
where $\lambda_C(A)$ is the Compton wavelength of the ion. 
$x_{max}$ is $4\times 10^{-3}$, $3\times 10^{-4}$, $1.4\times 10^{-4}$  
for O, Sn, Pb ions, respectively. Here and
also throughout the rest of the paper we use natural units, i.e.,
$\hbar=c=1$.

The collisions of $e^+$ and $e^-$ has been the traditional way to
study $\GG$ collisions. Similarly photon-photon collisions can also be
observed in hadron-hadron collisions, see Fig.~\ref{fig_collisions}. 
Since the photon number scales
with $Z^2$ ($Z$ being the charge number of the nucleus) such effects
can be particularly large. This factor of $\gtrsim 6000$ (corresponding
to Au, $Z=79$) is the reason
why the name ``gold flashlight'' \cite{KleinS97a} has been used to
describe very peripheral (AuAu) collisions at RHIC. 

Similarly the strong electromagnetic field can be used as a source
of photons to induce electromagnetic reactions in the second ion, 
see Fig.~\ref{fig_collisions}.
Since the ion, which is hit by these photons, is moving in the 
collider frame the photon-hadron invariant masses can become 
very high. In the rest frame of one of the ions (sometimes called 
the "target frame") the Lorentz factor of the other ion is given by  
$\G_{ion}=2 \G_{lab}^2 -1$, where $\G_{lab}$ is the Lorentz factor in 
the collider (cm) frame.
The maximum photon energy in this frame is 500~TeV for the LHC and 
600~GeV for RHIC.

This high equivalent photon flux has already found many useful applications in
nuclear physics \cite{BertulaniB88}, nuclear astrophysics
\cite{BaurR94,BaurR96}, particle physics \cite{Primakoff51} (sometimes
called the ``Primakoff effect''), as well as, atomic physics 
\cite{Moshammer97}. Previous  reviews of the present topic
can be found in \cite{BaurHT98,KraussGS97,BaurHT98b,HenckenSTB99}.

The theoretical tool to analyze very peripheral collisions is 
the equivalent photon method. This method is described in 
Chap.~\ref{sec_lum}. 
The equivalent photon method is often used for high energy reactions, e.g.,
in the description of $\EPEM$ and $e$p collisions. We put most emphasis here 
on the peculiarities which arise due to the finite size and the strong 
interactions of the ions. These equivalent photons can collide with the other
nucleus and with each other. This leads to the possibility of doing 
photon-hadron and photon-photon physics at relativistic heavy ion 
accelerators, with high fluxes of equivalent photons in hitherto 
inaccessible invariant mass regions.

Thus there will be new possibilities to study photon-hadron interactions
at RHIC and LHC. It extends the $\G$p interaction studies at HERA/DESY to 
$\G$A interactions. The photon-hadron invariant mass range at RHIC will be 
somewhat below the one at HERA, whereas at LHC one will reach higher invariant 
masses than those possible at HERA. As was mentioned above, these 
photons can be regarded as quasireal and the freedom to vary the 
four-momentum transfer $Q^2$ is not given in the heavy ion case. 
According to \cite{KleinN99,Klein00b} relativistic heavy ion colliders
can be regarded as {\it ``vector meson factories''.} 
This is discussed in detail in chapter~\ref{sec_photonhadron}.

This vector meson production is essentially a diffraction process: 
the (equivalent) photon emitted by one of the nuclei
is diffractively excited to a vector meson on the 
other nucleus. This can be viewed as a photon-Pomeron interaction. 
In addition there are also Pomeron-Pomeron processes, which are interesting
processes of their own \cite{EngelRR97}. Their importance as a possible 
background will be studied in chapter~\ref{sec_diffractive}. 

The physics potential of photon-photon collisions is
discussed in chapters~\ref{sec_ggqcd} and~\ref{sec_newphysics}. These 
chapters extend our previous studies for the  CMS Heavy Ion Programme
\cite{BaurHTS98,BaurHTS99}, see also \cite{Felix97}. It ranges from 
studies in QCD (Chap.~\ref{sec_ggqcd}) like meson production 
and the total $\GG \rightarrow \mbox{hadron}$ cross section to the 
search for new particles (Chap.~\ref{sec_newphysics}) like a light Higgs 
particle.

Chapter \ref{sec_leptons} is devoted to lepton pair production.
Due to their low mass especially the electrons have a special status and 
EPA is not always a good approximation. Effects of the strong fields 
manifest themselves essentially in multiple $\EPEM$ pair production.
Besides the free pair production we also discuss some important related
processes like bremsstrahlung from these leptons, the production of muons with
a large transverse momenta and bound-free pair production.

Central collision events are characterized by a very high multiplicity.
Therefore all major heavy ion detectors are tuned to deal with the large
amount of data in this case. On the other hand, the multiplicity in 
very peripheral collisions is comparatively low (therefore the name 
{\it ``silent collisions''} has been coined for them). The ions do not 
interact strongly with each other and move on essentially undisturbed in 
the beam direction. The only possible interactions are due to the long 
range electromagnetic interaction and diffractive processes. Still some of 
these "silent events" are not so silent, e.g., quite a few particles will 
be produced in a 100~GeV on 100~GeV $\GG$-interaction leading to a final 
hadron state. The background coming from especially grazing collisions needs
to be taken into account. Also a way to trigger on 
very peripheral collisions is needed. This will be discussed in 
chap.~\ref{sec_eventsel}.

\subsection{A Short History of Very Peripheral Collisions}
\label{sec:intro:history}

Relativistic heavy ion collisions have been suggested as a general
tool for two-photon physics about twelve years ago and the field has 
grown rapidly since. Yet the study of a special case, the production 
of $\EPEM$ pairs in nucleus-nucleus collisions, goes back to
Landau and Lifschitz in 1934 \cite{LandauL34} and to Racah in 1937
\cite{Racah37}. In those days one thought more about high energy cosmic 
ray nuclei than about relativistic heavy ion colliders.
The general possibilities and characteristic features of two-photon
physics in relativistic heavy ion collisions have been discussed in
\cite{BaurB88}. The possibility to produce a Higgs boson via
$\GG$ fusion was suggested in \cite{GrabiakMG89,Papageorgiu89}. 
In these papers the effect of strong absorption in heavy 
ion collisions was not taken into account. Absorption is a feature, which 
is quite different from the two-photon physics at $\EPEM$ colliders. The
problem of taking strong interactions into account was solved by using
impact parameter space methods in \cite{Baur90d,BaurF90,CahnJ90}. Thus
the calculation of $\GG$ luminosities in heavy ion collisions was put
on a firm basis and rather definite conclusions were reached by many
groups working in the field, as described, e.g., in 
\cite{KraussGS97,BaurHT98}. This opens the way
for many interesting applications.

\begin{figure}[tbh]
\begin{center}
\resizebox{6cm}{!}{\includegraphics{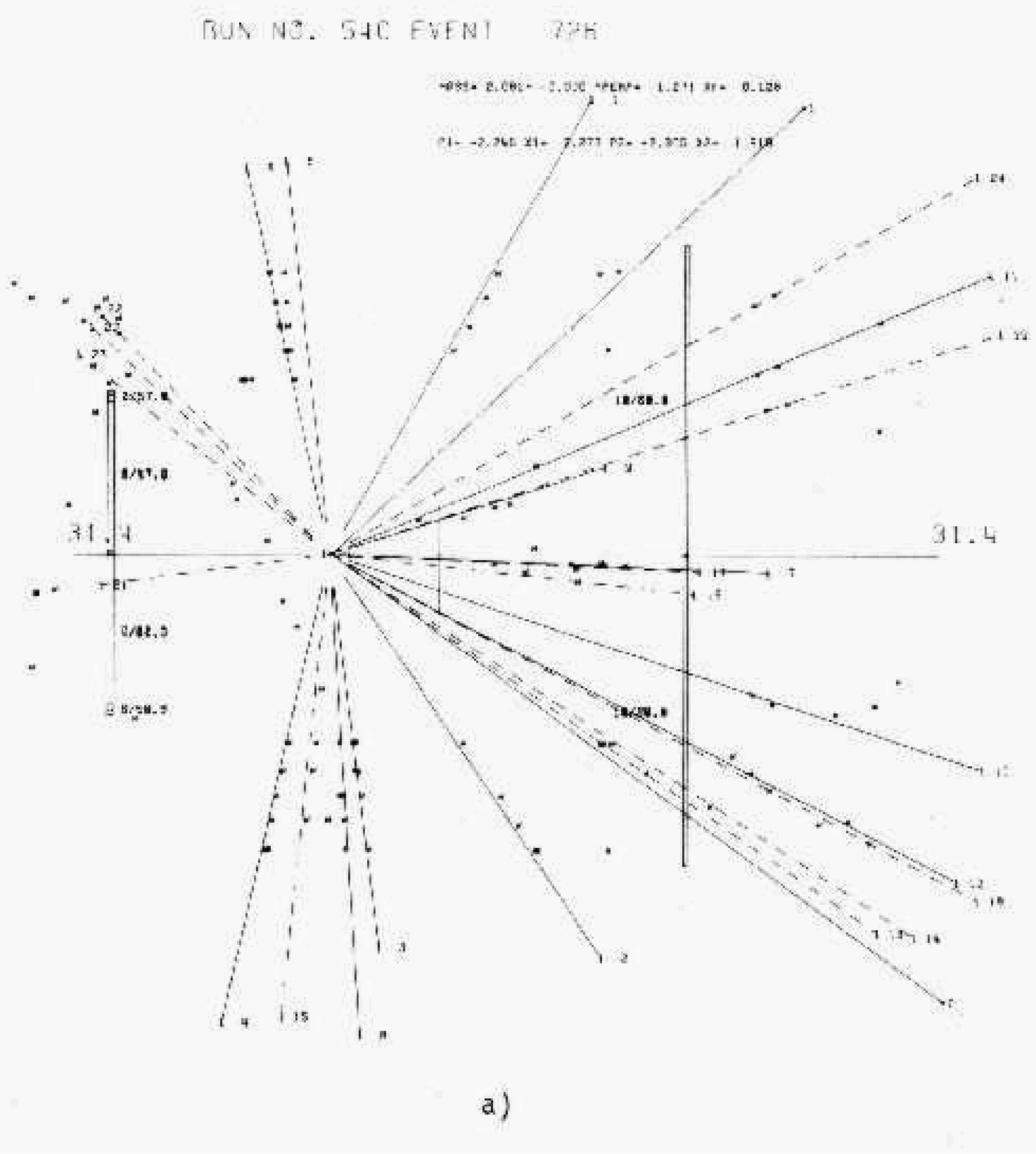}}
(a)~~
\resizebox{6cm}{!}{\includegraphics{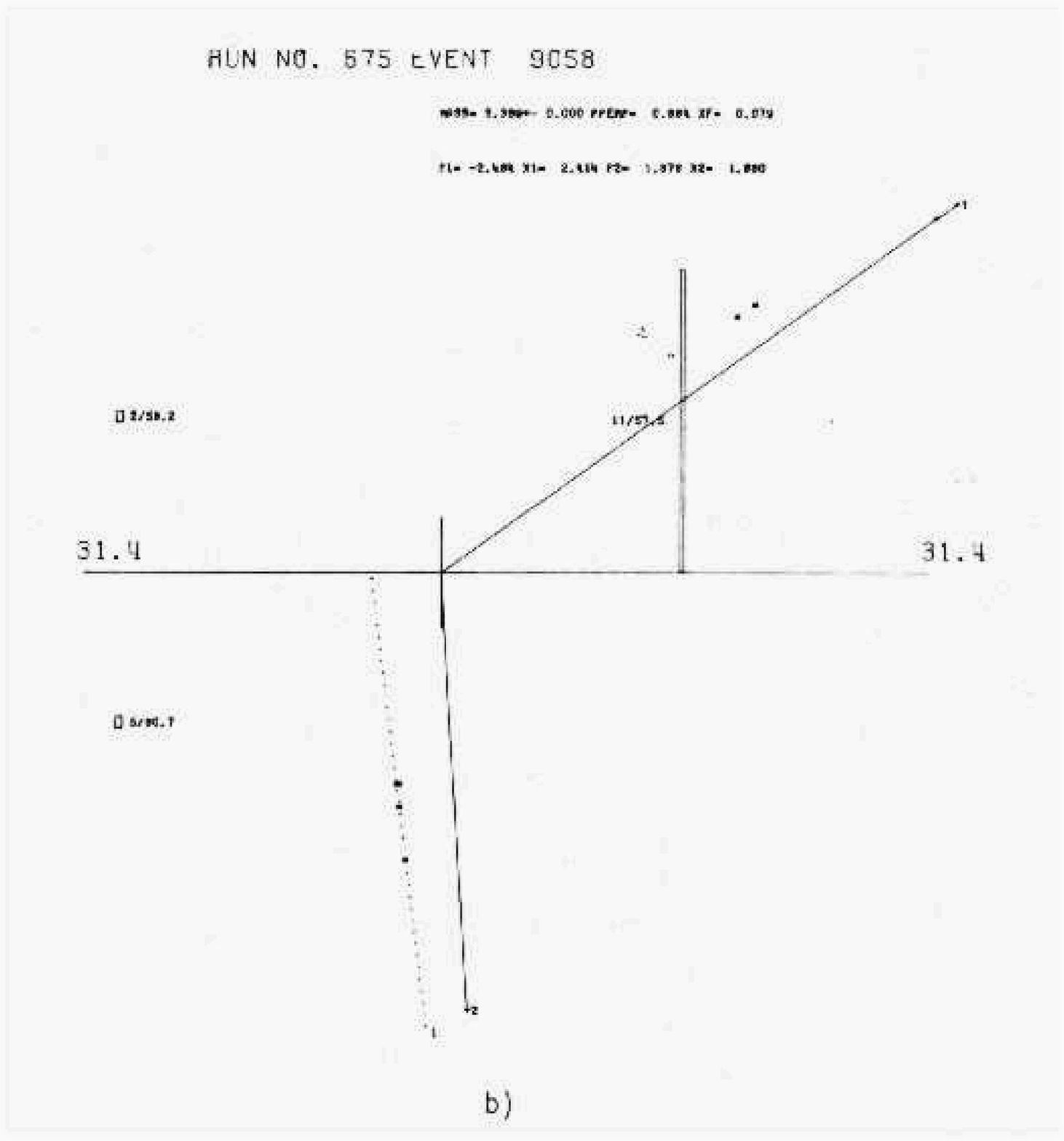}}
(b)
\end{center}
\caption{The production of a muon pair at the ISR with (a) and without
the production of hadrons (b) is shown. Typical characteristics 
of very peripheral collisions are seen: low particle multiplicity and a small 
sum of transverse momenta. Reproduced from Fig.~2, p. 244 of 
\protect\cite{Vannucci80} with kind permission of Springer-Verlag and the 
author.}
\label{fig_vannucci}
\end{figure}

Up to now hadron-hadron collisions have rarely been used for
two-photon physics. The work of Vannucci et al. \cite{Vannucci80} in 1980
may be regarded as its beginning, see Fig.~\ref{fig_vannucci}.
In this work, the production of $\mu^+\mu^-$ pairs was studied at the ISR.
The special class of events was selected, where no
hadrons are seen associated with the muon pair in a large solid angle
vertex detector. In this way one makes sure that the hadrons do not
interact strongly with each other, i.e., one is dealing with
very peripheral collisions; the photon-photon collisions manifest themselves 
as ``silent events'', that is, with only a  relatively small multiplicity.
It seems interesting to quote from Vannucci \cite{Vannucci80}:
{\it ``The topology of these events, their low transverse momentum,
and the magnitude of the cross section can be most naturally
be interpreted by the $\GG$ process. This effect, which is still 
a small background at the ISR compared with the Drell-Yan mechanism,
grows with energy and could trigger enough interest to be studied 
for itself, possibly at the $\mathrm{p}\bar\mathrm{p}$ collider, 
probably at the ISABELLE machine.''\/} Although ISABELLE has never 
been built, we have now the Relativistic Heavy Ion Collider (RHIC) in this 
ring, where $\GG$ processes will be studied with the ``gold flashlight''
in this way.
Another example is the search for a magnetic monopole in photon-photon
elastic scattering in $\mathrm{p} \bar\mathrm{p}$ collisions at the 
Tevatron following the idea in \cite{GinzburgS98}.
This is discussed in chapter~\ref{sec_newphysics}. Dimuons with a very 
low sum of transverse momenta (i.e., coming from $\GG$-interactions) are 
also considered as a luminosity monitor for the ATLAS detector at LHC 
\cite{ShamovT98}. A more recent discussion can also be found
in \cite{Martin01}.

The physics potential of relativistic heavy ion collisions for photon-hadron
studies was gradually realized during the last decade. Nuclear collisions 
without nuclear contact have been used very successfully over the past 
50 years to study nuclear structure, especially low lying collective 
excitations. Coulomb excitation, mainly at energies below the Coulomb barrier 
has been reviewed by Alder et al. \cite{AlderBHM56,AlderW66,AlderW75}.
The theory of relativistic electromagnetic excitation was given 
by Winther and Alder in \cite{WintherA79}.
This seminal paper started a whole new era of investigations.
It was recognized for some time that the cross section for the excitation of 
the giant dipole resonance is huge. This is the reason why also new phenomena 
like the double giant dipole resonance excitation could be studied by 
electromagnetic excitation. The excitation of the giant dipole resonance is 
also an important loss process in the relativistic colliders, as the giant 
dipole resonance decays predominantly by neutron emission. The decay neutrons 
are also a useful tool to measure the collider luminosity. 

In addition it was 
recognized in \cite{BaurB89} that there is a sizeable cross section for 
photon-nucleus interactions beyond 1~GeV. Thus a field of studies similar to 
the one at HERA will be opened. However, as mentioned already above, in the 
heavy ion case the photon is restricted to be quasireal, and the study of 
interactions of virtual photons with high $Q^2$ with hadrons is not possible.
On the other hand, the photon-nucleus interaction (rather than the 
photon-proton interaction) can be studied with RHIC and LHC. This is 
described in \cite{KleinN99,Klein00b}. Since the vector meson production 
cross section is very big --- of the order of the geometric cross section 
--- relativistic heavy ion colliders may justly be called ``vector meson 
factories'' \cite{KleinN99}.
Photon-gluon processes were discussed in \cite{HofmannSSG91} and later in 
\cite{BaronB93,GreinerVHS95}. The very interesting possibility to produce 
$t\bar t$ in this way was recently described in \cite{KleinNV00}.

At the STAR (Solenoidal Tracker At RHIC) detector at RHIC a program to study
photon-photon and photon-nucleus (especially diffractive ``photon-Pomeron'')
interactions in very peripheral collisions exists 
\cite{KleinS97a,KleinS97b,KleinS95a,KleinS95b,NystrandK98}.
First experimental results are just becoming available 
\cite{Klein01b,Meissner01}. 
At RHIC the equivalent photon spectrum goes up to about 3 GeV. 
Therefore the available $\GG$-invariant mass range is up to about the mass 
of the $\eta_c$. At the  RHIC/INT workshop at the LBNL (Berkeley), the 
physics of very peripheral collisions was discussed
by Klein \cite{Klein99} and Brodsky \cite{Brodsky99}.
The experimental feasibility of making these measurements with STAR
were described, expected backgrounds along with the techniques
and triggering algorithms to reject these signals are discussed, see also
\cite{KleinS97a,KleinS97b,KleinS95a,KleinS95b,NystrandK98}.

When the ``Large Hadron Collider'' will be scheduled to begin taking data
in 2006/2007, the study of these reactions can be extended to both
higher luminosities but also to much higher invariant masses, hithero
unexplored, see, e.g., \cite{HenckenKKS96,Sadovsky93,BaurHTS98}.

To conclude this introduction we quote Bjorken \cite{Bjorken99}:
{\it It is an important portion (of the FELIX program at LHC
\cite{Felix97}) to tag on
Weizs\"acker Williams photons (via the non-observation of completely
undissociated forward ions) in ion-ion running, creating a high
luminosity $\GG$ collider.}
Although the FELIX detector will not be realized in this form at the LHC,
these photon-photon and photon-ion collisions can and will be studied at other 
LHC detectors (ALICE and CMS).

\section{From Impact-Parameter Dependent Equivalent Photon Spectra to
{$\GG$-Luminosities}}
\label{sec_lum}

\begin{figure}[tbh]
\centerline{
\resizebox{28mm}{!}{\includegraphics{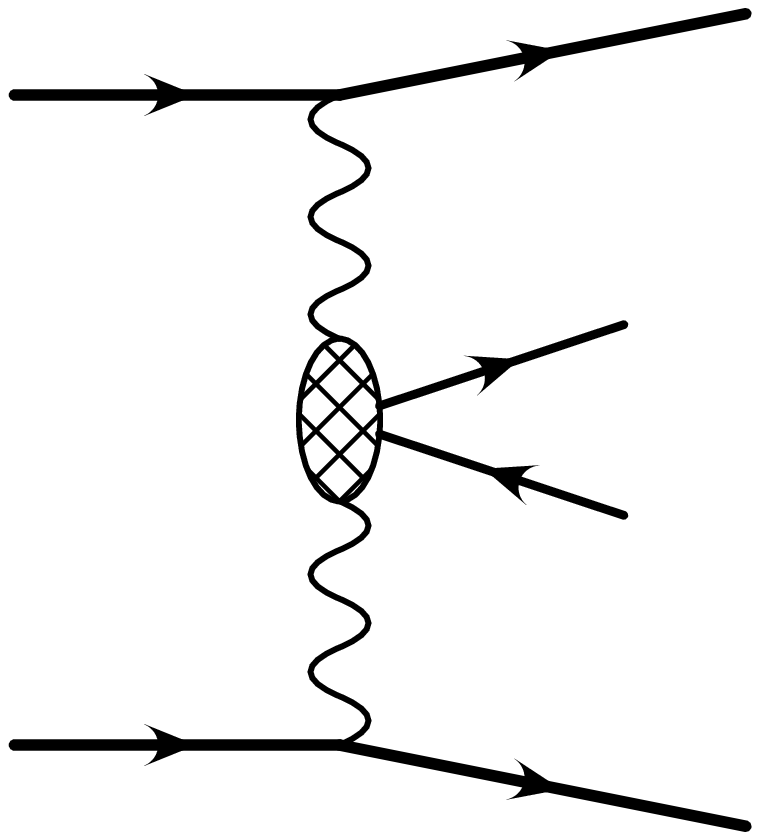}}
(a)~~~~~~
\resizebox{28mm}{!}{\includegraphics{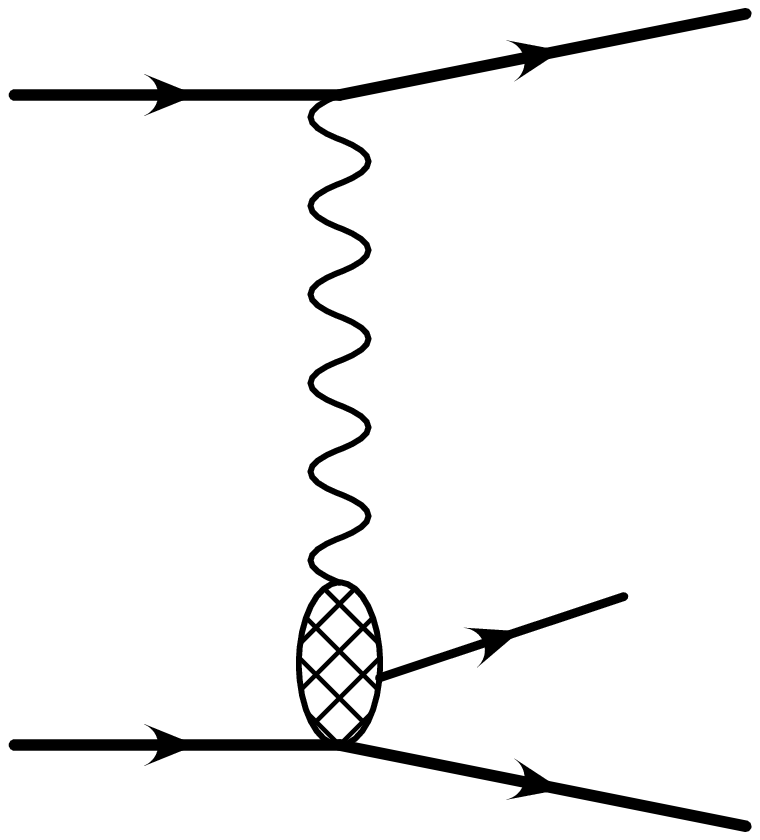}}
(b)
}
\centerline{
\resizebox{28mm}{!}{\includegraphics{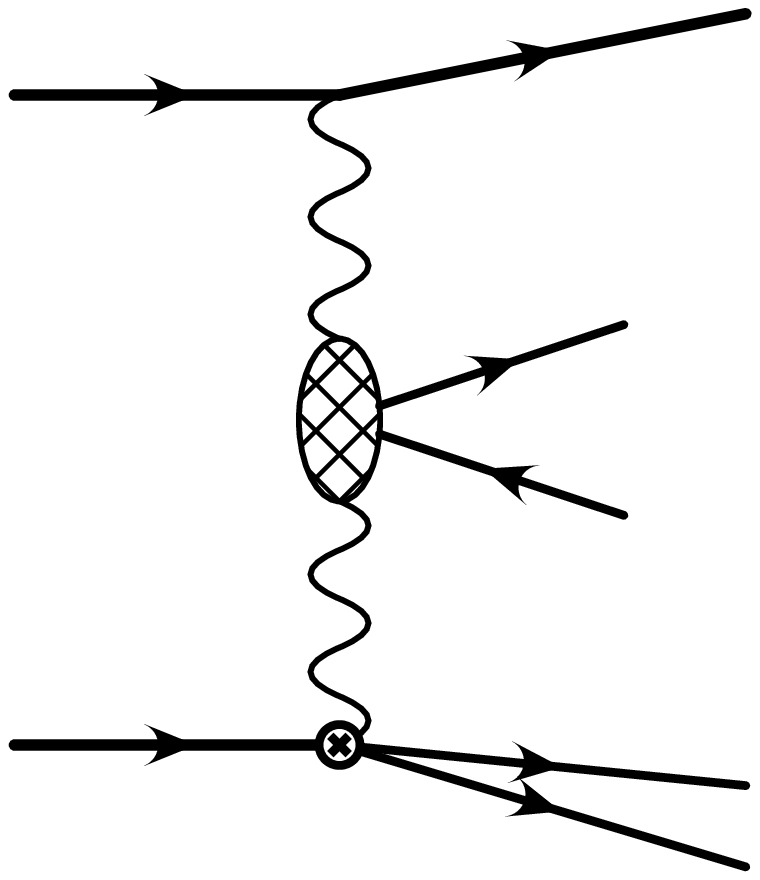}}
(c)~~~~~~
\resizebox{28mm}{!}{\includegraphics{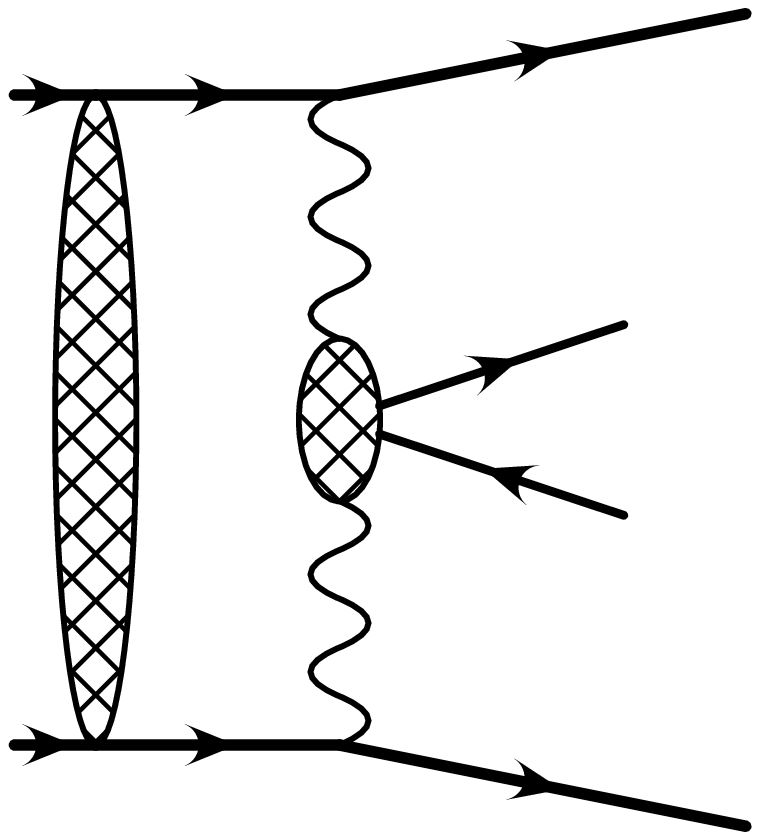}}
(d)
}
\centerline{
\resizebox{28mm}{!}{\includegraphics{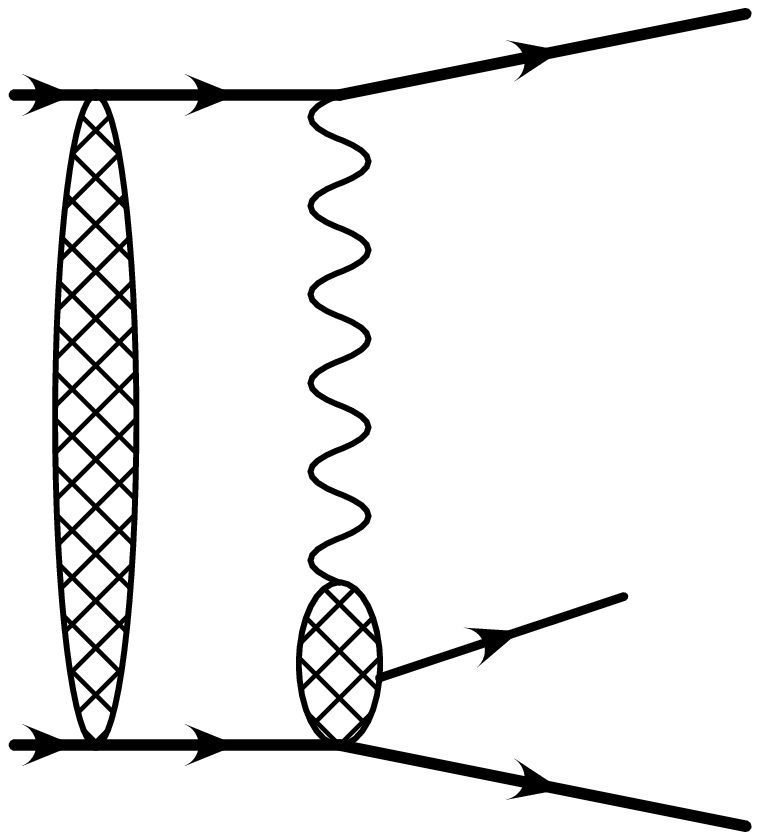}}
(e)~~~~~~
\resizebox{28mm}{!}{\includegraphics{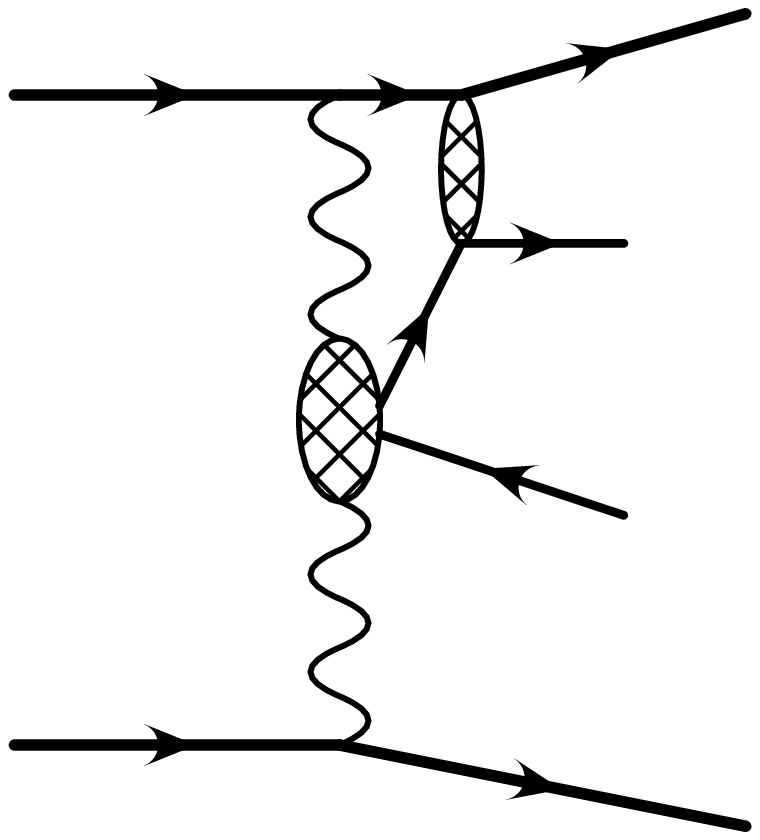}}
(f)
}
\caption{In the collision of either leptons with hadrons or hadrons with
hadrons photon-photon (a) and photon-hadron collisions (b) can be studied.
The principal diagrams are shown 
schematically here. In the collisions of hadrons additional
effects need to be taken into account: inelastic photon emission processes
(c), ``initial state interaction'' (d) and (e), as well as 
final state interaction (f).}
\label{fig_ggee}
\end{figure}

Electron-positron colliders have been and still are the basic tool to
study $\GG$ physics, see, e.g., the rich physics program at LEP
\cite{Freiburg99,Ambleside00}. The lowest order graph of two-photon (TP) 
processes is given in Fig.~\ref{fig_ggee}(a).
The plane-wave description is very accurate for $e^+$ and $e^-$,
which only interact electromagnetically. This is, for example,
described in great detail in \cite{BudnevGM75}. For lepton-hadron 
and lepton-ion collisions (e.g., for $ep$, or $e$A)
and hadron-hadron, hadron-ion, and ion-ion collisions (like $pp$, $p$A, 
and AA) such kind of 
two-photon interactions occur as well. 
If both particles are nuclei with charge $Z$ (symmetric AA 
collisions), the cross section of a TP process is enhanced by a 
factor of $Z^4$. For $e$A and $p$A collisions the enhancement is a factor 
$Z^2$. This has been called ``{\em the power of coherence}'' by Brodsky 
\cite{Brodsky99}. In these later cases there is also the interesting 
possibility of photon-ion processes, see Fig.~\ref{fig_ggee}(b), which we will 
call one-photon (OP) processes in the following.

Under rather general circumstances these processes
can be described by the concept of equivalent photons.
In this way a great simplification is achieved. The cross section 
factorizes into an equivalent photon spectrum $n(\omega)$ and the 
cross section for the photon-ion interaction process $\sigma_\gamma(\omega)$
in the case of one-photon processes:
\begin{equation}
  \sigma_{OP} = \int \frac{d\omega}{\omega} n(\omega  )
  \sigma_{\gamma}(\omega).
\label{eq:sigmaih}
\end{equation}
For the general case of two-photon collisions the cross section for the 
reaction $A_1 A_2\to A_1' A_2' X_f$ factorizes into
the photon-photon luminosity $L_{\gamma\gamma}$ and the cross section 
of the photon-photon interaction process $\GG \to X_f$
where $A_1$, $A_2$ are the initial particles, $A_1'$, $A_2'$ their final state
after the photon emission, and $X_f$ the final state produced in the 
photon-photon 
collision, see also Fig.~\ref{fig_epamostgeneral} and Eq.~(\ref{eq:TPcross})
below:
\begin{eqnarray}
  \sigma_{TP} &=& \int \frac{d\omega_1}{\omega_1} n_1(\omega_1)
  \int \frac{d\omega_2}{\omega_2} n_2(\omega_2)
  \sigma_{\GG\to X_f}(\sqrt{4 \omega_1\omega_2}) \\
  &=& \int \frac{dW}{W} \int dY \frac{dL_{\GG}}{dWdY} \sigma_{\GG \to X_f}(W)\\
  &=& \int \frac{dW}{W} \frac{dL_{\GG}}{dW} \sigma_{\GG\to X_f}(W).
\label{eq:sigmaiip}
\end{eqnarray}
The cross sections for the corresponding subprocesses are given by 
$\sigma_{\gamma}$ and $\sigma_{\GG \to X_f}$; the equivalent
photon spectra and the $\GG$ luminosity are denoted by $n(\omega)$ and 
$dL_{\GG}/dWdY$, respectively, where $Y$ is the rapidity of the produced 
system $X_f$. The invariant mass of the $\GG$ system is denoted by 
$W =\sqrt{4\omega_1\omega_{2}}$. The $\GG$ luminosity is obtained 
by a folding of the equivalent photon spectra, as will be explained below.

As compared to the point-like and non-strongly interacting
electrons and positrons there are in the case of incoming hadrons or nuclei
additional initial and final state interaction effects. They must be taken 
into account in the calculation of the equivalent photon spectra and $\GG$ 
luminosities. We distinguish three conceptually different kinds:
\begin{enumerate}
\item For extended objects one has an extended charge distribution, as well as,
an excitation spectrum. This is taken care of by  elastic and inelastic 
form factors, see Fig.~\ref{fig_ggee}(c).
\item For strongly interacting particles (like in $pp$, $p$A, or
AA systems) there are effects like absorption (nuclear interaction), 
which modify the $\GG$
luminosities. They may be called ``initial state interactions'', 
see Fig.~\ref{fig_ggee}(d) and~(e).
In the case of heavy ions the ``initial state interaction'' consists of
both the nuclear absorption and the Coulomb interaction. For the Coulomb
interaction the Sommerfeld parameter $\eta\approx Z_1 Z_2 \alpha$ is larger 
than one and the semiclassical description can be used.
The main initial state interaction is then the nuclear absorption for 
trajectories with $b<2R$. This can be very conveniently taken into account by
introducing the appropriate cutoff for the impact parameter dependent 
equivalent photon spectrum. It would be much more 
cumbersome to take these strong interaction effects into account in a
purely quantal description. 
\item  Particles produced in $\GG$ interactions, which have strong 
interactions, can have final state interactions with the incident particles,
see Figs.~\ref{fig_ggee}(f).
\end{enumerate}

\subsection{Equivalent Photon Numbers}
\label{pt_of_GG}

Before looking in more detail at the equivalent photon approximation
for the different cases of interest, let us discuss here first some of the 
main characteristics in a qualitative way. Here we make use of the plane
wave approximation, neglecting the ``initial state interaction'' between
the two ions. One should remember that therefore the statements made are only 
rather qualitative concerning the heavy ion cases, where the semiclassical 
straight line approximation is the relevant one, as the Sommerfeld parameter 
is always larger than one, see below and should be used instead.

For the elastic emission from a nucleus with spin 0,
characterized by an elastic form factor $F_{el}(q^2)$ and using the plane wave 
approximation, the equivalent photon number of a nucleus with charge $Z$ is 
given, see Eq.~(\ref{eq_nepael}) below, by
\begin{equation}
n(\omega) = \frac{Z^2\alpha}{\pi^2} \int d^2q_\perp 
\frac{q_\perp^2}{\left[
\left(\frac{\omega}{\gamma}\right)^2+ q_\perp^2 
\right]^2} 
F_{el}^2\left(\left[\frac{\omega}{\gamma}\right]^2+ q_\perp^2\right).
\label{eq:nepamainprop}
\end{equation}
The elastic form factor $F_{el}$ is characterized by the property, that it is
$\approx 1$ up to values of $Q^2=-q^2$ of about $1/R^2$ and quickly falls of 
for larger values of $Q^2$. This leads to the two main characteristics already
discussed in the introduction: for $\omega>\gamma/R$ the
equivalent photon number quickly decreases due to the fall-off of the form 
factor, giving a maximum ``usable'' photon energy of 
$\omega_{max} \approx \gamma/R$.

In addition also the transverse momentum distribution of the photon is 
limited to rather small values. This is again due to the fall-off of the 
elastic form factor, restricting the transverse momenta essentially to 
$q_\perp\lesssim 1/R$. But in addition the
integrand of the expression Eq.~(\ref{eq:nepamainprop}) peaks at values of 
$q_\perp\approx\omega/\gamma$. Most equivalent photons will therefore have
even smaller transverse momenta. This will also apply to
the TP case. Also here the produced system will be characterized by rather
small transverse momenta with $P_\perp\lesssim1/R$.

Approximating rather crudely the form factor by $\Theta(1/R^2-Q^2)$, where 
$Q^2=-q^2$, the integration can be done analytically and the equivalent 
photon spectrum in the leading logarithmic approximation is given by
\begin{equation}
n(\omega)= \frac{2 Z^2 \alpha}{\pi} \ln\left(\frac{\gamma}{\omega R}\right).
\label{eq:epaapprox}
\end{equation}
There are more refined formulae for the equivalent photon approximation 
than Eq.~(\ref{eq:epaapprox}).

For heavy ions the relevant equivalent photon number is the one obtained in 
the impact parameter approach given below, see Chap.~\ref{ssec:semiclassic}.
The approximate form of Eq.~(\ref{eq:epaapprox}) can be obtained also in this
case from Eqs.~(\ref{eq:nepab}) and~(\ref{eq:nomegafromb}) below. 
This equation is useful for quick estimates and also to see in a simple way 
the general properties of the spectrum. 
We see that the spectrum falls off logarithmically with $\omega$ up to
$\omega_{max}=\gamma/R$. Note also that the $1/\omega$ dependence, 
characteristic of the equivalent photon spectra, has been taken out in the 
definition of $n(\omega)$ in Eq.~(\ref{eq:sigmaiip}).

Equivalent photon spectra of point-like particles like the electron 
require a further discussion. The integral Eq.~(\ref{eq:nepamainprop}) is 
divergent, if the form factor is constant. The choice of the cutoff depends 
then on the specific process to be considered. This is thoroughly discussed
in \cite{BudnevGM75}, and it is sufficient here to refer to this reference.

The photon-photon luminosity can be found (again neglecting possible initial 
state effects)from  Eq.~(\ref{eq:sigmaiip}). One obtains, see also 
Eq.~(\ref{eq:fcahnjackson}) and Eq.~(\ref{eq:dl/dwdy}) below,
\begin{equation}
\frac{dL_{\GG}}{dWdY} = \frac{2}{W} n\left(\frac{W}{2}e^Y\right)
n\left(\frac{W}{2} e^{- Y}\right).
\end{equation}
Due to the decrease of the equivalent photon number with $\omega$, this
luminosity has a maximum at rapidity $Y=0$. It falls as a function
of $|Y|$. The width $Y_{max}$ of the distribution can be estimated to be
\begin{equation}
Y_{max}=\ln( 2\omega_{\max}/W).
\end{equation} 
The width decreases with larger invariant mass $W$.

\subsection{The Quantum Mechanical Plane Wave Formalism}

Let us first deal with the effects due to the finite size of the
particles. This leads to two effects: we need to introduce a form factor 
for the elastic photon emission and in addition have to take the 
inelastic photon emission into account, where the particle makes a 
transition to an excited state. Both can be dealt within the general 
plane wave framework, we are showing here. We will in the following deal 
only with the derivation
of the equivalent photon approximation in the two-photon case;
the one-photon case then follows along the same line.

\begin{figure}[tbh]
\centerline{
\resizebox{45mm}{!}{\includegraphics{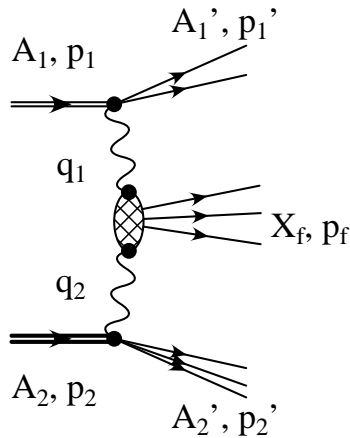}}
}
\caption{The general form of a two-photon process is shown. The two
incoming particles $A_1$ and $A_2$ either stay in their ground states
or undergo a transition to exited states $A_1$ and $A_2'$, while each
emitting a photon. The two photon fuse to a final state $X_f$. Also shown
are the momenta of all particles involved}
\label{fig_epamostgeneral}
\end{figure}

Let us look at the most general case, where two incoming particle $A_1$ and 
$A_2$ undergo a transition to $A_1'$ and $A_2'$ while emitting a photon each 
and where the collision of these photons then produces a final state $X_f$, 
see Fig.~\ref{fig_epamostgeneral}. We follow in our derivation here 
\cite{BudnevGM75}, but are more general in treating both the elastic and 
inelastic case and using a somewhat different notation, see also 
\cite{HenckenTB95}. The cross section of this two-photon process can be 
written as
\begin{eqnarray}
  d\sigma_{A_1A_2\to A_1' A_2' X_f} &=& 
\frac{(2\pi)^4 (4\pi\alpha)^4}{8E_1 E_2}
\int  \delta^{(4)}(q_1+q_2-p'_f) \left|T\right|^2 \nonumber \\
&& d^4q_1 d^4q_2 dR_1(p_1-q_1) dR_2(p_2-q_2) dR_f(q_1+q_2),
\label{eq:TPcross}
\end{eqnarray}
where $p_1$,$p_2$ are the four-momenta of the initial particles, $E_1=p_1^0$, 
$E_2=p_2^0$
their energies; $q_1$,$q_2$ are  the four-momenta of the exchanged photons and 
$dR_i(p'_i)$ denotes the integration over the final state $i$ (the summation 
and averaging over the final and initial spin states is assumed to be 
implicitly included):
\begin{equation}
dR_i(p'_i) = \prod_{k\in i} \frac{d^3p_k}{(2\pi)^3 2 E_k} 
\delta^{(4)}\left(p'_i-\sum_{k\in i} p_k\right).
\end{equation}
The square of the matrix element $|T|^2$ is given by
\begin{equation}
|T|^2 = \frac{1}{(q_1^2)^2} \frac{1}{(q_2^2)^2} \Gamma_1^\mu \Gamma_1^{\mu' *}
\Gamma_2^\nu \Gamma_2^{\nu' *} M_{\mu\mu'\nu\nu'},
\label{eq:matrx}
\end{equation}
where $M_{\mu\mu'\nu\nu'}$ describes the production of the system $X_f$ in 
the photon-photon interaction and $\Gamma_i$ are the electromagnetic 
transition currents. Integrating over all possible final state with momentum 
$p_i'=p_i-q_i$ ($i=1,2$), we can reexpress these electromagnetic currents 
in terms of 
the electromagnetic tensor $W_i^{\mu\mu'}$, see, e.g., \cite{HalzenM84}
\begin{equation}
W_i^{\mu\mu'} = \frac{1}{2\pi 2 m_i} \int dR_i(p-q) (2\pi)^4
\Gamma_i^\mu \Gamma_i^{\mu'}.
\label{eq:wmumup}
\end{equation}

From Lorentz invariance and current conservation considerations the general 
form of the tensor $W^{\mu\mu'}$ can be written as
\begin{eqnarray}
W_i^{\mu\mu'} &=& \left(-g^{\mu\mu'} + \frac{q_i^\mu q_i^{\mu'}}{q_i^2}\right) 
W_{i,1}
\nonumber\\&
+& \left(p_i^\mu - \frac{p_i\cdot q_i}{q_i^2} q_i^\mu\right)
\left(p_i^{\mu'} - \frac{p_i\cdot q_i}{q_i^2} q_i^{\mu'}\right) 
\frac{W_{i,2}}{M^2},
\label{eq:wmunu}
\end{eqnarray}
where $W_{i,1}$ and $W_{i,2}$ are two scalar functions of the two independent 
invariants $q_i^2$ and $\nu_i=-p_i\cdot q_i/m_i$, 
which characterize generally the electromagnetic structure of the 
electromagnetic currents. Please keep in mind that we are treating the
photons as being emitted here, whereas in electron scattering, the photons
are normally assumed to be absorbed by the particle. The cases 
differ only by the sign of $q$.

At high energies the main contribution to this cross sections comes from
small values of $q^2$ and where the three-momentum of the photon is almost 
aligned to
the beam axis. In this case only the transverse part (with respect to the
momentum of the photon) of the tensor $W^{\mu\nu}$ is important. Assuming
in addition that $M_{\mu\mu'\nu\nu'}$ in this case essentially only depends 
on the energies of the two photons ($\omega_i=q_{0,i}$) and not on the 
photon virtualities $q_i^2$, it can be related to the cross section 
$d\sigma_{\gamma\gamma}$ of the corresponding process for two real photons.
For a more thorough discussion of the applicability of the EPA, we refer
to \cite{BudnevGM75}.

Under these assumptions the cross section can be written as
\begin{equation}
d\sigma_{A_1A_2\to A_1' A_2' X_f} = \int \frac{d\omega_1}{\omega_1}
\int \frac{d\omega_2}{\omega_2} n_1(\omega_1) n_2(\omega_2)
      d\sigma_{\GG\to X_f}(\omega_1,\omega_2),
\end{equation}
with the equivalent photon number given by
\begin{eqnarray}
n_i(\omega_i) &=& \int d^3q \ \frac{\alpha}{\pi^2} \frac{\omega_i^2 m_i}
{E_i \left(q_i^2\right)^2}
\left[2 W_{i,1} + \frac{q_{i\perp}^2 P_i^2}{\omega_i^2 m_i^2} W_{i,2} 
\right]\\
&=& \frac{\alpha}{\pi^2} \int d^2q_{i\perp} \int d\nu_i 
\frac{1}{\left(q_i^2\right)^2} \left[2 \frac{\omega_i^2 m_i^2}{P_i^2} W_{i,1} 
+ q_{i\perp}^2 W_{i,2} \right],
\label{eq:epaW1W2}
\end{eqnarray}
where $E_i\approx P_i$ are energy and momentum of the incoming particle.
This equation corresponds to Eq.~(D.4) of \cite{BudnevGM75}.

One important feature of the description of the electromagnetic structure (and
therefore of the equivalent photon spectrum) in terms of $W_1$ and $W_2$ is 
the fact that they are Lorentz scalars and can therefore be calculated in 
any frame --- e.g., in the rest frame of the projectile (nucleus).

\subsection{Elastic Photon Emission}

Let us first discuss the most relevant case: the elastic photon emission.
In this case the two invariants are not independent of each other but are
related via $2 m \nu = - q^2$. Therefore the two structure function will be
of the form
\begin{equation}
W_{1,2}(\nu,q^2)= \widehat W_{1,2}(q^2) \delta(\nu+q^2/2m_i)
\end{equation}
and the integration over $\nu$ can be done trivially. In many cases ---
and especially in the heavy ion case --- we have $Q^2=-q^2 \ll m_i^2$;
we can then neglect the recoil and use to a good approximation $\nu\approx0$. 
We then get $q_z=\omega/\beta$, where $\beta$ is the velocity of the particle
and
\begin{equation}
-q^2 =  \left(\frac{\omega}{\beta\gamma}\right)^2 + q_\perp^2
\end{equation}

For the case of a nucleus with a  $J^P=0^+$ ground state  (an 
even-even nucleus) we have $\widehat W_1=0$ and $\widehat W_2=Z^2 
\left|F_{el}(-q^2)\right|^2$, with the elastic form factor normalized to 
$F_{el}(0)=1$. In this case the equivalent photon number is given by
\begin{equation}
n(\omega) = \frac{Z^2\alpha}{\pi^2} \int d^2q_\perp \frac{q_\perp^2}{(q^2)^2}
F_{el}^2(q^2).
\label{eq_nepael}
\end{equation}

It should be clear that in this equation coming from the plane wave 
approximation the important strong interaction effects (which correspond 
to a cutoff in impact parameter space, as will be explained below) are not 
taken into account. For a collection of some other relevant cases see, e.g., 
Table~8 of 
\cite{BudnevGM75} (with the relation of $C$ and $D$ there to $\widehat W_1$ 
and $\widehat W_2$ given by $C=\frac{4 m^2}{(-q^2)} \widehat W_1$, 
$D=\widehat W_2$). 
In many cases the dominant contribution comes from $W_2$ alone and $W_1$ 
can be neglected.

For nuclei various degrees of sophistication for the elastic form factor can be
applied: Gaussian form factors or form factors corresponding to a 
homogeneously charged sphere have been used in the literature. A comparison of
the different results can be found, e.g., in \cite{KraussGS97}. 

One can include also the effects of electromagnetic moments like 
$M1$ and $E2$, with their corresponding contributions to $\widehat W_1$ and 
$\widehat W_2$, see also Eqs.~(\ref{eq_w1temc}) and~(\ref{eq_w2temc}) below. 
E.g., the projectile 
$^{197}$Au --- which is used at RHIC --- has a magnetic moment of 
$\mu=0.14485$ n.m. and an electric quadrupole moment of $q=+0.6$~barn 
\cite{LedererS78}. A classical version of an $E2$ equivalent photon spectrum 
is given in \cite{Bertulani93}. Perhaps at some level of accuracy of the 
RHIC experiments the inclusion of such effects will become necessary.

Quite general spectra of the equivalent photons in the plane wave approximation
are given explicitly 
in \cite{KraussGS97} (with their definition of $n(\omega)$ including an
extra $1/\omega$):
\begin{equation}
n(\omega) = \frac{2 Z^2 \alpha}{\pi} \int_{\omega/\gamma}^{\infty}
d\kappa \frac{\kappa^2 - \left(\frac{\omega}{\gamma}\right)^2}{\kappa^3}
F_{el}^2(\kappa^2),
\label{eq:nomegawithF}
\end{equation}
for any elastic form factor $F_{el}$. The elastic form factor for extended 
charge distributions is in general characterized by a radius $R$. 
Defining a maximum photon energy $\omega_0=\gamma/R$, the equivalent photon 
number can then be cast into the form
\begin{equation}
n(\omega) = \frac{2 Z^2 \alpha}{\pi} \widehat f(\omega/\omega_0),
\end{equation}
where $\widehat f$ is a universal function, independent of $\gamma$ and given by
\begin{equation}
\widehat f(x) = \int_x^\infty dz \frac{z^2-x^2}{z^3} F_{el}^2\left( \frac{z^2}{R^2}
\right).
\end{equation}
Three different cases are given in \cite{KraussGS97} point particle, 
homogeneous charged sphere and a Gaussian form factor. The exact result for 
the equivalent photon spectrum of the proton can be found in \cite{Kniehl91}, 
where the usual dipole parameterization of the form factors is  used.

\subsection{Inelastic Form Factors}

The structure of the electromagnetic current in Eq.~(\ref{eq:wmunu}) is
equally valid for inelastic photon emission processes. In this case 
$W_1(\nu,q^2)$ and $W_2(\nu,q^2)$ are functions of both invariants $\nu$ and
$q^2$. At low excitation energies, discrete states will dominate ---
resonances of either the nucleus or the proton. These resonances will be
at well defined discrete energies (neglecting their finite width), and ---
as we can again neglect recoil effects in most cases --- $\nu$ has to be 
equal to
$\Delta$, where $\Delta$ is the energy of the excited state in its rest frame.
This can be seen by evaluating the relation $(p-q)^2=(m+\Delta)^2$ and 
neglecting terms quadratic in $q$ and $\Delta$. Therefore, similar to
the elastic case, $W_1$ and $W_2$ will be of the form
\begin{equation}
W_{1,2}(\nu,q^2)= \widehat W_{1,2}(q^2) \delta(\nu-\Delta),
\end{equation}
and the integration over $d\nu$ in Eq.~(\ref{eq:epaW1W2}) can be done trivially. 
The effect of the 
finite excitation energies on $q^2$ can be taken into account 
\cite{HenckenTB95}, with $q^2$ in this case given by
\begin{equation}
- q^2 = \frac{\omega^2}{\gamma^2} + 2 \frac{\omega \Delta}{\gamma}
 + \frac{\Delta^2}{\gamma^2} + q_\perp^2.
\end{equation}
One clearly sees that as long as $\omega > \gamma \Delta$ the 
modification of $q^2$ is small and can be neglected in these cases.

In the case of nuclei it is more convenient to express $\widehat W_1$ and 
$\widehat W_2$ in terms of the more familiar (to nuclear physicists) Coulomb, 
transverse electric or transverse magnetic matrix elements 
\cite{BlattW79,Walecka83,deForestW66,HenckenTB95}:
\begin{eqnarray}
\widehat W_1 &=& 2 \pi \left[|T^{\mathrm{e}}|^2 + |T^{\mathrm{m}}|^2 \right]\\
\label{eq_w1temc}
\widehat W_2 &=& 2 \pi \frac{q^4}{(\Delta^2 - q^2)^2} 
\left[ 2 |M^{\mathrm{C}}|^2 - \frac{\Delta^2-q^2}{q^2} 
\left(|T^{\mathrm{e}}|^2 + |T^{\mathrm{m}}|^2 \right) \right].
\label{eq_w2temc}
\end{eqnarray}

Inelastic photon processes on nuclei are dominated by the giant dipole 
resonance (GDR). For them the Goldhaber-Teller model gives \cite{HenckenTB95}
\begin{eqnarray}
\widehat W_1 &=& \frac{N Z }{A} \frac{\Delta}{2 m_N} 
\left|F_{el}(\Delta^2-q^2)\right|^2\\
\widehat W_2 &=& \frac{N Z}{A} \frac{-q^2}{2 m_N \Delta} 
\left|F_{el}(\Delta^2-q^2)\right|^2,
\end{eqnarray}
where $F_{el}$ is the elastic form factor of the nucleus, $m_N$
the nucleon mass and the excitation energy $\Delta$ is given experimentally by
\begin{equation}
 \Delta=E_{GDR} = \frac{80 \mbox{MeV}}{A^{1/3}}.
\end{equation}

For protons the lowest excited state is the $\Delta$-resonance. For this 
transition the contribution to the photon spectrum was estimated in 
\cite{BaurHT98}, using the parameters given in \cite{Chanfray93}. A
contribution of the order of 10\% was found. This is in contrast to the
similar situation in heavy ion collisions, where the contribution of the
most important excited state, the GDR, is in general less than 1\% 
\cite{HenckenTB95}.

For higher excitation energies and momentum transfers on nuclei
the quasielastic emission of nucleons is dominating and the quasifree
approximation can be used. For an estimate the Fermi gas model can be 
used. One obtains a simple expression in terms of the elastic nucleon 
form factors 
\cite{Tsai74,Moniz71}:
\begin{equation}
W_i^{qe} = C(t) \left[ Z W_i^{p} + (A-Z) W_i^{n} \right],
\end{equation}
with 
\begin{equation}
C(t) = \left\{
\begin{array}{ll}
1 & Q_{rec}> 2 P_F \\
\frac{3Q_{rec}}{4P_F} \left[1 - \frac{1}{12} \left(\frac{Q_{rec}^2}{P_F} 
\right)^2 \right]
& \mbox{otherwise},
\end{array}
\right.
\end{equation}
with $Q_{rec}^2=(-q^2)^2/(2 m_p)^2 + (-q^2)$ and the Fermi momentum 
$P_F=0.25$~GeV.

Finally at very large momentum transfer ($|-q^2| \gg 1 \mbox{GeV}^2$)
the parton model applies and $W_1$ and $W_2$ can be expressed in terms of
the quark structure functions $f_i(x)$
\cite{DreesGN94,OhnemusWZ94}:
\begin{eqnarray}
W_1 &=& \frac{1}{2 M} \sum_i e_i^2 f_i(x) \\
W_2 &=& \frac{1}{\nu} \sum_i e_i^2 x f_i(x),
\end{eqnarray}
that is, $W_1$ and $W_2$ are related by $2 M W_1 = \nu W_2 / x$ and 
the Bjorken variable is $x= -q^2/(2m\nu)$. Of course in this case
the integration over $d\nu$ can be replaced by one over $dx$.
In the case of the proton, the equivalent photon spectra were calculated
in \cite{DreesGN94,OhnemusWZ94}. Whereas the integration over $q^2$ in
the elastic case is limited due to the form factor to small values of $q^2$,
here the integration goes up to the kinematical limit ($-q^2<4 E^2$).

\subsection{Integrated Photon-Photon Luminosity}
\label{ssec:ggluminosities}

Let us now calculate the luminosity function $dL_{\GG}/dW$ as defined in 
Eq.~(\ref{eq:sigmaiip}). Neglecting  possible ``initial state 
effects'' between the collision partners, we can fold 
the equivalent photon spectra 
$n_1(\omega_1)$ and $n_{2}(\omega_{2})$ to obtain \cite{CahnJ90}
\begin{equation}
\tau \frac{dL}{d\tau} = \int_\tau^1 \frac{dx}{x} f\left(x\right)
f\left(\frac{\tau}{x}\right),
\label{eq:fcahnjackson}
\end{equation}
where $\tau=W^2/s = 4 \omega_1 \omega_2 / s$ is the square of the fraction
of the total energy carried away by the photon-photon system with 
$s=(p_1+p_2)^2$. The photon distribution function $f$ here is related to the 
equivalent photon number via
\begin{equation}
f(x) = \frac{E}{\omega} \ n(x E) = \frac{n(x E)}{x}
\end{equation}
and $E$ is the total energy of the initial particle in a given reference frame.
$x=\omega/E$ is as before the ratio of the photon energy $\omega$ to 
the energy of the incoming particle $E$.

Explicit calculations of $\GG$ luminosity functions exist for the
$pp$ case. The strong interaction between the protons are not taken
into account. Several approximate results are given in 
\cite{OhnemusWZ94}. The simplest one (in leading order of $\ln(\tau)$) 
is given by
\begin{equation}
\tau \frac{dL}{d\tau}=\left(\frac{\alpha}{\pi}\right)^2 \frac{2}{3}
 \ln^3\left(\frac{1}{\tau}\right).
\end{equation}
A more detailed result is given by Drees et al. \cite{DreesEZ89}
based on the equivalent photon spectra of the proton given in
\cite{DreesZ89}. 

Such a formula for the $\GG$-luminosity (Eq.~(\ref{eq:fcahnjackson})), in 
which the initial state interactions are neglected, can be applied to 
$\GG$ interactions in $ep$ and $e$A collisions. For $e$A collisions this was 
done by Levtchenko \cite{Levtchenko96}.
Eq.~(\ref{eq:nomegawithF}) was used for the equivalent photon spectrum 
of the nucleus. In $ep$ and $eA$collisions initial state interactions are 
weak and can be safely neglected.

In the case of protons calculations have been made with either one or both
protons breaking up \cite{DreesGN94,OhnemusWZ94}.
This leads to $\GG$ luminosities 
which are larger than the ones in the elastic-elastic case. Since 
the charges of the quarks ($2/3,-1/3$) are comparable to the proton charge, 
this is not unexpected. In the case of a heavy nucleus on the
other hand the charges of the constituents (protons) are much smaller compared
to the total charge $Z$ of the ion; therefore incoherent effects, although 
present, are generally less important. Possible strong interaction effects in
the initial state are not taken into account in these calculations. 

The interesting possibility to study photon-photon events in $pp$ collisions
at the LHC by tagging the two final protons was studied in 
\cite{Piotrzkowski00}. Only protons emitting a photon with a substantial
part of their energy (corresponding to $x>0.1$) can be tagged in this way,
as their path after the interaction deviates sufficiently from the 
non-interacting ones. Photon-Photon luminosities under these conditions
(and also including transverse momentum cuts, which are less important)
were calculated.

\subsection{Semiclassical Impact Parameter Description}
\label{ssec:semiclassic}

For heavy ions (as opposed to, e.g., $pp$ collisions) the semiclassical 
description works very well, as the Sommerfeld parameter 
$\eta=\frac{Z_1 Z_2 e^2}{\hbar \beta}$ is much larger than one, see, e.g., 
the discussion in Chap.~2 of \cite{BaurHT98}. Therefore an impact parameter
dependent equivalent photon number $N(\omega,b)$ can be defined.
It is most important to note that in this semiclassical impact parameter 
description it is very easy to take strong absorption effects into account. 
This is done by introducing a cutoff for those impact parameters
where the ions interact strongly with each other.
The calculation of the impact parameter dependent equivalent photon
spectra is explained in \cite{JacksonED} for the case of $E1$
(electric dipole)
excitations. The generalization of the equivalent photon spectrum
to all electromagnetic multipoles was given by \cite{WintherA79}. 
For the $E1$ case
\begin{equation}
N_{E1}(\omega,b) = \frac{Z^2 \alpha}{\pi^2} 
\left(\frac{\omega}{\gamma\beta}\right)^2 \left[ K_1^2(u) + \frac{1}{\gamma^2} 
K_0^2(u) \right],
\end{equation}
with $\beta$ and $\gamma$ the velocity and Lorentz factor of the ion and
where $u=\omega b /(\gamma\beta)$. Integrating from some $R_{min}$ to infinity 
gives
\begin{equation}
n_{E1}(\omega) = \frac{2}{\pi} Z^2 \alpha \left[ \xi K_0(\xi) K_1(\xi) - 
\frac{\xi^2}{2} \left( K_1^2(\xi) - K_0^2(\xi) \right) \right],
\end{equation}
where $\xi=\omega R_{min} / \gamma\beta$.
Using the expression for $K_0$ and $K_1$ for $\xi\to 0$, it can easily be seen,
that this expression agrees in logarithmic approximation with 
Eq.~(\ref{eq:epaapprox}).
Impact parameter dependent equivalent photon spectra $N(\omega,b)$ for 
extended charge distributions are calculated in \cite{BaurF91}. They are
\begin{equation}
N(\omega,b) = \frac{Z^2 \alpha}{\pi^2} \frac{1}{\beta^2 b^2}
\left| \int_0^\infty dv \ v^2 J_1(v) \frac{F_{el}\left(-\frac{u^2+v^2}{b^2}
\right)} {u^2 + v^2} \right|^2,
\label{eq:nepab}
\end{equation}
where $u=\omega b/\gamma\beta$ as above. In the case of a monopole or 
dipole form factor analytical results can be found \cite{Baron94,HenckenTB94}.

\subsection{Effects of Strongly Interacting Particles} 

Let us now turn to the effects of strong interactions between the
colliding particles (``initial state interactions'').
For the case of nucleus-nucleus (AA) collisions
we can use the semiclassical method: to a very good approximation the 
nuclei move on classical straight line trajectories 
with an impact parameter $b$.
Collisions with $b<R_{min}=R_1+R_2$ are dominated by strong interactions and 
electromagnetic effects (even though they still occur) are generally 
completely swamped by those violent processes. As an example for the
presence of these electromagnetic effects in central collisions we mention
Ref.~\cite{HenckenTB00}. In this paper $\EPEM$ production due to the 
electric fields of the ions is calculated. Their effect was found to be 
much smaller than $\EPEM$-pair production due to hadronic processes and 
from meson decay \cite{Lenkeit98}. Usually, the experimental conditions 
imposed on the study of electromagnetic processes ($\GG$, $\gamma h$ and 
$\gamma$A) exclude these collisions, as they cannot be extracted from 
hadronic events. Of course one has to be careful to exclude them in the 
same way in the theoretical study of these processes as well. One generally 
imposes a (sharp) cutoff on the impact parameter 
\begin{equation}
b > R_{min} = R_1 + R_2.
\end{equation}
In the one-photon case one then gets the equivalent photon number as
\begin{equation}
n(\omega) = \int_{R_{min}}^\infty 2 \pi b db N(\omega,b).
\label{eq:nomegafromb}
\end{equation}
\begin{figure}[tbh]
\centerline{
\resizebox{45mm}{!}{\includegraphics{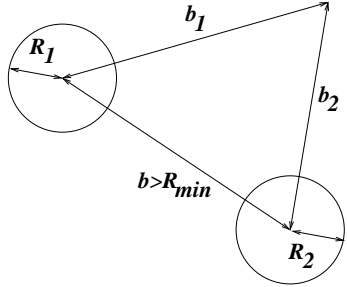}}
}
\caption{In the semiclassical picture, which is valid in the case
of heavy ion collisions, initial state interactions between the two
ions take place if the impact parameter between the two ions is smaller
than $R_{min}=R_1+R_2$. Final state interaction can occur if the individual 
$b_i$ are smaller than $R_i$.}
\label{fig_semiclassic}
\end{figure}

Let us look first at the one-photon case (``$\gamma$A collisions'') between 
two ions. It is a 
good assumption that the total nuclear charge is contained inside the sphere 
with radius $R_i$. As it is well known from electrostatics, the electric 
field outside of a spherical symmetric charge distribution is the same as 
if the total charge is concentrated in the center. The two ions cannot come
closer than $R_{min} = R_1 + R_2$. Impact parameter dependent equivalent 
photon spectra for extended charge distributions can in principle be 
calculated. But from the argument given above it is clear that the form of 
the charge distribution does not enter
in this case in Eq.~(\ref{eq:nomegafromb}). It only matters that the 
nuclear charge $Z$ is entirely contained inside the nuclear radius $R_i$.

The corresponding electromagnetic process can be written as
\begin{eqnarray}
\sigma_{OP} &=& \int_{R_{min}}^\infty 2 \pi b db 
\int \frac{d\omega}{\omega} N(\omega,b)
\sigma_{\G}(\omega)\\
&=& \int_{R_{min}}^\infty 2 \pi b db P_{OP}(b),
\end{eqnarray}
see Eq.~(\ref{eq:sigmaih}), allowing to define the impact parameter dependent 
probability for the one-photon process in this case.
In numerous studies of nuclear excitations of fast (relativistic) heavy ions 
this has been proven to be a good approximation. 

The sharp cutoff assumption
may be relaxed using a smooth absorption profile $T(b)$, see, e.g.,
Chap.~2 of \cite{BaurHT98}. The importance of
such uncertainties depends on each case: for dipole excitations the
dependence on $R_{min}$ is only logarithmic, for higher
multipolarities it depends on an inverse power of the cutoff
$R_{min}$. So in the former case the uncertainties in the choice
of a cutoff will be small. For RHIC the effect of the surface thickness was 
calculated and found to contribute to about 5--10\% \cite{NystrandK98}.
An interesting question is the choice of a cutoff in the case of deformed
nuclei. Depending on the orientation of the nuclei, the minimum impact 
parameter varies. It would be interesting to study this effect in the future.
For a related problem we refer to \cite{FaeldtG90}.

Let us mention the attempt by Benesh, Haynes and Friar \cite{BeneshHF96}
to include the effects of the $R_{min}$-cutoff by using plane waves, 
a nuclear form factor, and some cutoff $q_{max}$ (corresponding loosely 
to a minimum impact parameter $R_{min}$). From the above it is clear that 
this cannot work quantitatively, as is explained in detail in a comment 
by Baur and Bertulani \cite{BaurB97}, see also \cite{BeneshHF97}. The 
equivalent 
photon spectrum used in \cite{BeneshHF96} corresponds to the "point-form" 
spectrum in Fig.~2.3 of \cite{KraussGS97}. 
In addition the results of \cite{BeneshHF96} were compared to
experimental data for giant resonance excitations in \cite{RubehnMT97}.

For the $\GG$ luminosity in the two-photon (TP) case for AA collisions, 
modifications due to the strong interaction are important especially for 
the high energy end of the luminosity function. Again we apply the 
condition that the minimum impact parameter between the two ions must be 
larger than $R_{min}$. The photon-photon luminosity can then be calculated as
\cite{Baur90d,CahnJ90}
\begin{eqnarray}
\frac{dL_{\GG}}{dWdY} &=& \frac{2}{W} 
\int d^2b_1 \int d^2b_2 \nonumber\\
& &\times N_1(\frac{W}{2}e^Y,b_1) N_2(\frac{W}{2}e^-Y,b_2) 
\Theta(|\vec b_1 + \vec b_2| - R_{min}).
\label{eq:dl/dwdy}
\end{eqnarray}
Here, especially for non-strongly interacting final states, one does not
necessarily have a condition for the individual $b_i$ to be larger then
$R_i$. Therefore form factor effects might become important in contrast to
the $\gamma$A collisions considered above.

This formula has also been derived ab initio within the semiclassical 
approximation in \cite{Baur92,BaurB93}. Effects due to the photon 
polarization are also included. They are neglected here for simplicity,
as they were found to be small.
A similar approach for the derivation of the impact parameter dependent
luminosity $L_\GG$ was done in \cite{VidovicGB93}. There an expression not
in terms of the individual $b_i$, but a direct impact parameter
dependent luminosity function $L_\GG(b)$ or $n(\omega_1,\omega_2,b)$ was
derived. (Of course this expression can also be transformed into a formula
dependent on the individual $b_i$, as explained in \cite{KraussGS97}).

For a calculation of the $\GG$ luminosities it is 
easiest to calculate first the $\GG$ luminosities without the 
restriction $b<R_{min}=R_1+R_2$, which can be done analytically, e.g., 
for point particles with the restriction $b_i>R_i$. The correct result will 
then be obtained by subtracting those cases, where $b<R_{min}$. 
One obtains \cite{CahnJ90}
\begin{equation}
\frac{dL_{\GG}}{dWdY} = \frac{2}{W} n\left(\frac{W}{2} e^{Y}\right) 
n\left(\frac{W}{2} e^{- Y}\right) - \frac{d\Delta L_{\GG}}{dWdY},
\end{equation}
with
\begin{eqnarray}
\frac{d\Delta L_{\GG}}{dWdY} &=& \frac{8 \pi}{W}
\int_{R_1}^{\infty} b_1 \ db_1 \int_{\max(b_1-R_{min},R_2)}^{b_1+R_{min}} 
b_2 \ db_2 \nonumber\\
&&\times      N\left(\frac{W}{2} e^{Y},b_1\right)
      N\left(\frac{W}{2} e^{-Y},b_2\right) \phi_{\mbox{crit}}.
\end{eqnarray}
The integral over $\phi$ goes only from $0$ to
$\phi_{\mbox{crit}}$ and from $2\pi-\phi_{\mbox{crit}}$  to
$2\pi$. Only in these cases do we have contributions from $b<R_{min}$. 
$\phi_{\mbox{crit}}$ is given by 
\begin{equation}
\phi_{\mbox{crit}} = \arccos 
\left( \frac{b_1^2 + b_2^2 - R_{min}^2}{2 b_1 b_2}\right).
\end{equation}

For $b_1 \rightarrow \infty$ we also have $b_2 \rightarrow b_1$ 
and $\phi_{\mbox{crit}} \rightarrow  0$. The integral converges 
rapidly.  
It was also noted in \cite{CahnJ90} that for symmetric collisions the
$\GG$ luminosity can be written in the form
\begin{equation}
\tau \frac{dL_{\GG}}{d\tau} = L_0 \xi(z),
\end{equation}
with
\begin{equation}
L_0= \frac{16 Z^4 \alpha^2}{3 \pi^2}
\end{equation}
and a universal function $\xi(z)$ of $z=2 M R\sqrt{\tau}=m R / \gamma$.
A parameterization of this universal function is given there as well.
The case $z \ll 1$  was first discussed in
\cite{BaurB88}.

The last modification concerns possible final state interactions
of the produced particles with the colliding particles.
If the produced particles are hadrons, they can interact with the ions, 
if they are produced ``within'' one of the ions $b_1<R_1$ or $b_2<R_2$. Such 
processes are excluded in the approach given above, as integration over
$b_1$ and $b_2$ are starting from $R_1$ and $R_2$, respectively. 
Since the electric field strength inside the nucleus decreases with 
decreasing radius such effects are expected to be small. 
They will be of
importance only at the very high end of the $\GG$ spectrum.
A calculation taking into account the final state interaction has to our 
knowledge not been performed. But one can obtain an estimate (or an upper 
bound) of the size of this effect by comparing $\GG$ luminosities with either
an integration over all $b_i$ or restricted to $b_i>R_i$ respectively. Of 
course for this one needs to use impact parameter dependent equivalent photon
spectra with an elastic form factor.

\subsection{Effective $\GG$-Luminosities and Perspectives for RHIC and LHC}

The effective $\GG$ luminosity $L_{\rm eff}$ in ion-ion collisions 
is defined in terms of the  
the beam luminosity $L_{AA}$ as
\begin{equation}
\frac{dL_{\rm eff}}{dM} = L_{AA}\frac{dL_{\GG}}{dM},
\end{equation}
where  the $\GG$ luminosity $L_{\GG}$ is given  by Eq.~(\ref{eq:dl/dwdy}). 
Luminosities of the heavy ion beams are several orders
lower compared to those of light ions and protons. One reason is  the
large cross sections of electromagnetic processes of heavy ions, which
either disintegrate the ions or change their charge state due to electron
capture. In \cite{BrandtEM94} the influence of different beam-beam
interaction processes on the beam lifetime is considered for the
LHC. The main processes, which contribute to the beam-beam interactions,
are hadronic nuclear interactions, electromagnetic dissociation, where
an ion is excited and subsequently decays, and bound free electron-position 
pair production, see below. Thus, the maximum 
ion luminosity is derived from the 
cross sections of the beam-beam interaction. The luminosities for different 
ion species at the RHIC and the LHC are given in \cite{RHIC:CDR} and 
\cite{Brandt00}, respectively.
\begin{table}[tbh]
\centerline{
\begin{tabular}{c|ccccc}\hline\hline
 & Projectile & $Z$ & $A$ & $\sqrt{s}$, A GeV & Luminosity,
                                        $\mbox{cm}^{-2}\mbox{s}^{-1}$ \\
                                                     \hline
L & $p$  &  1 &   1 & 14000 & $1.4\cdot10^{31}$ \\
H & $Ar$ & 18 &  40 & 7000  & $5.2\cdot10^{29}$ \\
C & $Pb$ & 82 & 208 & 5500  & $4.2\cdot10^{26}$ \\   \hline
R & $p$  &  1 &   1 & 500   & $1.4\cdot10^{31}$ \\
H & $Cu$ & 29 &  63 & 230   & $9.5\cdot10^{27}$ \\
I & $Au$ & 79 & 197 & 200   & $2.0\cdot10^{26}$ \\  
C &      &    &     &       &                   \\   \hline\hline
\end{tabular}
}
\caption{Average luminosities at LHC and RHIC
for $pp$, medium and heavy ion beams.}
\label{tab:beamlumi}
\end{table}
In Table~\ref{tab:beamlumi} we quote the average
luminosities at RHIC and LHC for $pp$, medium and heavy ion
collisions.
The specific experiments may also apply their requirement
to lower the luminosities to satisfy the detector load
\cite{Morsch2001}. 

\begin{figure}[tbh]
\parbox{0.48\hsize}{
\resizebox{\hsize}{!}{\includegraphics{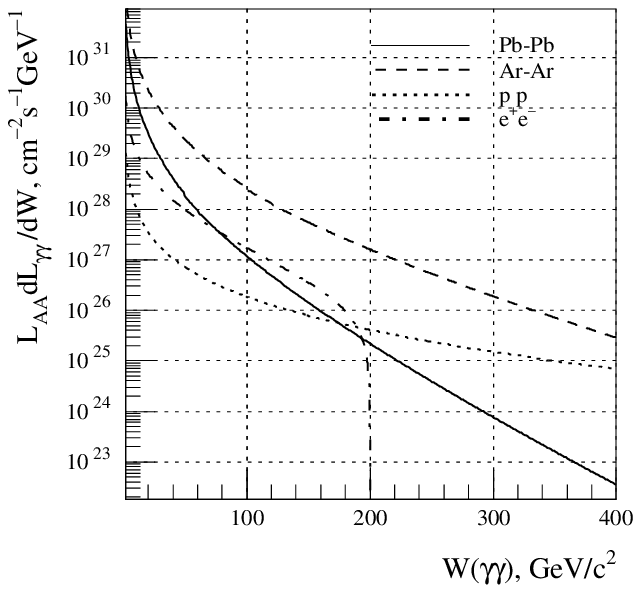}}
}
\hfill
\parbox{0.48\hsize}{
\resizebox{\hsize}{!}{\includegraphics{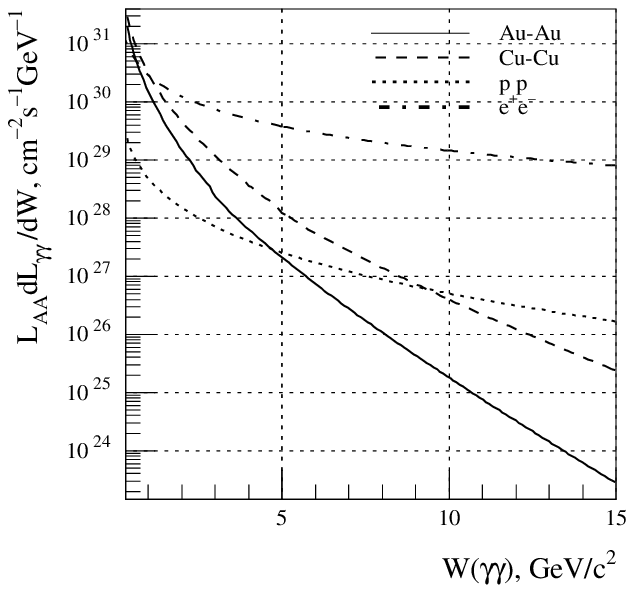}}
}
\caption{Effective $\GG$ luminosity at LHC (left) and RHIC (right) for different
  ion species and protons as well as at the $\EPEM$ collider LEP-II.
  The $L_{AA}$ in the last two cases of course corresponds to the $pp$ or $ee$
  beam luminosity.}
\label{fig:gglum}
\end{figure}
Two-photon luminosities for various ion species and protons as a
function of $\GG$ mass are shown in Fig.\ref{fig:gglum} (left) for LHC
and (right) for RHIC. The $\GG$ luminosities are calculated by the Monte
Carlo program TPHIC \cite{HenckenKKS96}. The luminosities at the ion
colliders are compared with the $\GG$ luminosity at LEP-II with a
c.m. energy of 200-GeV and an 
$\EPEM$ luminosity of $5\cdot10^{31}~\mbox{cm}^{-2}\mbox{s}^{-1}$
\cite{Evans1993}.

Although the heaviest ions generate the highest $\GG$ luminosities due
to their large charge, the beam luminosities are lower for these ions.
As a result, the medium-weight ions (Argon at LHC, Copper at RHIC) are
the most prominent source of two-photon processes. RHIC can compete
with LEP in two-photon studies at low $\GG$ masses while LHC will be
the best machine to study two-photon physics at all ranges of $\GG$
masses.

%
%
\section{$\GG$ Physics at Hadron Colliders, General Considerations}
\label{sec_ggqcd}

We will now give a general discussion of possible photon-photon physics 
at relativistic heavy ion colliders. Two-photon interactions allow to test 
the main properties of the standard model in both the electroweak sector
and in QCD, as well as, physics beyond the Standard Model. In contrast to 
hadronic processes, many of the processes here can be calculated in principle 
to some degree of accuracy. Below we review QCD interactions  and also 
physics related to electroweak interactions, which are of 
interest to be studied in two-photon interactions
in heavy ion collisions. This covers the range starting from exclusive meson
production, meson pair production up to the total hadronic cross section.
In the next chapter we will then be concerned with the potential for the 
discovery of new physics, which seems to be possible in principle
due to the high two-photon invariant masses of up to about 100~GeV, which are 
available at the LHC. An interesting topic in itself is the electron-positron 
pair production. The fields are strong enough to produce multiple pairs in a 
single collision. A discussion of this subject will be given in 
Chap.~\ref{sec_leptons} below.

Up to now photon-photon scattering has been mainly studied at $\EPEM$
colliders. Many reviews \cite{BudnevGM75,KolanoskiZ88,BergerW87}
as well as conference reports
\cite{Amiens80,SanDiego92,Sheffield95,Egmond97,Freiburg99,Ambleside00}
on this subject exist. Whereas in the past the range of invariant masses
has been the region of mesons, ranging from $\pi^0$ ($m_{\pi^0}=135$~MeV) 
up to about $\eta_c$ ($m_{\eta_c}=2980$~MeV), the higher invariant masses
at LEP2 have allowed to study an invariant mass range up to about 
185~GeV \cite{L3:01}, where one interesting subject has been the
study of the total $\GG\rightarrow$~hadron cross-section.

We are concerned here mainly with the invariant mass region relevant for 
the LHC; see the $\GG$-luminosity in Fig.~\ref{fig:gglum}.
At RHIC $\GG$ physics can be done in the $\GG$ invariant mass region
up to several GeV. This topic has also been studied  experimentally now
by the ``Peripheral Collision Group'' at STAR  and this will be discussed 
below in Chap.~\ref{sec_eventsel}.
Apart from the production of $\EPEM$
(and $\MUPM$) pairs, the photons can always be considered as
quasireal. The cross section for virtual photons deviates from the one 
for real photons only for $Q^2$, which are much larger then the coherence 
limit $Q^2\lesssim 1/R^2$, see also the discussion in \cite{BudnevGM75}.
Therefore photon-photon processes are restricted to {\em quasireal} photons.
Tagging of the elastic scattered beams is very difficult in the heavy ion
case, but for the $pp$ case at the LHC such a possibility has been investigated
in \cite{Piotrzkowski00}. The standard detector techniques used for 
measuring very forward proton scattering will allow a reliable separation 
of interesting $\GG$ interactions

\subsection{The Photon at High Energy}

Real photons have a complicated nature. In a first approximation the
photon is a point-like particle, although in field theory it may fluctuate 
also into a fermion pair \cite{SchulerS97}. 
Fluctuations into a quark-antiquark pair
$\gamma \to q \bar{q}$ can interact strongly and give a large
contribution to the two-photon hadronic cross section, while fluctuations 
into a lepton pair $\gamma \to l \bar{l}$ interact electromagnetically only
and do not influence the total $\gamma\gamma$ 
cross section very much (see also the rates predicted for the different
contributions below for the LHC). Lepton fluctuations can be calculated 
perturbatively, which is not true for the quark pairs, as
low-virtuality fluctuations need to be described by nonperturbative QCD. 
Therefore, the spectrum of the photon fluctuation
can be separated into low-virtuality and high-virtuality parts,
according to \cite{SchulerS97}. To describe the first part, a
phenomenological model of vector-meson dominance (VMD) is used, according 
to which the photon fluctuates into a sum of vector meson states. Quark 
pair with high virtuality are described by the perturbative theory. As a
whole, the photon wave function can be written then as follows 
\begin{figure}[tbh]
\parbox{0.48\hsize}{
  \centerline{\resizebox{3cm}{!}{\includegraphics{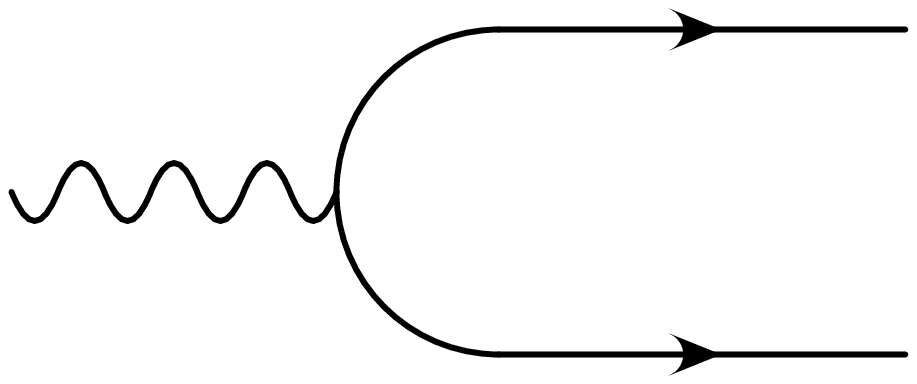}}(a)}
  ~\\
  \centerline{\resizebox{3cm}{!}{\includegraphics{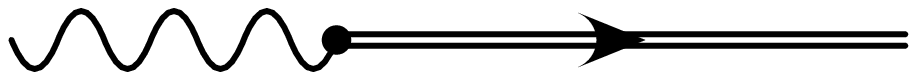}}(b)}
}
\hfill
\parbox{0.48\hsize}{
  \caption{The photon can fluctuate into other states, among them into
a pair of fermions (leptons, quarks, (a)), but also into vector mesons
(b). The total cross section in photon-photon collisions is dominated by
the fluctuations into two quarks and their strong interaction.}
\label{fig_gqqbar}
}
\end{figure}
\begin{equation}
\begin{array}{rcl}
\mid\gamma\rangle = c_0 \mid\gamma_0\rangle & + &
 \sum_{V=\rho^0,\omega,\phi,J/\psi,\Upsilon}{c_V\mid V\rangle} +
 \sum_{q=u,d,s,c,b}{c_q \mid q\bar q \rangle} + \\
 & & \sum_{l=e,\mu,\tau}{c_l \mid l^+ l^- \rangle},
\end{array}
 \label{wfun}
\end{equation}
where $\mid\gamma_0\rangle$, $\mid V\rangle$, $\mid q\bar q\rangle$
and $\mid l^+ l^- \rangle$ are wave functions of the point-like photon,
a vector meson, a quark- and a lepton pair, respectively. The coefficients
$c_i$ in general depend on the scale $\mu$, which probes the photon.
The coefficients for the contribution to a lepton pair are well known to be
equal to
$$
c_l \approx \frac{2}{3}~\frac{\alpha_{\rm em}}{2\pi}~\ln(\mu^2/m_l^2).
$$
To separate the low- and high-virtuality quark fluctuations one 
introduces a parameter $p_0$. For high-virtuality fluctuations 
one sets
$$
c_q \approx (\alpha_{\rm
  em}/2\pi)2Q^2_q\ln(\mu^2/p_0^2).
$$  
The $q\bar q$ fluctuations with virtuality below $p_0$ are described by 
the vector meson dominance part, which does not depend on the scale $\mu$. 
As a rule of thumb the scale $\mu$ is taken equal to the
transverse momentum of the parton process. The value of $p_0$, which
provides the best description of the total cross section, is found to be
$p_0=0.5$~GeV \cite{SchulerS97}.
The $c_V$ are defined as 
$$
c^2_V = 4\pi \alpha_{\rm em} /f^2_V, 
$$
where the decay constants $f^2_V/4\pi$ are determined from experimental data
on $V\rightarrow l^+l^-$;
they are $f^2_V/4\pi = 2.20$ for $\rho^0$, 23.6 for $\omega$, 18.4 for $\phi$, 
11.5 for $J/\psi$. Finally the $c_0$ is then defined from unitarity, and 
is usually close to one.

\begin{figure}[tbh]
\begin{center}
\resizebox{7.5cm}{!}{\includegraphics{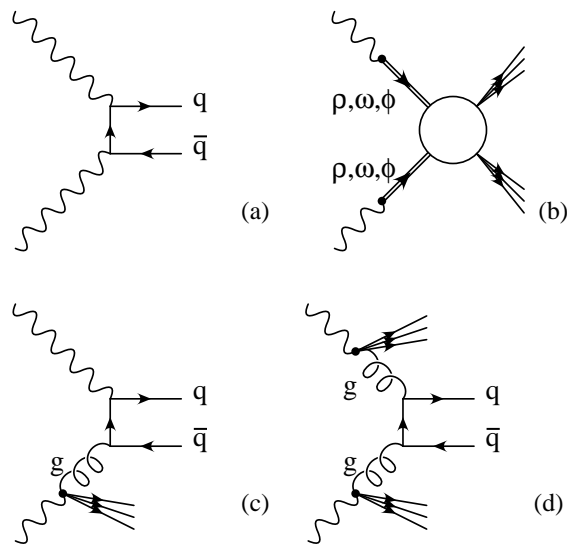}}
\end{center}

\caption{\it
Diagrams showing the contribution to the $\GG\rightarrow$hadron
reaction: direct mechanism (a), vector meson dominance (b), single (c)
and double (d) resolved photons.
}
\label{fig_resolved}
\end{figure}

According to the photon wave function representation of Eq.~(\ref{wfun}), the
following partonic processes can take place in two-photon interactions, see 
also Fig.~\ref{fig_resolved}:
\begin{itemize}  
\item Purely electromagnetic processes of order ${O}(\alpha_{\rm em}^2)$ 
involving quarks $q$ and leptons $l$, i.e.,  $\gamma\gamma \to q\bar q$, 
$\gamma l \to \gamma l$, $ \gamma\gamma \to l \bar{l}$, $l l' \to l l'$,
as well as, the production of $W^+W^-$ pairs and of pairs of particles
beyond the Standard Model.
\item Processes with one electromagnetic and one strong vertices of order
${O}(\alpha_{\rm em} \alpha_{\rm s})$ such as  $\gamma q \to gq$ and 
$\gamma g \to q\bar q$.
\item Strong processes involving quarks and gluons of order 
${O}(\alpha_{\rm s}^2)$ $qq' \to qq'$, $q\bar q \to q'\bar q'$,
  $q\bar q \to gg$, $qg \to qg$, $gg \to q\bar q$, $gg \to gg$.
\item Nonperturbative QCD processes (elastic, diffractive scattering,
normally at low transverse momentum).
\end{itemize}

To describe those two-photon processes where quarks and gluons
interact, one uses the structure function of the photon, which has
been measured experimentally (see, e.g. \cite{DreesG85,SchulerS95}).
The structure functions of the vector mesons are less well known,
therefore the approximation $|\rho^0\rangle \approx |\pi^0\rangle
\approx (|\pi^+\rangle + |\pi^-\rangle)/2$ is often used, where the
structure functions of pions are known \cite{Gluck92}.  Below we give
the cross section formulae of the elementary processes within the
leading order of the perturbative theory and describe some
nonperturbative models, which involve photons.

\subsection{Total $\GG$ Cross Section}
The main contributions to the total $\gamma\gamma$ collision comes
from the strong interactions. The total cross section
$\GG \rightarrow \mbox{hadrons}$ can be
parameterized in the form predicted by Regge theory
\begin{equation}
\sigma^{\gamma\gamma}_{\mbox{\scriptsize tot}} \approx A
\left(s/s_0\right)^\epsilon + B \left(s/s_0\right)^{-\eta},
    \label{eq:ggtotal}
\end{equation}
where $s$ is the square of the invariant $\gamma\gamma$ mass, $s_0= 1 GeV^2$, 
and the exponents $\epsilon$ and $\eta$ are universal parameters,
that is, they are identical to those in $\gamma p$, $pp$, but also $\pi p$
total cross sections, corresponding to Pomeron and Reggeon exchange, 
respectively. The universal values for these exponents from a combined fit of 
all cross section are $\epsilon=0.093(2)$ and $\eta=0.358(15)$ \cite{PDG00}.

This total cross section is an important input in the study of high energy
$\EPEM$ collisions, see, e.g., the review of Peskin \cite{Peskin00}.
It is also interesting to understand what part comes from point-like processes 
and what part from soft processes involving the hadronic constituents of the 
photon.  Eventually theory should explain, as well as, be constrained by 
the data of both $\sigma(\GG)$ and $\sigma(\gamma p)$.
As also done in \cite{L3:01} there are in general two fits to the total
cross section data: one with PHOJET and the other one with the PYTHIA 
event generators. They differ substantially,
and this is due to the fact that about
40\% of the cross-section is unobserved at LEP and that theoretical 
models differ considerably in the size of the contribution from these very soft
events.

The best fitted values {\em before LEP2} were as follows \cite{SchulerS97}:
\begin{equation}
A\approx 211 \mbox{nb}, \quad B\approx 297 \mbox{nb}, \quad
\epsilon \approx 0.0808, \quad \eta \approx 0.4525.
\end{equation}

The L3 collaboration recently made a measurement of the total hadron
cross-section for photon-photon collisions in the interval $5 GeV <
W_{\GG} < 185 GeV$ \cite{L3:97,L3:01}. Fitting the data up to 65~GeV,
it was found that the $\GG \rightarrow$hadrons cross-section is consistent 
with the universal Regge behavior of total hadronic cross-sections.
Values of 
\begin{equation}
A=173\pm7 \mbox{nb}, \quad B=519\pm125 \mbox{nb}, \quad \epsilon=0.0790, \quad 
\eta=0.4678
\end{equation}
were given in \cite{L3:97} (where the values of $\epsilon$ and $\eta$ had
been held fixed). Their cross section were found to be in agreement with the
universal values of $\eta$ and~$\epsilon$.

Using the larger invariant mass range up to 185 GeV in \cite{L3:01}
a deviation from the universal value for $\epsilon$ was found.
Fitting the data with $\eta=0.358$ fixed gave values of 
\begin{equation}
A=59\pm10 \mbox{nb}, \quad B=1020\pm146 \mbox{nb}, \quad
\epsilon=0.225\pm0.021, 
\end{equation}
that is a value for $\epsilon$
more than twice as large as the universal one.
This large exponent was also found to be independent of the Monte Carlo 
model used to correct the data.

\subsection{Charged Fermion Pair Production}

For the invariant  mass of the $\gamma\gamma$ system above the threshold
$\sqrt{s} > 2m_f$ a fermion (lepton or quark) 
pair can be produced in two-photon collision. 
In lowest order of QED 
\begin{figure}[tbh]
\centerline{\resizebox{30mm}{!}{\includegraphics{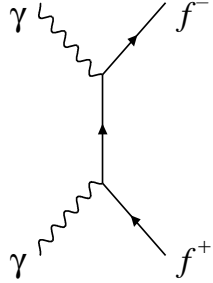}}}
\caption{Process $\gamma\gamma \to f^+f^-$ in the lowest order QED.}
 \label{diag:ggff}
\end{figure}
this process is shown in Fig.\ref{diag:ggff}. The 
differential cross section of the process $\gamma\gamma \to f^+ f^-$
in the c.m. system is given by the equation \cite{Barklow90,BrodskyKT70}
\begin{equation}
\frac{d\sigma}{d\cos\theta}
(\gamma\gamma \rightarrow f^+f^-) =
\frac{e^4\beta Q^4_f N_c}{8\pi s}
\frac{1+2\beta^2(1-\beta^2)(1-\cos^2 \theta) - \beta^4 \cos^4 \theta}
     {(1 - \beta^2 \cos^2 \theta)^2}, 
\label{dxsgg-ff}
\end{equation}
where $Q_f$ is a fermion charge in units of electron charge, $N_c$ in
a number of colors, $\beta = \sqrt{1-4m^2_f/s}$ is the velocity of the 
fermion in the $\GG$ rest frame.
The integral cross section of this process is
\begin{equation}
\sigma(\gamma\gamma \rightarrow f^+ f^-)
    = \frac{4\pi\alpha^2 Q^4_f N_c}{s}\beta
    \left[\frac{3-\beta^4}{2\beta}\ln\frac{1+\beta}{1-\beta} -
    2 +\beta^2\right]. \label{xsgg-ff}
\end{equation}
The dependence of the integral cross section on the ratio of the fermion
mass to the beam energy in the c.m. system, $2m_f/\sqrt{s}$ is shown in
Fig.\ref{ggff_beta}.
\begin{figure}[tbh]
\parbox{0.48\hsize}{
  \centerline{\resizebox{7.0cm}{!}{\includegraphics{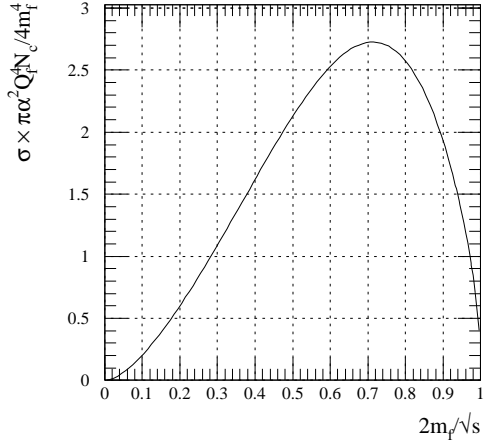}}}
}
\hfill
\parbox{0.48\hsize}{
  \caption{Dependence of the process $\gamma\gamma \to f^+ f^-$ cross
    section on $2m_f/\sqrt{s}$.}
  \label{ggff_beta}
}
\end{figure}

The production of fermion pairs in two-photon collisions can be used to
study the coupling of $s$-, $c$- and $b$-quarks to photons and to
study the fragmentation of these quarks into $K$, $D$ and $B$ mesons,
because these processes are clean from hadronic background. To
normalize the two-photon luminosity the process $\gamma\gamma \to
\mu^+ \mu^-$ can be used as it is easy to observe and simple to calculate
\cite{ShamovT98,BudnevGMS72,BudnevGMS73}.
In Table~\ref{xs-ff} production cross section of 
$\mu^+ \mu^-$, $s\bar s$, $c\bar c$ and $b\bar b$ are shown in different 
invariant mass intervals in collisions of various species of ions at the RHIC
and LHC. Figure~\ref{fig_overviewC} shows cross section and event rates for
the LHC.
\begin{table}[tbh]
\centerline{
\begin{tabular}{c|c|c|c|c} 
\hline\hline
$f^+f^-$ & Mass range, GeV  & 
\multicolumn{3}{c}{$\sigma(AA \to AA +f^+f^-)~\mu$b}\\
              &                   & AuAu     & CaCa & PbPb\\
\hline
$\mu^+\mu^-$  & $1 < W < 10$      & 280       & 730   & $1.1\cdot10^5$\\
$s\bar s$     & $1 < W < 10$      & 7.6       & 20    & $3.1\cdot10^3$ \\
$c\bar c$     & $3.7 < W < 10$    & $1.5\cdot10^{-2}$ & 6.2 & 790      \\
$b\bar b$     & $10.6 < W < 20$   & ---       & $1.3\cdot10^{-2}$ & 1.2\\ 
\hline\hline
\end{tabular}}
\caption{Production cross sections of fermion pairs in two-photon
    interactions in AuAu collisions at $\protect\sqrt{s_{AA}} =
    100$~GeV per nucleon at the RHIC, as well as in CaCa and PbPb 
    collisions at $\protect\sqrt{s_{AA}} =
    7.2$~TeV per nucleon $\protect\sqrt{s_{AA}} =
    5.5$~TeV per nucleon at the LHC.} 
\label{xs-ff}
\end{table}
\begin{figure}[tbh]
\begin{center}
\resizebox{12cm}{!}{\includegraphics{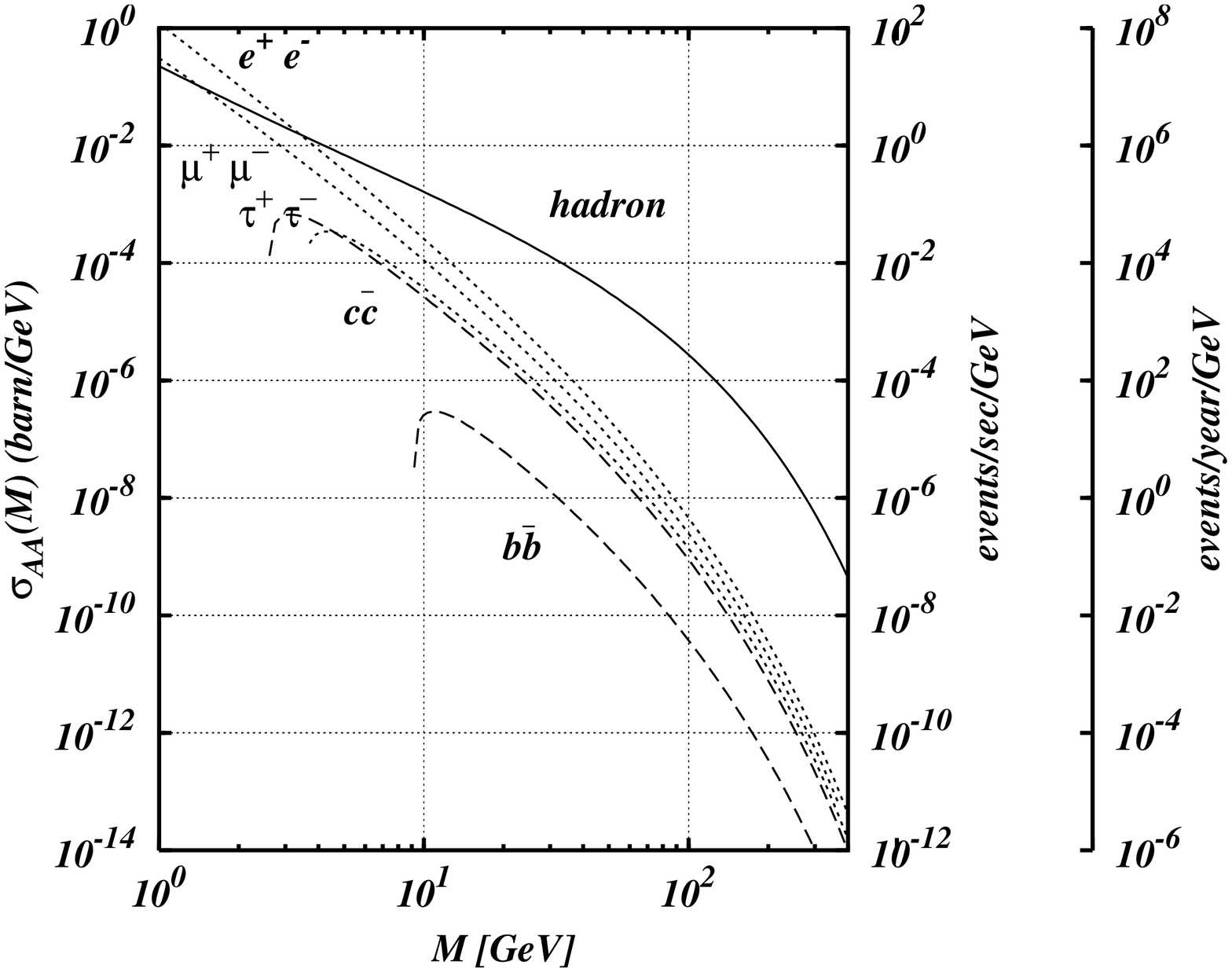}}
(a)\\
\resizebox{12cm}{!}{\includegraphics{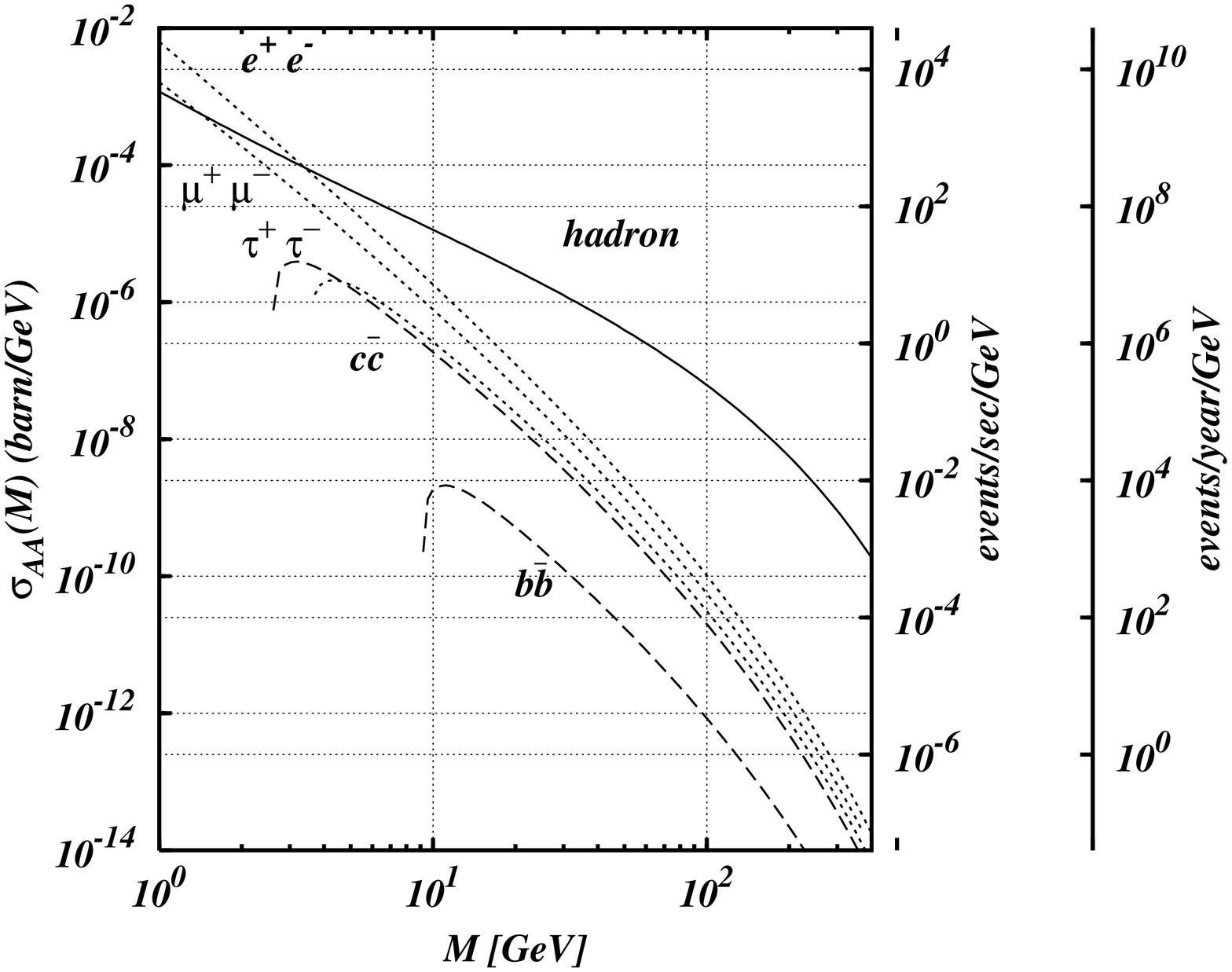}}
(b)\\
\end{center}
\caption{Cross section per GeV for different charged fermion pair production 
at the LHC for PbPb (a) and CaCa (b) collisions, using the lowest order QED
cross section. Also shown is the total rate for $\GG \to$hadrons, see 
Eq.~(\protect\ref{eq:ggtotal}). Also shown are event rates per s and per
($10^6$) year.
}
\label{fig_overviewR}
\end{figure}

\subsection{ $W^+W^-$-Pair Production}

\begin{figure}[tbh]
\centerline{\resizebox{0.5\hsize}{!}{\includegraphics{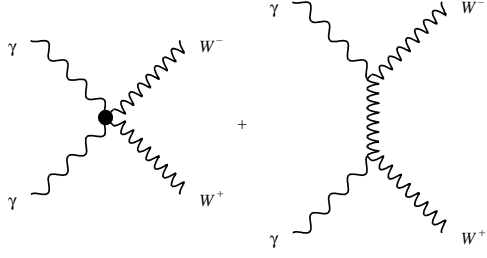}}}
\caption{Process $\gamma\gamma \to W^+W^-$ in the lowest order.}
\label{diag:ggww}
\end{figure}
If the energy of the two-photon collision is higher than twice the  mass of
$W^\pm$, the process $\gamma\gamma \to W^+W^-$ is possible. This 
process involves the gauge couplings of these gauge bosons . The 
lowest order graphs are shown in Fig.~\ref{diag:ggww}. The 
differential cross section of this process is given by
\begin{equation}
\begin{array}{l}
\displaystyle
\frac{d\sigma}{d\cos\theta}(\gamma\gamma \rightarrow W^+W^-) = \\[1mm]
\displaystyle
\qquad \frac{\pi \alpha^2\beta}{s} \frac
{19 - 6\beta^2(1-\beta^2) + 2(8-3\beta^2)\beta^2 \cos^2 \theta +
 3\beta^4 \cos^4 \theta}{(1 - \beta^2 \cos^2 \theta)^2},
\label{dxs-ww}
\end{array}
\end{equation}
with the same notations as those in Eq.~(\ref{dxsgg-ff}). The integral cross
section is \cite{Katuya83}
\begin{equation}
\sigma(\gamma\gamma \rightarrow W^+W^-)
       = \frac{\pi\alpha^2}{s}\beta\left[
       -3\frac{1-\beta^4}{\beta}\ln\frac{1+\beta}{1-\beta} +
        2\frac{22-9\beta^2+3\beta^4}{1-\beta^2}\right].
\label{gg-ww}
\end{equation}
The magnetic moment, as well as, the quadrupole moment of the $W$ 
are fixed in the standard model, for the more general case of a magnetic
moment different from the one given by the standard model see also 
\cite{TupperS81}.

\begin{figure}[tbh]
\parbox{0.48\hsize}{
  \centerline{\resizebox{7cm}{!}{\includegraphics{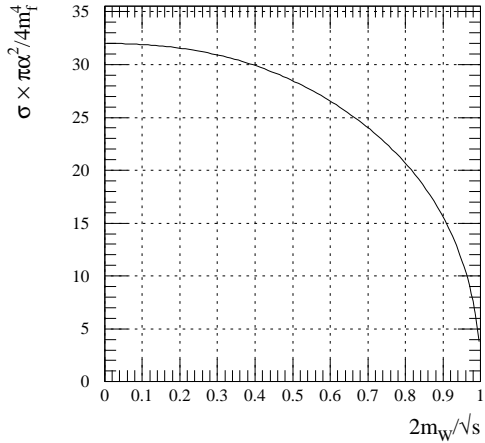}}}
}
\hfill
\parbox{0.48\hsize}{
  \caption{The process $\gamma\gamma \to W^+ W^-$ cross
    section as function of $2m_W/\sqrt{s}$.}
  \label{ggww_beta}
}
\end{figure}

Fig.~\ref{ggww_beta} shows the dependence of the integral cross section
on the ratio of a $W^\pm$ mass to the beam energy in the c.m. system,
$2m_W/\sqrt{s}$.
For example, the cross section estimates of the $W^+W^-$ production in 
CaCa and PbPb collisions at the LHC energies are 
$\sigma(AA \to AA + W^+W^-) = 29$~pb and~190~pb, respectively.

\subsection{Vector Meson Pair Production}
\label{sec:vector-meson-pair}

Among the nonperturbative processes in photon-photon interactions
there is the exclusive vector-meson pair production. 
There are various mechanisms to produce hadrons in such collisions,
recall the discussion about the nature of the photon above.  According
to this idea, vector meson pair production is explained to happen in
the following way: both photons fluctuate first into a vector meson
($\rho$, $\omega$, $\phi$, $J/\psi$ and so on, the VMD component).
These vector mesons then scatter elastically to produce the (real)
vector meson pair in the final state. The study of these
processes allows a test of the Pomeron-exchange factorization
relation, see \cite{GribovP62,GribovP62b}. If the exchange of a single
Pomeron dominates in these processes, the factorization relation among 
$pp$, $p\gamma$ and $\gamma\gamma$ collisions predicts
\begin{equation}
  \left[\sigma_{\rm tot} (\gamma p)\right]^2 = \sigma_{\rm tot} (pp) \times 
  \sigma_{\rm tot} (\gamma \gamma).
  \label{ggvv-factor}
\end{equation}

As for the available experimental data there exist measurements of the
production cross section of vector meson pairs $\rho^0\phi$ in
$\GG$-interactions by the ARGUS \cite{ARGUS1994} and $\rho\rho$ by the
L3 \cite{Soldner1998,Vogt1999} collaboration. These data were compared with the
cross section estimate of the reaction $\GG\to VV'$ ($V,V' = \rho^0,
\phi$) based on the Pomeron factorization model and all possible
combinations of the existing sets of data on the reactions $\gamma p
\to V p$ and $pp \to pp$ \cite{Achasov1995,Achasov1999}.  A strong
discrepancy between the existing experimental data (by order of
magnitude) and the factorization model prediction was found. These
estimates are in good agreement with independent calculations in
\cite{SchulerS97}.  Thus we come here to the puzzling situation: 
why does one of the most well-grounded phenomenological models predict 
cross sections, which are larger by order of magnitude in the 
reaction $\GG\to VV'$ compared to those measured in the experiments? 
One explanation proposed in \cite{Achasov1999} is that we face here a
new ``{\it defiant phenomenon in the formation mechanism of the Pomeron
exchange for quasi-two-body reactions}''.

The production of vector meson pairs can well be studied at RHIC with
high statistics in the region of up to several GeV \cite{KleinS97a}.
In connection with the above mentioned puzzle the results from the STAR would
be quite desirable and interesting.  For the
possibilities at LHC, we refer the reader to \cite{Felix97} and 
\cite{BaurHTS98}, where also experimental details and simulations are
given.

\subsection{Resonance Production}
\label{sec_resproduction}

One may say that photon-photon collisions provide an independent view
on meson and baryon spectroscopy. They provide powerful
information on both the flavor and spin/angular momentum internal
structure of the mesons. Much has already been done at
$\EPEM$ colliders. Light quark spectroscopy is very
well possible at RHIC, benefitting from the high
$\GG$-luminosities. Detailed feasibility studies exist
\cite{KleinS97a,KleinS97b,KleinS95a,KleinS95b}.  In these studies, $\GG$
signals and backgrounds from grazing nuclear and beam gas collisions
were simulated with both the FRITIOF and VENUS Monte Carlo codes. 
The possibilities to
produce these mesons at the LHC have been discussed in detail in the 
FELIX LoI \cite{Felix97}. Rates are given and possible
triggers are discussed. The general conclusion is that all these processes 
are very promising tools for meson spectroscopy.

In two-photon collisions with real photons general
 symmetry requirements restrict the possible final states, 
as is well known from the Landau-Yang theorem \cite{Yang48}. Especially
it is impossible to produce spin 1 final states. Only resonance states 
with positive $C$-parity can be produced, such as $J^{PC}= 0^{-+}$, $0^{++}$, 
$2^{++}$, $\cdots$.
Two photon collisions therefore give access to most of the $C=+1$ mesons.
In $\EPEM$ annihilation on the other hand only states with $J^{PC}=1^{--}$, 
that is, with odd $C$-parity can be produced directly. 

In principle $C=-1$ vector mesons can be produced by the fusion of
three (or, even less important, five, seven, \dots) equivalent
photons. This cross section scales with $Z^6$. But it is smaller than
the contribution coming from $\G$-A collisions, as discussed in 
Chap.~\ref{sec_photonhadron}, even for nuclei with large $Z$ 
(see also \cite{BaurHT98} and the corresponding discussion for the 
positronium in Sec.~\ref{ssec:dlbstates}).

The production cross section of the resonance is given by
\begin{equation}
\sigma_{\gamma\gamma \rightarrow R} = 8\pi(2J+1)
   \frac{\Gamma_{\gamma\gamma}\Gamma_{\mbox{\scriptsize tot}}}
   {(W^2-M^2_R)^2+M^2_R\Gamma^2_{\mbox{\scriptsize tot}}}, \label{gg-r}
\end{equation}
where $W$ is the $\GG$ invariant mass , $M_R$ is the
mass of the resonance $R$, $\Gamma_{\gamma\gamma}$ and
$\Gamma_{\mbox{\scriptsize tot}}$ its two-photon and total width.
For sufficiently narrow state ($\Gamma_{\mbox{\scriptsize
tot}} \ll M_R$) the expression (\ref{gg-r}) can be approximated by
\begin{equation}
\lim_{\Gamma_{\mbox{\scriptsize tot}}/M_R \to 0}
 {\sigma_{\gamma\gamma \rightarrow R}} =
    4\pi^2 (2J+1)\frac{\Gamma_{\gamma\gamma}}{M_R^2} \ \delta(W-M_R).
    \label{gg-r-delta}
\end{equation}
This makes it easy to calculate the production cross-section 
$\sigma_{AA\rightarrow AA+R}$ of a particle in terms of its basic properties.

In this case the production cross section of a resonance in heavy ion 
collisions factorizes, according to equation (\ref{eq:sigmaiip}), to 
\begin{equation}
\sigma(AA \to AA+R) = 4\pi^2 (2J+1)\frac{\Gamma_{\gamma\gamma}}{M_R^2} 
    \left. \frac{d{ L}_{\gamma\gamma}}{dW} \right|_{W=M_R},
    \label{AA-AAR}
\end{equation}
i.e., it is given only by its two-photon width and the value of the
$\gamma\gamma$-luminosity function at its resonance mass.

\begin{figure}[tbh]
\begin{center}
\resizebox{7cm}{!}{\includegraphics{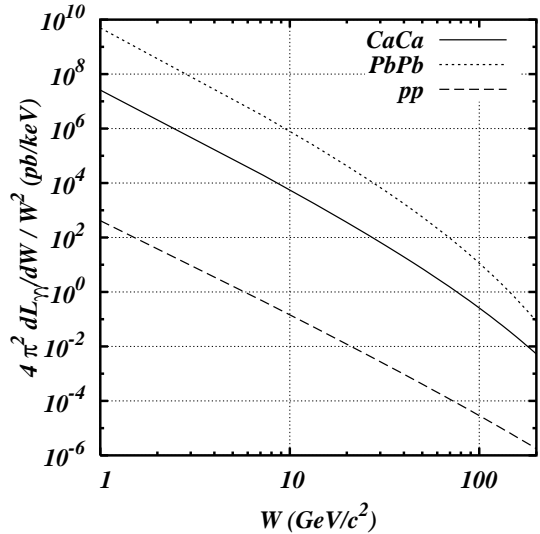}}
\end{center}
\caption{\it
The universal function $4\pi^2 dL_{\GG}/dW /W^2$ is
plotted for different ion species at LHC. We use $R=1.2 \times A^{1/3}$~fm and
$\gamma=2950$, $3750$ and $7000$ for PbPb, CaCa and $pp$ collisions, 
respectively.
}
\label{fig_sigmagamma}
\end{figure}

\begin{figure}[tbh]
\begin{center}
\resizebox{10cm}{!}{\includegraphics{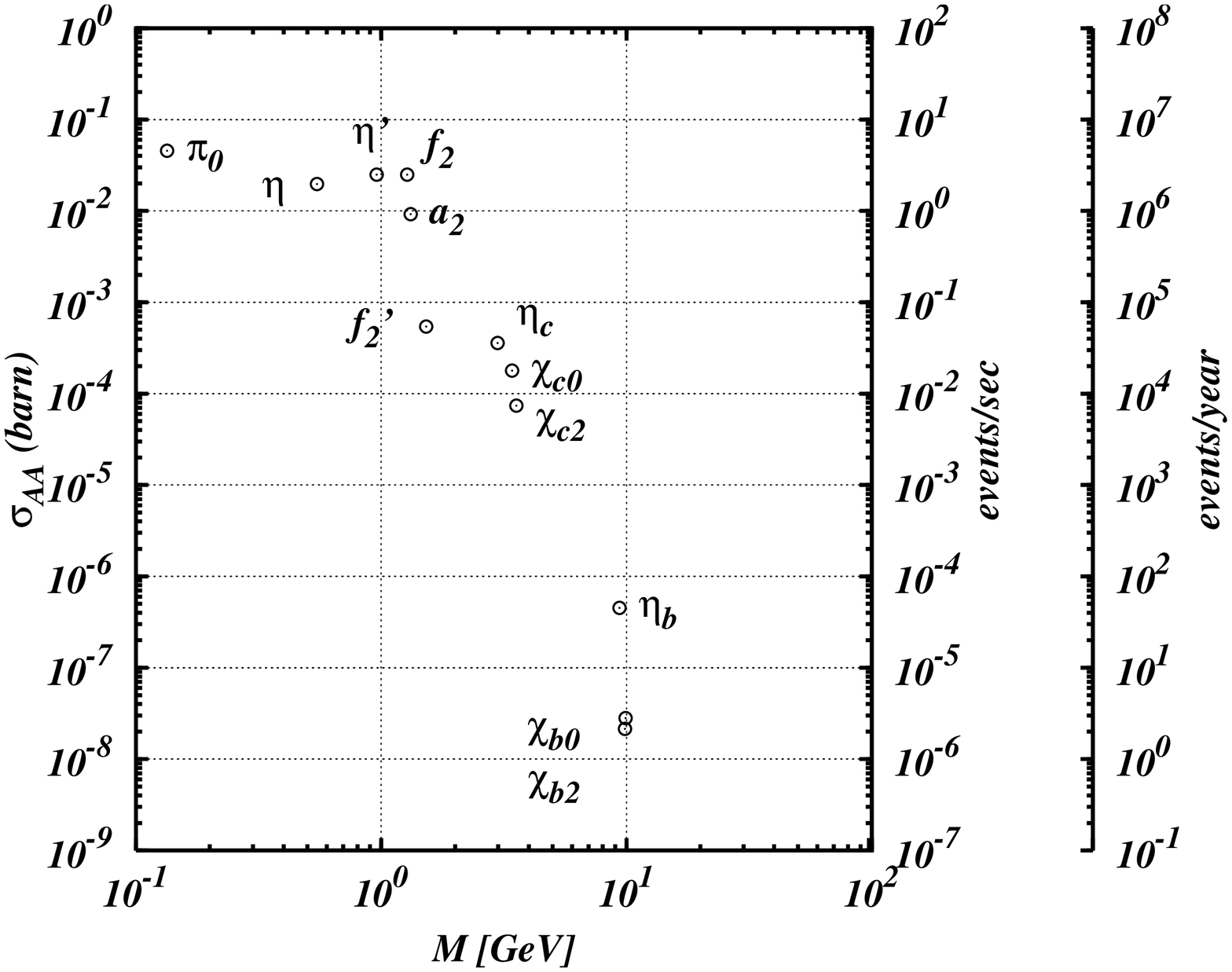}}
(a)\\
\resizebox{10cm}{!}{\includegraphics{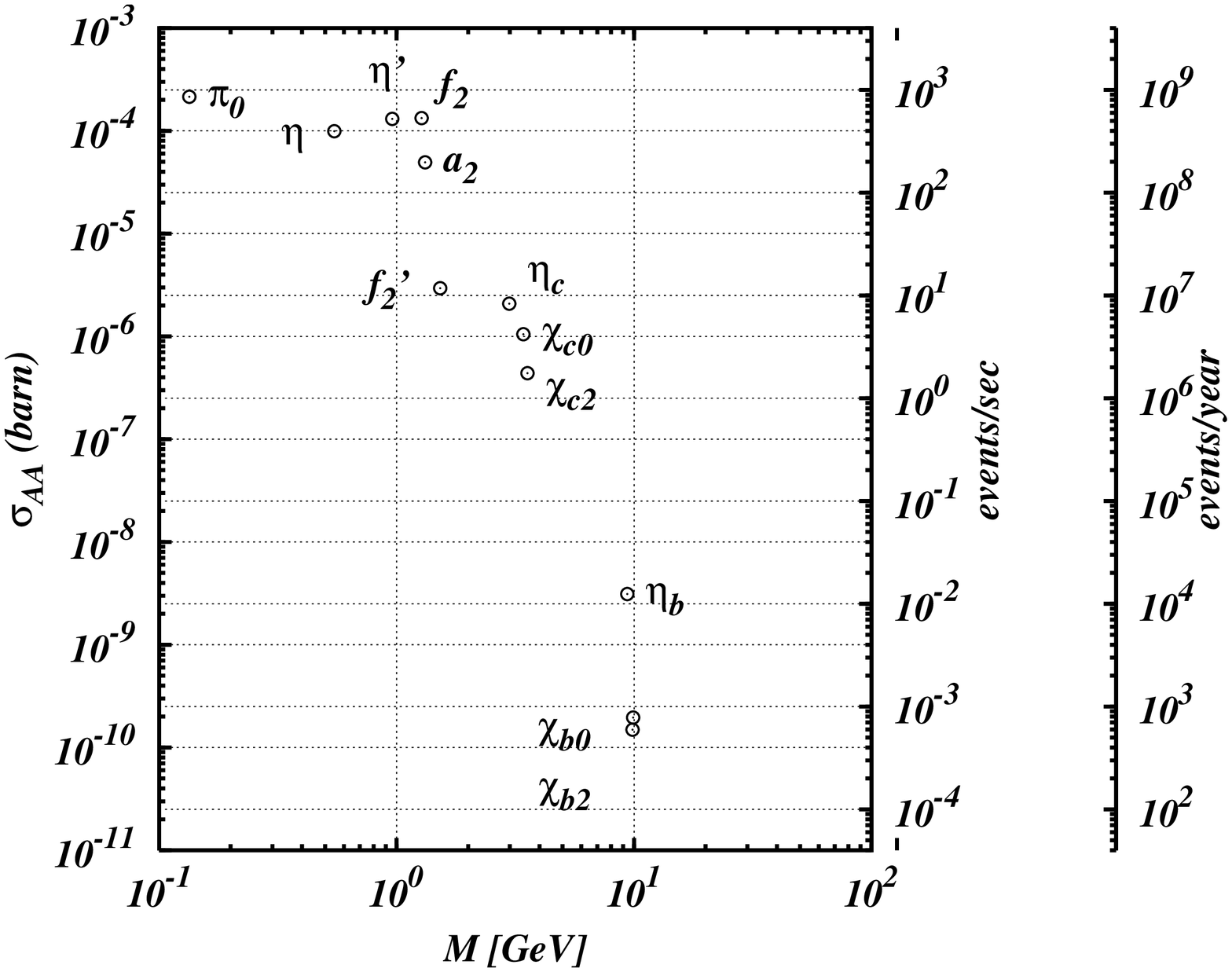}}
(b)
\end{center}
\caption{Cross section for the production of different resonances at the
LHC for PbPb (a) and CaCa (b) collisions, using the parameters of 
Table~\protect\ref{xs-qq}. Also shown are the event rates per s and per
($10^6$s) year.}
\label{fig_overviewC}
\end{figure}

In Fig.~\ref{fig_sigmagamma} the function $4\pi^2 dL_{\GG}/dW /W^2$,
which is universal for a produced resonances, is plotted for various
systems. It can be directly used to calculate the cross-section for
the production of a resonance $R$ with Eq.~(\ref{AA-AAR}).

\subsubsection{Quarkonia}
Among the interesting processes for resonance production is the study of 
the quarkonium states $c\bar c$ and $b\bar b$ such as $\eta_{c(b)}$ ($^1S_0$),
$\chi_{c(b)0}$ ($^3P_0$), $\chi_{c(b)2}$ ($^3P_2$) with the aim of
measuring their two-photon widths, and therefore to test
quarkonium models, as well as, studying their decay modes. Production
of light $q \bar q$ states, like $\pi^0$, $\eta$ and $\eta'$, whose widths 
are well-known, can be used for calibration.

In the nonrelativistic quarkonium model of \cite{Kwong88}, which takes
into account first-order QCD terms, analytical expressions for gluon,
lepton and photon widths of various quarkonium states are obtained.
\begin{table}[tbh]
\centerline{
\begin{tabular}{ccc} \hline\hline
Process & Width & Correction ${ O}(\alpha_{\rm s})$ \\ \hline
$^1S_0 \to \gamma\gamma$ & $12\pi Q^4\alpha_{\rm em}^2 |\Psi(0)|^2 / m^2_q$ &
    $1 - 3.4\alpha_{\rm s}/\pi$ \\
$^1S_0 \to gg$ & $8\pi\alpha^2_{\rm s} |\Psi(0)|^2 / 3m^2_q$ &
    $1 + 4.8\alpha_{\rm s}/\pi \quad (\eta_c)$ \\
 && $1 + 4.4\alpha_{\rm s}/\pi \quad (\eta_b)$ \\[2mm]
$^3S_1 \to e^+e^-$ & $16\pi Q^2\alpha_{\rm em}^2 |\Psi(0)|^2 / M^2$ &
    $1 - 16\alpha_{\rm s}/3\pi$ \\
$^3S_1 \to ggg$ & $40(\pi^2-9)\alpha^3_{\rm s} |\Psi(0)|^2 / 81m^2_q$ &
    $1 - 3.7\alpha_{\rm s}/\pi \quad (J/\psi)$ \\
 && $1 - 4.9\alpha_{\rm s}/\pi \quad (\Upsilon)$ \\[2mm]
$^3P_0 \to \gamma\gamma$ & $27 Q^4 \alpha^2_{\rm em} |R'_{nP}(0)|^2 / m^4_q$ &
    $1 + 0.2\alpha_{\rm s}/\pi$ \\
$^3P_0 \to gg$ & $6 \alpha^2_{\rm s} |R'_{nP}(0)|^2 / m^4_Q$ &
    $1 + 9.5\alpha_{\rm s}/\pi \quad (\chi_{c0})$ \\
 && $1 + 10.0\alpha_{\rm s}/\pi \quad (\chi_{b0})$ \\[2mm]
$^3P_2 \to \gamma\gamma$ & $36 Q^4 \alpha^2_{\rm em} |R'_{nP}(0)|^2 / 5m^4_q$ &
    $1 - 16 \alpha_{\rm s}/3\pi$ \\
$^3P_2 \to gg$ & $8 \alpha^2_{\rm s} |R'_{nP}(0)|^2 / 5m^4_q$ &
    $1 - 2.2 \alpha_{\rm s}/\pi \quad (\chi_{c2})$ \\
 && $1 - 0.1 \alpha_{\rm s}/\pi \quad (\chi_{b2})$ \\ \hline\hline
\end{tabular}}
\caption{Quarkonia decay widths in the nonrelativistic model. Adapted from
\cite{Kwong88}.}
    \label{qqrates}
\end{table}

In Table~\ref{qqrates} these expressions for states with even
total angular momentum are given, as well as, those for the vector state 
$^3S_1$ for normalization. In these expressions $Q$ denotes the electric 
charge 
of a quark, constituting the quarkonium, $M$ is the quarkonium mass, and $m_q$
the quark mass. For the quark masses we use $m_c = 1.5$~GeV, 
$m_b = 4.8$~GeV. The strong coupling constant $\alpha_{\rm s}$ is
derived from the relation between the total and partial widths of the decay
of the known states. Comparison with experimental data gives
\cite{Kwong88} $\alpha_{\rm s}(m_c) = 0.19$, $\alpha_{\rm s}(m_b) =
0.17$. The unknowns are the value of the wave functions at the origin 
$\Psi(0)$ and $R'_{nP}(0)$.

Assuming that the wave function $\Psi(0)$ is the same for all $1S$
states, and equal to those in Table~\ref{qqrates}, the width ratio of
$J/\psi(\Upsilon)$ ($1^3S_1$) to $\eta_c$ ($\eta_b$) ($1^1S_0$) can be
calculated:
\begin{equation}
 \frac{\Gamma(1^3S_1 \to e^+e^-)}{\Gamma(1^1S_0 \to \gamma\gamma)} = 
   \frac{1}{3Q^2} \left( \frac{2m_q}{M}\right)^2
   \left(1 - 1.9\frac{\alpha_{\rm s}}{\pi}\right).
\label{wid-eegg}
\end{equation}
This ratio for $J/\psi$ and $\eta_c$ is equal to $0.74$, in perfect 
agreement with the experimental result $0.71 \pm 0.14$ \cite{PDG00}.
Therefore, one can predict two-photon width of the $\eta_b$ from the lepton 
width of the $\Upsilon(1S)$. The ratio of the total and two-photon width of 
the $\eta_b$ is obtained analogously, assuming that the two-gluon decay 
dominates:
\begin{equation}
\frac{\Gamma(\eta_b \to gg)}{\Gamma(\eta_b \to \gamma\gamma)} = 
   \frac{2\alpha^2_{\rm s}}{9\alpha^2_{\rm em} Q^2}
    \left(1 + 7.8\frac{\alpha_{\rm s}}{\pi}\right).
\label{wid-totgg}
\end{equation}
Using the experimental lepton decay width $\Gamma(\Upsilon(1S) \to
e^+e^-) = 1.32$~keV \cite{PDG00}, one obtains from Eqs.~(\ref{wid-eegg}) 
and~(\ref{wid-totgg}) the total and the two-photon widths of
the $\eta_b$:
\begin{equation}
\Gamma_{\mbox{\scriptsize tot}}(\eta_b) = 6~\mbox{MeV}, \qquad
\Gamma_{\gamma\gamma}(\eta_b) = 0.43~\mbox{keV}.
\label{wid-etab}
\end{equation}

$P$-states of bottomonium are not yet studied, therefore predictions of
their total and partial widths cannot be obtained from a comparison of
experimental data. In \cite{Bodwin94} lattice calculations
give a value for the wave function of the $1P$-state of bottomonium
at the origin
 $|R'_{1P}(0)|^2 = 0.75~\mbox{GeV}^5$. These calculations
allows to define total and two-photon widths of the $\chi_{b0}$ and the
$\chi_{b2}$ from the analytical expression of Table~\ref{qqrates}:
\begin{equation}
\Gamma_{\mbox{\scriptsize tot}}(\chi_{b0}) = 0.24~\mbox{MeV}, \qquad
\Gamma_{\gamma\gamma}(\chi_{b0}) = 25~\mbox{eV},
\label{wid-chib0}
\end{equation}
\begin{equation}
\Gamma_{\mbox{\scriptsize tot}}(\chi_{b2}) = 65~\mbox{keV}, \qquad
\Gamma_{\gamma\gamma}(\chi_{b2}) = 6.7~\mbox{eV}.
\label{wid-chib2}
\end{equation}

\begin{sidewaystable}[tbh]
\centerline{
\begin{tabular}{c|c|c|c|c||c|c|c||c|c|c} \hline\hline
$J^{PC}$ & particle & $M$, & $\Gamma_{\gamma\gamma}$, &
  $\Gamma_{\mbox{\scriptsize tot}}$, & 
     \multicolumn{3}{c||}{$\sigma(AA \to AA + R)$} &
     \multicolumn{3}{|c}{Production rate per $10^6$-sec run} \\ \cline{6-11}
 & & MeV & keV & MeV & 
   AuAu,   & PbPb,    & CaCa,    & AuAu,   & PbPb,    & CaCa,    \\ 
 & &           &     &     &
   $\gamma=100$ & $\gamma=3000$ & $\gamma=3700$ & $\gamma=100$ & $\gamma=3000$ & $\gamma=3700$ \\ \hline
$0^{-+}$ & $\pi^0$      & 135 & $8 \cdot 10^{-3}$ & $8 \cdot 10^{-6}$   & 
    5.0 mb & 46 mb & 210 $\mu$b    & $1.0 \cdot 10^6$ & $4.6 \cdot 10^6$ & $8.4 \cdot 10^8$ \\
$0^{-+}$ & $\eta$       & 547  & $0.46$            & $1.2 \cdot 10^{-3}$ &
    0.85 mb & 20 mb & 100 $\mu$b & $1.7 \cdot 10^5$ & $2 \cdot 10^6  $ & $4.0 \cdot 10^8$ \\
$0^{-+}$ & $\eta '$     & 958  & $4.2$             & $0.2$               & 
    0.59 mb & 25 mb & 130 $\mu$b & $1.2 \cdot 10^5$ & $2.5 \cdot 10^6$ & $5.2 \cdot 10^8$ \\
$2^{++}$ & $f_2(1270)$  & 1275 & 2.4               & 185                 &
    0.41 mb & 25 mb & 133 $\mu$b & $8.2 \cdot 10^4$ & $2.5 \cdot 10^6$ & $5.2 \cdot 10^8$ \\
$2^{++}$ & $a_2(1320)$  & 1318 & 1.0               & 107                 &
    0.14 mb & 7.7 mb& 49 $\mu$b  & $2.8 \cdot 10^4$ & $7.7 \cdot 10^5$ & $2.0 \cdot 10^8$ \\
$2^{++}$ & $f_2'(1525)$ & 1525 & 0.1               & 112                 &
    6.6 $\mu$b & 0.45 mb& 2.9 $\mu$b & $1.3 \cdot 10^3$ & $4.5 \cdot 10^4$ & $1.2 \cdot 10^7$ \\
$0^{-+}$ & $\eta_c$     & 2979 & $7.4$             & $13.2$              & 
    1.8 $\mu$b & 0.54 mb & 3.7 $\mu$b & 360 & $5.4 \cdot 10^4$ & $1.6 \cdot 10^7$ \\
$0^{++}$ & $\chi_{c0}$  & 3415 & $4.0$             & $14.9$              & 
    0.38 $\mu$b & 0.17 mb &1.2 $\mu$b & 76 & $1.7 \cdot 10^4$ & $4.8 \cdot 10^6$ \\
$2^{++}$ & $\chi_{c2}$  & 3556 & $0.46$            & $2.0$               & 
    0.17 $\mu$b & 85 $\mu$b & 0.59 $\mu$b & 34 & $8.5 \cdot 10^4$ & $2.4 \cdot 10^6$ \\
$0^{-+}$ & $\eta_b$     & 9366 & $0.43$            & $6$                 & 
                & 0.32 $\mu$b & 2.8 nb    &     & 32               & $1.1 \cdot 10^3$ \\
$0^{++}$ & $\chi_{b0}$  & 9860 & $2.5 \cdot 10^{-2}$ & $0.24$            &
                & 15 nb       & 0.15 nb   &     & 1.5              & 600 \\
$2^{++}$ & $\chi_{b2}$  & 9913 & $6.7 \cdot 10^{-3}$ & $6.5 \cdot 10^{-2}$&
                & 20 nb & 0.18 nb         &     & 2.0              & 720 \\ \hline\hline
\end{tabular}
}
\caption{Prediction of production cross section and rates for a $10^6$~sec-run
        of some mesons at RHIC and LHC.
        Masses and widths for mesons except bottomonia are taken from
        \cite{PDG00}, bottomonium widths are given from equations
        (\ref{wid-etab}), (\ref{wid-chib0}) and (\ref{wid-chib2}). The
        beam luminosities used are $2\cdot 10^{26}$cm${}^{-2}$s${}^{-1}$ for
        AuAu at RHIC, $10^{26}$cm${}^{-2}$s${}^{-1}$ for PbPb
        and $4\times10^{30}$cm${}^{-2}$s${}^{-1}$ for CaCa at LHC}
\label{xs-qq}
\end{sidewaystable}

For charmonium production, the two-photon width $\Gamma_{\GG}$ of the
$\eta_c$ (2960 MeV, $J^{PC} = 0^{-+}$) is known from experiment
\cite{PDG00}. But the two-photon widths of the $P$-wave charmonium states
have been measured only with modest accuracy. Two photon widths of
$P$-wave charmonium states can be estimated following the PQCD
approach of \cite{Bodwin94}. Similar predictions of the bottomonium two-photon 
widths can be found in \cite{Kwong88}. For RHIC the
study of $\eta_c$ is a real challenge \cite{KleinS97b}; the
luminosities are falling strongly with increasing $\GG$ mass
and the branching ratios to experimentally interesting channels are small. 

In Table \ref{xs-qq} (adapted from Table~2.6 of \cite{Felix97})
properties of some $q\bar q$ states are given, and their production cross
sections are predicted, where possible. 
Similarly an overview of different rates is also given in 
Fig.~\ref{fig_overviewR} for both PbPb and CaCa collisions at the LHC.
The results for AuAu collisions at
RHIC and PbPb collisions at LHC are done on the basis of \cite{BaurF90} and 
from calculations of the program {\sc Tphic} \cite{HenckenKKS96}. Mass values 
and known widths are taken from \cite{PDG00}, bottomonium widths are used from 
Eqs.~(\ref{wid-etab}) to~(\ref{wid-chib2}).  Also given is the number of 
events in a $10^6$s run.  Ion luminosities of 
$2\times 10^{26}$cm${}^{-2}$s${}^{-1}$ for AuAu collisions at RHIC
and $4\times 10^{30}$cm${}^{-2}$s${}^{-1}$ for CaCa and
$10^{26}$cm${}^{-2}$s${}^{-1}$ for PbPb collisions at LHC are used. Millions
of $C$-even charmonium states will be produced in coherent two-photon
processes during a standard $10^6$s heavy ion run at the LHC. The
detection efficiency of the charmonium events has been estimated to be about
5\% for the forward-backward FELIX geometry \cite{Felix97}, i.e., one can
expect the detection of about $5\times 10^3$ charmonium events in PbPb
and about $10^6$ events in CaCa collisions. This is two to three
orders of magnitude higher than what is expected during the five years of
LEP200 operation. Experiments with a well-equipped central detector
like CMS on the other hand should have a much better efficiency.
Further details --- also on experimental cuts, backgrounds and the
possibilities for the study of $C$-even bottomonium states --- are given in
\cite{Felix97}.

\subsection{Exotic Mesons}

The two-photon width of a
resonance is a probe of the charge of its constituents, so the
magnitude of the two-photon coupling can serve to distinguish quark
dominated resonances from glue-dominated resonances (``glueballs'').
The absence of meson production via $\GG$ fusion would therefore 
be one signal of great interest for glueball search. 
In $\GG$ collisions a glueball can only be produced via the
annihilation of a $q\bar q$ pair into a pair of gluons, whereas a
normal $q\bar q$-meson can be produced directly. Therefore we expect
the ratio for the production of a glueball $G$ compared to a normal
$q\bar q$ meson $M$ to be
\BE
\frac{\sigma(\GG \rightarrow G)}{\sigma(\GG \rightarrow M)}
=
\frac{\Gamma(G \rightarrow \GG)}{\Gamma(M \rightarrow \GG)}
\sim
\alpha_s^2,
\EE
where $\alpha_s$ is the strong interaction coupling constant. On the
other hand glueballs are most easily produced in a glue-rich
environment, for example, in radiative $J/\Psi$ decays, $J/\Psi
\rightarrow \gamma gg$. In this process we expect the ratio of the
cross section to be
\BE
\frac{\Gamma(J/\Psi \rightarrow \G G)}{\Gamma(J/\Psi \rightarrow \G
M)} \sim \frac{1}{\alpha_s^2} .
\EE
A useful quantity to describe the gluonic character of a mesonic state
X is therefore the so called ``stickiness'' \cite{Cartwright98}, defined as
\BE
S_X = \frac{\Gamma(J/\Psi \rightarrow \G X)}{\Gamma(X \rightarrow
\G \G)} .
\EE
One expects the stickiness of all mesons to be comparable, while for
glueballs it should be enhanced by a factor of about $ 1/\alpha_s^4 \sim 20$.
In the recent work of \cite{Godang97} results for the search for $f_J
(2220)$ production in two-photon interactions were presented. A
very small upper limit for the product of $\Gamma_{\GG} B_{K_sK_s}$
was given, where $B_{K_s K_s}$ denotes the branching ratio of
its decay into $K_s K_s$.  From this it was concluded that this is a
strong evidence that the $f_J(2220)$ is a glueball.

Two-photon processes can also be used as a tool to observe mesons
beyond the quark model. The $a_0(980)$ and $f_0(980)$ could be
four-quark states $q\bar{q}q\bar{q}$ (or ``quarktets'')
\cite{Achasov1991}.  The ARGUS collaboration observed a $I^G(J^{PC}) =
2^+(2^{++})$ peak in the reaction $\GG\to\rho\rho$ near threshold
\cite{ARGUS1991}. This state was called by PDG $X(1600)$ and also
interpreted as a four-quark state.

For a rather general  discussion of glueballs but also other (even more) 
exotic mesons like ``quarktets'',
``hybrids'' (or ``centauros'', made of a quark, an antiquark and a gluon) 
we refer to \cite{Wilczek00}. The QCD studies in $\GG$ collisions at the 
CLEO detector were summarized recently by Savinov in \cite{Savinov01}.
Also the anti-search for glueballs is described.

%
%
\section{Two-Photon Collisions as a Tool for the Search of New Physics}
\label{sec_newphysics}

The large photon-photon luminosity and the high invariant mass range make
two-photon collisions at heavy ion colliders also of interest for the
search for new particles and new physics. This includes the possible
production of the Higgs-boson in the $\GG$-production channel 
(unfortunately, this possibility is only marginal, see below) or new
physics beyond the standard model, like supersymmetry or
compositeness. Many studies for different extensions
of the standard model have been performed. In the following the
conclusions of some of these studies will be summarized, further discussions
can also be found in \cite{KraussGS97}.

Before doing this we mention the plans to build a future $\EPEM$ linear 
collider. Such a linear colliders will be used for $\EPEM$, $e\G$
and $\GG$-collisions (PLC, ``photon linear collider''). 
The photons will be obtained by the scattering of
laser photons (of eV energy) on high energy electrons (up to the TeV
region), see \cite{Telnov95}. These photons in the TeV energy range
will be roughly monochromatic and polarized. The physics program at such
future machines was discussed , e.g., in \cite{ginzburg95,Ginzburg00};
it includes Higgs
boson and gauge boson physics and the discovery of new particles.
A recent review can be found in Parts 3 and 6 of \cite{TESLATDR01}.
The physics topics reach from Higgs boson physics to supersymmetry 
and extra dimensions. Such a collider will provide  a rather clean environment 
for physics 
in a new  energy region. This is of interest for the
(far) future, whereas the LHC and its detectors are built at
present. 

In \cite{DreesGN94,OhnemusWZ94} $\GG$-processes at $pp$
colliders (LHC) are studied. It is observed there that non-strongly
interacting supersymmetric particles (sleptons, charginos,
neutralinos, and charged Higgs bosons) are difficult to detect in
hadronic collisions at the LHC. The Drell-Yan and $gg$-fusion mechanisms
yield low production rates for such particles. Therefore the
possibility of producing such particles in $\GG$ interactions at
hadron colliders is examined. Since photons can be emitted from
protons which do not break up in the radiation process, clean events
can be generated which should compensate for the small production number. 
In
\cite{DreesGN94} it was pointed out that at the high luminosity of
$L=10^{34}$cm${}^{-2}$s${}^{-1}$ at the LHC($pp$), one expects about
16 minimum bias events per bunch crossing. Even the elastic $\GG$
events will therefore not be free of hadronic debris. Clean elastic
events will be detectable at luminosities below
$10^{33}$cm${}^{-2}$s${}^{-1}$. This danger of ``overlapping events''
has also to be checked for the heavy ion runs, but it will be much
reduced due to the lower luminosities.

Similar considerations for new physics were also made in connection
with the proposed $eA$ collider at DESY (Hamburg). Again, the coherent
field of a nucleus gives rise to a $Z^2$ factor in the cross-section
for photon-photon processes in $eA$ collisions \cite{Krawczyk96,herafuture96}.

\subsection{Supersymmetry Particle Pair Production}
Two-photon collisions allow to study the production of particles beyond
the Standard Model. Here we consider the processes of chargino,
slepton and charged Higgs pair production in the frame of the minimal
supersymmetric standard model (MSSM) \cite{HaberK85}.

Charginos are coupled to photons as standard fermions of spin $1/2$,
and therefore the cross section of the process $\gamma\gamma \to
\tilde\chi_1^+ \tilde\chi_1^-$ is described by Eqs.~(\ref{dxsgg-ff}) 
and~(\ref{xsgg-ff}) with $Q_f = 1$ and $N_c = 1$. Following \cite{FengS95} 
we assume  $R$-parity
conservation, i.e. continuity of supersymmetric lines. We take the
chargino $\tilde\chi_1^\pm$ as the lightest observable supersymmetric
particle and  the neutralino $\tilde\chi_1^0$ as the lightest
supersymmetric particle. Thus, it is  not observable. Sleptons, squarks
and gluinos are assumed to be much heavier than the
$\tilde\chi_1^\pm$. All parameters of the supersymmetry breaking can
be taken as real, i.e., $CP$-parity violation does not take place in
the chargino processes. All these assumptions result in the fact that the
$\tilde\chi_1^\pm$ decays to neutralino $\tilde\chi_1^0$ and a weak
duplet of fermions via a virtual $W^\pm$, or via a sfermion $\tilde f^\pm$ or
Higgs $H^\pm$: $\tilde\chi^+_1 \to (\tilde\chi^0_1 W^{+*}, \tilde f^*_i
\bar{f_j}, \tilde\chi^0_1 H^{+*}) \to \tilde\chi^0_1 f_i \bar{f_j}$.

Experimental search for supersymmetric particles impose limits on
their masses \cite{PDG00}. The lightest neutralino
$\tilde\chi^0_1$ was not found at the mass $m < 32.5$~GeV. The
chargino $\tilde\chi^\pm_1$ mass is bound by $m > 67.7$~GeV. This
restricts the possibility to search for the processes of the chargino
pair production in two-photon interactions at heavy ion colliders,
since the invariant mass of the two-photon system must be above
$130$~GeV, while the spectrum of the $\gamma\gamma$-system at LHC
is limited essentially to $W_{\gamma\gamma} \approx 200$~GeV (see 
Eq.~(\ref{eq_wmax})).

\begin{figure}[tbh]
\parbox{0.48\hsize}{
  \caption{Production cross section of $\tilde\chi^+_1 \tilde\chi^-_1$
    in $Pb$-$Pb$ and $Ca$-$Ca$ collisions at $\protect\sqrt{s_{AA}} =
    5.5$ and $7.2$~TeV per nucleon respectively.} \label{chipair}
}
\hfill
\parbox{0.48\hsize}{
  \centerline{\resizebox{7cm}{!}{\includegraphics{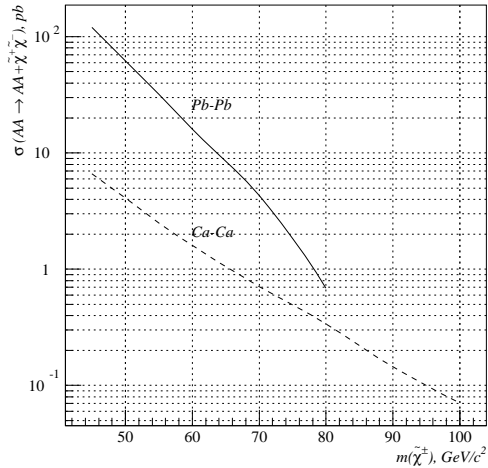}}}
}
\end{figure}

Cross sections of the chargino pair production vs. their mass in
two-photon interactions in $Pb$-$Pb$ and $Ca$-$Ca$ collisions at LHC  
are shown in Fig.~\ref{chipair}. The results are obtained with the 
program {\sc Tphic} \cite{HenckenKKS96}.
\begin{figure}[tbh]
\centerline{\resizebox{0.5\hsize}{!}{\includegraphics{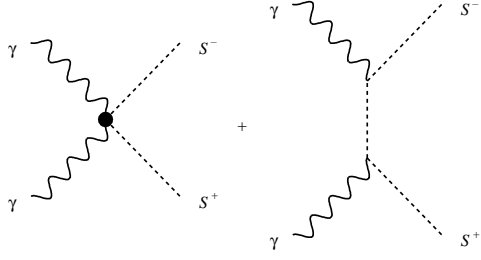}}}
\caption{Process $\gamma\gamma \to S^+S^-$ in the leading order, $S^\pm$
is either a slepton $\tilde l^\pm$, or charged Higgs $H^\pm$.}
\label{diag:ggss}
\end{figure}

\begin{figure}[tbh]
\parbox{0.48\hsize}{
  \centerline{\resizebox{7cm}{!}{\includegraphics{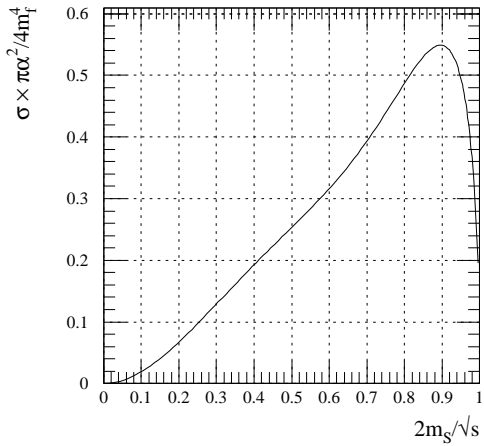}}}
}
\hfill
\parbox{0.48\hsize}{
  \caption{The  $\gamma\gamma \to S^+ S^-$ cross
    section as a function of  $2m_S/\sqrt{s}$.}
  \label{ggss_beta}
}
\end{figure}

When the mass of sleptons $\tilde l^\pm$ and charged Higgs $H^\pm$ are
not too high, their pairs can also be produced in the two-photon
interactions. Sleptons and Higgs are scalar particles and the differential 
cross sections of the processes $\gamma\gamma \to \tilde l^+ \tilde l^-$ 
and $\gamma\gamma \to H^+ H^-$: 
\begin{equation}
\frac{d\sigma(\gamma\gamma \rightarrow S^+S^-)}{d\cos(\theta)} 
= \frac{\pi\alpha^2}{s} \beta 
\left[ 1 - \frac{2 \left(1-\beta^2\right)}{1 - \beta^2 \cos^2(\theta)}
 - \frac{2 \left(1-\beta^2\right)^2}{\left(1 - \beta^2 \cos^2(\theta)\right)^2}
\right],
\label{gg-ss_diff}
\end{equation}
as well as, the total cross section
\begin{equation}
\sigma(\gamma\gamma \rightarrow S^+S^-)
       = \frac{4\pi\alpha^2}{s}\beta\left[
        2 - \beta^2 - \frac{1-\beta^4}{2\beta}\ln\frac{1+\beta}{1-\beta}
        \right]. 
\label{gg-ss}
\end{equation}
are well known, see \cite{AchieserB62,OhnemusWZ94}.
Fig.~\ref{ggss_beta} shows how the integral cross section depends
on the ratio of a $S^\pm$ mass to the beam energy in c.m. system,
$2m_S/\sqrt{s}$.
Comparing this plot with the cross section behavior of fermions
(Fig.~\ref{ggff_beta} for $N_c=1$, $Q_f=1$) shows that the yield of
supersymmetric scalars is lower than that of charginos.

The search for sleptons and charged Higgs imposes a lower limit on their
mass $m(H^\pm) > 69.0\mbox{~GeV}$, $m(\tilde l^\pm) > 67.1
\mbox{~GeV}$ \cite{PDG00}.

Recent (unpublished) studies done for FELIX and ALICE show that the
chargino pair production can be detectable, if the lightest chargino
would have a mass below 60~GeV. Unfortunately the chargino
mass limits set recently by LEP experiments already exclude the existence of 
charginos below 67.7~GeV with 95\% confidence level \cite{PDG00}.
Therefore the observation of MSSM-particles in $\GG$-interactions in heavy ion 
collisions seems to be hard to achieve.
CMS on the other hand should be more suited for these observations.

\subsection{Higgs Search}

One of the most important tasks in present high energy 
physics is the discovery and the study of the properties of the Higgs
boson. In this respect, the $\GG$ production of the Higgs boson is
of special interest.

There are a  number of calculations of the $\GG$ production
of  a medium heavy standard
model Higgs \cite{DreesEZ89,MuellerS90,BaurF90,Papageorgiu95,%
Norbury90}. For
masses $m_H < 2 m_{W^\pm}$ the Higgs boson decays dominantly into
$b\bar b$. 
In \cite{Papageorgiu95} a comparison of $gg \rightarrow H$ and
$\GG \rightarrow H$ emphasized the favorable signal to background ratio
for the latter process. Unfortunately, at the 
LHC a heavy ion year consists only of the order of  $10^6$ s
instead of the assumed $10^7$ s. Chances of finding the standard model
Higgs in this case are marginal \cite{BaurHTS98}.

The search for anomalous Higgs couplings in 
peripheral heavy ion collisions at the LHC was studied by Lietti 
et al. \cite{LiettiNRR01}. They consider corrections to the standard
model from new physics associated with a high energy scale $\Lambda$.
It is concluded that ``{\it limits for anomalous Higgs couplings
which can be obtained in peripheral heavy ion collisions at the LHC via 
electromagnetic processes are one order of magnitude tighter
than the limits that can be obtained in the upgraded Tevatron
and comparable to limits coming from the $pp$ mode of the LHC.\/}
For further details we refer to this paper.

An alternative scenario with a light Higgs boson was, e.g., given in
\cite{ChoudhuryK97} in the framework of the ``general two Higgs
doublet model''. Such a model allows for a very light particle in the
few GeV region. With a mass of 10~GeV, the $\GG$-width is about 0.1
keV. The authors of
\cite{ChoudhuryK97} proposed to look for such a light neutral Higgs
boson at the proposed low energy $\GG$-collider. 
We want to point out
that the LHC CaCa heavy ion mode would also be very suitable for such
a search.
A systematic parameter study of the production of Higgs bosons in the minimal 
supersymmetric extension of the standard model (MSSM) was done in 
\cite{GreinerVS93}.
For certain values of the free parameters in enhancement of the cross section
of up to a factor 10 was found.
 
In \cite{Piotrzkowski00}
it was shown how to tag two-photon production in $pp$ collisions
at the LHC, see Fig.~4 there. It was concluded that ``{\it the significant 
luminosity of the tagged two-photon collisions opens an exciting 
possibility of studying the exclusive production of the Higgs boson
as well as search for new phenomena.\/}''

Last year all 4 LEP experiments have reported on their Higgs search in
the data collected during the LEP-II run. All of them set their new
lower limits on the Higgs mass which are $M_H > 107.0$~GeV (L3),
109.7~GeV (OPAL), 110.6~GeV (ALEPH), 114.3~GeV (DELPHI) at
95\% CL \cite{Abrue2001,Achard2001,Abbiendi2001,LEPWGH2001}.
However, the likelihood analysis shows a preferences for a Higgs
boson with a mass of 115.6~GeV.

In the Standard Model the arguments of self-consistency of the theory
can be used to obtain upper and lower limits on the Higgs mass. The
upper limit is obtained from arguments of perturbative continuation to
the GUT scale up to $\Lambda_{GUT}=10^{16}$~GeV \cite{Cabibbo1979},
i.e., from the requirement that the electroweak interaction would be still
"weak" up to this energy scale. The high-energy behavior of the weak 
interaction was discussed also in a more general framework 
\cite{Pomeran1970,Logunov1978}. A lower limit for the Higgs mass is obtained
from quantum corrections involving top-loops to the Higgs interaction potential
\cite{Linder1989,Sher1993,Sher1994,Altarelli1994,Casas1995}.
Thus it appears that the requirement of self-consistency of the SM up to
$10^{16}$~GeV leads to the theoretical limits for the Higgs mass of 
$130 \mbox{~GeV} < M_{H^o} < 190$~GeV (to be compared with the 
lower mass limits of $M_H > 107.0-114.3$~GeV at LEP, see above). 
This range is just the mass range acceptable for the Higgs
production in two-photon processes of CaCa collisions at the LHC.

\subsection{Search for Extra Dimensions}

Graviton production in very peripheral 
heavy-ion collisions was recently studied theoretically in \cite{AhernNP00}.
Such a possibility has recently become of interest due to a new class of 
theoretical models, for further explanations see, e.g., Chap.~6 of 
\cite{Peskin00}. In these models quantum gravity and string 
physics may become accessible to experiment; they may
even appear directly in the realm of LHC and the linear colliders.
Graviton production in $\GG$ collisions can be  calculated in such models. 
It depends on a fundamental gravity scale M  which is related 
to the scale  $R$ at which the Newtonian inverse-square law is expected to fail
(see, e.g., Eq.~(25) of \cite{Peskin00}).
Cross sections for graviton production in heavy ion 
collisions were found to be substantially greater
than for graviton production in $\EPEM$ collisions \cite{AhernNP00}.
A value of $M=1TeV$ was assumed in those calculations. The signature
for such a process would be a large missing mass in the reaction.
In the heavy ion case this is hard to realize experimentally
and the authors conclude that ``{\it a definite experimental signature 
cannot be predicted\/}''. For further details we refer to this paper.

\subsection{Search for Heavy Point-like Dirac Monopoles}
In \cite{GinzburgS98}
it was proposed to search for heavy magnetic monopoles in $\GG$
collisions 
at hadron colliders like the Tevatron and LHC. The idea is that 
photon-photon 
scattering 
$\gamma\gamma\rightarrow\gamma\gamma$
below the monopole production threshold 
is enhanced due to the strong coupling of magnetic monopoles to 
photons. The magnetic coupling strength $g$ is given by
$ g= \frac{2\pi n}{e} $ where $n=\pm1, \pm2,...$. Since 
$\frac{e^2}{4 \pi}=1/137$ the magnetic coupling strength is
indeed quite large. In this reference differential cross sections
for $\GG$ scattering via the monopole loop are calculated for
energies below the monopole production threshold. The result depends strongly
on the assumed value of the spin of the monopole. With this elementary
cross section as an input, the cross section for the process 
$ pp\rightarrow \GG +$anything is calculated. Elastic, i.e., anything$=pp$
and inelastic contributions are taken into account. The signature of 
such a process is the production of two photons where the transverse
momentum of the pair is much smaller than the transverse momentum of
the individual photons. 
At the Tevatron such a search was performed. They looked at a pair of photons
with high transverse energies. No excess of events above background was found 
\cite{Abbott98}.
Thus a lower limit on the mass of the magnetic monopole could 
be given. A mass of 610, 870, or 1580~GeV was obtained, for
the assumed values of a monopole spin of $0,~1/2,$ or $1$ respectively.

\subsection{Tightly Bound States}

One can also speculate about new particles with strong coupling to the
$\GG$-channel. As seen in Sec.~\ref{sec_resproduction} above, large 
$\Gamma_{\GG}$-widths will directly lead to large $\GG$ production 
cross sections. The two-photon width of quarkonia, for example, is 
proportional to the wave function 
squared in the center of the system, see Table~\ref{qqrates}. Thus we 
can expect, that if a system is very tightly bound it should have a
sufficiently large two-photon width due to the factor $|\Psi(0)|^2$
which is large in these cases.
Examples for such tightly bound systems are discussed,
e.g., in \cite{Renard83,BaurFF84}.
Composite scalar bosons at $W_{\GG}\approx 50$~GeV are expected to have 
$\GG$-widths of several MeV \cite{Renard83,BaurFF84}. 
The search for such kind of resonances in the $\GG$-production channel 
will be possible at LHC.
In Part 3 (p.~110ff) of the TESLA Design Report \cite{TESLATDR01}
the reader can find some interesting 
remarks about the "agnostic" approach to compositeness. 
From Eq.~(\ref{AA-AAR} and Fig.~\ref{fig_sigmagamma}
one can easily obtain a value for the production cross-sections
of such states and the corresponding rates in the various collider
modes. Of course, such ideas are quite speculative. However due
to the high flux of equivalent photons such searches seem worth-
while, and a possible discovery would be quite spectacular.

Certainly, many aspects in the present section are quite speculative,
and one will have to wait for the future $\EPEM$ and $\GG$ colliders
to do the physics in the region of several 100 GeV. However,
it may be possible to have a glimpse into this region with the 
very peripheral collisions at LHC which will be taking data in a few
years from now.

%
%

\section{Photon-Nucleus and Photon-Hadron Interactions}
\label{sec_photonhadron}

Let us start this section with a few rather qualitative remarks which 
may serve as a guideline. Two-photon processes in relativistic ion collisions 
have a relatively simple nature:  both ions interact by means of the 
quasireal photons, which they both emit. Theses photons can be thought of as 
being away, well separated from the nuclei ($b>R_1+R_2$, $b_i>R_i$). 
The nuclei remain  
almost intact after such a collision. The coherence of the ion interaction is 
taken into account in terms of ion form factors, see Chap.~\ref{sec_lum}. And 
thus we have virtually pure  $\GG$-interactions and the rich $\GG$-physics as 
discussed above.

Photon-nucleus processes in relativistic ion collisions are in general more 
complicated. The photon radiated by one (``spectator'') ion interacts with 
the other (``target'') ion in a wide range of photon energies. Besides 
interacting electromagnetically it can also have a hadronic component,
due to its fluctuation into a vector meson (``vector meson dominance'', etc.).
This leads to quite different interaction mechanisms in photon-ion reactions.
If the photon is of low energy we observe the excitation of the giant dipole
or other multipole resonances. As these are relatively low energy
excitations, they can occur at rather large distances. When the photon 
(being a non-strongly interacting particle) interacts inside the target ion 
with a single nucleon we observe the excitation of nucleon resonances and 
related  phenomena. 

When the photon fluctuates either to a $q\bar q$-pair or a vector meson before
interacting with the nucleons of the target, diffractive processes are 
possible. These can also be called ``photon-Pomeron'' processes. This
signature is similar to the photon-photon or Pomeron-Pomeron processes.
An important case is the diffractive vector meson production. Due to the
production with a nuclear target, both the coherent and incoherent production
can take place. As the vector meson produced inside the nucleus has to go
through the nuclear medium, new phenomena will appear. They are  widely  
discussed in the literature in terms of  color transparency of the ions,
of formation- and coherence-lengths of hadrons etc., see, e.g., the  
review \cite{Jain1996}.

Below we give an overview of all these quite interesting phenomena, we present 
basic formulae and discuss also practical aspects, which are --- as it appears 
now --- important for experiments at ion colliders.  
      
\subsection{Photonuclear Processes: Giant Dipole Excitation, 
Beam Loss and Luminosity Monitor}
\label{photonuclear}

At relativistic heavy ion accelerators in the region up
to several $A$GeV the study of the electromagnetic  excitation 
of the giant dipole resonance (GDR) and other multipole resonances 
has been a subject of considerable physics interest.
Even the excitation of the double phonon giant dipole resonance
(DGDR) was observed and
studied. This is made  possible due to the strong electromagnetic fields 
in these collisions, for recent reviews see 
\cite{Emling95,AumannBE98,BertulaniP99}.
In the fixed target experiments at SPS/CERN with energies in the 100~$A$GeV
region, fragmentation processes have been investigated like the fission of
Pb in PbPb interactions observed by the NA50 collaboration \cite{Abreu98}.
Such experiments also help to extrapolate to the 
collider energies (A similar situation is also present in the case of 
bound-free electron-positron production, see Sec.~\ref{ssec:boundfree} 
below).
While in general a good agreement with theoretical expectations is found 
\cite{Pshenichnov01}, a puzzling discrepancy still exists. 
In \cite{Datz97} a measurement of the total dissociation cross section of 
Pb for different targets was done. Beside the dominant electromagnetic
interaction (proportional to $Z_T^2$) and the nuclear interaction,
a component proportional to $Z_T$ was found.
This contribution is in addition to the contribution 
due to the target electrons, which is sometimes called "incoherent". 
The size of this extra 
contribution in the case of a Pb target is 9.74~barn, larger than the 
hadronic/nuclear cross section of 7.86~barn. This large value precludes 
explanations in terms of purely nuclear effects. A possible interpretation 
in terms of incoherent electromagnetic processes on protons in the target 
nucleus was 
investigated in \cite{BaurHT98}. But the size of this effect was found to be 
too small to explain the results. 

The dominant contribution to the electromagnetic excitation of the ions comes
from the excitation of the giant dipole resonance (GDR). This is a 
strongly collective nuclear state which exhausts the Thomas-Reiche-Kuhn
sum rule to a very large extent. 
The impact parameter dependent probability of the one-phonon GDR excitation 
is given to a good approximation by
\begin{equation}
P_{GDR}(b)= S/b^2 \exp\left(-S/b^2\right)
\label{eq:PGDR}
\end{equation}
for $b>R_{min}=R_1+R_2$. The quantity S denotes an area which is given by
\cite{BertulaniB88}
\begin{equation}
S=5.45 \times 10^{-5} Z_1^2 N_2 Z_2 A_2^{-2/3} \mbox{fm}^2.
\end{equation}
The total cross section is obtained by integrating over the impact parameter 
from a minimum value of $R_{min}$ up to the adiabatic cutoff radius
$R_{max} \sim \gamma_{ion} c / \omega$. 
The energy of the GDR is given phenomenologically by
$\hbar \omega =80~MeV / A^{1/3}$.
This and the assumption that the classical Thomas-Reiche-Kuhn sum rule 
is exhausted to 100\% by the GDR enter into the expression for $S$ given above.
To lowest order in $S/R_{min}^2$ the cross-section is given by
\begin{equation}
\sigma_{GDR} = 2 \pi S \ln \left(\frac{\gamma_{ion} c}{\omega R_{min}} \right).
\end{equation}

The interest in these (low energy) excitations at the heavy ion colliders
is mainly motivated by  more  practical considerations:
the huge cross section for electromagnetic interaction (of the order of 100's
of barn) leads to beam loss and --- as was only noticed very recently 
\cite{Klein01} --- to a local beam pipe heating, with the possible danger of
quenching of the superconducting magnets.
This is due to the effect that most of the nuclear excitations (to the GDR)
are followed by the emission of a few neutrons. The remaining 
nucleus (with a change in the  $Z/A$ ratio) will 
hit the beam pipe and deposit its energy in a 
rather small area. An even more serious problem is the formation of 
one-electron atoms in bound-free electron-positron pair creation
(see below, Sec.~\ref{ssec:boundfree}).

On the other hand it was noticed also that the neutrons coming from the
GDR decay can conveniently be detected in the zero degree calorimeter (ZDC). 
These low energy neutrons in the rest frame of the nucleus are  
high energy neutrons in the collider frame (with energies in the 
TeV range in the case of the LHC). This leads to the possibility of using 
them as a luminosity monitor.
Especially the use of mutual GDR excitation has been studied in detail
as a possible luminosity monitor for RHIC, as well as, LHC \cite{Baltz98}. 

The mutual excitation probability is to a good approximation
given by the product of the probabilities for the single excitations
\cite{Baur90d}. For symmetric collisions ($Z_1=Z_2$, $A_1=A_2$) we have
\begin{equation}
P_{\mbox{\scriptsize mutual}}(b) =P_{\mbox{\scriptsize GDR}}^2(b)=
\frac{S^2}{b^4} \exp\left(-2 S/b^2\right)
\end{equation}
and the total cross section is given  by the integration over the impact 
parameter from $R_{min}$ to $\infty$:
\begin{equation}
\sigma_{\mbox{\scriptsize mutual}} =
\frac{\pi S}{2} \left[1 - \exp\left(- 2 \frac{S}{R_{min}^2}\right)\right]
\approx
\frac{\pi S^2}{R_{min}^2},
\end{equation}
where the last equation is valid if $S/R_{min}^2$ is small.

This formalism can be extended  to include also the excitation of
the double giant dipole resonance (DGDR) and even higher phonon states
with phonon numbers $N_1$ and $N_2$ respectively.
Taking into account the Bose character of these excitations one obtains  
\begin{equation}
P_{N_1,N_2}(b) = \frac{1}{N_1! N_2!} \left(\frac{S}{b^2}\right)^{N_1+N_2}
\exp\left(-2 S/b^2\right).
\end{equation}
The total cross section is found again by integration over the impact parameter
from $R_{min}$ to $\infty$:
\begin{displaymath}
\begin{array}{rcl}
\displaystyle
\sigma_{N_1,N_2} & = &
     \displaystyle\frac{(N_1+N_2-2)! \pi S}{N_1! N_2! 2^{N_1+N_2-1}}
     \left[ 1 - \exp(-2 S/R_{min}^2) \sum_{k=0}^{N_1+N_2-2} 
     \left\{\frac{2 S}{R_{min}^{2}}\right\}^2 \frac{1}{k!} \right]\\
&\approx&
     \displaystyle\frac{\pi S}{N_1! N_2! (N_1+N_2-1)} 
     \left[\frac{S}{R_{min}^2}\right]^{N_1+N_2-1},
\end{array}
\end{displaymath}
where again the last equation is valid to lowest order in $S/R_{min}^2$.
Even for the heaviest systems  $S/R_{min}^2$ is less than one, but not very 
much: with $R_{min}=2 R_A$) we obtain 
$S/R_{min}^2=0.53$ and 0.48 for the system PbPb ($A=208$)
and AuAu ($A=197$), respectively.
The probability to excite a one phonon GDR in close collisions
is thus of the order of 30\%.
Detailed calculations using photonuclear data (instead of the gross 
properties of the GDR as discussed above) as an input were 
performed by Vidovic et al. \cite{VidovicGS93} and Baltz et al. 
\cite{BaltzRW96}.
An improved value of $S=(17.4\mbox{ fm})^2$, including photonuclear excitation 
beyond the GDR, was given in \cite{BaltzS98} for the PbPb collisions at LHC,
based on \cite{BaltzRW96}.

Close collisions of ions contribute most to the mutual excitation.
This leads  to a stronger  dependence of the mutual excitation cross section 
on the detailed form of the cutoff, as compared to the single neutron 
production, with its rather weak logarithmic dependence on the cutoff radius.
On the other hand the measurement of the mutual excitation process is less 
affected by
possible background processes --- like beam-gas interactions --- than the 
single GDR excitation. This makes it more suitable as a luminosity monitor. 
Mutual excitation is  sensitive to some extent to
nuclear effects. The theoretical calculations \cite{Baltz98} and especially 
\cite{Pshenichnov98,Pshenichnov01}
show that nevertheless a good  accuracy can be achieved.
In \cite{Pshenichnov01} it is  concluded  that ``{\it Good description of
CERN SPS experimental data on Au and Pb dissociation gives confidence in the
predictive power of the model for AuAu and PbPb collisions at RHIC and 
LHC\/''}. This method to monitor the luminosity is currently used at RHIC and
plans are also underway to use it at LHC/CERN. In \cite{Gallio00}
it was concluded that this method is possible also for the ALICE experiment,
see also the discussion in Chap.~\ref{sec_eventsel} below.

\begin{figure}[tbh]
\centerline{
\epsfxsize=0.6\hsize
\epsfbox[42 165 521 645]{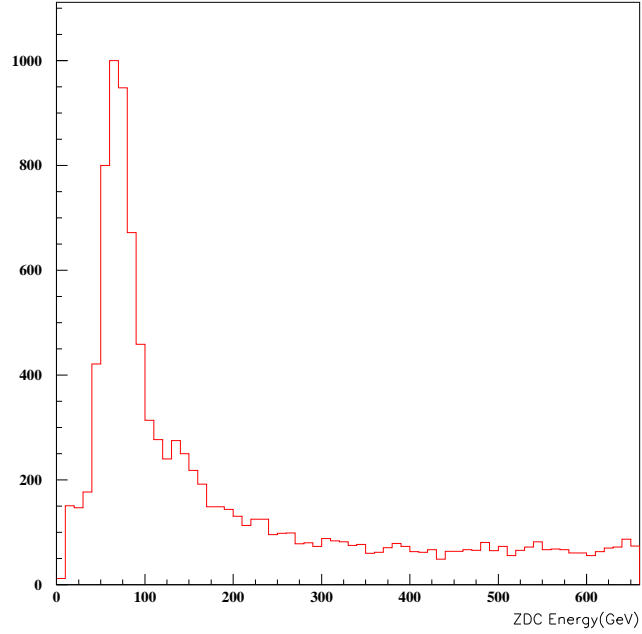}
}
\caption{Online ZDC energy distribution of neutrons obtained during RHIC 
colliding beam operation with beam energy 65 GeV per nucleon. 
Extracted from \protect{\cite{AdlerDCM00}.}}
\label{En_in_ZDC}
\end{figure}

\begin{figure}[tbh]
\centerline{
\epsfxsize=0.6\hsize
\epsfbox{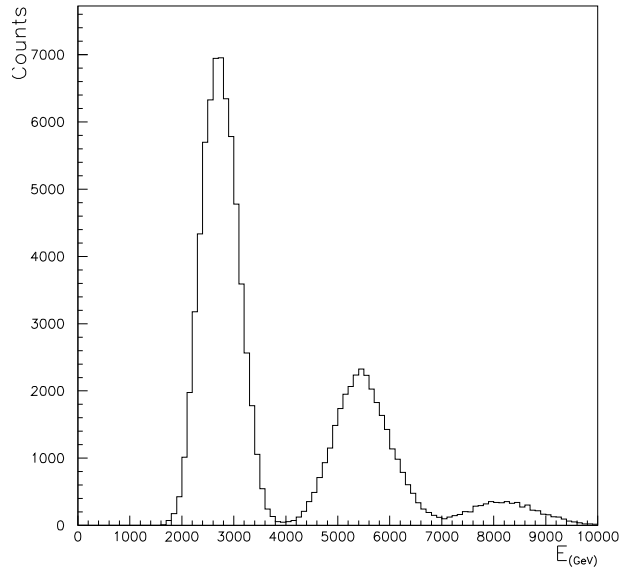}
}
\caption{Expected ZDC energy distribution of neutrons from e.m. of giant 
dipole resonance in the ALICE experiment at LHC. The ZDC energy resolution 
is included in the calculations. Extracted from 
\protect{\cite{AliceZDC_TDR}.}}
\label{AliceZDC}
\end{figure}

Very recently the first experimental results have become available from the
ZDC at RHIC \cite{White01}. The ratio of the total cross section (mutual 
Coulomb plus nuclear interaction) and nuclear interaction alone was measured 
to be $0.64$ in good agreement with the expected theoretical result of 0.66.
Also ratios of cross sections for the emission of different numbers of 
neutrons from each ion were compared with theoretical predictions. 
The energy spectrum in the ZDC measured in the experiment at RHIC is shown 
in Fig.~\ref{En_in_ZDC}, taken from \cite{AdlerDCM00},   
where the dominant peak is caused by high-energy neutrons in the ZDC. At 
LHC 
energies a clear structure due to multiple-neutron emission of the ions is 
predicted, see Fig.~\ref{AliceZDC} \cite{AliceZDC_TDR}, where the expected 
energy distribution for neutrons in the ZDC for the ALICE experiment is shown. 
Another application of the neutron emission processes in heavy ion 
collisions is to use them as a trigger for the electromagnetic interaction
(only) of the ions in the collisions by observing neutrons in both of the 
ZDC's. This opens the possibility to select $\GG$-processes and suppress other
peripheral processes in ion collisions --- like photon-Pomeron and 
Pomeron-Pomeron processes in experiments with an open geometry, i.e., where 
it is impossible to identify the $\GG$-processes by measuring directly the 
very small $P_t$ of the system produced in $\GG$-processes. This will be
discussed in more detail in Chap.~\ref{sec_eventsel} below.

For both of these purposes the cross section needs to be known very well. 
Especially the different decay channels resulting in a different number of 
neutrons emitted have therefore been studied in a detailed investigation in 
\cite{Pshenichnov98,Pshenichnov01}. For the simulation the internuclear 
cascade in the excited nucleus was used.

\subsection{Excitation of Nucleon Resonances}

The electromagnetic excitation of the $\Delta$ resonance in 
peripheral heavy ion collisions was already observed at the 
SPS \cite{PriceGW88}. 
Cross sections for the electromagnetic 
excitation of the nucleon resonances are sizeable at the collider energies.
This was first studied in \cite{BaurB89}, where the following values
are given: 17~b for UU collisions at RHIC and 25~b for PbPb
collisions at LHC. These numbers are based on the folding
of the equivalent photon spectra with an experimental
input for the $\gamma$-nucleon cross-sections up to $E_{\gamma}=2$~GeV
(in the nucleon rest frame).
For further details we refer to that reference.  
These resonances predominantly decay by the emission of mesons,
mainly pions. These pions will be emitted in a rather narrow cone in the 
beam direction. Maybe they are useful? 
The cross section for nucleon resonances is 25~b
for PbPb collisions at LHC, compared to 8~b for the total nuclear cross 
section. With the branching ratio of about 33\% for the decay of the
$\Delta$ to $n\pi^+$ and of about 66\% for the decay to $p\pi^0$, and
relatively small decay energy, a beam of pions with a narrow energy
range will be produced. For a recent reference of 
photoproduction of mesons we refer to \cite{KruscheABM01}. Some time ago
the production of beams of pions and other strongly interacting particles 
by photons was considered by Drell in \cite{Drell60}.
A detailed study of the pion emission in heavy ion collisions is performed in 
\cite{ChikinKKS00}. Various mechanisms are considered along with their 
characteristics. The electromagnetic excitation of nucleon resonances is found
to be one of the important channels. 

\subsection{Photonuclear Reactions to the Continuum 
above the Nucleon Resonances}

The continuum above the nucleon resonances 
is also strongly excited at collider energies. A first  estimate 
of this effect was given
in \cite{BaurB89}. In this exploratory calculation
a constant value (independent of the photon
energy) of $\sigma=100\mu$b was assumed for photon energies
above 2~GeV (in the nucleon rest frame). This gives a value of 2.6~b for 
UU at RHIC and 19~b for PbPb at LHC. More detailed calculations were done
in \cite{BaronB93}
and  \cite{VidovicGS93}. In this last work experimental values
of photonuclear cross-sections 
for Pb up to 80~GeV were used as an input into the calculations.
For PbPb collisions at LHC a cross section of 53.7~b was
given for the region between 40~MeV and 2~GeV. From 2~GeV to 80~GeV the
cross section is 18.7~b in good agreement with the estimate above. For AuAu
collisions at RHIC the corresponding experimental photonuclear data 
points only exist up to 9.5~GeV. Scaling
the experimentally known  cross section for Pb 
to the case of Au a value of 5.6~b was given for the region
of 2--80~GeV. For the region above 80~GeV, where no experimental data
were available, several parameterizations have been used. Only a small
 cross section of 0.7~barn was found for AuAu at RHIC, in contrast to
16.3~barn for PbPb collisions at LHC. This is due to the much 
harder equivalent photon spectrum at the higher beam energies at the LHC.

As has been mentioned already before, the electromagnetic breakup of the
nuclei is one of the main loss processes in PbPb collisions at LHC. 
Therefore a precise knowledge of these cross sections also at higher
photon energies is of interest.

\subsection{Coherent Vector Meson Production in Photon Diffraction}
\label{sec_photodiffractive}

A lot of theoretical studies of coherent, as well as, incoherent 
photoproduction of vector mesons in nuclei exist. With the 
hard equivalent photon spectra present at the relativistic 
heavy ion colliders new experimental perspectives are opened up
for these studies. As was emphasized above the photon virtuality
$Q^2$ is essentially restricted to zero, as opposed to the situation
in electron scattering, where this quantity can be varied. Let us 
mention here some basic concepts of this vast field, where we 
essentially stick to the case of quasireal photons.

Among the many possible channels, the coherent production of vector mesons
(that is, when the nucleus stays in the ground state), as well as, other 
$C$-odd mesons in photon-Pomeron processes is a very interesting
one to be studied in peripheral heavy ion collisions. From an experimental 
point of view these processes are quite similar to the two-photon processes 
discussed above. The main difference lies in the quantum number of the produced
system, which in a photon-Pomeron collisions corresponds to $C =-1$.
The low value of the transverse momentum $P_\perp$ of the mesons is a very 
clear 
experimental signature for this, comparable to the similar one in the 
photon-photon case. In fact
the production of $\rho^0$ mesons was already observed at STAR/RHIC 
\cite{Klein01b} by looking for its decay into
$\pi^+ \pi^-$ pairs, see Sec.~\ref{ssec:experimentsHI} below. The invariant 
mass distribution
for pairs with  $P_t< $ 0.1~GeV shows a clear peak at the $\rho$ mass.

Let us describe the main ideas to treat coherent vector meson production, 
where we follow essentially the work of \cite{KoelbigM68} and
\cite{KleinN99}. 
The calculation is done in the rest frame of one of 
the nuclei. The minimum momentum transfer is given to a good accuracy by 
$t_{min}=(M_V^2/2k)^2$, where $M_V$ is the vector meson mass and $k$ the 
photon momentum. This quantity (it corresponds to 
the minimal $q_L^2$ defined in Eq.~(\ref{eq:diffractiveql}) below) 
is very small (essentially zero) for high photon momenta. The coherence 
length $l_c =1/q_L$ is larger than the size of the nucleus.
This means that the photon will fluctuate into a vector meson pair already
outside the nucleus and this virtual meson is then put on-shell in the
interaction with the nucleus.

We are interested in the total cross section for the reaction
$\gamma A \rightarrow VA$. To get to this several steps have
to be done. One starts with the experimentally known cross section for
$\gamma p \rightarrow V p$ as an input. This cross section can usually
be parameterized in the Regge form \cite{Crittenden97}
\BE 
\left. \frac{d\sigma(\gamma p \rightarrow Vp)}{dt} \right|_{t=0}
= b_V \left( X W^\epsilon + Y W^{-\eta}\right),
\EE
where W is the center-of-mass energy and the constants $b_V$, $X$, $Y$, 
$\epsilon$ and $\eta $ are determined from fits to the data 
(compare also with the parameterization of the total $\GG$ cross section 
Eq.~(\ref{eq:ggtotal})). A major simplification comes from the use of
vector meson dominance \cite{BauerSYP78}, which allows to relate this
photoproduction cross section to the cross-section for forward elastic 
$V p \rightarrow V p$ scattering:
\begin{equation}
\left. \frac{d\sigma(\gamma p \rightarrow Vp)}{dt} \right|_{t=0}
= \frac{2 \pi \alpha}{f_V^2}
\left. \frac{d\sigma(V p \rightarrow Vp)}{dt} \right|_{t=0},
\end{equation}
where $f_V$ denotes the vector-meson--photon coupling.
The further calculations are simplified if one assumes that 
$\mbox{Re} f(0)/ \mbox{Im} f(0) \ll 1$, i.e., the $Vp$ scattering is 
mainly absorptive. This assumption is, 
e.g., relaxed in \cite{PautzS98} at the expense of having to introduce a new 
parameter. This assumption is certainly better fulfilled for the $\rho$
meson than for the $J/\Psi$.
With this the forward scattering cross section is proportional
to the square of the imaginary part of the forward scattering amplitude.
The optical theorem can be used to relate this forward elastic 
scattering cross section to the total cross section $\sigma_{Vp}$ for vector 
meson nucleon scattering:
\begin{equation}
\sigma_{Vp}^2 = \left .16 \pi 
\frac{d\sigma(V p \rightarrow V p)}{dt} \right|_{t=0}.
\end{equation}

Using this elementary cross section $\sigma_{Vp}$ and the nuclear 
density distribution $\rho_A(r)$ 
we can obtain in a Glauber calculation \cite{KoelbigM68}
the  total elastic scattering cross section as well as the
total cross section for VA scattering:
\begin{equation}
\begin{array}{lcr}
\sigma_{\rm elastic}(VA) &=& \displaystyle\int d^2b 
\left[ 1 - \exp\left(-\sigma_{Vp} T_A(b)/2 \right)\right]^2
\label{eq:diffVMelastic}\\
\sigma_{\rm total}(VA) &=& 2 \displaystyle\int d^2b
\left[ 1 - \exp\left(-\sigma_{Vp} T_A(b)/2 \right)\right],
\end{array}
\end{equation}
where $T_A$ is the nuclear thickness function
\begin{equation}
T_A(b) = \int dz \rho_A(\sqrt{b^2+z^2}).
\end{equation}
VMD allows us now to relate the cross section for elastic $VA$ 
scattering back to the coherent production cross section
\begin{equation}
\sigma(\G A \rightarrow V A ) = \frac{2 \pi \alpha}{f_V^2} 
\sigma_{\rm elastic}(VA).
\end{equation}
This expression allows us to discuss the dependence of the cross section on
$A$ in two limiting cases. In the transparent limit, we expand the exponent
to first order. Using a nuclear density distribution corresponding to 
a homogeneous charged sphere, integration
of Eq.~(\ref{eq:diffVMelastic}) gives us $\sigma(\G A \rightarrow V A )\sim
A^{4/3}$. In the black disc limit on the other hand we find
$\sigma(\G A \rightarrow V A )\sim R^2 \sim A^{2/3}$.
If only a finite range of $t$
needs to be taken into account, the differential scattering cross
section $d\sigma/dt$, see \cite{KoelbigM68} can be used.

A slightly different path is used in \cite{KleinN99}.
The optical theorem for nucleus $A$ and vector meson dominance is used to get
\begin{equation}
\left.
\frac{d\sigma(\G A \rightarrow V A)}{dt} \right|_{t=0}
= \frac{\alpha \sigma_{\rm total}^2(VA)}
{4 f_V^2} .
\end{equation}
From this equation one can directly see the $A$-dependence of the 
coherent forward photoproduction amplitude for the two limiting cases:
in the transparent limit there is a $A^2$ behavior 
(typical for coherent processes) and in the black disc limit we have an 
$A^{4/3}$
rise with nucleon number $A$.  

The $t$-dependence of the coherent photoproduction is 
modeled by the nuclear form factor $F_{el}(t)$. A convenient parameterization
of this form factor is given in Eq.~(14) of \cite{KleinN99}. This
form factor falls off very  fast with $t$, the angular distribution is 
peaked very much in the forward direction (up to a $t_{max}$ of the order of
$1/R^2$). Integration over $t$ in order to obtain the total coherent
cross section yields a factor of $A^{-2/3}$ . Thus the total coherent 
photoproduction scales as $A^{4/3}$ for the transparent and as  $A^{2/3}$
for the black nucleus case, in agreement with the discussion above.
The above description  illustrates an important idea,
 which goes back to the work of Drell 
and Trefil \cite{DrellT66}: using the nucleus as a laboratory, one can 
determine otherwise unaccessible vector-meson--nucleon scattering 
cross sections. Several groups studied such processes already in the 1960s and 
1970s. This is reviewed in \cite{BauerSYP78}. More recent experiments
using virtual photons from muon or electron scattering
are less comparable because the photons are more virtual (compared to $1/R_A$).
There are some experimental results on coherent cross sections: older data 
exist from DESY and Cornell. In Figs.~5,6,7 and~9 of \cite{PautzS98} there 
are good fits to the Cornell 
data. Photons of $E_{\gamma}=6.1$ and 8.8 GeV were scattered on 
complex nuclei like C, Mg, Cu, Ag, Pb and U. At RHIC and LHC one has the 
possibility to study these processes in great detail, for vector mesons up 
to the Upsilon and over a much wider energy range.
With the forthcoming new detailed results, the theoretical analysis
can be improved, for a detailed formalism see, e.g., \cite{KoelbigM68} 
or \cite{PautzS98}. 

Further interesting ideas are advanced in \cite{KleinN99,Klein00b,KleinN00}.
Since either nucleus can emit the photon, there is an 
interference between these two indistinguishable processes. Such 
interference effects are discussed theoretically in \cite{Klein00b,KleinN00}.
We refer the reader to this work for further details.
Interference effects are sensitive to amplitudes. Therefore it will  
be interesting to see  whether such effects can 
tell us more about the real and imaginary parts of the amplitudes involved.

Another possibility is multiple vector meson production. There is 
an analogy  to the double phonon giant dipole excitation, see, e.g.,
\cite{BertulaniB88,Emling95}: in both cases identical bosons are produced and 
symmetrisation has to be done. This is very important in the case of the
double phonon states, since the dipole phonons are only labeled by their 
spin-projection quantum numbers $M=0,\pm 1$. On the other hand for   
the produced vector mesons, there is  in addition 
a continuous quantum number --- namely their momenta --- which can be 
quite different in general. Therefore  one can expect that boson effects are 
less important in this case. However, it would be most interesting to 
observe the production of two identical vector mesons with close enough 
momenta and study such Boson effects. For details and further references,
see \cite{Klein00b}.

It is noted in
\cite{Klein00b} and \cite{KleinN00} that coherent vector meson production 
cross sections are huge. E.g., the $\rho$ production in AuAu collisions
at RHIC is 10~\% of the total hadronic cross-section, in PbPb collisions 
at LHC it is about equal to the total hadronic cross-section. So heavy ion 
colliders can act as vector meson factories with rates comparable to
those of $e^+e^-$ vector meson machines. Up to $10^{10}$ $\Phi$ mesons
will be produced in $10^6$~s with Ca beams at LHC.
This can be compared to the expected rates at the dedicated $\Phi$-factory
DA$\Phi$NE \cite{DaphneHandbook}:
at the beginning, a number of $2.2 \times 10^{10}$ yearly produced $\Phi$ 
is expected. This will allow for searches of rare decay modes,
CP violation and the like.
Also there is the possibility to do vector meson spectroscopy:
mesons like the $\rho(1450)$, $\rho(1700)$ and 
$\Phi(1680)$ will be copiously produced. 
The current situation of vector meson spectroscopy is outlined in 
\cite{Donnachie01}. It is concluded there that many interesting questions 
are still not understood. Hopefully relativistic heavy ion colliders can
contribute to these QCD studies.
Of course, it will have to be seen
how well such experiments can be done in the hostile environment
of the violent central collisions with their extremely high 
multiplicities.

\subsection{Incoherent Production of Vector Mesons}

In the incoherent vector meson production process the photon interacts with 
a nucleon to produce a vector meson leaving the nucleus in an excited state. 
We follow here mainly the formalism of \cite{HKN1996}.

The coherence length is defined as $l_c =1/q_L$, where 
$q_L$ is the difference between the longitudinal 
momenta of the photon and the produced vector meson. 
It is given by:
\begin{equation}
q_L=\frac{M_V^2+p_T^2}{2\omega},
\label{eq:diffractiveql}
\end{equation}
where $p_T$ denotes the transverse momentum of the produced vector 
meson with mass $M_V$, and $\omega$ is the photon energy. 
The incoherent cross section for the production of a vector 
meson in a reaction, where the nuclear state changes from
$|0\rangle$ to $|f\rangle$ ($f\not=0$) in Glauber theory is 
given by:
\begin{equation}
 \frac{d\sigma^{\gamma V}_{inc}(0\to f)} {d^2p_T} =
 \left |
 \int\frac{d^2b}{2\pi}\ \exp(-i\ \vec p_T\ \vec b)\
 \left\langle f
 \left|\Gamma^{\gamma V}_A(\vec b) \right |0\right\rangle
 \right |^2,
\label{HKN3}
\end{equation}
where the transition operator $\Gamma^{\gamma V}_A(\vec b)$ 
is defined as 
\begin{equation}
\begin{array}{l}
 \Gamma^{\gamma V}_A(\vec b;\{\vec s_j,z_j\}) = \\
 \sum_{j=1}^A
 \Gamma^{\gamma V}_N(\vec b-\vec s_j)\ e^{iq_Lz_j}\
 \prod_{k(\not=j)}^A\left[1-
 \Gamma^{VV}_N(\vec b-\vec s_k)\
 \Theta(z_k-z_j)\right]\
\end{array}.
\label{HKN1}
\end{equation}
It consists of the vector-meson production amplitude 
$\Gamma^{\gamma V}_N$ and an amplitude $\Gamma^{VV}_N$, which 
describes the elastic scattering of the vector meson on the nucleon. 
By taking $\left|f\right>=\left|0\right>$, we can treat also the
elastic/coherent case, above. Of course the equations there are related to
the one here by making use of the ``cumulant expansion'' \cite{HuefnerN81}.

Because of the phase factor $e^{iq_Lz_j}$ in this equation, the photon 
production 
amplitude on two nucleons with positions $|z_i-z_f|< l_c$ adds up 
coherently. E.g., with $M_V=0.8~GeV$ and $\omega=8~GeV$ we have 
$l_c=4$~fm (for $p_T\approx 0$). The equivalent photon spectrum at 
RHIC extends up to several hundred GeV (as viewed from an orbiting nucleus), 
therefore this coherence condition will be fulfilled in many cases. 

The low and high energy limits of incoherent photoproduction are given in 
Eq.~(12) of \cite{HKN1996} as
\begin{equation}
\label{HKN12}
\begin{array}{c}
 \sigma^{\gamma V}_{inc} = \sigma(\gamma^*N\to VN)\
 \displaystyle\int d^2b\ 
 \displaystyle\int\limits_{-\infty}^{\infty} dz\ \rho(b,z)\ \times \\
 \times \left\{
\begin{array}{ccc}
 e^{-\sigma^{VN}_{in}T_{z}(b)} & (\mbox{low energy}) \\
 e^{-\sigma^{VN}_{in}T(b)}     & (\mbox{high energy})
\end{array}
 \right\}.
\end{array}
\end{equation}
This formula contains the elementary vector meson production cross section 
and a shadow correction factor. In the low energy limit, with $l_c\ll R$, the 
attenuation of the outgoing vector meson is governed by the thickness 
$T_z(b)$, which 
is the thickness experienced by the vector meson from its point of creation
$(\vec b,z)$ till its exit from the nucleus. In the high energy limit the
attenuation is governed by $T(b)=T_{z\to-\infty}(b)$, the thickness of the 
nucleus along the total path. This corresponds to the interpretation already
given above, that at high energies the photon has already converted to a 
virtual vector meson long before hitting the nucleus. The meson is then put 
on the mass shell by the interaction with a target nucleon. This 
interpretation is familiar from the vector dominance model.

\begin{figure}
\begin{center}
\resizebox{7cm}{!}{\includegraphics{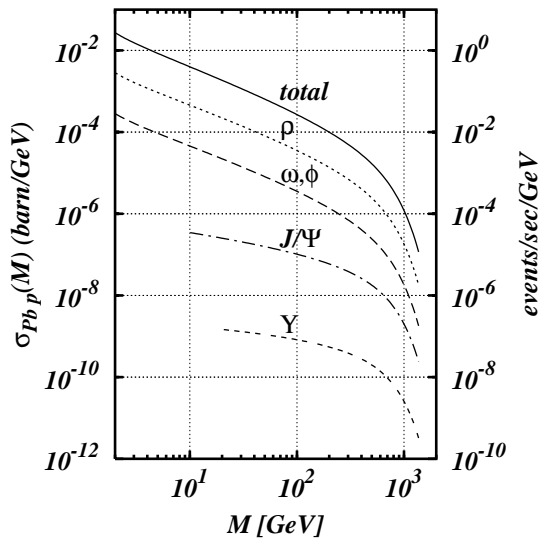}}
\end{center}
\caption{
The cross sections for the diffractive vector meson production of
$\rho$,$\omega$,$J/\Psi$, and $\Upsilon$, as well as, the total diffractive
cross section is shown for the collision of a virtual photon coming from
a Pb ion at LHC energies with a proton
as a function of the invariant mass $M$ of the photon-proton system.
For the completely incoherent cross
section, this cross section needs to be multiplied by $A$. Parameter for
the different cross section are taken from \protect\cite{Crittenden97}.}
\label{fig:vmdiffprod}
\end{figure}

An interesting quantity is the nuclear transparency. It is defined as 
the ratio of the incoherent production cross section on a nucleus
$A$ to $A$ times the elementary photoproduction cross section on a nucleon.
In Fig.~\ref{fig:vmdiffprod} we show the cross sections for incoherent 
vector meson production at the LHC. A nuclear transparency of 1 
was assumed. This should (of course) be scaled with some realistic 
value of the transparency factor, which can, e.g., be read off from Fig.~1
of \cite{HKN1996}. For a Pb nucleus those authors find a value of about 0.1 
for $\omega> 10$~GeV ($Q^2=0$).
   
Both the coherent and incoherent production of vector mesons have been
studied at HERMES/HERA in $eA$ collisions \cite{Ackerstaff99}.
The photon virtuality was $0.4~GeV^2 < Q^2 <5~GeV^2 $which
is much larger than the $Q^2 < 1/R^2$ in  AA collisions.
A theoretical discussion is given in  \cite{RenkPW00} 
where a systematic multiple scattering formalism is developed for 
vector meson electroproduction from nuclei. Also formation-, propagation-, 
and hadronization-scales for quark-gluon fluctuations of the virtual 
high-energy photon are considered.

\subsection{Photon-Gluon Fusion and the Gluon Structure Function
in a Nuclear Environment}

The production of $c \bar c$ and $b\bar b$ pairs in photon-gluon
processes has been suggested as a method to measure the gluon
distribution in a nucleon by Frixione et al. \cite{Frixione1993}.
Such processes are under study experimentally at HERA
\cite{Freiburg99} (especially \cite{Daum99}), and \cite{Ambleside00}
(especially \cite{Hochman2000}).  Also there has been much progress in
the theoretical study of these processes and the investigation of
higher order effects. A recent reference is \cite{Frixione2001}.  With
the high photon fluxes at relativistic heavy ion colliders such
studies will also be possible for nuclei instead of nucleons. This is
reviewed in \cite{KraussGS97} and we give only a brief update here.

In these photon-gluon fusion processes, one of the ions provides the photon,
which hits the gluon of the other ion to produce a $q \bar q$ pair.
Such processes are coherent for one ion and incoherent for the other. 
This leads to a special kind of event topology:
there is one nucleus, which remains intact (or at
least almost, the ``spectator'', which provides the 
equivalent photons), and the ``target'' (which provides the gluon), will
be broken up (often called the ``remnant'') plus a $q \bar q$ jet. 

The first estimate of the production of 
heavy quarks in such collisions was done in \cite{SchneiderGS92}. 
Central collisions were not excluded in their approach. More refined 
calculations focusing on the production of $c\bar c$ and $b\bar b$ quarks were
done in \cite{HofmannSSG91,BaronB93,GreinerVHS95,KraussGS97}. 

The flavour content of the quark jets can be determined by detecting the 
corresponding $D$ or $B$ mesons. This of course needs to be included in
the specification of the triggering. In this context the study of 
inclusive processes with heavy flavour production is also of 
interest. We mention the measurement of inclusive $D_s^{\pm}$ photoproduction
at HERA \cite{Breitweg00}. 
The photon-proton c.m. energies were in the range of
$130<W< 280$~GeV. Photoproduction events were selected by the 
requirement that no scattered positron was identified.
The c.m. energies of the photon-proton system can therefore not be 
determined directly from the momentum of the scattered positron, but needs to 
be extracted from the transverse momentum of the jet from the two produced 
quarks. They were determined with the Jacquet-Blondel estimator of $W$, 
see also Sec.~\ref{sec:equimuon} and especially Eq.~(\ref{eq:yjblondel}).

Modification of parton distribution functions inside the nucleus is one
of the interesting topics of QCD \cite{Arneodo94,EskolaKR98,EskolaKR99}. These 
modifications are expected to be present especially at small value of $x$,
where the partons distributions of the individual nucleons start to overlap
with each other. In a recent work \cite{GelisP01}
the gluon distribution in this low-$x$ region was assumed to be given by 
the ``color glass condensate''. In this model, the photoproduction of a 
$q\bar q$ pair in very peripheral
$AA$ collisions was calculated. The authors conclude that
``{\it it appears that this process is very sensitive to the properties of 
the gluon distribution in the region where saturation might 
play an important role}''. 

Quite recently
the production of top quarks at the LHC was studied in \cite{KleinNV00}.
It was found there that 210 $t\bar t$ pairs will be produced
in a $10^6$s OO run at LHC (with an assumed luminosity of 
$1.4 \times 10^{31} \mbox{cm}^{-2}\mbox{s}^{-1}$). Top quark production 
has been studied before in \cite{SchneiderGS92}.
In \cite{KleinNV00} 
the photon spectrum excluding central collisions and
a more modern version of the
gluon distribution function is used. The results are sensitive
to the large $Q^2$ behavior of the distribution function.
Shadowing is also included. For more details we refer to this
work. It will be an interesting channel to be studied at the 
forthcoming LHC.

%
%
\section{Diffractive Processes}
\label{sec_diffractive}

Diffractive processes are of interest in $pp$ and AA collisions
for their own sake and as a possible background to $\gamma\gamma$ 
collisions. At HERA diffractive photon-proton interactions have been 
studied in great detail, see \cite{Abramowicz01,Cartiglia97,Levy97}.
The processes of diffractive photoproduction of vector mesons on
nuclei were discussed in subsections \ref{sec_photodiffractive} and
\ref{sec:vector-meson-pair}.
Elastic proton proton scattering will be studied at RHIC \cite{pp2pp98}.

In this subsection we discuss double diffractive (or double Pomeron) 
processes in ion-ion collisions. This case is certainly even more 
difficult than the diffraction in the $\gamma\gamma$ and $\gamma p$
interactions, as discussed above. The double diffractive processes have been 
studied already at the ISR in $pp$-collisions and at the TEVATRON in $p\bar p$ 
collisions \cite{GoulianosCDF01,AffolderCDF01,Santoro01}. These processes 
are also one of the main topic in the current COMPASS experiment
at the SPS \cite{COMPASS}.  

Generally one could define diffractive processes as hadronic or ion
interactions by means of exchange of one or several Pomerons. The Pomeron
itself is regarded in QCD as a colorless object consisting of 2 (correlated)
gluons $gg$. Also discussed now are three-gluon objects $ggg$, i.e., 
Odderons. The simplest diffractive process meeting this definition is the 
elastic scattering and there are several special experimental programs 
devoted to this question in $pp$ collisions \cite{TOTEM99,PP2PP98,Martin01}. 
For the elastic scattering in AA collisions, see also Chap.~2 in 
\cite{BaurHT98}.
 
We are interested in diffraction mainly as a possible background for 
photon-photon and photon-hadron processes. We will therefore look only at 
those processes where particles are produced in the central region
together with no breakup of the two ions, as they correspond to the tagging 
conditions for photon-photon and photon-hadron processes, see 
Chap.~\ref{sec_eventsel}. Experimentally these double diffractive 
processes (``double Pomeron exchange'') have been studied for alpha-alpha 
collisions already at the ISR in 1985 \cite{CavasinniPMS85}. 
Results for double diffraction in $p\bar p$ collisions at the TEVATRON have 
also
been studied recently \cite{AffolderCDF01}. Diffractive photon-nucleus 
processes could also be studied with nuclear targets in a future option at 
HERA \cite{herafuture96}. For an overview of diffractive processes we refer 
to \cite{Bjorken92,Bjorken93,Felix97}.

\begin{figure}[tbh]
\begin{center}
\resizebox{8cm}{!}{\includegraphics{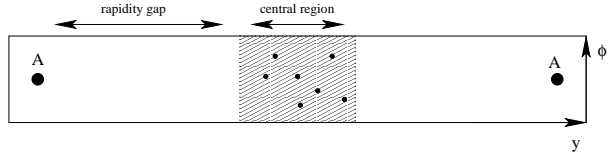}}
\end{center}
\caption{Coherent double diffractive processes are characterized by particles
produced in the central rapidity region and the two ions leaving the 
interaction with almost the initial rapidity. Between them is a large empty 
region, defining the ``rapidity gap''. Adapted from \protect\cite{Bjorken92}}.
\label{fig_rapgaps}
\end{figure}

These double diffractive processes are characterized by hadrons produced
in a region of rapidity $y$ well separated by so-called ``rapidity gaps''
from the rapidity of the initial ions, see Fig.~\ref{fig_rapgaps}.
If a colored object (e.g., a single quark or gluon) would be transfered from
one of the ions, the probability for the existence of such a gap is suppressed
exponentially with $y$ \cite{Bjorken92,Martin01}, therefore these events will
be mainly due to the exchange of colorless objects. This characteristics
coincides also with 
the condition for finding the two ions in the ground 
state (``coherent case''). This process will therefore be due to the
exchange of either neutral mesons or Pomerons, where these two contributions 
were already present in the total cross section 
$\gamma\gamma\rightarrow$~hadrons, 
discussed in Chap.~\ref{sec_ggqcd} above.
At high energies, as it takes place in relativistic ion collisions,
only the Pomeron exchanges are dominant in the cross sections.   

In the phenomenological approach of Donnachie and Landshoff 
\cite{DonnachieL88,DonnachieL88b,DonnachieL88c,Landshoff01}
the Pomeron is described as a single Regge pole with a ``propagator''
\begin{equation}
D_\P(t,s) = \frac{(s/m^2)^{\alpha_\P(t)-1}}
{\sin \left( \frac{1}{2}\pi \alpha_\P(t) \right) }
\exp\left(-\frac{i}{2}\pi \alpha_\P(t)\right),
\end{equation}
where $s$ is the square of the total c.m. energy, $t=-q^2$ the square of
the momentum transfer and the Regge trajectory of the Pomeron given by
\begin{equation}
\alpha_\P(t)=1+\epsilon+\alpha_\P' t,
\end{equation}
with $\epsilon=0.085$ and $\alpha_\P'=0.25$~GeV$^{-2}$ 
(see also the parameterization mentioned above for 
$\gamma\gamma\rightarrow$~hadron). The coupling of the Pomeron to
the proton is given by
\begin{equation}
\beta_{p\P}(t) = 3 \beta_0 F_1(-t),
\end{equation}
with $\beta_0=1.8$~GeV$^{-1}$ and $F_1$ is 
taken as the isoscalar electromagnetic
form factor, 
for which the usual dipole parameterization is assumed.
How the Pomeron couples with the whole nucleus is an interesting
question. Very little seems to be known about this at present. This will introduce
some uncertainty into the calculations discussed here. In general it is assumed
to be of the form
\begin{equation}
\beta_{A\P}(t) = 3 A \beta_0 F_A(-t),
\end{equation}
that is, a coherent coupling to all nucleons $\sim A$ with the elastic form
factor guaranteeing that the Pomeron emission does not lead to a breakup of 
the ion. The coherence will be destroyed rather quickly with increasing $t$,
$t$ being restricted essentially by  
$|t|\lesssim 1/R^2 \approx (60$~MeV$)^2$, compared to the ``natural range''
given by Regge theory $|t|\lesssim 1/r_0^2 \approx (500$~MeV$)^2$ with 
$r_0^2= \alpha_\P' \ln(s/m^2)$.

The use of the elastic form factor corresponds to taking into account the
finite size of the nucleus, see the discussion in Chap.~\ref{sec_lum}. 
The additional hadronic ``initial state interaction'' between
the other nuclei need to be taken into account. Whereas their effect was
small in most cases in the photon-photon  
collisions, it will be much more severe
here, due to the short range nature of the Pomeron. Also the ``final state
interaction'', the third effect discussed in Chap.~\ref{sec_lum}, will be more 
important in this case. Whereas the electromagnetic field decreases inside
the nucleus and therefore the number of equivalent photons there is small, 
this is not the case for the equivalent Pomerons inside the nucleus.
Both the finite size and the initial state interaction
effects are assumed to lead to a considerable decrease of the 
(effective) coupling so that we finally expect only an increase of the cross 
section with 
$A^{\delta}$, where $\delta<1$.

\begin{figure}[tbh]
\begin{center}
\resizebox{3cm}{!}{\includegraphics{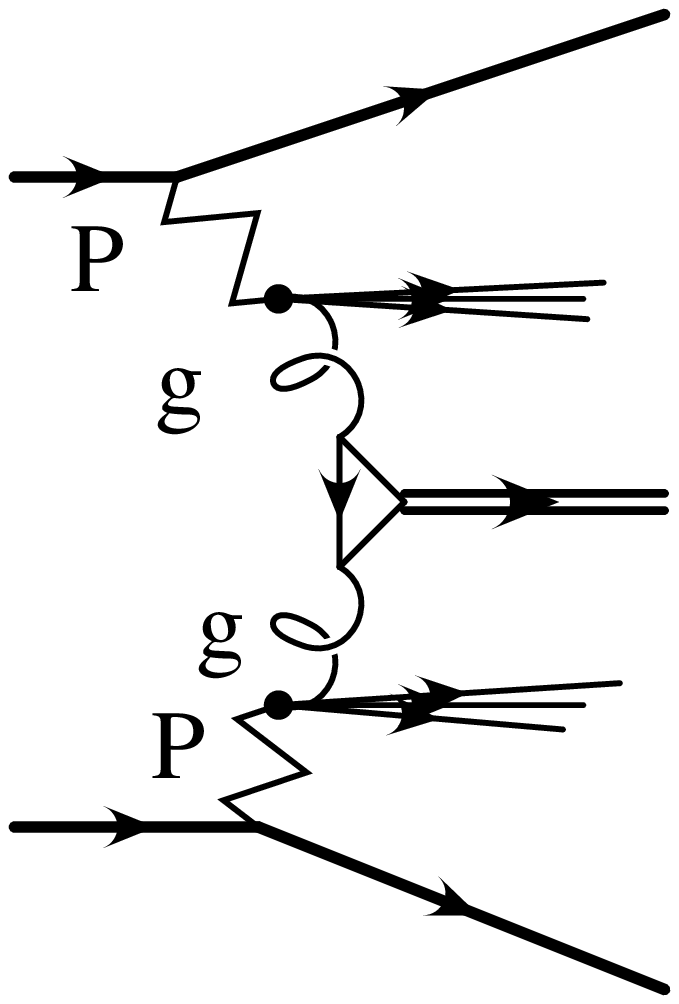}}
(a)~~~
\resizebox{3cm}{!}{\includegraphics{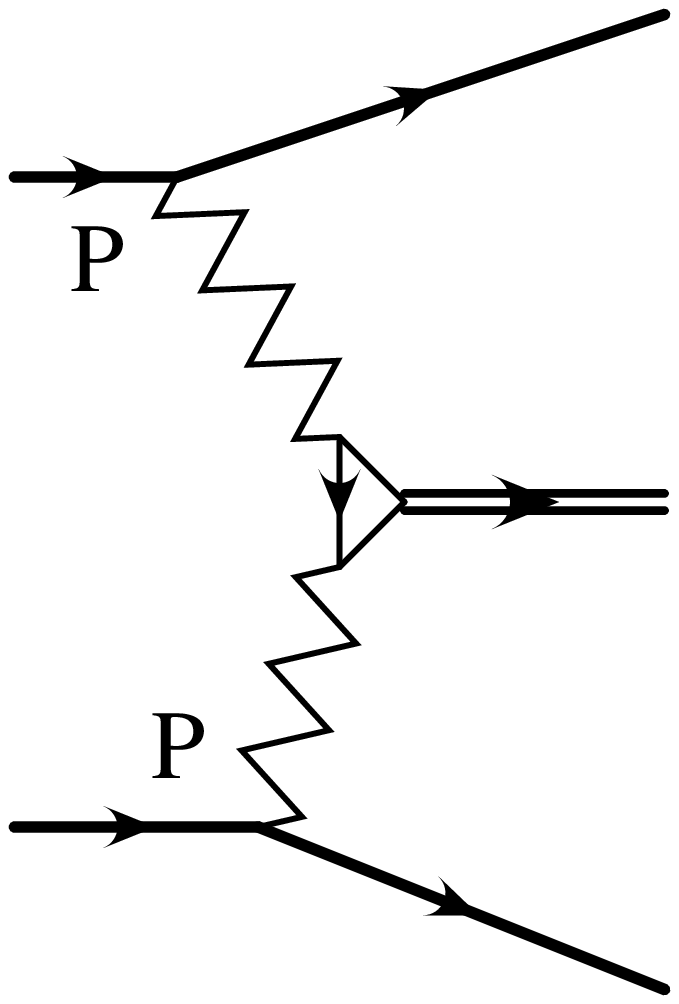}}
(b)~~~
\end{center}
\caption{The inclusive (a) and exclusive (b) production of, e.g., a Higgs
boson through Pomeron-Pomeron interaction. The dominant process in both cases
comes from the coupling of the Higgs boson with the two Pomerons or gluons
through a heavy quark loop.
}
\label{fig_diffhiggs}
\end{figure}

This model for the Pomeron has been used as the basis in a number of studies 
within the
``equivalent Pomeron approximation'' \cite{MuellerS91}. The inclusive
production of the Higgs boson ($\P\P \to H +$anything) 
was studied for $pp$ collisions in
\cite{SchaeferNS90}, see Fig.~\ref{fig_diffhiggs}(a). It was assumed that
the major contribution for the Higgs boson production comes from the 
gluon-gluon fusion via a heavy-quark loop into the Higgs. The gluon distribution
function inside the Pomeron is not fully settled, but it was assumed that the 
Pomeron consists mainly of gluons. A gluon distribution of the form 
\begin{equation}
G_\P(x)=6/x(1-x)^5
\end{equation}
was used.

This approach was extended to AA collisions in \cite{MuellerS91}. Also an estimate
for the exclusive Higgs boson production ($\P\P\to H$,
where the Higgs boson alone without any other hadronic particles is produced, 
see Fig.~\ref{fig_diffhiggs}(b)) was given.
The Pomeron-Pomeron-Higgs coupling was modeled by assuming that the Pomeron
is a correlated two gluon exchange \cite{DonnachieL88}, and the coupling to
the Higgs boson was then done again via a heavy quark loop,
see Fig.~\ref{fig_diffhiggs}(b). In contrast to the photon the Pomeron,
being not a real particle but a phenomenological description, has a finite
size. Therefore the exclusive production of some heavy mass resonances 
(corresponding to an object with a small size) is highly suppressed.

Central collisions were excluded in \cite{MuellerS91} by using only the 
impact parameter range $b>R_1+R_2$. This exclusion of the central collisions 
leads to a 
suppression of the inclusive production cross section by six orders of 
magnitude. It was concluded that both the exclusive and inclusive 
production are small compared to the photon-photon case.
The exclusive production of $\eta$ mesons was studied similarly in 
\cite{SchrammR97}. Again it was found to be only a small background compared
to the dominant photon-photon process.

\begin{figure}[tbh]
\begin{center}
\resizebox{4cm}{!}{\includegraphics{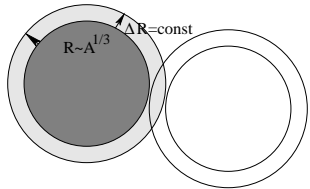}}
\end{center}
\caption{Due to the strong interaction of the nucleons in both ions and
the short range of the Pomeron, essentially only a narrow ring around
the nuclear radius, contributes to the diffractive interactions without
nuclear breakup. As the nuclear radius increases with $R\sim A^{1/3}$ this
explains the increase of the Pomeron-Pomeron cross section with $A^{1/3}$.}
\label{fig_ringzone}
\end{figure}

The calculation of the production of a central cluster of hadrons with a given
invariant mass was done in \cite{EngelRR97} within the Dual Parton Model
\cite{CapellaSTT94}.
By assuming that only one of the nucleons produces a central cluster and
all other nucleons scatter only elastically, central collisions were removed.
Within a Glauber-like calculation of this, the dependence of the
cross section on $A$ is given by $\sigma\sim A^{1/3}$.
This $A^{1/3}$ can be understood by assuming that there is only a 
``ring zone'' surrounding each ion, which is active for Pomeron-Pomeron 
interactions, see Fig.~\ref{fig_ringzone}, corresponding to the black disc 
limit.
It was concluded that the cross-section for Photon-Pomeron and Pomeron-Pomeron 
interactions for this case are larger than the cross section for photon-photon 
events for almost all ion species except for the very heavy ones (Pb), were 
they are comparable.

Whereas the central collisions were effectively removed in \cite{EngelRR97},
the coherence condition, which essentially guarantees that the ion does not
break up due to the Pomeron emission, was not included in this approach.
As mentioned above the Pomeron is a short ranged object, it would normally 
give a large momentum transfer of the order of 500~MeV to the nucleus.
Therefore it is very likely that the nucleus will break up. An elastic form
factor for the whole ion was not taken into account in this approach. 
Some estimates 
were made \cite{Engel98}, indicating that including this would lead
to a further suppression of the diffractive events, favoring again the 
photon-photon case.

This further decrease of the diffractive production cross section
due to coherence effects
was confirmed also in \cite{RoldaoN00} within the equivalent
Pomeron approach, see above. Three different types of final states 
were considered and compared to the photon-photon case: the exclusive 
production of resonances (mainly different $\eta$ mesons) was studied, as well 
as, the (continuum) production of a $\pi$ pair. The contribution of 
Pomeron-Pomeron interaction compared to photon-photon collisions in these 
cases is found to be only 1\% or less for both RHIC and LHC for all ions. 
This is again due to the fact that the exclusive process is highly suppressed.
As a third type of reaction the production of a central hadronic cluster is 
studied. Here it is found that for light ions (Ca,Ag) at LHC diffractive 
processes can become of the same size and even dominate at an invariant mass 
of the cluster below 5~GeV. Also at RHIC the diffractive processes will 
dominate over the photon-photon processes. No results are given for the 
photon-Pomeron interactions in \cite{RoldaoN00}.

A useful expression for the production of a central cluster of hadrons
in Pomeron-Pomeron processes is given in the framework of the simple Regge
model, see, i.e., \cite{Chung91,Felix97}. The cross section of these 
processes are given approximately by
\begin{equation}
   d \sigma_{PP}/dm_X^2 = \frac{\sigma_{0}}{m_X^2}~(A^{1/3}+\delta)~
   \ln \frac{s_{NN}}{(m_X m_N R)^2},
\label{DPdm}
\end{equation}
where $\sigma_{0}= 1.3~\mu b$, $m_X$ is the mass of the produced hadron
cluster in a central rapidity region, $s_{NN}$ is the
square of the c.m. energy of the nucleon-nucleon collisions, $R$ the radius
of the nucleus, $m_N$ is the nucleon mass and $\delta$ is the width of the 
nuclear surface. After integration of Eq.~(\ref{DPdm}) over $dm_X^2$ an 
estimate of the cross section for Pomeron-Pomeron processes can be obtained. 
One sees again that this cross section roughly increases with $A^{1/3}$, that
is, also with $Z^{1/3}$. This should be compared with the factor $Z^4$ in the 
cross section for
two-photon processes. Thus in heavy ion collisions we can expect the
predominance of two-photon processes, whereas in light ion- and
especially in $pp$-collisions the contribution of Pomeron-Pomeron
processes will be much more essential. These more qualitative cross section
estimates are fully confirmed by detailed calculations, see above and
Sec.~\ref{sssec:physback}.

Due to strong shadowing of the Pomeron inside the nucleus only the surface
region will contribute to peripheral diffractive processes. Photon-photon
and Pomeron-Pomeron processes will therefore scale approximately with the 
ion species as $Z^2$ and $A^{1/3}$, respectively \cite{Felix97}.
Thus for heavy ions, like Pb, we may expect dominance
of the photon-photon processes whereas, say in $pp$ collisions, the
Pomeron-Pomeron processes will dominate in coherent collisions. 

Diffractive processes are an additional field of studies 
in  peripheral ion collisions. They essentially have the same triggering 
conditions and therefore one should be able to record them at the same time 
as photon-photon and photon-hadron events. One may think also about
going beyond the coherent case (``no breakup''). Processes with a large 
rapidity gap have been proposed \cite{ChatrchyanJZ97,AlberiG81} as an 
interesting 
physics topic and many interesting questions, like the probability for a gap 
as a function of its width $y$ have been posed.

As a potential background for coherent photon-photon and photon-hadron
collisions in very peripheral collisions we conclude that they are in 
most cases only a small background for the heaviest ions. Without the 
triggering on ``no breakup'' for the two ions, they dominate the production 
of hadrons in the central region. The same is also the case for $pp$ 
collisions at the LHC.

\section{Special Aspects of Dilepton Pair Production}

\label{sec_leptons}

Due to their small mass, electrons (positrons) and to some 
extent also muons play a special role in
very peripheral heavy ion collisions. They are produced 
much more easily than other heavier particles, and
--- especially in the case of $\EPEM$
pair production --- some  new phenomena, like multiple pair
production, will occur. A small mass means that the  Compton wave
length --- for electrons this is
$\lambda_C=\frac{1}{m_e}\approx386$~fm --- is large, larger than 
the nuclear radius $R$ of several fm. The equivalent 
photon approximation has to be modified when applied in this case. 
The cutoff radius is not given by the nuclear radius anymore but 
by the Compton wavelength of the particle.
The reason is that in the cross section for the
$\GG\rightarrow e^+e^-$ subprocess the dependence on the
virtualities $Q_1^2$ and $Q_2^2$ of the (quasireal) photons can no
longer be neglected. As was explained in Chapter~\ref{sec_lum}
the photon virtuality in heavy ion collisions is limited by the
nuclear radius up to $Q^2 \lesssim 1/R^2$, which is much larger than $m_e^2$.
Especially for invariant masses of the order of several $m_e$ ---
which contribute mostly to the total cross section ---  the
photon-photon cross section deviates from the one for real photons for
virtualities  $Q_1^2$ and/or $Q_2^2 \gtrsim m_e^2$, (see App.~E and 
Eq.~(6.25) of \cite{BudnevGM75}). For the muon --- with a Compton
wavelength of about 2~fm --- we expect the standard equivalent photon
approximation (EPA) to be applicable with smaller corrections. For $e^+e^-$
pair production using EPA 
together with the cutoff radius equal to the electron Compton wave length
gives a fair approximation for the total cross section. On the other
hand EPA completely fails for certain regions of the phase space and also 
for the calculation of pair production probabilities at small
impact parameters (as discussed in detail below). 

The calculation of pair production in heavy ion collisions dates back to
the work of Landau \& Lifschitz \cite{LandauL34} and Racah \cite{Racah37}
in the thirties, as already mentioned in the introduction. In connection
with the relativistic heavy ion colliders there has been some renewed interest
and cross section calculations have been done using the semiclassical 
approximation in \cite{BertulaniB88} and using Feynman Monte Carlo 
techniques in \cite{Bottcher89}. For lower energies 
(about 1--2~GeV per nucleon) coupled channel calculations have been done 
also by a number of groups \cite{MombergerGS91,RumrichSG93}.

\begin{figure}[tbh]
\begin{center}
\resizebox{5.5cm}{!}{\includegraphics{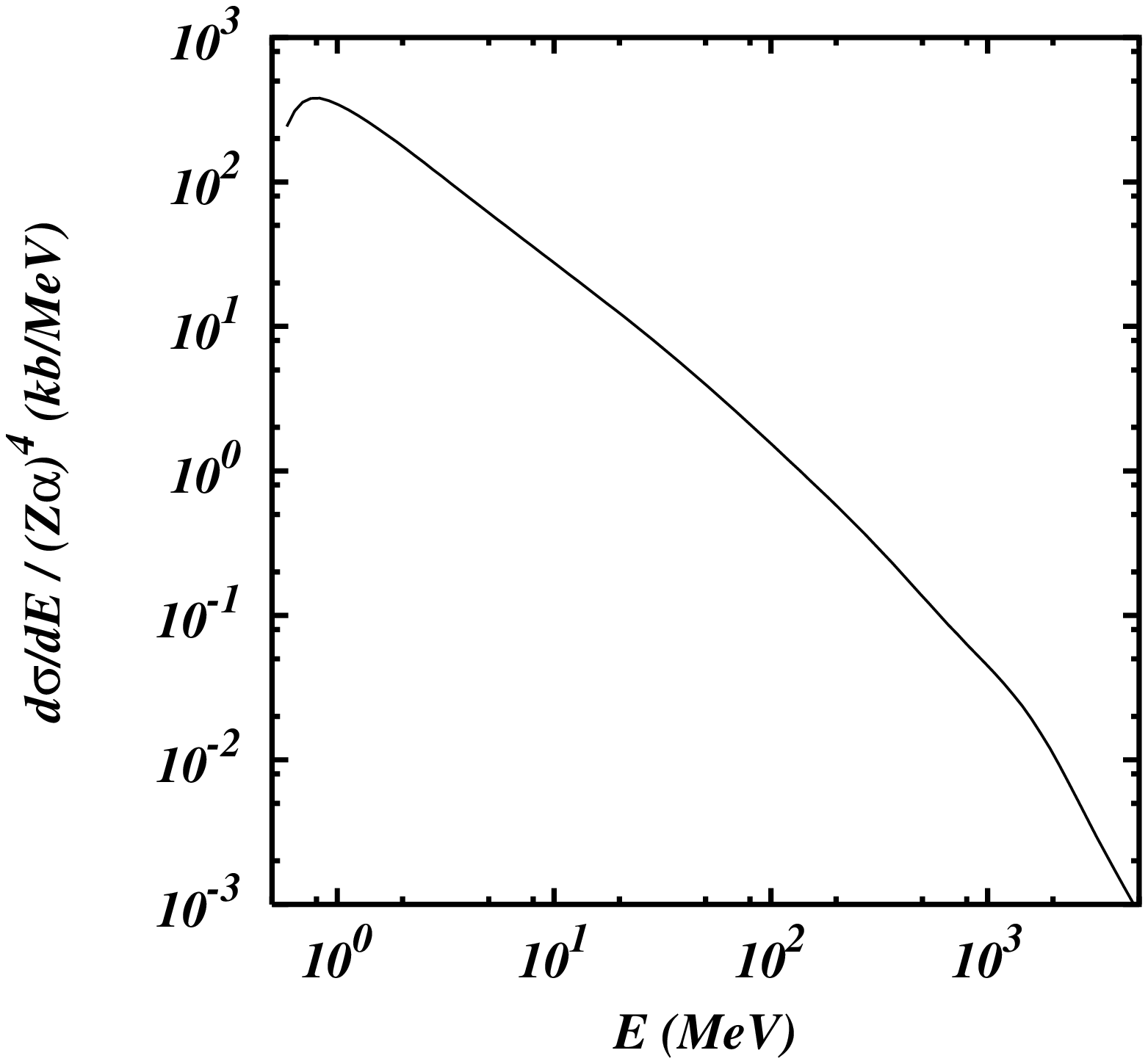}}
(a)~~~
\resizebox{5.5cm}{!}{\includegraphics{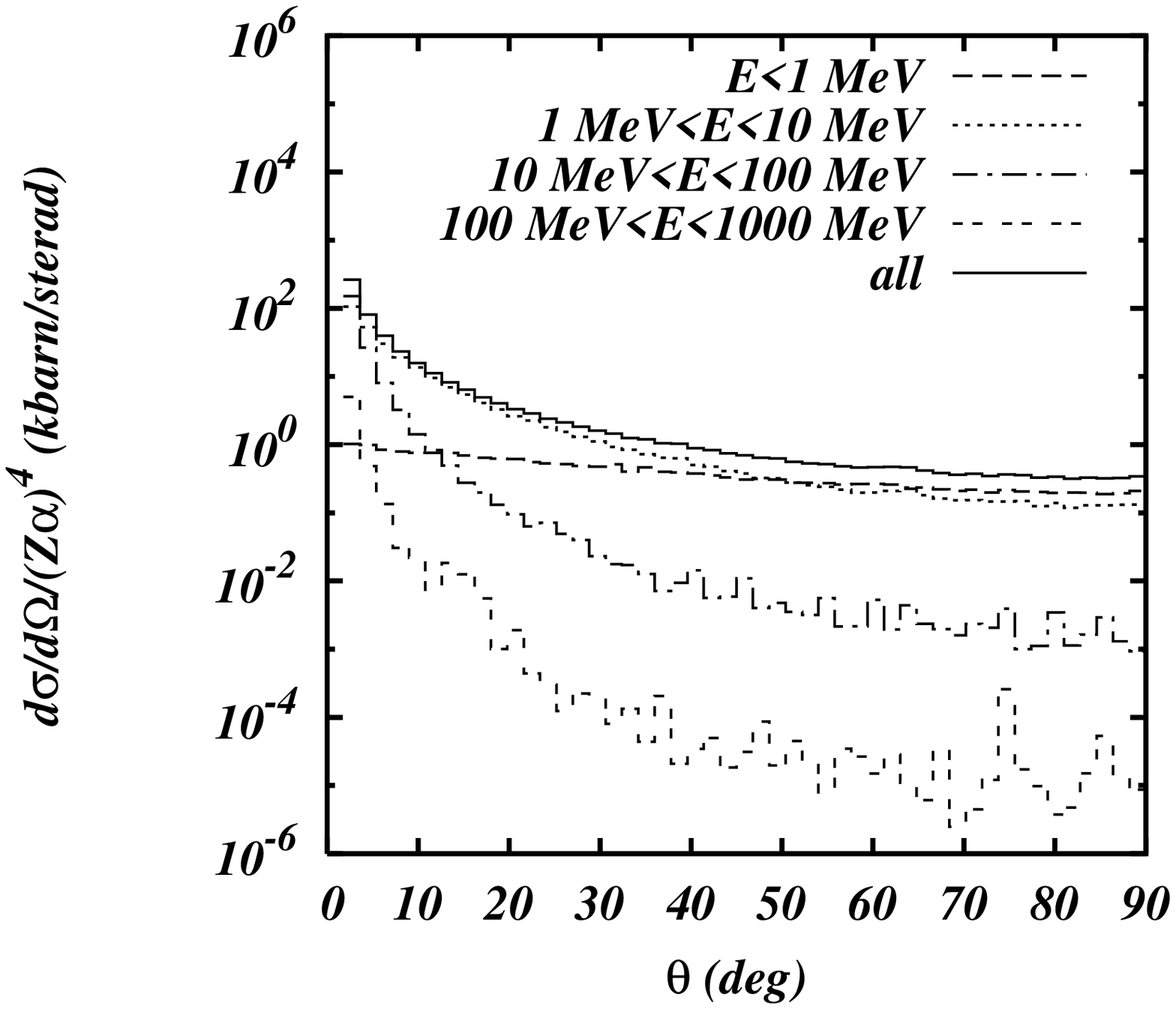}}
(b)
\end{center}
\caption{
Single differential cross sections for the $e^+e^-$ 
pair production at LHC as a function of
the energy (a) of either electron or positron and as a function of the
angle of the electron or positron with the beam axis (b). 
Most pairs are produced with energies between 2--5 MeV in the Lab frame
and in the very forward or backward direction.
}
\label{fig_eee}
\end{figure}

The total cross section for pair production is huge 
(about 200~kbarn for PbPb at LHC, 30~kbarn for AuAu at RHIC 
\cite{AlscherHT97}). Electron-positron pairs therefore present a possible
background. The differential cross section show that most of the particles are
produced at low invariant masses (below 10 MeV)
and into the very forward direction, see Fig.~\ref{fig_eee}. The highly 
energetic electrons and positrons tend
to be even more concentrated along the beam pipe, therefore most of them
remain unobserved. On the other hand, the cross section for highly energetic
electrons and positrons is still of the order of barns. These QED pairs 
constitute a potential hazard for the detectors, see below in 
Chap.~\ref{sec_eventsel}. On the other hand they can also be useful as a
possible luminosity monitor, as is discussed, e.g., in 
\cite{Felix97,ShamovT98}.

Due to the strong electromagnetic fields of short duration new phenomena, 
especially multiple pair production, but also Coulomb-corrections, will appear
in relativistic heavy ion collisions. They are  of 
interest for the  study of QED of strong fields. This will
be discussed in the next two sections. In addition there are also a number of 
processes related to  pair production, like bremsstrahlung from produced 
pairs, or the production of electrons and muons not only as free particles 
but either produced
into an atomic state bound to one of the ions (``bound-free pair
production'') or as a bound state of the pair (``positronium'',
``dimuonium'', or even ``ditauonium'' production). This will be
discussed then in the rest of this chapter. For an introduction to aspects
of atomic physics of relativistic heavy ion collisions, we refer the reader 
to \cite{Eichler90,EichlerM95}.

\subsection{Strong Field Effects in Electron Pair Production:
Multiple Pair Production}
\label{sec_leptons_mpairs}

The special situation of the electron pairs can already be seen from
the formula for the impact parameter dependent probability for
(single) pair production $P^{(1)}$ in lowest order.
(For heavy nuclei 
the semiclassical approximation is valid and the impact parameter $b$
can be considered to be  an observable quantity,
 see, e.g., Chapter~2 of \cite{BaurHT98}
and further references given there.) Using the
double equivalent photon approximation (valid for $1/m_e\ll b\ll \gamma/m_e$)
with the cutoff radius set to $1/m_e$ one obtains \cite{LeeM00,LeeMS01}
\begin{eqnarray}
P^{(1)}(b) &\approx& \frac{28}{9\pi^2} \frac{Z_1^2 Z_2^2 \alpha^4}{m_e^2 b^2}
\left[
2 \ln \gamma_{lab}^2 - 3 \ln (m_e b) \right]
\ln(m_e b) \quad\mbox{for $1\ll m_e b \ll \gamma_{lab}$} \nonumber\\
           &\approx& \frac{28}{9\pi^2} \frac{Z_1^2 Z_2^2 \alpha^4}{m_e^2 b^2}
\left[
\ln \frac{\gamma_{lab}^2}{m_e b}
\right]^2 \quad\mbox{for $\gamma_{lab}\ll m_e b \ll \gamma_{lab}^2$}
\label{eq_pbapprox}
\end{eqnarray}
As this calculation 
underestimates the probability in the range of small impact parameters, 
one might expect that also probabilities larger than
one are possible. As mentioned already in the introduction,
the use of the equivalent photon approximation is not justified here. 
Therefore (more) exact calculations need to be done. 

The impact parameter dependent probability in lowest 
order was calculated numerically in \cite{HenckenTB95b,Guclu95}.
The quantity $P^{(1)}(b)$ found
there is larger than the one given by Eq.~(\ref{eq_pbapprox}), 
and values larger
than one are possible at RHIC and LHC. As an example $P^{(1)}(b=0)=3.9$ (1.6) 
\cite{HenckenTB94}, and $P^{(1)}(1/m_e)=1.5$ (0.6) \cite{HenckenTB95b} for LHC 
(RHIC) are found. It means that a new class of phenomena appears here.

The fact that the cross section for the electron-pair production in 
lowest order rises too quickly with beam energy and will eventually 
violate unitarity at very high energies was already mentioned by Heitler 
in \cite{Heitler34}, and speculations about possible
remedies were made there. Whereas it was assumed to be an {\it 
``academic problem''} at that time, the fact that unitarity is violated 
in the lowest order calculations at RHIC and LHC energies for the heavy 
systems has generated some renewed interest \cite{BertulaniB88}. It led 
to a series of studies starting almost a decade ago. It was found that the 
production of multiple pairs in a single collision restores unitarity 
\cite{Baur90}.

\begin{figure}
\begin{center}
\resizebox{!}{4cm}{\includegraphics{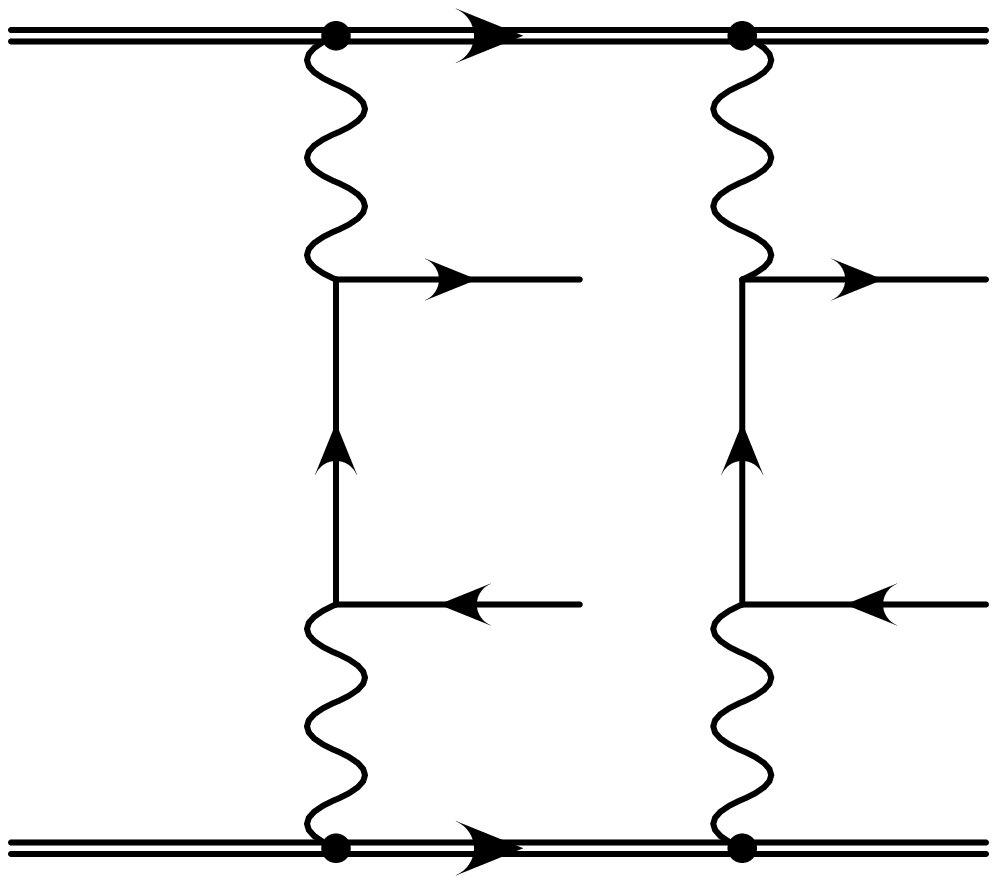}}
(a)~~~
\resizebox{!}{4cm}{\includegraphics{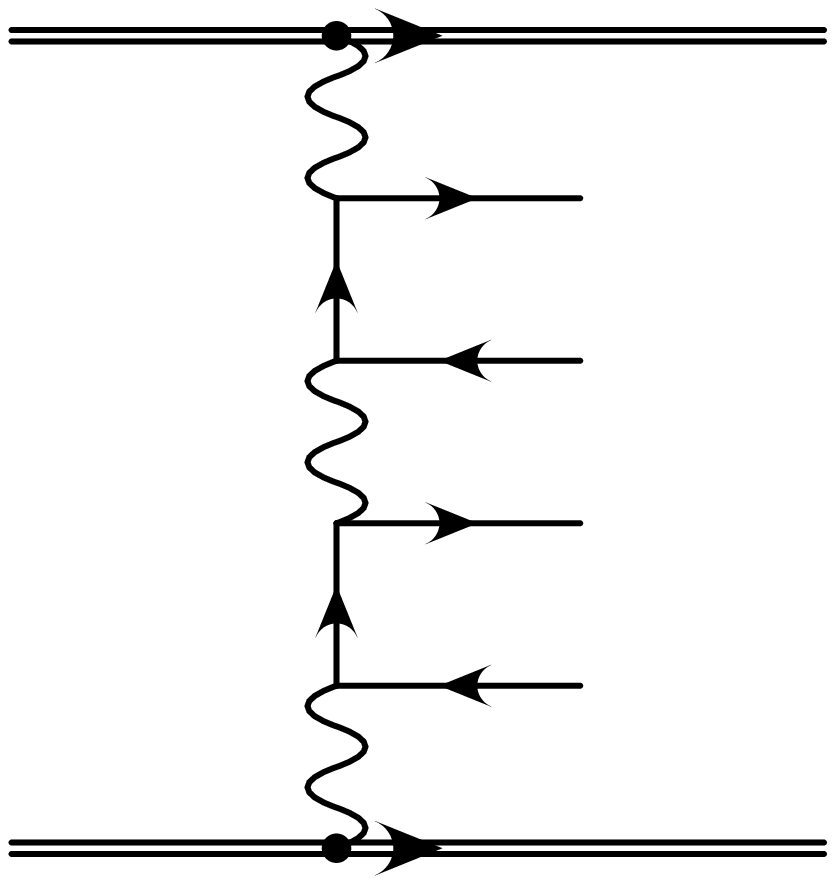}}
(b)
\end{center}
\caption{The dominant two pair production process at the energies of
RHIC and LHC is shown in (a). Due to $Z\alpha\sim1$ this process is favored,
giving a cross section of $\sim Z^8\alpha^8\ln^2\gamma_{lab}^2$. 
On the other hand
the process of (b) is proportional to $\sim Z^4\alpha^6\ln^4\gamma_{lab}^2$. It
will therefore take over at extremely high $\gamma_{lab}$.}
\label{fig_mpairsdiag}
\end{figure}

For $Z\alpha\sim1$ the interaction of the leptons with several photons 
coming from the external field will be much more likely than the exchange 
of photons among the singly charged leptons, see 
Fig.~\ref{fig_mpairsdiag}. Therefore we will neglect the lepton-lepton 
interaction, treating it as a pure external field problem.

This was first studied in \cite{Baur90,Baur90c}. Hereby, the $e^+e^-$ pair 
was treated
as a ``quasiboson''. Neglecting rescattering terms a closed expression was
obtained to all orders. A similar approach was also followed in 
\cite{RhoadesBrownW91}. In \cite{BestGS92} another approach was used, 
starting from a 
description within the Dirac sea picture. All studies essentially make
use of the ``quasiboson'' approximation, that is, treating the pair as an 
``unbreakable unit''. All find the probability
to produce $N$ pairs $P(N,b)$ to be given by a Poisson distribution:
\begin{equation}
P(N,b) = \frac{P^{(1)}(b)^{N}}{N!} \exp\left[-P^{(1)}(b)\right],
\label{eq_poisson}
\end{equation}
where $P^{(1)}(b)$, the probability for pair production calculated in 
lowest order 
(which, as shown before, can become larger than one), is actually
the ``average number of $e^+e^-$ pairs'' produced in a single ion collision: 
\begin{equation}
\left<N(b)\right> = \sum_N N P(N,b) = P^{(1)}(b).
\label{eq_averageN}
\end{equation}

One should keep in mind that the two-pair production process as shown in 
Fig.~\ref{fig_mpairsdiag}(a) is rising approximately with $\sim Z^4 \alpha^8
\ln^2(\gamma_{lab}^2)$, whereas the one shown in Fig.~\ref{fig_mpairsdiag}(b)
rises with $\sim Z^4 \alpha^6 \ln^4(\gamma_{lab}^2)$ \cite{Serbo70,LipatovF71,%
BudnevGM75} and will finally take over at extremely high values of 
$\gamma_{lab}$. We are still far away from this regime. On the other hand,
as mentioned in \cite{BudnevGM75} the multiple pair production at large
invariant masses is again dominated by Fig.~\ref{fig_mpairsdiag}(b).

The use of the Poisson distribution Eq.~(\ref{eq_poisson}) allows us to 
interpret the lowest order calculation as a calculation of  multiple pair 
production. Multiple pair production cross sections can be defined by 
integrating Eq.~(\ref{eq_averageN}) over $b$, for example:
\begin{equation}
\sigma_T=\int d^2b \left<N(b)\right> = \sum_N N \sigma(N).
\label{eq_sigmaNav}
\end{equation}

\begin{figure}[tbh]
\begin{center}
\resizebox{4.5cm}{!}{\includegraphics{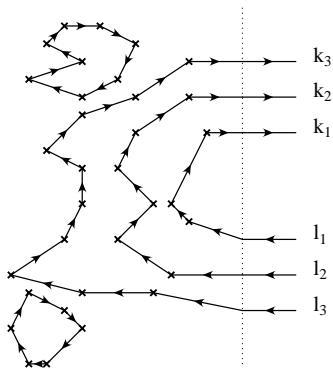}}
\end{center}
\caption{Graphical illustration of the general form of the $N$-pair production
process. The interaction with the external field is shown as crosses. The
production of a pair is described by a fermion line coming from and leaving
to the future, interacting an arbitrary number of times 
with the external field. Such a line corresponds to $s^{+-}$ in
Eq.~(\protect\ref{eq_mnpairs}). The vacuum-vacuum amplitude
$\left<0\right|S\left|0\right>$ corresponds to the sum of all closed
fermion loops. For details we refer to \protect\cite{HenckenTB95a}.}
\label{fig_pairproduction}
\end{figure}

It was shown in \cite{HenckenTB95a,AlscherHT97} that the matrix element for 
$N$-pair production $S_N$ factorizes quite generally into an antisymmetrized 
product of individual pair production amplitudes $s^{+-}$ 
(corresponding to individual fermion lines) and the vacuum-vacuum amplitude 
(corresponding to all closed fermion loops),
see also Fig.~\ref{fig_pairproduction}:
\begin{equation}
S_N = \left<0\right|S\left|0\right> \sum_\sigma \mbox{sgn}(\sigma)
s^{+-}_{k_1l_{\sigma(1)}} \cdots s^{+-}_{k_Nl_{\sigma(N)}},
\label{eq_mnpairs}
\end{equation}
where $k_i$,$l_i$ are the quantum numbers (momenta and spin projection)
of electron and positron, respectively, and $\sigma$ denotes a permutation of
$\{1,\cdots,N\}$. The vacuum-vacuum amplitude is present in all QED 
calculations,
but whereas it is of absolute value one without external fields or for static
fields (and is 
therefore factored out and dropped in the calculation), it is  $<1$ here, 
as pair creation occurs out of the initial vacuum state and the probability 
for no-pair production $\left|\left<0\right|S\left|0\right>\right|^2$ 
has to be smaller than one. This result was confirmed recently 
in \cite{BaltzGMP01}. The amplitude $s^{+-}$ corresponds to the one of 
single pair
production neglecting the vacuum-vacuum amplitude; it is the one that is
calculated, for example, in lowest order Born approximation.

\begin{figure}[tbh]
\begin{center}
\resizebox{3cm}{!}{\includegraphics{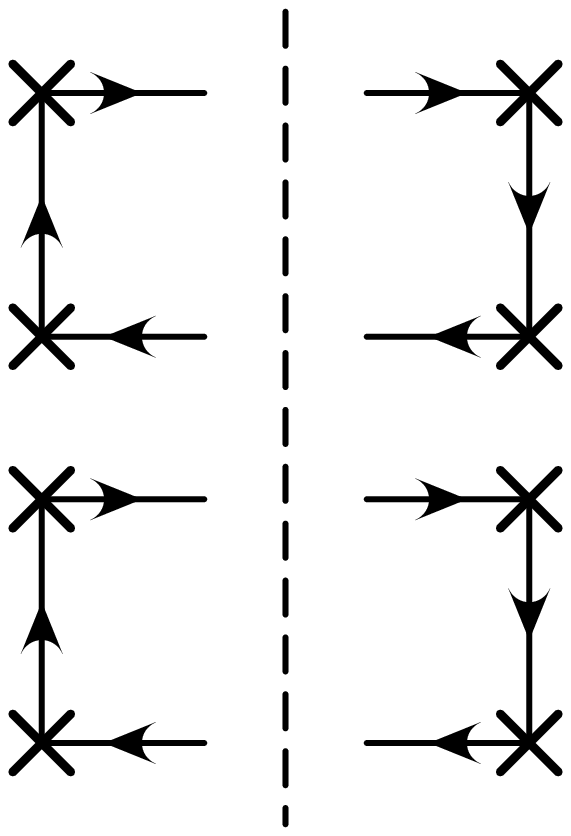}}
(a)~~~
\resizebox{3cm}{!}{\includegraphics{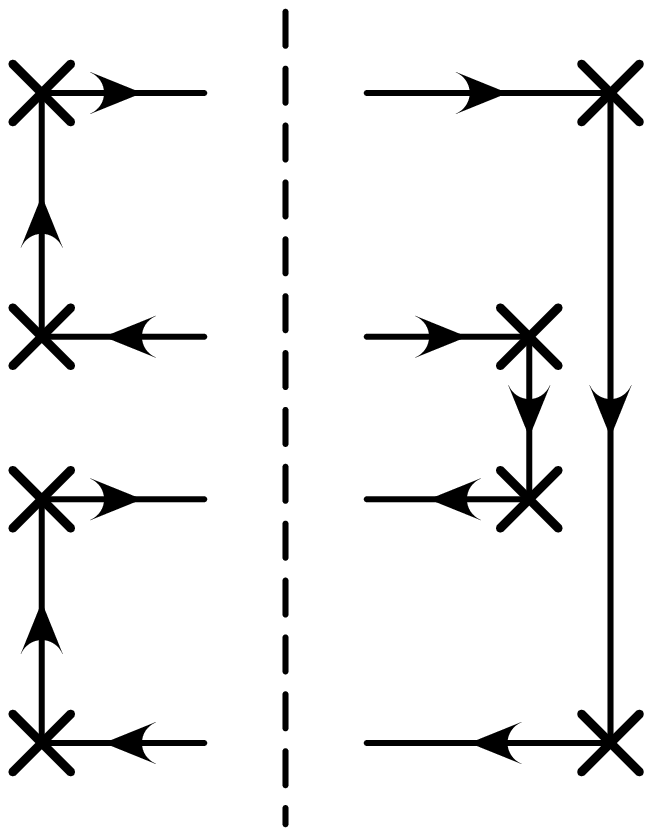}}
(b)
\end{center}
\caption{
The two pair production probability consists of two types of diagrams.
The class of diagrams (a), where electron and positron are always matched
in $S_N$ and $S_N^+$ leads to the Poisson distribution. Diagrams of type (b)
are assumed to be small due to the correlation of the momenta of
electrons and positrons.
}
\label{fig_2pairDXdiagrams}
\end{figure}

Calculating the total $N$ pair probability 
\begin{equation}
P(N,b) = \frac{1}{(N!)^2} \sum_{k_1,\cdots,k_N,l_1,\cdots,l_N} 
\left|S_N\right|^2
\end{equation}
one gets two different types of
contributions. In the first type --- see Fig.~\ref{fig_2pairDXdiagrams}(a) 
---
the same permutation $\sigma$ is present in $S_N$ and $S_N^+$, whereas
different ones are used in the second type 
--- see Fig.~\ref{fig_2pairDXdiagrams}(b). Neglecting this second class
--- and therefore the antisymmetrization in the final
state --- one recovers again the Poisson distribution Eq.~(\ref{eq_poisson}).

\begin{figure}[tbh]
\begin{center}
\resizebox{7cm}{!}{\includegraphics{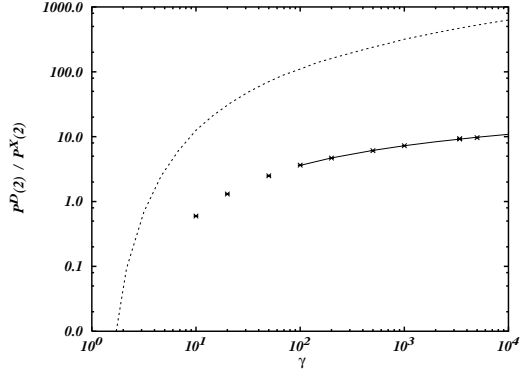}}
\end{center}
\caption{
The contribution of diagram~(a) and~(b) of 
Fig.~\protect\ref{fig_2pairDXdiagrams} 
to the two-pair production probability
at impact parameter $b\approx0$ is shown as a function of the Lorentz factor 
$\gamma_{lab}$ of the collision. The dotted line shows contributions from
the diagram of Fig.~\protect\ref{fig_2pairDXdiagrams}(a), the numerical
data points and 
solid line those of Fig.~\protect\ref{fig_2pairDXdiagrams}(b). Taken from
\protect\cite{Hencken94}.
}
\label{fig_dxcontribution}
\end{figure}

The fact that the second class of diagrams gives in general a small 
contribution as compared
to the first one can be understood due to the correlation of the electron
and positron momenta. They are not produced completely independently of each
other. Especially their transverse momenta are correlated; this is obvious in
the equivalent photon approximation where the sum of the transverse momenta 
is close to zero, which is also true in the exact calculation in lowest 
order. Therefore the 
``quasiboson'' approximation, that has been used in the early studies is 
justified. In addition the deviation from the Poisson distribution in the 
case of two pair production was calculated explicitly for small 
impact parameters $b\approx 0$ \cite{HenckenTB95a,Hencken94}. 
Figure~\ref{fig_dxcontribution} shows the contribution from the two diagrams 
of Figure~\ref{fig_2pairDXdiagrams}. At LHC energies the deviation is found 
to be only of the order of 1\%.

\begin{figure}[tbh]
\begin{center}
\resizebox{5.5cm}{!}{\includegraphics{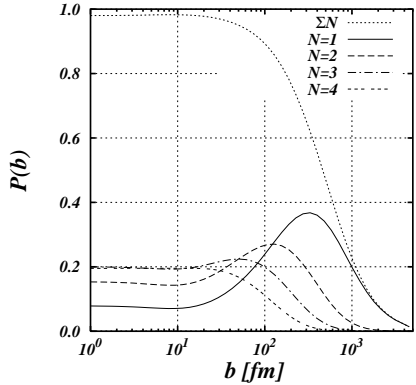}}
\end{center}
\caption{
The impact parameter dependent probability to produce $N$
$\EPEM$-pairs ($N=1,2,3,4$) in one collision is shown for the
 LHC ($\G_{lab}=2950$,PbPb). Also shown is the total probability to 
produce at least one $\EPEM$-pair $\sum_N P(N,b) =  1-P(0,b)$. 
One sees that at small impact parameters multiple pair production 
dominates over single pair production.
}
\label{fig_pbee1}
\end{figure}

Using Eq.~(\ref{eq_poisson}) one can use the impact parameter dependent 
probability $P(b)$ of \cite{HenckenTB95b,Guclu95} to obtain the 
probabilities for $N$-pair production $P(N,b)$, see Fig.~\ref{fig_pbee1}. 
One can see that for impact parameters $b \approx 2R$ up to about $1/m_e$
on the average 3--4 pairs will be produced in PbPb collisions at the LHC.
This means that each photon-photon event --- especially those with high 
invariant mass, which occur predominantly at impact parameters close to 
$b \gtrsim 2 R$ --- is accompanied by the production of several (low-energy)
$\EPEM$ pairs (most of them however will remain unobserved experimentally).
Integrating over the impact parameter the total multiple pair production cross
section were given in \cite{AlscherHT97}.
The single pair production probability $P(b)$ falls off essentially 
as $1/b^2$. The total cross section 
is dominated by large impact parameters and not very sensitive to the 
multiple pair production effects at small $b$. In \cite{LeeMS01} 
the reduction of the exclusive one-pair production cross section $\sigma(N=1)$
was estimated to be -6.4\% for RHIC and -4.7\% for LHC, using the numerical
results of \cite{HenckenTB95a,AlscherHT97}.
Multiple pair production cross section on the other hand are dominated by
impact parameters, around $b\approx\lambda_c =1/m_e$. Measuring them
would be of interest. At the SPS (CERN) the effect was looked for, see
\cite{VaneDDD97}, but 
only an upper bound could be given, which is still above the theoretical 
prediction.

\subsection{Strong Field Effects in Electron Pair Production:
Coulomb Corrections}
\label{sec_leptons_higherorder}

The calculation of the impact parameter dependent probabilities described 
above have been done in lowest order perturbation theory, even though 
$Z\alpha\approx0.7$ is not small. One could 
therefore ask how reliable these calculations are. Higher order effects 
(in addition to multiple pair production discussed in the previous 
section) could become important. This will be discussed now.

For low beam energies (around 1--2~GeV per nucleon) coupled
channel calculations have been made by a number of people 
\cite{MombergerGS91,RumrichSG93}.
They found a rather large increase of the cross section. For the higher
energies at RHIC and LHC such an approach is no longer viable due to the
large number of channels to consider \cite{MombergerBS95}
(For a calculation at $\gamma=200$, see \cite{ThielHGS95}).
Since the review in \cite{BaurHT98} a number of papers appeared treating 
higher order effects in ultrarelativistic heavy ion collisions in the 
high energy limit.

In the following we are concerned only with higher order corrections to the 
single fermion line, that is to $s^{+-}$ of Eq.~(\ref{eq_mnpairs}), 
assuming that the Poisson distribution can then be used 
to account for multiple pair production. These effects are generally 
called  ``Coulomb corrections'', as they mostly deal with the Coulomb 
distortion of both the electron and positron wave functions due to 
additionally exchanged
photons. Formally two kinds of classes can be distinguished. In the first one
only Coulomb rescattering will be considered, that is, only one pair will be
present at all intermediate time steps. The second one, which we call 
``multiple particle corrections'', are those processes, where more than one 
pair is present at an intermediate time step, with an electron and a positron
from two different pairs annihilating at a later step.
Such a distinction is less useful in the usual Feynman approach, as 
intermediate
lines describe both electrons and positrons, but it is well
defined in a retarded boundary condition approach, corresponding to the
Dirac sea picture.

The second class of diagrams was studied in \cite{HenckenTB95a} for small
impact parameters $b\approx0$ within the framework of the Magnus theory and
neglecting all Coulomb rescattering terms. The effect was found to be rather 
small, at most 5\% for PbPb collisions at LHC. As these effects are expected
to be largest for the small impact parameters, one has to conclude that their
effects will be rather small in the cross section.
We therefore consider the first class in the following.

A classical result for this first class of higher-order effects 
can be found in the Bethe-Heitler formula for 
the process $\gamma+Z \rightarrow e^+ + e^- +Z$ (corresponding to the 
highly asymmetric case $Z_1\alpha\rightarrow 0$, $Z_2 \alpha\approx 1$):
one obtains for an unscreened nucleus
\BE
\sigma = \frac{28}{9} \frac{(Z\A)^2}{m_e^2} \left[ \ln \frac{2 \omega}{m_e} -
\frac{109}{42} - f(\ZA) \right] ,
\EE
with the higher-order term given by
\begin{equation}
f(\ZA) = (\ZA)^2 \sum_{n=1}^{\infty} \frac{1}{n(n^2+(\ZA)^2)}  
= \gamma + \mbox{Re} \psi(1+i \ZA)
\end{equation}
with the Euler constant $\gamma\approx0.57721$ and $\psi$ the Psi (or Digamma)
function. As far as total cross sections are concerned the higher-order 
contributions tends to a constant for $\omega \rightarrow \infty$. 

\begin{figure}[tbh]
\begin{center}
\resizebox{3cm}{!}{\includegraphics{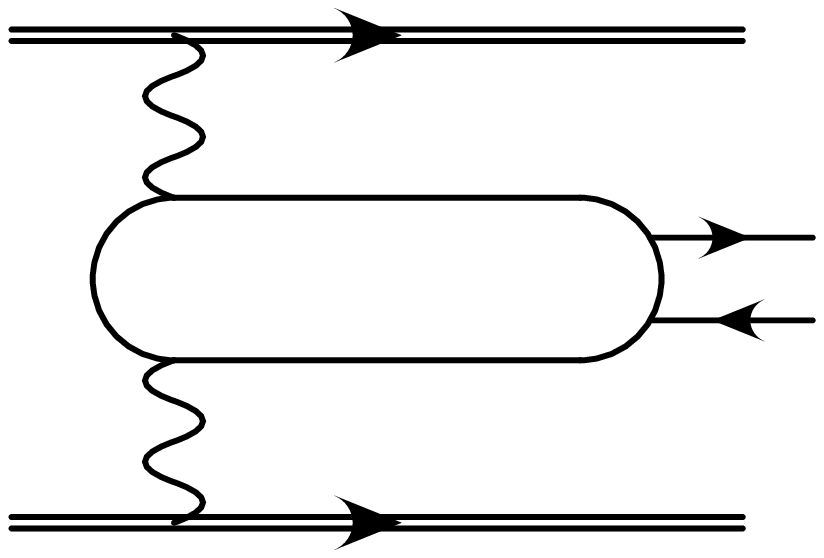}}
(a)~~~~
\resizebox{3cm}{!}{\includegraphics{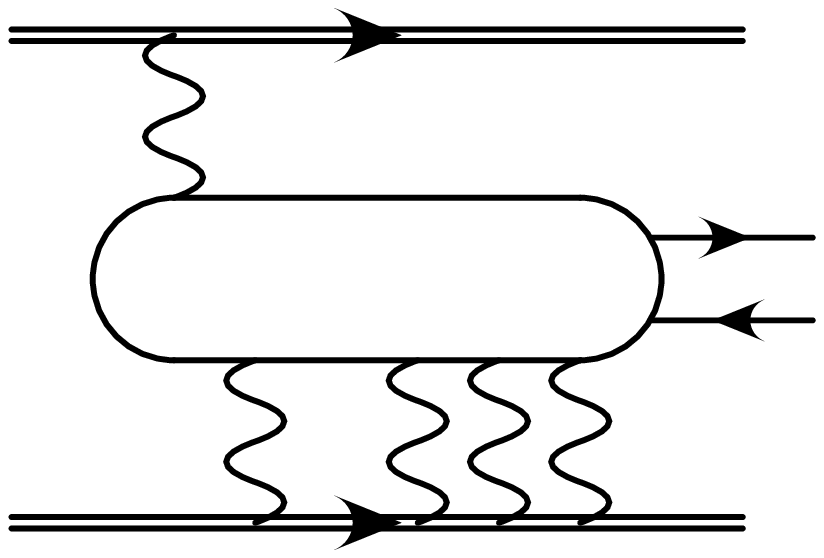}}
(b)
\\
\resizebox{3cm}{!}{\includegraphics{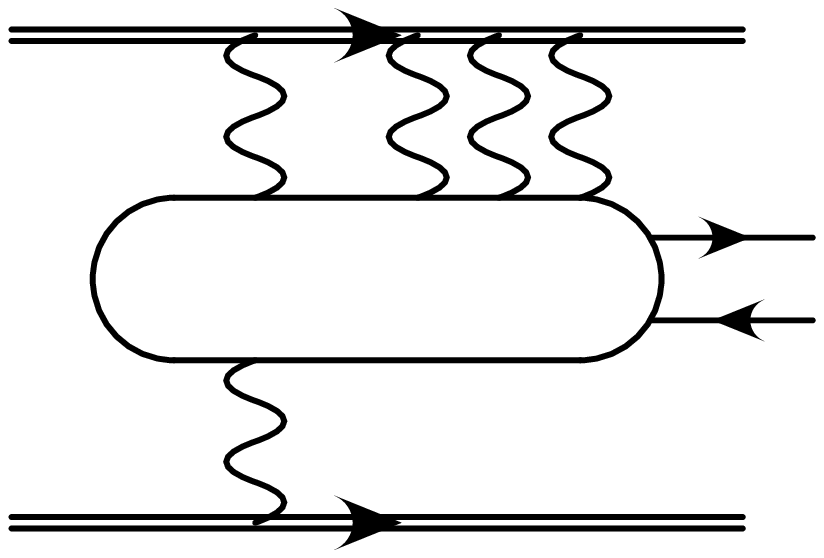}}
(c)~~~~
\resizebox{3cm}{!}{\includegraphics{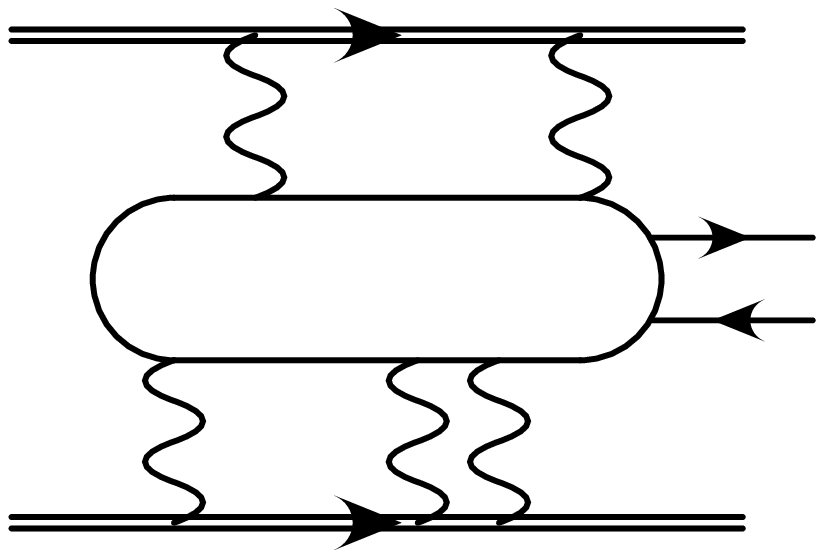}}
(d)
\end{center}
\caption{Contributions to the single pair production cross section can
be classified by the number of photons that is emitted by each ion.
If each ion emits only one photon (a) the cross section has a $\ln^3\G$
dependence. If only one of the ions emits only one photon ((b) and (c)),
only a $\ln^2\G$ dependence is found. Finally with both ions emitting several
photons (d), only a $\ln\G$ dependence is expected.}
\label{fig_serbo}
\end{figure}

This is used as the basis for a detailed calculation of higher order effects
on the (single pair) cross section in \cite{IvanovSS99}. The higher order 
effects were classified according to the number of photons $n$ and $n'$
exchanged with each ion, see Fig.~\ref{fig_serbo}. The dominant term is 
given by the $n=n'=1$ contribution (Fig.~\ref{fig_serbo}(a)), due to the 
$1/b^2$ behavior of the single photon exchange. It leads to the famous 
$ln^3\gamma$ rise of the (total one pair production) cross section. 
The next important terms are those with $n=1$, $n'>1$ and $n>1$, $n'=1$
(see Fig.~\ref{fig_serbo}(b) and (c)). They are enhanced proportional to
$\sim \ln^2\gamma$. The last class of diagrams will only be 
$\sim \ln \gamma$ and is therefore neglected in \cite{IvanovSS99}.
A systematic way to take the leading terms into 
account in $\EPEM$ pair production was used in \cite{IvanovM97,IvanovSS99},
leading to a result which corresponds to the Bethe-Heitler result above
and is therefore in accord with the Bethe-Maximon corrections 
\cite{BetheM54,DaviesBM54,LandauLQED}.
The authors find corrections to the total Born cross sections,
which are negative and equal to -25\% and -14\% for RHIC and LHC.
Therefore the Coulomb corrections in the single pair production cross section
(better $\sigma_T$ as defined in 
Eq.~(\ref{eq_sigmaNav}), as this is what is really calculated) 
seem to be well understood.
A negative correction was also found in the work of \cite{BertulaniB88}. 
These theoretical developments are further supported in \cite{LeeM00}.

Please note that such a classification according to $n$ and $n'$,
resulting in different powers of $\ln\G$ is not helpful for multiple pair 
production, as at least two photons must be emitted from each nucleus 
in this case (corresponding always to the case $n>1,n'>1$ above) and one will 
not have a cross section proportional to $\ln^3\gamma$ from the beginning. 
A different approach, which is good in the small impact parameter region, is 
therefore needed.

\begin{figure}
\begin{center}
\resizebox{!}{6cm}{\includegraphics{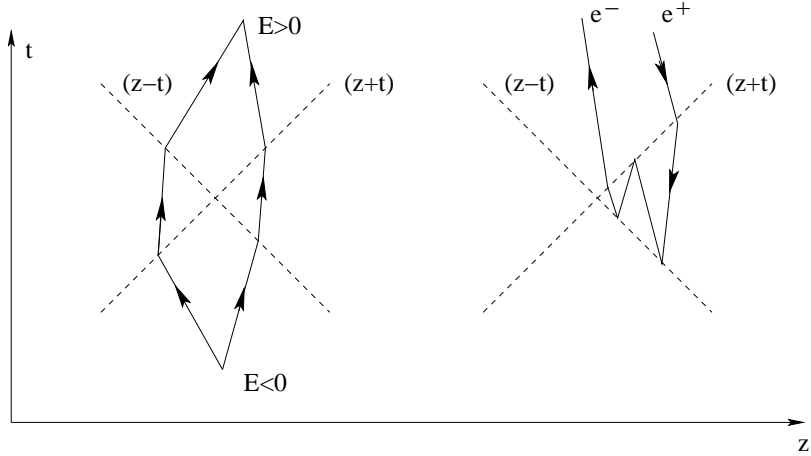}}
\end{center}
\caption{The special structure of the electromagnetic interaction can be
seen in this $t$-$z$ plot. Due to the Lorentz contraction the electromagnetic
fields are localized in two sheets corresponding to $z=\pm t$. In the 
``retarded'' or ``Dirac sea'' approach (left) an electron with negative 
energies comes from $t=-\infty$, crosses the field of each ions
only once before leaving as an electron with positive energy. In the ``Feynman''
approach, the positron comes from $t=+\infty$ and can go forward and backward
in time therefore interacting a number of times with the ions.}
\label{fig_retarded}
\end{figure}

In a series of papers the exact solution of the Dirac equation of an 
electron in the field of the two nuclei in the limit of the Lorentz 
factor $\gamma_{lab}\rightarrow\infty$ was studied.
The interest started after an article showed how the summation to all orders
in the high energy limit can be done in the related problem of bound-free pair
production \cite{Baltz97}, see also below. In a first article \cite{SegevW98}
it was shown that also in the free pair production case the summation to
all orders can be done analytically. Due to the form of the 
electromagnetic field of the two ions in the high energy limit 
--- both of them being essentially localized on two different sheets 
corresponding to $z = \pm t$, where $z$ is the direction of the beams
 --- only a certain class of diagrams was 
found to be dominant in this case, see Fig.~\ref{fig_retarded}. 
The sum of the interactions with one of the ions in the highly energetic case 
gives the typical eikonal phase. It was shown that this leads to a matrix 
element very similar to the one in lowest order, but where the photon 
propagator is replaced with
\begin{equation}
\frac{1}{q_\perp^2+q_l^2} \rightarrow \frac{1}{\left(q_\perp^2+q_l^2\right)^{1-iZ\alpha}}.
\label{eq_ieta}
\end{equation}
The authors of \cite{BaltzM98} come to the same conclusion; they also show
that by integrating over the impact parameter the cross section becomes 
identical to the lowest order Born result, as only the absolute values squared
of the modified photon propagators Eq.~(\ref{eq_ieta}) are needed. The same 
result is also found in \cite{SegevW98b}. In \cite{EichmannRSG99} the 
scattering of electrons is studied, which is then related via crossing 
invariance to the pair production process.

A numerical calculation of the impact parameter dependent probability was
done in \cite{HenckenTB99} and also in \cite{Guclu00} (see also the two
comments following \cite{Guclu00}). Calculating the impact
parameter dependent probability in this approach a reduction of up to
50\% was found for small impact parameters, leading to the same reduction for
the multiple pair cross section. Therefore Coulomb corrections are expected
to be important also at small impact parameters.

The fact that the cross section to all orders corresponds to the lowest order 
Born result
is obviously in contradiction to the Bethe Maximon corrections, as described
above. This can be seen from Eq.~(7.4.3) of \cite{BertulaniB88}
and  it was  pointed out in \cite{IvanovSS99} and later also in
\cite{LeeM00,EichmannRG00}. In \cite{EichmannRG00} it was 
noted that in the limit of $\gamma\rightarrow\infty$, different types of 
diagrams are of importance in the electron scattering case and in the pair 
production case, that is, taking only the dominant diagrams in both cases
will violate crossing symmetry. In \cite{LeeM00} the question of how to do 
this limit is studied in more detail. By making a careful transition to 
$\gamma\rightarrow\infty$ the Bethe Maximon corrections were found 
to be present also in this case.
The failure of the eikonal technique to reproduce the Bethe Maximon results
was already noted in \cite{BlankenbeclerD87}.

The equality of the nonperturbative cross section to all orders 
with lowest order result can therefore be seen as 
an artifact of the regularization of the modified photon propagator.
Eq.~(\ref{eq_ieta}) is found by taking the limit $\gamma\rightarrow\infty$
first; in this limit the modified photon propagator is 
$1/(q_\perp^2)^{1-i\eta}$. Taking this limit --- which essentially 
corresponds to the sudden approximation --- is justified only
for small impact parameters. In fact integrating this expression over 
$b$ would lead to a diverging cross section, essentially due
to the singular behavior of the propagator for $q_\perp \rightarrow 0$. 
Therefore in \cite{BaltzM98} the form of the propagator Eq.~(\ref{eq_ieta}) 
was suggested. It is mainly this form of the propagator which
in the end leads to the equality of the
two cross sections. A more careful regularisation of the propagator
would lead to an expression in agreement with the Bethe Maximon 
theory \cite{LeeM01}. On the other hand this means that the probabilities
for small impact parameter, which get a large part of their contribution from
large $q_\perp$ should be expected to be reliable.

In all derivations of the pair production in the high energy limit 
\cite{SegevW98,BaltzM98,EichmannRSG99} the pair production are not 
calculated using Feynman boundary conditions, that is,
treating the positrons as particles going backward in time 
\cite{Feynman49,Schwinger54,ItzyksonZ80,BialynickiB75}. Instead a retarded 
approach, that is, essentially the ``Dirac sea'' picture is used:
the problem is treated as the scattering of a negative energy electron into
a positive energy state, see, e.g., \cite{Feynman49,BialynickiB75}.
In this approach the structure of the interaction with the external field can
be exploited easily, see Fig.~\ref{fig_retarded}.

This has two consequences: the total cross section as well as the differential 
cross section with respect to either the electron or the positron should be 
reliable, but not the differential cross section with respect to both 
leptons, as the ``rescattering of the hole'' is not taken into account 
\cite{LeeM00}. In addition in \cite{BaltzGMP01} it is shown that the cross 
section calculated in this way does not correspond to the exclusive one-pair 
production $\sigma_1$ but to $\sigma_T$, as defined in 
Eq.~(\ref{eq_sigmaNav}). 
This result was already found there to be true for the calculation of
the single-pair production cross section, neglecting the vacuum-vacuum 
amplitude. But whereas the Poisson distribution
was needed there, it is found in \cite{BaltzGMP01} to be exact in the general 
case as well.

Single and multiple pair production in charged particle (heavy ion) collisions
without nuclear contact is a QED process and should in principle
be calculable precisely with the methods of QED. Although the calculations 
in detail are quite complicated, the theoretical concepts are well understood 
by now. Especially the importance of higher order corrections due to the 
multiple
pair production cross section would be an interesting test. First experimental 
observations of $e^+e^-$ pairs has been reported already by STAR at RHIC 
\cite{Spencer01NEW}. A comparison with experiments will hopefully be 
possible soon.

\subsection{Electromagnetic Production 
of Electron Pairs and Other Particles in Central Collisions
}
\label{sec_electrons_central}

Differential production probabilities for dileptons from $\GG$ processes
in central 
relativistic heavy ion collisions are calculated using the equivalent
photon approximation in an impact parameter formulation and are compared 
to Drell-Yan and thermal ones in \cite{Baur92,BaurB93b,Baur92b}. The very 
low $p_\perp$ values and the angular distribution of the pairs give a handle 
for their discrimination to those coming from other 
sources, e.g., from meson decay.

In \cite{HenckenTB00} it is shown that particle production 
in central collisions via the $\GG$-mechanism is so small that it 
can be neglected in practice. The probability to produce a 
final state with invariant mass M and rapidity Y is given by 
\BE
\frac{d^2P}{dMdY}=\frac{Z_1 Z_2 \alpha^2}{\pi^3 R^2}
\left(1-a^2 \exp(-2\left|Y\right|)\right) 
\frac{2\sigma_{TT}(M^2)}{M},
\EE
where $\sigma_{TT}$ is the corresponding $\GG \rightarrow f$
fusion cross-section and $a=\frac{RM}{2\gamma}$ ($R$
is the larger one of the nuclear radii). This probability 
is very small, since we typically have $\sigma_{TT}\ll R^2$. 
Electromagnetically produced electrons were also measured 
in distant $S-Pt$ collisions in  the CERES/NA45 experiment
\cite{BaurCERES94}. The authors also conclude
that ``{\it the rate of $e^+e^-$ pairs from QED production in collisions with 
nuclear overlap for the CERES acceptance is negligible compared to
$e^+e^-$ pairs from meson decays}''. For further details we refer
to \cite{HenckenTB00,BaurCERES94}.

\subsection{Electrons and Muons as Equivalent Particles,
Equivalent Electron (Muon) Approximation}
\label{sec:equimuon}

The equivalent photon approximation can be generalized to 
other particles \cite{ChenZ75,BaierFK73}. A light or zero
mass particle moving with high energy splits into a pair of light or zero mass
 particles. The constituents of this daughter pair 
subsequently react as "equivalent beams" with the target system.
This subprocess is easier to describe than the original reaction.
Here we are interested in those processes, where the 
(equivalent) photon splits into a lepton pair and the 
(equivalent) leptons induce reactions with the target,
e.g., deep inelastic lepton scattering.
(The discussion here has some similarity to the 
more general one in Chap.~\ref{sec_ggqcd} above,
where also vector mesons and quark pairs are considered.)

The production of lepton pairs from (real or virtual) photons producing
a pair by inelastic scattering from a nuclear target was first studied in 
\cite{DrellW64}, see also \cite{Tsai74}, where it was mostly regarded as an
additional contribution to the (elastic) pair production.
Now we suggest that the equivalent photons may also be used as a 
lepton beam to study deep inelastic lepton scattering from a nucleus.

\begin{figure}[tbh]
\begin{center}
\resizebox{!}{4cm}{\includegraphics{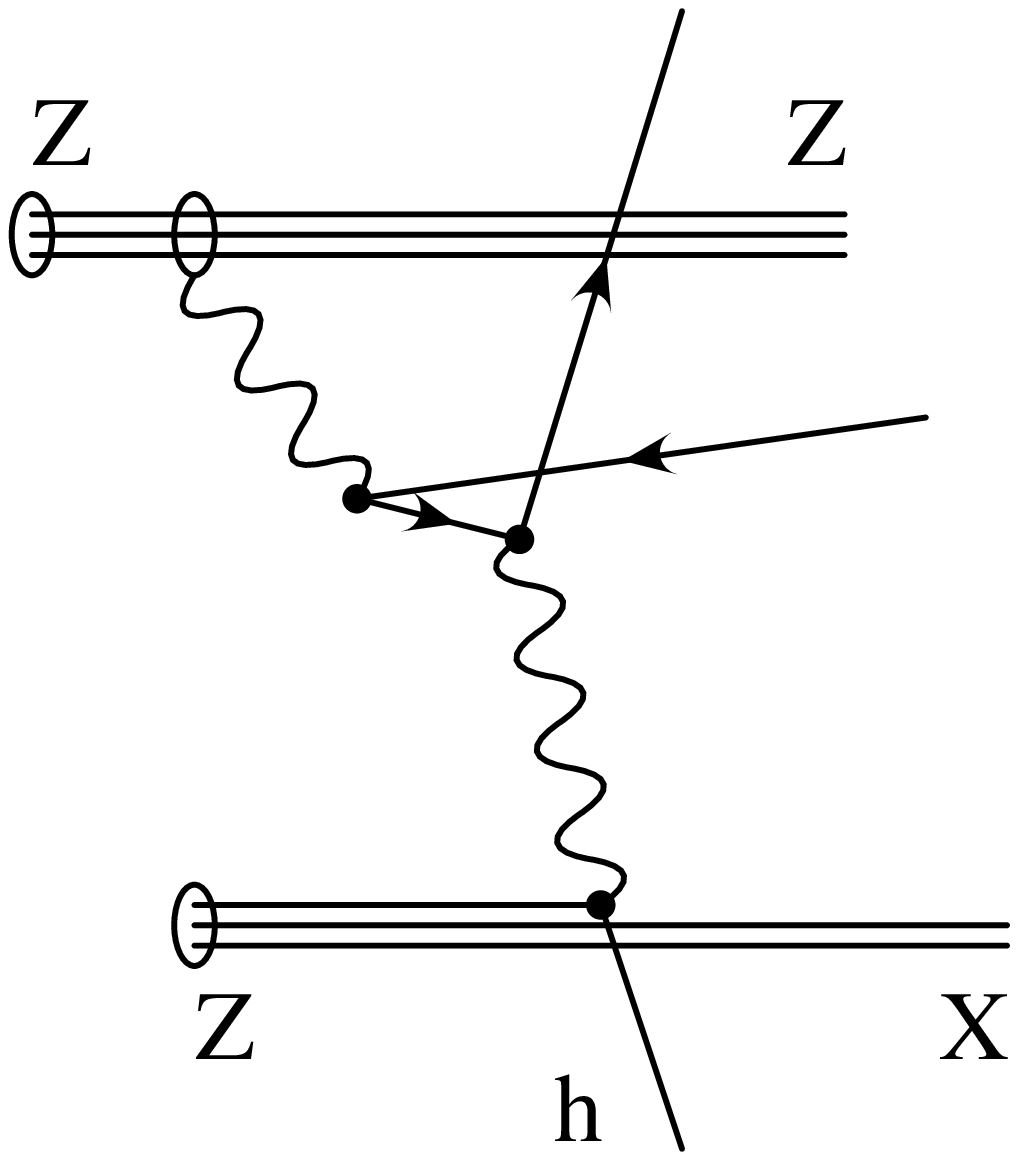}}
~~~
\resizebox{!}{4cm}{\includegraphics{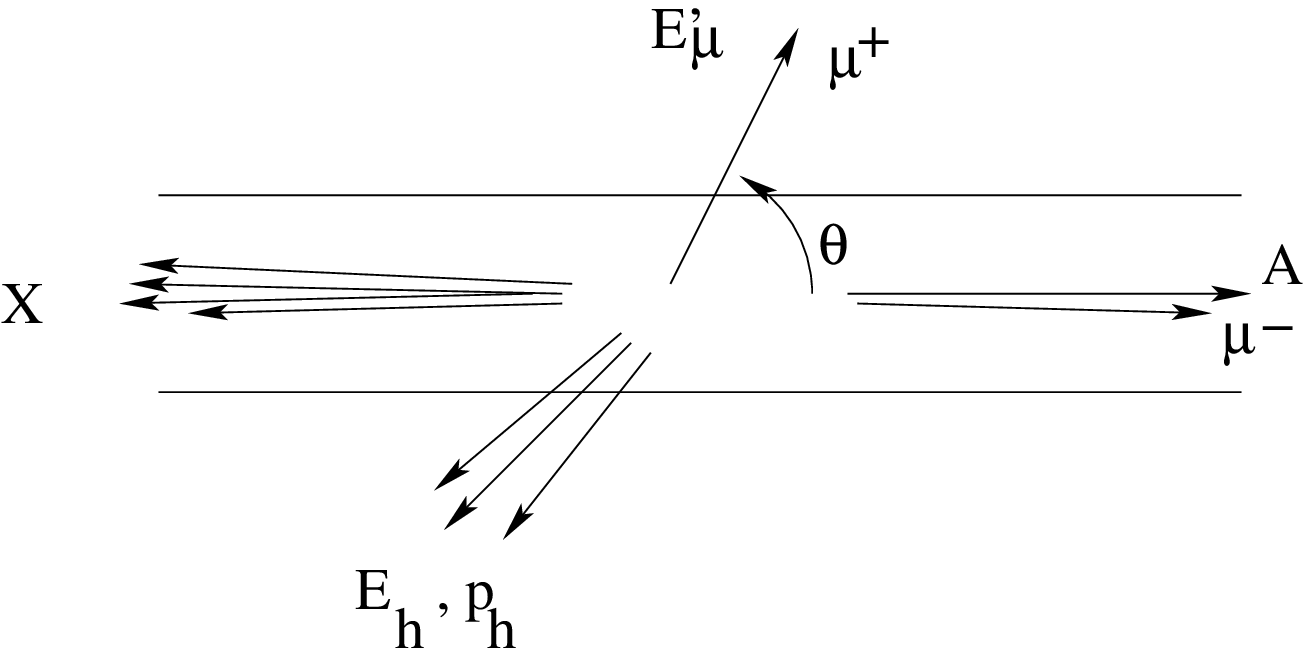}}
\end{center}
\caption{
Events where the ``equivalent muon'' is scattered deep inelastically on
the other ion are characterized by one muon being emitted with a large
transverse momentum, whereas the other muon escapes unobserved 
(together with the remnants of the struck ion)
 in the forward direction. Energy and momentum
of the struck parton can be used to infer the energy of the (initial)
equivalent muon.}
\label{fig_emadis}
\end{figure}

Up to now only the production of dileptons in heavy ion collisions was 
considered, for which the four-momentum $Q^2$ of the photons was less than 
about $1/R^2$ (coherent interactions). There is another class of processes, 
where one of the interactions is coherent ($Q^2 \le 1/R^2$) and the other
one involves a deep inelastic interaction ($Q^2\gg 1/R^2$), see
Fig.~\ref{fig_emadis}. These processes are much rarer than the events 
with two coherent interactions. But they may also be of 
interest in the future. They are readily described theoretically using the
equivalent electron-- (or muon--, or tau--) approximation, as given,
e.g., in \cite{ChenZ75,BaierFK73}. They are characterized by
the fact that the lepton is almost on-shell (i.e. the propagator 
of the virtual particle will become very large, since its momentum squared
$|p^2| \simeq m^2$).
In the following we will speak of muons, but the same considerations 
apply also to electrons.

The equivalent photon can be considered as containing 
muons as partons. The equivalent muon number is given by \cite{ChenZ75}:
\BE
f_{\mu/\G} ( \omega,x) = \frac{\alpha}{\pi}
\ln\left(\frac{\omega}{m_\mu}\right) \left[ x^2 + (1-x)^2 \right],
\EE
where $m_\mu$ denotes the muon mass. The energy $E_\mu$ 
of the equivalent muon is given by
$E_\mu=x \omega$, where $\omega$ is the energy of the equivalent photon.
This spectrum has to be folded with the equivalent photon spectrum given 
approximately by, see Eq.~(\ref{eq:epaapprox}):
\BE
f_{\G/Z}(u) = \frac{2\alpha}{\pi} \frac{Z^2}{u} \ln\left[ \frac{1}{u
m_A R}\right].
\EE
For $u<u_{max}=\frac{1}{R m_A}$ we then get:
\BE
f_{\mu/Z} (x_1) = \int_{x_1}^{u_{max}} du f_{\G/Z} (u) f_{\mu/\G}(uE_A,x_1/u).
\EE
The energy of the heavy ion is denoted by $E_A$, the photon energy is given
by $\omega=u E_A$,and the muon energy by $E_\mu=x_1 E_A=uxE_A$.
This expression can be calculated analytically.

The deep inelastic lepton-nucleon scattering can now be calculated in
terms of the structure functions $F_1$ and $F_2$ of the nucleon.
It would be interesting to see how the nucleon structure functions 
$F_1 $ and $F_2$ are modified in the nuclear medium. (It remains however
unclear at present how accurate such measurements could be.)
The inclusive cross section for the  deep-inelastic scattering of the
equivalent muons is  given by
\BE
\frac{d^2\sigma}{dE_\mu' d\Omega} = \int dx_1 f_{\mu/Z}(x_1)
\frac{d^2\sigma}{dE_\mu' d\Omega} (x_1),
\EE
where $\frac{d^2\sigma}{dE_\mu' d\Omega} (x_1)$ can be calculated from the 
usual invariant variables in deep inelastic lepton scattering (see, e.g., 
Eq.~(35.2) of \cite{PDG96}). Here the scattering angle of the lepton is 
$\theta$ and its energy $E'$. The accompanying muon of opposite charge,
as well as the remnants of the struck nucleus, will scatter to small
angles and generally remain unobserved, see Fig.~\ref{fig_emadis}.
The hadrons scattered to large angles
can be observed, with total energy $E_h$ and momentum in the beam
direction of $p_{zh}$. One can use the Jacquet-Blondel variable 
$y_{JB}$ defined as 
\BE
y_{JB}=\frac{E_h-p_{zh}}{2E_\mu}.
\label{eq:yjblondel}
\EE
 The missing hadrons which do not enter the detector 
(which we assume to be almost $4\pi$) have a small $p_T$, and thus their
contribution to the numerator $E_h-p_{zh}$ is small
(ultrarelativistic particles are assumed). One gets a
good estimate of $y$ using this method. Combining this with Eq.~(3.27) 
of \cite{Levy97} the initial energy of the equivalent muon $E_\mu$ can in 
principle be reconstructed as
\BE
E_\mu = \frac{1}{2} \left[ E_h - p_{zh} + E_\mu' (1-\cos\theta)\right].
\EE

This is quite similar to the situation at HERA, with the difference
that the energy of the lepton beam is continuous, and its energy has
to be reconstructed from the kinematics (How well this can be done in
practice remains to be seen).
Of course the struck parton also has to go through
the nucleus, and there is the possibility of FSI with it. This could
introduce some uncertainty in the reconstruction.

Quite similar also the excitation of individual nuclear states is
possible in this way. As these states can decay electromagnetically, they
are a possible source of high energetic photons. Recently this excitation
was discussed \cite{KorotkikhC01} for a related process, where pairs are
produced first and then excite the nuclear levels in a second step. It is
astonishing that the authors found a quite large cross section of 5.1~barn
for this process in the case of CaCa collisions at the LHC.

\subsection{Radiation from $\EPEM$-Pairs}
The bremsstrahlung in very peripheral relativistic heavy ion collisions was
found to be small, both for real \cite{BertulaniB88} and virtual
\cite{MeierHTB98} bremsstrahlung photons. This is due to the large
mass of the heavy ions.
Since the cross section for $\EPEM$ pair production is so large, one
may  expect to see sizeable effects from the radiation of these light
mass particles. For $e^+e^-$ colliders it was found that
{\it``at large values of $s$ the photon emission process in electroproduction
of the $e^+e^-$ pair \dots becomes very important (and in a number of cases,
decisive)''} \cite{FadinK73}.

\begin{figure}[tbh]
\begin{center}
\resizebox{3cm}{!}{\includegraphics{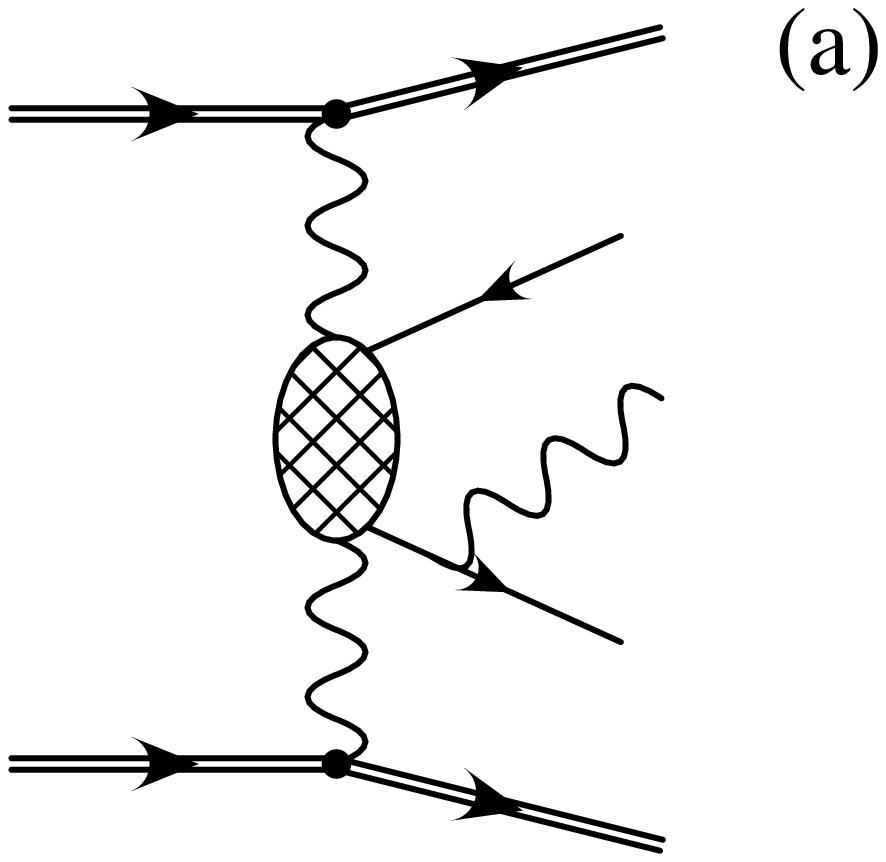}}
~~~~
\resizebox{3cm}{!}{\includegraphics{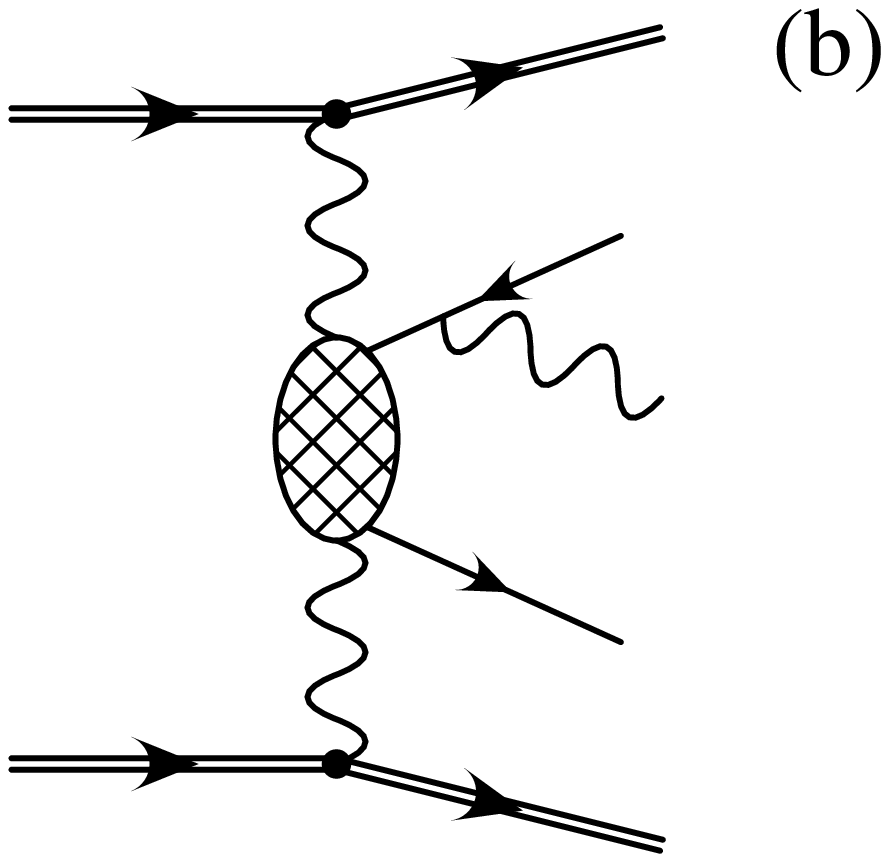}}
~~~~
\resizebox{3cm}{!}{\includegraphics{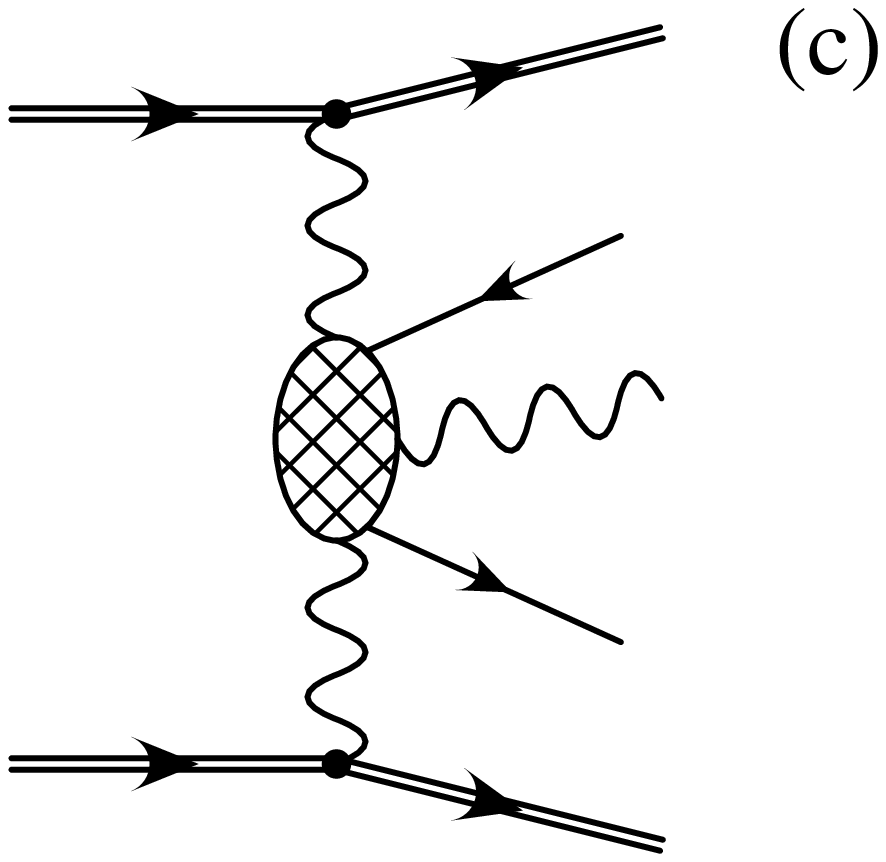}}
\end{center}
\caption{Emission of a bremsstrahlung photon from pairs produced 
electromagnetically. Process (a) and (b) are dominant in the infrared regime.
For the calculation either the IR approximation together with a full 
calculation of the pair production was used or a full calculation of
all three diagrams together with  photon spectra from double equivalent photon
approximation (DEPA). Similar diagrams, where the photon line is attached to 
the ions are generally small due to the heavy mass of the ions. They are
neglected here.}
\label{fig_eebrems}
\end{figure}

In the soft photon limit (see, e.g., \cite{Weinberg97}) one
can calculate the cross section for $e^+e^-$ pair production 
accompanied by soft
photon emission
\BE
Z + Z \rightarrow Z + Z + e^+ + e^- + \gamma
\EE
as
\BE
d\sigma(k,p_-,p_+) = 
- e^2 \left[ \frac{p_-}{p_- k} - \frac{p_+}{p_+ k} \right]^2
\frac{d^3k}{4 \pi^2 \omega} d\sigma_0(p_+,p_-),
\EE
where $d\sigma_0$ denotes the cross section for the $e^+ e^-$ pair
production in heavy ion collisions
(without soft photon emission). The corresponding 
Feynman graphs are given in Fig~\ref{fig_eebrems}(a)+(b). 
An alternative approach is 
to  use  the double equivalent photon approximation  and
calculate  the (lowest order) matrix element for the QED subprocess
$$
\gamma + \gamma \rightarrow e^+ + e^- + \gamma.
\label{eq_ggbrems}
$$

\begin{figure}[tbh]
\begin{center}
\resizebox{6cm}{!}{\includegraphics{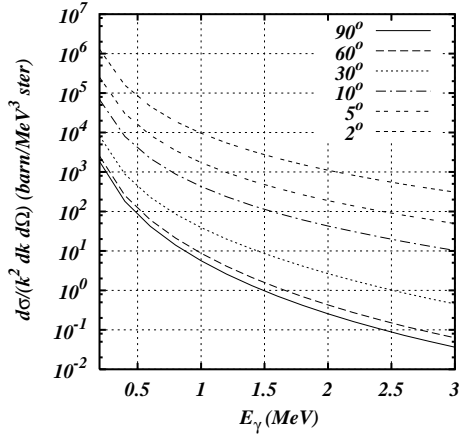}}
\end{center}
\caption{
The energy-dependence of bremsstrahlung-photons from
$e^+ e^-$ pair production is shown for different angles. We show
results for PbPb collisions at LHC.} 
\label{fig_brems}
\end{figure}

In Fig.~\ref{fig_brems} we show results of calculations for soft 
photon emission. The cross section for the QED process 
(Eq.~(\ref{eq_ggbrems}) above) in lowest order was used and folded with the 
corresponding  (double) equivalent photon spectra \cite{HenckenTB99b}.

These low energy photons might constitute a background for the
detectors. Unlike the low energy electrons and positrons, they are 
not bent away by the magnets. The angular distribution of the
photons also peak at small angles, but again a substantial amount is
still left at larger angles, even at $90^o$. The typical energy of
these low energy photons is of the order of several MeV, i. e., much
smaller than the expected level of the energy equivalent noise in the
CMS ECALs \cite{CMS-ECAL-TDR}.

\subsection{Bound-free Pair Production}
\label{ssec:boundfree}

The bound-free pair production, also known as ``electron-pair production
with capture'', is a process which is also of practical importance for 
relativistic heavy ion  colliders. An electron-positron pair is produced,
where the electron is not  free  but bound in an atomic state of one
of the ions:
\BE
Z_1+Z_2 \rightarrow (Z_1+e^-)_{1s_{1/2},\cdots} + e^+ + Z_2.
\label{eq_boundfree}
\EE

 As this changes the charge state of the nucleus, it is
lost from the beam in the collider.
 Together with the electromagnetic dissociation of
the nuclei (see Sec.~\ref{photonuclear} above) these two processes are the
dominant loss processes for heavy ion colliders. It has only been
realized recently \cite{Klein01}, see also \cite{Brandt00,Jeanneret00},
that this process can also result in a localized beam-pipe heating: 
the atomic states are produced with a perpendicular momentum of the order 
of $m_e$, which is rather small and therefore leads to a narrow singly-charged
ion beam. These beams with altered
magnetic rigidity will deposit their energy in a localized region of 
the beam pipe and  cause a  localized heating. This can lead to
the quenching of the superconducting magnets. 
The energy deposited per unit time is of the order of Watts.
This limits the luminosity of the PbPb collider at LHC \cite{Brandt00}. 
Due to the lower beam energy at RHIC, the energy deposit there is
smaller and therefore less important.  

From a simple kind of coalescence model, a scaling law for the cross-section 
$\sigma_n$ for bound-free pair production of the form
\BE
\sigma_n \sim  \frac{Z_1^5 Z_2^2}{n^3}
\label{eq_sigmascaling}
\EE
can be derived, where $n$ is the principal quantum number of the bound state
and only $s$-states are populated \cite{MungerBS94,Nemenov85}.
It is easiest to see this in a system where the nucleus $Z_1$
is at rest. 
\begin{figure}[tbh]
\begin{center}
\resizebox{4cm}{!}{\includegraphics{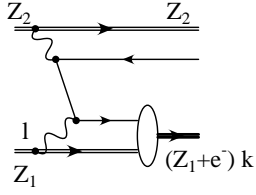}}
\end{center}
\caption{
A typical graph describing the bound-free pair production process, see 
\protect\cite{MungerBS94}.}
\label{fig_bfreediag}
\end{figure}

A typical graph is shown in Fig.~\ref{fig_bfreediag}.
An atom $(Z_1+e^-)$ with momentum $\vec k$ 
is produced. The bound state wave function is a strongly peaked function of 
the relative momentum $\vec\kappa_{rel}=\vec l-(M_{Z_1}/M_{(Z_1+e^-)})\vec k
\sim \vec l-\vec k$ where $\vec l$ is the three-momentum of the photon 
exchanged 
with the target. This momentum is of the order of $m_e$, while the
momentum $\kappa_{rel}$ is of the order of $Z_1 \alpha m_e$
(i.e. $Z_1$ times the Bohr momentum $p_{Bohr}=\alpha m_e$.). 
Since this momentum $l$ is usually much larger
than $\kappa_{rel}$ the loop integration over the relative momentum 
factors out and the factor $\Psi(0)\sim Z^{3/2}$ is found, where $\Psi$
is the bound state $s$-wave function in coordinate space. 
The cross section to produce free pairs is proportional to
$Z_1^2 Z_2^2$; so the scaling law Eq.~(\ref{eq_sigmascaling}) above is 
obtained directly from the scaling properties 
of the hydrogen-like wave-functions. 

\begin{figure}[tbh]
\begin{center}
\resizebox{6cm}{!}{\includegraphics{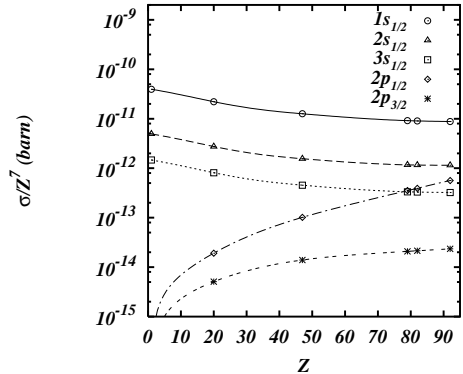}}
\end{center}
\caption{
The cross section for the bound-free pair production process are shown
for LHC energies (using $\gamma=3400$) as a function of the charge of the 
ion $Z$. The cross sections for capture to a given state
with quantum numbers $n_f$, $\kappa_f$ are scaled by the 
factor $Z^7$, see Eq.~(\protect\ref{eq_sigmascaling}).
}
\label{fig_zdepbfree}
\end{figure}

The calculations to be discussed
in the following are not done along these lines.
Instead the matrix elements involving the 
full wave functions of the bound and free lepton in the field of 
nucleus $Z_1$ are evaluated directly. As will be seen below,
these scaling rules still hold to a good degree of accuracy. 
Not unexpectedly deviations are found for the large values of $Z_1$ ,
where  $Z_1 \alpha \lesssim 1 $ . In these cases the argument given above
is not so well fulfilled ($\kappa_{rel} \lesssim m_e$ but not $\ll m_e$). 
But still, the $1/n^3$ scaling is also very well fulfilled
in this case, see below (and Fig.~\ref{fig_zdepbfree}).

Recently, the cross-section for the process Eq.~(\ref{eq_boundfree})
was calculated  in PWBA using exact Dirac wave functions for a 
point nucleus \cite{MeierHHT01}. The formalism of \cite{MeierHHT98} was 
extended in a straightforward way to the heavy ion case.  The cross-section 
for capture to the $n,\kappa$ state can be written as
\begin{eqnarray} 
\sigma^{(bfpp)}_{n_{f},\kappa_{f}} & = & 16\pi^2\,\left(
\frac{Z_P \alpha_{em}}{\beta_{ion}}\right)^2
\sum_{\kappa_i}\int_{m-E_i}^\infty d\omega \int_0^\infty \frac{d(k_\perp^2)}
{k_\perp^2+k_z^2}
\nonumber\\&&
\times\left\{\frac{1}{k_\perp^2+k_z^2}T_l
+\frac{\beta_{ion}^2}{2}\,\frac{k_\perp^2}
{\left[k_\perp^2+\left(k_z/\gamma_{ion}\right)^2\right]^2} T_\perp\right\}.
\label{eq_capturexsection}
\end{eqnarray}
Here $\beta_{ion}$ and $\gamma_{ion}$ are the velocity and the Lorentz factor
of the projectile ion in the rest frame of the target ion,
$k_\perp$ denotes the perpendicular momentum of the exchanged photon, 
$k_z= \omega/\beta_{ion}$  is its component in the beam direction and 
the energy is given by $\omega$. The quantities $T_\perp$ and $T_l$
are functions of $\omega$ and $Q^2$, independent of $\gamma$. For the 
integrations over $k_\perp$ and $\omega$ a logarithmic grid in $\omega$ 
and $k_\perp$, with energies and momenta up to 500~$m_e$ and angular momenta 
up to $l=200$ was used.

In addition to $1s_{1/2}$ capture the capture to the 
$2s_{1/2}$, $2p_{1/2}$, $2p_{3/2}$ and $3s_{1/2}$ states 
was also calculated explicitly. 
The only approximation in these calculations is the neglection of higher 
order effects (in $Z_2$). In Fig.~\ref{fig_zdepbfree} we show the dependence 
of the cross-section (divided by the approximate scaling factor $Z^7$) as a 
function of $Z$ for various bound states (We consider the symmetric case  
$Z_1=Z_2=Z$; as the dependence on the projectile charge (in lowest order)
is given by $Z_2^2$ --- see the scaling law Eq.~(\ref{eq_sigmascaling}) ---
the asymmetric cases can be found as well.)
We especially note the rise of the
cross-section for $p_{1/2}$ states. This is due to the $s$-wave 
character of the small component of the bound-state wave-function.
We also see a deviation from the $Z^7$ scaling, which is due  
to the  behavior of the Dirac wave
functions for small values of $r$ \cite{BetheS57}. 
While the bound state wave function is further increased  
as compared to the nonrelativistic $Z^3$ rise, the Coulomb repulsion
of the positron causes the $s_{1/2}$ curves in Fig.~\ref{fig_zdepbfree}
to decrease with increasing $Z$.  
The $1/n^3$ scaling with the principal quantum number $n$ for the $s$-states 
is very well fulfilled in these calculations for all values of $Z$.    

The cross section per target in Eq.~(\ref{eq_capturexsection}) can be shown 
to be of the form 
\BE
\sigma= A \ln \gamma_{lab} + B = C \ln \gamma_{ion} + D.
\label{eq_lnAB}
\EE
This form has been found to be a universal one at sufficient high values 
of $\G$. The constant $A$ and $B$ (and alternatively $C$ and $D$) then 
only depend on the $Z$ value of the target.
Parameterizations for $A$ and $B$ \cite{MeierHHT01} for typical
cases are given in Table~\ref{tab_capture}.
\begin{table}[tbh]
\begin{center}
\begin{tabular}{|c|c|c|c|c|}
\hline
Ion & $A$ & $B$ & $\sigma$,$\gamma_{lab}\approx 100$ & $\sigma$, 
$\gamma_{lab}\approx 3400$\\
\hline
Pb & $35.5$barn & $-22.1$barn       & 146 barn     & 272 barn \\
Au & $28.7$barn & $-17.8$barn       & 114 barn     & 212 barn \\
Ca & $4.46$mbarn & $-2.88$mbarn     & 18.7 mbarn     & 33.9 mbarn\\
\hline
\end{tabular}
\end{center}
\caption{
Parameters $A$ and $B$ (see Eq.~(\ref{eq_lnAB})) 
as well as total cross sections for the bound-free pair production per target
for RHIC and LHC. The parameters are taken from \protect\cite{MeierHHT01}.
Capture to the $1s$-, $2s$-, $2p1/2$-, $2p3/2$-, and $3s$-states are taken 
into account.}
\label{tab_capture}
\end{table}

The calculation of bound-free pair production has a  long
history. Let us mention here the papers 
\cite{BaltzRW91,BaltzRW93,BeckerGS87,AsteHT94,AggerS97,RhoadesBrownBS89}.
Most of these calculations have either not been done  at the high
beam energies relevant for RHIC and LHC or are based on the EPA.
In \cite{BertulaniB88} an analytical formula for bound-free
pair production was given, starting from approximate wave functions for 
electron and positron. For large values of $Z$ this formula underpredicts 
the cross-section. The reason for this was recently discussed in 
\cite{BertulaniD01}. All other results were found to be in agreement with 
each other \cite{MeierHHT01}.

Most calculations were done in first order in the interaction with the
projectile only. For a long time the effect of higher order 
processes have been under investigation.  Such higher order processes 
would violate the $Z_2^2$ scaling in Eq.~(\ref{eq_sigmascaling}). At lower 
beam energies, in the region of a few GeV per nucleon, coupled channel 
calculations have indicated, that these give large contributions, especially 
at small impact parameters, increasing the cross section by orders of 
magnitude. 
Newer calculations tend to predict considerably smaller values at higher 
energies, indicating that the number of channels used in the original 
calculations was not sufficiently large enough \cite{MombergerBS95}. 
In a recent
calculation such effects were studied at high beam energies \cite{Baltz97}
in the high energy limit $\G\rightarrow \infty$. In these calculations the
higher orders were found to be quite small (less than 1\%) and even tend to
reduce the cross section. Higher order effects are more important for
the smaller impact parameters $b \sim 1/m_e$, where the fields are largest. 
It seems interesting to note that very large (almost macroscopic) 
impact parameters contribute: for $\omega  = 1~\mbox{MeV} \sim 2m_e$
and $\gamma_{ion}=2 \times 10^7$ (relevant for PbPb collisions at LHC)
one has $b_{max}=\frac{\gamma_{ion}}{\omega} \sim 4 \mu m$. 

There are experimental results of bound-free pair production for fixed targets
at the Bevalac, AGS, and SPS/CERN.
At Bevalac energies, the experimental results \cite{Belkacem93,Belkacem94,%
BelkacemGF97} are not quantitatively reproduced by lowest order theories. 
Higher order effects are present for the systems with $Z \alpha \lesssim 1$,
see the conclusion of \cite{BelkacemGF97}.
At the higher AGS energies \cite{ClaytorBD97,BelkacemCD98}, higher order 
effects become smaller and  there is good overall agreement with theory. 
A cross-section of 8.8~b was found for AuAu collisions at 
$\gamma_{ion} =12.6 $, in agreement with the theoretical results 
of \cite{MeierHHT01,BeckerGS87,RhoadesBrownBS89}. Screening effects
due to the target electrons will be small, since only small impact parameters 
contribute appreciably. At the even higher SPS energies, the effects of the 
target electrons (screening, as well as, ``antiscreening'') should be taken 
into account \cite{VoitkivGS00,Sorensen98,AnholtB87};
for further details see the references given above.
The bound-free pair production was measured  
at the SPS for  $\G=168$ with a fixed target \cite{Krause98,Krause01}. 
There is an overall agreement with theoretical expectations, for details 
we refer to these papers.

In principle the same process is also possible for muon-production. Due to
their larger mass and therefore smaller Bohr radii, muon-capture is much more
sensitive to the finite size of the nucleus. But the cross section for this
process are rather small.

The process corresponding to Eq.~(\ref{eq_boundfree})
with one of the ions replaced by an antiproton $\bar p$ was used to produce 
fast antihydrogen
\cite{MungerBS94,BaurO96,Blanford98}. In \cite{Blanford98}
a value of $1.12\pm0.14\pm0.09$~pb was given for the cross section 
for $Z_2=1$. The momentum of the antiproton beam was in the range 
between 5203 and 6323~MeV.
Due to experimental limitations all capture into the $1s$-state and 98\%
of those into the $2s$-state but no higher states were measured.
This experimental value is in 
agreement with the theoretical ones  of 0.89~pb \cite{MeierHHT98} and 
1.02~pb \cite{BertulaniB98} for $\gamma_{ion}=6$. (The contribution
of the capture to the $2s_{1/2}$-state is given by the $1/n^3$ scaling law,
as discussed above.) 

Finally we discuss a  practical aspect.
When passing through the  magnetic fields in the collider, it is 
possible in principle that the atom formed in  the interaction region
will be subsequently ionized in the magnetic fields of the system. Since 
this would alleviate the problems related to beam pipe heating 
(see the discussion above) we give now a simple estimate, which shows 
that this effect will be negligible in practice. Let us assume that the 
$(Z+e^-)$ atom moves through a constant magnetic field of strength $B$. 
In its rest-frame this corresponds to an electric field of strength 
$E=\gamma_{lab} \beta_{lab} B$, i.e., a linear term $e E x$ is added to 
the Coulomb 
potential of the electron in the field of nucleus $Z$. This could cause 
ionization by tunneling through this barrier, see, e.g., \cite{BetheS57}. 
For a given binding energy $E=1 \mbox{Ry}\  \frac{Z^2}{n^2}$, with 
$1\mbox{Ry}=13.6eV$, this escape of the electron from the Coulomb
barrier is even allowed classically for a certain critical field. 
For a given $Z$ and electric field $E$ a critical value $n_{crit}$
is found, above which the atom ionizes in this classical
approach. Such field ionization is well known for the case of the loosely 
bound Rydberg atoms. This ionization will happen very fast, it is 
of the order of the classical orbiting time of an electron
in the corresponding state with principal quantum number $n$. 
We find 
\BE
n^4_{crit}= 3.2 \times 10^8 Z^3/E,
\EE 
where $E$ is the field strength given in units of $V$/cm. 
See also Eq.~(54.2) of \cite{BetheS57}.
As an example we take $Z=82$, $E=10^{10} V/$cm which corresponds
to a magnetic field of $B=1$~Tesla and $\gamma=3000$. We
have $n_{crit} \sim 12$, i.e. only states with very high 
principal quantum number will be ionized. According
to the $1/n^3$ scaling law, their contribution can safely 
be neglected. 

\subsection{Formation of Dilepton Bound States: Positronium and Muonium}
\label{ssec:dlbstates}

Finally we note that the electron and the positron can also form a bound state:
positronium. This is in analogy to the $\GG$-production of mesons
($q\bar q$ states) discussed in Sec.~\ref{sec_resproduction}.
With the known width of the parapositronium, 
\begin{equation}
\Gamma^{(0)}(n {^1}S_0) = \frac{\alpha^5 \, m_e}{2\,n^3}\,,
\label{eq:positroniumGamma}
\end{equation}
the photon-photon
production of this bound state was calculated in \cite{Baur90b}.  The
production of orthopositronium, $n=1$ ${}^3S_1$ was calculated recently
\cite{Ginzburg97} and \cite{KotkinKSS99}, see also \cite{BertulaniN01}
where the three-photon production of vector mesons is treated. 
In \cite{Ginzburg98} an effective parameter $\rho$ is introduced,
which controls the relative importance of two- and three-photon-
processes (see also p.~1677 of \cite{BaurHT98}):
\BE
\rho=\left(\frac{Z\alpha \Lambda}{m_{\mu \mu}}\right)^2,
\EE
with $1/\Lambda^2=1/6\left<r^2\right>$ and $\left<r\right>^2$ is the 
mean square radius of the charge distribution and $m_{\mu\mu}$ is the 
mass of dimuonium (or any other produced particle).
This factor $\rho$ is always much smaller than 1 except for 
positronium, where $\rho $ is given by $Z\alpha$ (in this case
$\Lambda \approx 1/R$ should be replaced by $\Lambda \approx 1/m_e$,
see the discussion at the beginning of this chapter). Therefore the 
production of orthopositronium is only suppressed by the factor 
$(\ZA)^2$ as compared to the two-photon production of parapositronium. 
Since $(\ZA)^2 \sim 0.5$ for the heavy systems like Au or Pb, one expects 
that both kinds of positronium are produced in similar numbers. 
Detailed calculations show that the three-photon process is indeed not much 
smaller than the two-photon process \cite{Ginzburg97,Gevorkyan98}.

The bound $\mu^+\mu^-$ system (``dimuonium'')
is an interesting example of a pure electromagnetic bound state, which
can be effectively studied in heavy-ion collisions.
Dimuonium has a mass of about two muon masses, $m_{\mu\mu} \approx
211$~MeV and the binding energy in the ground state is of the 
order of $1.4$~keV. \hbox{Although} this system is well known theoretically, 
see \cite{Karsh1,Jents,Karsh2,Nemenov,Malenfant,Bilen}, dimuonium 
has not been observed experimentally up to now.   
Like in the case of positronium, one should distinguish para- and ortho-
dimuonium with $S^C = 0^+$ and $1^-$ respectively, which can be 
produced in electromagnetic decays of light mesons, i.e. orthodimuonium 
in radiative decays of $\eta$- \cite{Nemenov,Kozlov} and $K_L$- mesons 
\cite{Malenfant}, while paradimuonium in decays of $\omega$- and 
$\phi$- mesons.   
The hyperfine structure and the decay rate of dimuonium are influenced 
by the electronic vacuum polarization in the far time-like asymptotic 
region, which does not yield any contribution in any other bound states 
\cite{mohr,jungmann}. In particular, a relative contribution of $1.6\,\%$
to the decay rate of orthodimuonium is predicted in \cite{Jents,Karsh2}
and this would allow a test of QED in a previously unexplored kinematic 
region.

\begin{figure}[tbh]
\centerline{\includegraphics[bb=58 600 612 750,width=0.7\hsize]{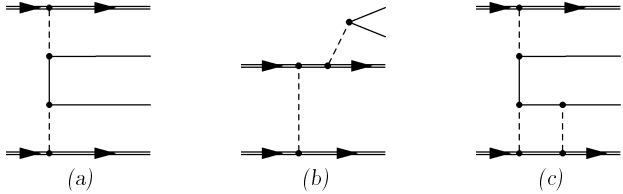}}
\caption{Diagrams for two-, one- (bremsstrahlungs) and three- photon 
production mechanism of 
a dimuon bound state in heavy ion collisions \protect{\cite{Ginsburg}}.}
\label{diag:3photon}
\end{figure}

Dimuonium production in heavy ion collisions is studied in \cite{Ginsburg}
in detail. Below we will mainly follow this paper. Paradimuonium is 
produced dominantly in two-photon processes, orthodimuonium in three-photon 
processes while its production by bremsstrahlung is suppressed, see 
Fig.\ref{diag:3photon}.
The paradimuonium production cross-section is determined according to 
Eqs.~(\ref{gg-r-delta}) and~(\ref{AA-AAR}) by the $\GG$-width, which in the 
leading order approximation is given by Eq.~(\ref{eq:positroniumGamma})
with $m_u$ instead of $m_e$. Orthodimuonium cross section is suppressed 
compared with that of 
paradimuonium by more than an order of magnitude 
see Eq.~A1 of \cite{Ginsburg}. The production cross sections 
are given in Table~\ref{tbl:dimu}.
\begin{table}[tbh]
\centerline{
\begin{tabular}{|l|c|c|} 
\hline\hline
               & $\sigma(AA \to AA + PDM),~\mu$b & 
                 $\sigma(AA \to AA + ODM),~\mu$b     \\
\hline
AuAu, RHIC  &  0.15            &  0.021   \\
PbPb, LHC~  &  1.35            &  0.089   \\
CaCa, LHC~  &  6.6$\times 10^{-3}$   & 6.9$\times 10^{-5}$    \\
\hline\hline
\end{tabular}}
\caption{Production cross sections for paradimuonium (PDM) and 
         orthodimuonium (ODM) in heavy ion collisions, as predicted in 
         \protect{\cite{Ginsburg}}.}
\label{tbl:dimu}
\end{table}
The main decay channels are the annihilation processes
into two photons for paradimuonium and $\EPEM$ for
orthodimuonium. Dimuonium can therefore be observed via these decay
channels. Further details, especially concerning background processes,
can also be found in \cite{Ginsburg}.

One might even think of the production of ditauonium 
(DT). With the ditauonium mass $m_{DT}= 3554~MeV$, its width $\Gamma_\GG$
(see Eq.~(\ref{eq:positroniumGamma})) is given by 0.018~eV.  From
these values we can immediately obtain the cross section of ditauonium
production in $\GG$-interactions in heavy ion collisions.
For example, in PbPb collisions at the LHC we obtain from
Eq.~\ref{AA-AAR} and Fig.~\ref{fig_sigmagamma} the production cross
section $\sigma_{DT}=0.7~nb$.
We note that the lifetime of the $\tau$ of $291 \times 10^{-15}$s
corresponds to the width of $\tau$-lepton $\Gamma_{\tau}=0.00226$eV,  
i.e., the ditauonium  
$\GG$-width is about 5 times larger than the width of its weak decay
given by $2\Gamma_{\tau}$ and thus the ditauonium can be really
produced as a bound state of two $\tau$-leptons.

%
%
\section{Methods of Detecting Very Peripheral Collisions in Experiments at 
Heavy Ion Colliders} 
\label{sec_eventsel}

In this chapter we discuss the study of $\GG$-events in heavy ion
collisions from an experimental point of view. We discuss the general 
characteristics of these kind of events and how they can be used as
basic signatures for the triggering and off-line selection of these events. 
We discuss further various background sources for $\GG$-processes, 
including single-photon, photon-Pomeron and double Pomeron processes,
their corresponding cross sections and expected trigger rates. 
We also give an overview of the experimental methods used for the selection 
of different final states produced in $\GG$ interactions, based on the 
experience of current $\EPEM$-collider experiments, as well as, on the first 
studies of very peripheral events in heavy ion collisions at RHIC and 
Monte-Carlo simulations done for the future heavy-ion experiments at LHC. 

\subsection{Signatures of $\GG$-Processes in Heavy Ion Collisions}

The experimental technique for the detection of $\GG$-processes 
is based on special observables (signatures) which make the $\GG$-processes 
different from other very peripheral processes, as well as, from strong 
peripheral interactions of the ions. In particular one of the main features 
of coherent electromagnetic interactions of relativistic ions is the 
small transverse momentum transfer from the ions. As a result the ions
remain intact after such an interaction and in general
they will be kept for some time in the beam pipes and thus cannot be detected
in an experimental setup. Therefore very clean events will be detected with 
particles produced mostly in the central rapidity region. 
(In Sec.~\ref{ssec:ggluminosities} we already mention the interesting 
possibility to tag the scattered protons in $pp$ collisions studied in
\cite{Piotrzkowski00}. In the case of ions this is probably impossible 
to achieve, due to the low limits on $x$, see Sec.~\ref{sec_intro} above.)

It is useful to mention in this context also the excitation of the Giant Dipole
(GDR) and other higher Multipole and/or Multiphonon Resonances (MR), which 
lead in their decay 
to the emission of one or several low-energy neutrons in the nucleus 
rest frame, see Sec.~\ref{photonuclear}. In the laboratory frame --- due to 
the very high Lorentz factor of ions and the low momenta of the neutrons in 
the ion frame --- one has almost monoenergetic neutrons, which can be 
detected in the very forward hadron calorimeters 
commonly called the Zero Degree Calorimeters (ZDC) 
of the experimental setup by the deposit of multiples of the energy 
$\gamma_{lab} m_n$ for each detected neutron. These almost monoenergetic signals 
in the ZDCs, see Figs.~\ref{En_in_ZDC} and~\ref{AliceZDC}, and their small 
number are therefore a good
signature of the electromagnetic nature of the interaction between the two
ions.

\begin{figure}[bth]
\parbox{0.5\hsize}{
\resizebox{\hsize}{!}{\includegraphics{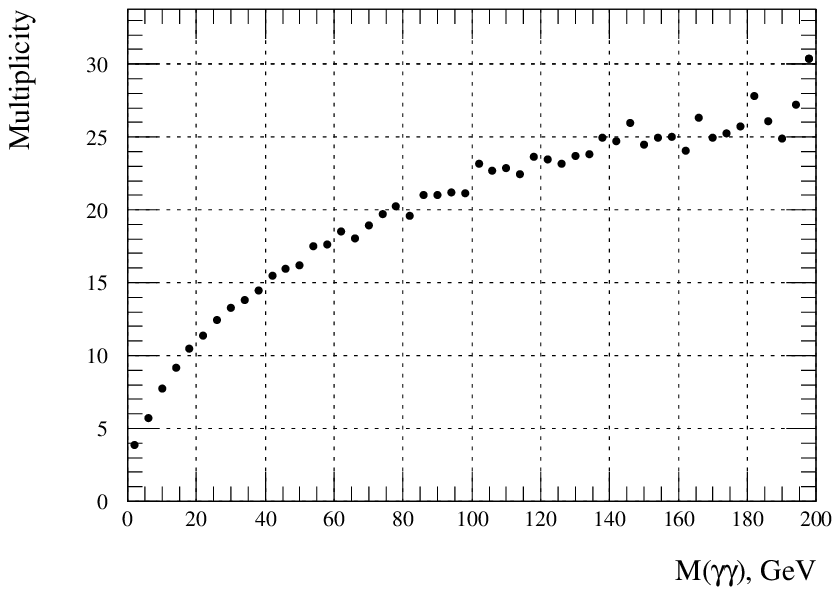}}
}
\hfill
\parbox{0.46\hsize}{
\caption{The average charged particle multiplicity in $\GG$-interactions 
  vs. the effective $\GG$-mass predicted by the PYTHIA model
  \protect{\cite{Pythia}.}}
\label{gg-mult}
}
\parbox{0.5\hsize}{
\resizebox{\hsize}{!}{\includegraphics{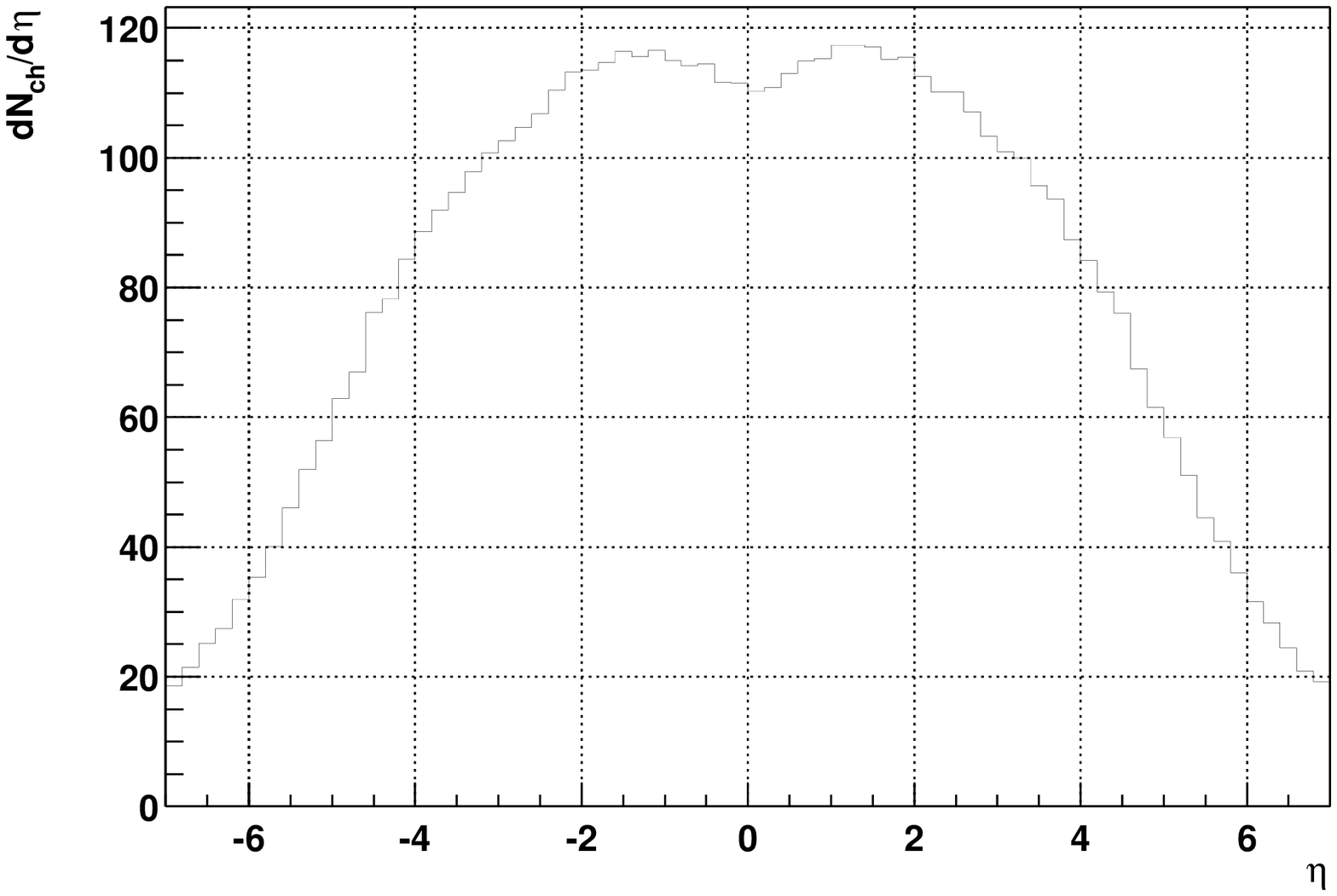}}
}
\hfill
\parbox{0.46\hsize}{
\caption{Charged particle multiplicity $dN_{\rm ch}/d\eta$ in hadronic
  peripheral PbPb collisions with impact parameter $1.8\times R_{Pb} <
  b < 2\times R_{Pb}$ calculated with the HIJING generator
  \protect{\cite{Hijing}}.}
\label{hh-mult}
}
\end{figure}

Another feature of $\GG$-processes in heavy ion collisions is a
relatively low multiplicity of secondary particles compared to those in
strong ion interactions. Fig.~\ref{gg-mult} shows the average charged
particle multiplicity for the process $\GG\to X$ as a function of the 
effective $\GG$-mass as predicted by PYTHIA \cite{Pythia}. This should be
compared with the multiplicity expected in strong peripheral ion
collisions. As an example the expected average charged particle
multiplicity distribution vs. pseudorapidity $dN_{\rm ch}/d\eta$ in
PbPb collisions with a large impact parameters $b$ between 
$1.8\times R_{Pb} < b < 2\times R_{Pb}$ is shown in Fig.~\ref{hh-mult} for LHC 
energy.  This multiplicity distribution has been calculated using HIJING
\cite{Hijing}. The total charged particle multiplicity in hadronic peripheral 
collisions is much higher than that in $\GG$-processes; even if one is 
using a restricted aperture, let's say, $|\eta|<1$, the multiplicity is still
expected to be of the order of hundreds and therefore remains very high.
The most important signature of $\GG$ processes in coherent ion-ion 
collisions is the quite low total transverse momentum $P_\perp$ of the produced 
system. As discussed in Chap.~\ref{sec_intro}, see Eq.~(\ref{pt_max}), 
the $\GG$ system has a transverse momentum less than about $\sqrt{2}/R$, 
where $R$ is the nuclear radius. But the average transverse momenta is even
smaller than this, namely
\begin{equation}
                      P_\perp \sim M_{\GG}/\gamma_{lab}.
\end{equation}
For the effective $\GG$ mass range at RHIC ($M_{\GG} < 3$~GeV), as
well as, at LHC ($M_{\GG} < 100$~GeV) we have $P_\perp \sim 30~MeV$.
It should be remembered however that transverse momentum distributions 
have not yet been calculated  explicitly within the semiclassical 
approximation. The problem is formulated in \cite{BaurB93}.
In order to use this feature for selecting $\GG$-processes the experimental 
setup has to provide a large enough aperture for the detection of both charged
and neutral particles, including also secondary photons. It should provide also
a sufficiently good transverse momentum resolution.

Many of the $\GG$ reactions, which are planned to be studied, are
exclusive processes. In this case the net charge of the detected particles or 
the number of charged tracks can be taken into account. This leads to one
more selection criteria for $\GG$-processes, requiring an even number of 
charged track of opposite signs.

We therefore may formulate the criteria for the selection of the two possible
classes of two-photon processes in collisions of relativistic ions:
\begin{itemize}
\item {\it The pure $\GG$-processes} are characterized by a small
  multiplicity of the charged particles in the central detectors, a small total
  transverse momentum of the produced system and the absence of any
  signal in both ZDCs;
\item {\it The $\GG$-processes with EM excitation of the secondary ions}
  are characterized as well by a small multiplicity of the charged
  particles in the central detectors, a small total transverse momentum 
  of the produced system and by neutrons coming from the GDR or MR decay in 
  one or both ZDCs.
\end{itemize}
Actually these are ideal criteria for the event identification, which in real 
experiments are often less clear due to different limitations of the 
experimental setup that lead to distortions and mixtures between the different
event classes, as well as, to the contamination of them by background 
processes.

\subsection{Background Processes}

In this section we discuss different sources of background for
very peripheral processes in relativistic heavy ion collisions. As
mentioned already in Chap.~\ref{sec_diffractive}, one should
distinguish between photon-photon, photon-Pomeron and Pomeron-Pomeron 
processes.
All of them are quite interesting in themselves and they also have similar
signatures. The physical nature of them is quite different, and each
of these processes should therefore be considered as a background for the 
others. All those processes belong to the class of very peripheral collisions,
taking place at impact parameter $b > 2R$. The strong interaction of the ions
with impact parameter $b \approx 2R$ is a common source of background for all 
very peripheral processes and should be taken into account. This is a physical
background, as it results from the interaction of the ions with each other.

On the other hand there is background coming from the apparatus, which is then
specific for each experimental setup. As potentially important sources of such
a background we discuss in the following one-photon processes, the photonuclear
excitation of the ions and multiple $\EPEM$ pair production.  Cosmic ray 
events and beam-gas interaction should be mentioned also in this respect 
as possible background sources of the apparatus.

\subsubsection{Physical Background}
\label{sssec:physback}
 
Two-photon processes in relativistic ion collisions are discussed in
detail in Chap.~\ref{sec_ggqcd} including the corresponding cross
sections and rates. Pomeron-Pomeron processes as a possible background
have been discussed in Chap.~\ref{sec_diffractive}. 

As was mentioned there the Pomeron-Pomeron cross section increases with
$A^{1/3}$ or equivalently with $Z^{1/3}$, compared to the $Z^4$ rise of
the photon-photon processes. Therefore in heavy ion collisions we can expect 
the predominance of two-photon processes, whereas in light ion and
especially in $pp$-collisions the contribution of Pomeron-Pomeron
processes will be much more essential. As shown in Chap.~\ref{sec_diffractive}
this is confirmed in detailed calculations, see, e.g., 
\cite{EngelRR97,RoldaoN00}.

The cross section for the photon-Pomeron processes in ion-ion collisions
can then be estimated with the help of the factorization relations
\cite{GribovP62,GribovP62b}, as explained in Sec.~\ref{sec:vector-meson-pair},
using the cross section for photon-photon and Pomeron-Pomeron interactions:
\begin{equation}
\sigma_{tot}^2(\gamma \P\to X) = \sigma_{tot}(\GG\to X)
~\sigma_{tot}(\P\P\to X),
\end{equation}
which is valid also for the corresponding differential cross sections.
This can be used to estimate the total cross section of the
photon-Pomeron processes in relativistic ion-ion collisions using the known
cross sections of the photon-photon and Pomeron-Pomeron processes as
 $\sigma_{\gamma \P} = \sqrt{\sigma_{\GG} \sigma_{\P\P}}$.

The cross sections of the three different processes for PbPb, ArAr and $pp$ 
collisions at the LHC, as well as, for AuAu, CuCu and $pp$ 
collisions at RHIC are shown in 
Table~\ref{Crss}. 

From this it is seen that at the LHC in the very peripheral PbPb
collisions we have dominantly two-photon processes, in $pp$ collisions
on the other hand double Pomeron processes dominate, while in the ArAr 
collisions cross sections of all three very peripheral processes are 
comparable. Similar cross sections ratios for different ion species are valid
also at RHIC.

Apart from the different $A$-dependence of the very peripheral processes  
the selection of one of the processes can be enforced by using the following 
additional signatures:
\begin{enumerate}
\item A different total $P_t$ distributions for very peripheral processes
  is expected because in the case of Pomeron-Pomeron processes the $P_t$ are 
  in general higher than in the case of photon-Pomeron processes, and these
  are again higher than those in photon-photon  processes, see \cite{Felix97}.
\item The different processes will lead to a different $C$-parity of the 
   (exclusive) system produced in the different very peripheral processes. 
  $C$-parity 
  should be even in the two-photon and double-Pomeron processes, while it is 
  odd in the case of photon-Pomeron processes,
\item The probabilities for the excitation of the GDR and MR of the ions,
  that is, of the additional electromagnetic excitation of the nuclei 
  should be a good sign for a very peripheral collision, discriminating them
  against, e.g., grazing collisions or diffractive processes. The probability 
  to excite a GDR or MR in the ions is much higher for an additional 
  photon-induced reaction than for a hadronic collision. The GDR favors 
  neutron over proton emission, whereas a direct hadronic nucleon emission 
  is insensitive to proton/neutron, and thus the emission of a soft proton in 
  the ion rest frame and detection of the almost monoenergetic proton in the 
  charged ZDC could be regarded as a sign of a Pomeron-ion interaction.
\end{enumerate}
\begin{table}[tbh]
\begin{center}
\begin{tabular}{l|l|c|c|c|c}
\noalign{\vspace{5pt}}
\hline\hline
\noalign{\vspace{2pt}}
& Reaction & $AA\to X$ 
          & $AA\to AAX$ 
          & $AA\to AAX$
          & $AA\to AAX$ \\ 
&         & (nuclear) & $({\P\P{\to \rm hadrons}})$ &
          $({\gamma\gamma{\to \rm hadrons}})$ & $({\gamma \P{\to \rm hadrons}})$ \\
\noalign{\vspace{2pt}}
\hline
 L & $\sigma$(PbPb), barn & 7.8 & $8.4\cdot10^{-4}$ &
                                  $1.5\cdot10^{-1}$ & $1.1\cdot10^{-2}$ \\
 H & $\sigma$(ArAr), barn & 2.1 & $6.7\cdot10^{-4}$ &
                                  $5.6\cdot10^{-6}$ & $6.1\cdot10^{-4}$ \\
 C & $\sigma(pp)$,   barn & 0.07& $5.2\cdot10^{-4}$ &
                                  $1.5\cdot10^{-8}$ & $2.8\cdot10^{-6}$ \\
\hline
 R & $\sigma$(AuAu), barn & 7.5 & $2.2\cdot10^{-4}$ &
                                  $1.8\cdot10^{-3}$ & $6.3\cdot10^{-4}$ \\
 H & $\sigma$(CuCu), barn & 3.1 & $2.1\cdot10^{-4}$ &
                                  $9.3\cdot10^{-5}$ & $1.4\cdot10^{-4}$ \\
 I & $\sigma(pp)$,   barn & 0.07& $2.2\cdot10^{-4}$ &
                                  $1.7\cdot10^{-9}$ & $6.1\cdot10^{-7}$ \\
 C &                      &                         &                   &                   \\
\hline\hline
\end{tabular}
\end{center}
\caption{Cross sections of the nuclear, Pomeron-Pomeron,  
photon-photon and photon-Pomeron processes ($M > 1~GeV$) in 
PbPb, ArAr, $pp$ collisions at the LHC and in   
AuAu, CuCu, $pp$ collisions at the RHIC.} 
\label{Crss}
\end{table}

\begin{figure}[hbt]
\parbox{0.32\hsize}{
\resizebox{\hsize}{!}{\includegraphics{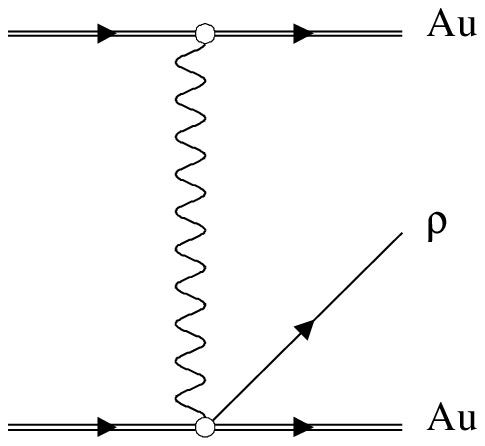}}
}
\hfill
\parbox{0.32\hsize}{
\resizebox{\hsize}{!}{\includegraphics{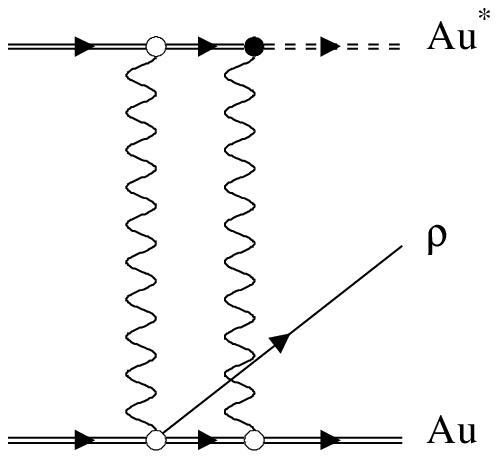}}
}
\hfill
\parbox{0.32\hsize}{
\resizebox{\hsize}{!}{\includegraphics{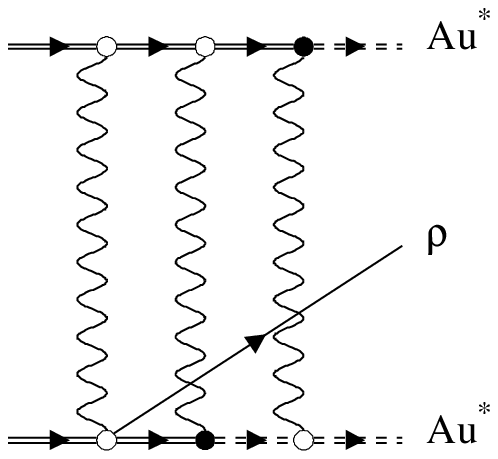}}
}
\caption{Diagrams of $\rho^0$ production in heavy ion collisions when
  neither, one or both nuclei are excited electromagnetically. Diagrams
  are show up to vertex permutations.}
\label{fig:rho_prod}
\end{figure}

Please note that in the case of heavy ion collisions, due to the large
probabilities for electromagnetic excitations, see Sec.~\ref{photonuclear},
additional electromagnetic processes, which we illustrate for the case of 
$\rho^0$-production in AuAu collisions at RHIC in Fig.\ref{fig:rho_prod},
are very likely.
In heavy ions, the probability for the excitation of the GDR is large for 
small impact parameter, but this is not essential for the selection of 
$\GG$-processes because for heavy ions $\GG$-processes dominate over the 
processes involving the Pomeron. The signature 3 above could
become important in the case of intermediate ion collisions, where on the one 
side the cross sections of photon-photon, photon-Pomeron and Pomeron-Pomeron 
processes are comparable, but on the other side the cross section of GDR 
excitation is low as well, and thus the analyzing power of signature 3 
becomes very high.

Concerning the strong interaction of the two ions, these events in general
are quite different in their properties from the very peripheral ones discusses
above, compare, for example,  Figs.~\ref{gg-mult} and~\ref{hh-mult}.  
Nevertheless in the case of detectors covering only a small aperture range,
like the central detector and especially the ZDC, the background
from these processes should be taken into account.

\subsubsection{Apparatus Backgrounds}

Since the cross sections of the interesting coherent processes in
relativistic very peripheral ion-ion collisions are much smaller than the
total cross sections, the key point in all experiments to study very 
peripheral ion-ion collisions is the organization of the primary or ``Level
Zero'' (L0) trigger, which is needed in order to start the Data 
Acquisition System (DAQ) of the experiment.

The Higher Level (HL) triggers are used then further only to stop 
or continue the data taking by the DAQ of the triggered event from the 
detectors. The main aim of the HL-triggers is to prevent storage of 
the physically uninteresting events. Due to the finite dead time of 
the DAQ the L0 trigger rate cannot be too high, otherwise the data
taking in the experiment will be fully blocked. Thus processes,
which give a too high a load on the L0 trigger detectors, should be 
considered as a most important source of background.

The first trigger on coherent processes in relativistic heavy ion
collisions was developed for the STAR experiment at the RHIC
\cite{STAR347}. The L0-trigger is based on the charged particle
multiplicity in the pseudorapidity interval $|\eta|<1$, measured in
the Central Trigger Barrel (CTB) and Multi-Wire Chambers (MWC) at the
endcaps of the TPC. The information from the trigger detectors is
divided into 4 bins in azimuthal angle which can be used to apply cuts
on event topology to select events with tracks lying in opposite
$\phi$-sectors. At the L1 and L2 trigger levels the improved
multiplicity information is available after refining kinematics. At
the L3-level the tracking information is available to apply momentum
cuts and use the charge of the tracks. The decision time of the trigger
levels 0, 1/2 and 3 were given by the authors as 1.7~$\mu$s,
100~$\mu$s and 10~ms respectively. Besides these triggers STAR also used
a minimum bias trigger to select events when both nuclei were
dissociated \cite{Klein01b}. This trigger is based on zero-degree
calorimeters (ZDC).  The presence of a signal in each of the ZDCs 
corresponding to one or more neutrons with the full beam energy defines 
events when the nuclei are electromagnetically dissociated.

The positioning of the main trigger detectors of an experiment is highly 
sensitive to the background processes. In this connection the low mass
single and multiple $\EPEM$-pair production in heavy ion collisions has
to be considered as one of the major background processes, because its
cross section is really huge (200~kbarn for PbPb at LHC and 30~kbarn
for AuAu at RHIC), the rapidity distribution is rather flat (and thus
covers all detectors) and all other signatures are quite similar to
those of the photon-photon processes, because the $\EPEM$-pair production
is also just a photon-photon process, see
Sec.~\ref{sec_leptons_mpairs} above. Electrons and positrons are produced 
preferably in the very forward or backward directions, see 
Fig.~\ref{fig_eee}(b), but these pairs do not constitute a background for 
the ZDCs as their energy is rather small, Fig.~\ref{fig_eee}(a), compared 
to the typical hadron energy in this pseudorapidity interval.
But the central trigger detectors (i.e., the CTB in the case of the
STAR experiment) can be essentially loaded by these pairs. Due to their
low transverse momentum $P_\perp$ an essential reduction of this background can 
be done, if one uses the central trigger detector at a sufficiently large 
radius, a strong solenoidal magnetic field, or both. (In the case of the CTB 
at STAR its radius, 2.2~m and $B=0.2$~T,
is sensitive to charged particle with $P_\perp>130$~MeV
\cite{STAR347}, which reduces the background from soft $\EPEM$-pair
significantly). On  the other hand if the central trigger detectors are too 
large, 
we may expect some background from cosmic ray events. As was shown in
\cite{STAR347} the rate of cosmic rays entering the TPC is due 
to the large size of the TPC (16~m$^2$) almost 3~kHz, which is rather high. In
\cite{NystrandK98} the cosmic ray flux through the STAR trigger
detectors was estimated to give an important percentage of all background
processes.

Another possible trigger detector, the ZDC, can be contaminated by 
the electromagnetic excitation of the ions alone to the Giant Dipole 
and other resonances, see Sec.~\ref{photonuclear}. 
We are studying here the following two types of processes: one-photon exchange
with the electromagnetic excitation of one of the ions and no particle detected
in the central rapidity region and the production of one or more low mass 
$\EPEM$-pair together with the electromagnetic excitation of one of the ions.
The cross sections of these processes in heavy ion collisions at RHIC and 
LHC are equal to $\sigma_{AuAu\to AuAu^*} =58$b,
$\sigma_{AuAu\to AuAu^*(e^+e^-)_{N>0}} = 13$b, and $\sigma_{PbPb\to PbPb^*} = 114$b, 
$\sigma_{PbPb\to PbPb^*(e^+e^-)_{N>0}} = 22$b, respectively, using for the 
GDR excitation Eq.~(\ref{eq:PGDR}) and for the $e^+e^-$ pair production
Eq.~(\ref{eq_poisson}).
The similar process but with the mutual excitation of both ions have almost
the same order of magnitude, as the probability for excite the GDR is not
much smaller than one, see Sec.~\ref{photonuclear}.
For intermediate ions (CuCu at RHIC and ArAr at LHC) the exchange of additional
soft photons leading to electromagnetic excitation is suppressed by a factor
of about $A^{10/3}$, see Eq.~(\ref{eq:PGDR}). Therefore the background load
on the ZDC as a potential trigger detector by electromagnetic excitation of
the ions alone is important only in the case of heavy ion collisions, 
but it is negligible small for the intermediate and low mass ions.

\subsubsection{Trigger for Very Peripheral Processes}

In this subsection we briefly summarize the most essential aspects,
which should be taken into account for a trigger to study the  
very peripheral ion collisions. 

The central trigger barrel (like the CTB in STAR) with endcups have 
to be considered as the major detectors for the L0-trigger to select 
very peripheral ion collisions. The trigger requirement is a low charged 
particle multiplicity $n_{ch}$ in the central trigger detectors, let's say 
$1<n_{ch}< 20$. The solenoidal magnet is an obligatory part in such 
type of experiments, because it prevents the overloading of the trigger 
barrel by background processes. Two background processes should be taken 
into account for an optimization of the trigger detector parameters: the 
processes of low mass multiple $e^+e^-$ production and the background from 
cosmic events. The basic parameters are the magnetic field 
in the solenoid $B$ and the 
trigger barrel radius $R_{bar}$ or the inner radius of endcups, $R_{inn}$.
For the optimization of the trigger one increases the value of $B R$ 
(where $R = \min\{R_{bar},R_{inn}\}$) to reduce the background from low mass 
multiple $e^+e^-$-pair production up to a value, at which the background from 
cosmic ray events is still not too high. The optimal detector parameters 
correspond probably to a detector configuration, when the background rates 
from the $e^+e^-$-pair production and from cosmic events are comparable. The 
signals from the ZDCs are not so important, due to their contamination with
processes of the electromagnetic excitation of the heavy ions. 
But it could be useful for the suppression of Pomeron-Pomeron processes in the
case of intermediate ion collisions. 

Another possibility for a very peripheral trigger in ion-ion
collisions is the barrel Electromagnetic Calorimeter (ECAL). This is
one of the most popular detectors used in experiments at $e^+e^-$ and
hadron colliders, see, for example, \cite{CMS-ECAL-TDR}. The
requirement that the total deposited energy in the ECAL is
sufficiently low compared with the one in strong interaction of the
ions, allows an effective selection of very peripheral events at a L0
trigger level. Because the energy in the ECAL from the low mass
multiple $e^+e^-$-pairs is quite small, these events do not provide
any background. The strong ion interactions in peripheral collisions
as a background process can be effectively suppressed by using
additional signals from the ZDCs, which for these events are very high
as was discussed in Chap.\ref{sec_photonhadron}.  Only cosmic ray
events provide in this case some background, which can be suppressed
at the HL trigger level, based on general event topology, see
\cite{Klein01b}.

\subsection{Detection of Some $\GG$-Processes}

Two-photon processes in heavy ion collisions have not yet been widely
studied experimentally with the first heavy ion collider RHIC starting its 
operation only a year ago. However, a group of experimentalists at the  STAR
detector have studied the possibility to observe $\GG$-processes during the  
last six years. Earlier $\GG$-physics has been studied at $\EPEM$-colliders
by almost all collaborations. The experience gained of these 
$\EPEM$-experiments to detect $\GG$ processes can be applied now to the heavy
ions experiments. Below we give an overview of the major experimental 
observations of $\GG$-processes at $\EPEM$-colliders and the studies done for 
heavy ion experiments.

\subsubsection{Experiments at $\EPEM$-Colliders}

All $\EPEM$-collider experiments which have been running so far, have 
accumulated a large experience for the detection of $\GG$ processes. 
Although the signatures of two-photon processes in electron-position 
collisions differ from those in heavy-ion collisions, they still have many 
common features which allows to take over their experience to the future 
heavy ion collider experiments. Many recent experimental data from
$\EPEM$-colliders (LEP-II, CLEO, KEKB) are for quasi-real photon
interactions, like those in heavy-ion collisions. Here we give a review of 
the $\GG$ processes detected by those experiments.

\paragraph*{\underline{$\GG\to hadrons$.}}
This reaction was studied by the L3 collaboration at LEP
\cite{L3:01} to measure the total cross section for the 
scattering of two real photons into hadrons. 
The two-photon cross section 
$\sigma(\GG\to$~hadrons) was measured in the invariant mass range of
$5< W_{\GG} < 185$~GeV. Hadronic two-photon events were selected by the 
following criteria:

\begin{itemize}
\item Events with the scattered beam electrons detected were rejected. This
  restricts the virtuality of the interacting photons to $Q^2<8$~GeV$^2$
  with an average value of $\langle Q^2 \rangle \sim 1.5\times 10^{-2}$%
  ~GeV$^2$.
\item The total energy in the electromagnetic calorimeter was required
  to be between 0.5 and 50~GeV to suppress beam backgrounds and
  exclude radiative events.
\item The detected multiplicity should be greater than six to exclude
  events containing $\tau$-pairs.
\end{itemize}
After the selection the visible effective mass of the event,
$W_{\rm vis}$ was measured. An unfolding procedure based on the Monte Carlo
event generators PHOJET and PYTHIA was used to calculate the 
two-photon mass $W_\GG$. The systematic uncertainties due to the Monte
Carlo generators were estimated to be negligible for $W_\GG<65$~GeV, but
are important for higher $\GG$-masses.

\paragraph*{\underline{\rm Production of light resonances.}}
The L3 collaboration studied also light resonance production in
$\GG$-collisions \cite{Acciarri:2000ev}. The resonances were studied
in the $K^0_S K^\pm \pi^\mp$ and $\eta \pi^+ \pi^-$ final states. The
$\eta(1440)$, $f_1(1420)$, $f_1(1285)$ were observed. The mass region
between 1200 and 1500~MeV is expected to contain several states. The
pseudoscalar state $\eta(1440)$ was observed in hadron collisions and
in radiative $J/\Psi$ decay, but not in $\GG$-collisions. Only a very low
upper limit of its two-photon width was measured, which supports the
interpretation of $\eta(1440)$ as a glueball candidate. Axial-vector
states $f_1(1420)$, $f_1(1285)$ are also present in this mass region. 

Since these are exclusive processes, their selection criteria differ
from those for the process $\GG\to X$ above:
\begin{itemize}
\item The $K^0\to\pi^+\pi^-$ was identified by the secondary vertex
  separated from the interaction point, the effective invariant mass
  of the two pions and the angle correlation between the $K^0$
  momentum and the two final tracks.
\item $K^\pm \pi^\mp$ was identified by the two tracks coming from the
  interaction point and their $dE/dx$ measurement being consistent with a
  $K\pi$ hypothesis.
\item The final state $\eta \pi^+ \pi^-$ was identified by the presence
  of two photons and two tracks of opposite charge and the $\GG$  invariant 
  mass.
\end{itemize}

\paragraph*{\underline{$\GG\to K^+K^-$, $\GG\to K^0_SK^0_S$.}} These
reactions were studied by the L3 collaboration at LEP
\cite{Acciarri:2000ex}, by the BELLE collaboration at KEKB
\cite{Huang2001} and by the CLEO at CESR \cite{Godang:1997hj}. The
events were selected according to usual criteria for exclusive
reactions, i.e., the requirement of two or four charged tracks for the $K^+K^-$
and $K^0_SK^0_S$ final states respectively, with a net charge of zero.
$K^0_S$ mesons were identified by secondary vertex reconstruction and
the invariant mass of the secondary track pairs. The vector sum of the
transverse momenta of all tracks should be small. These experiments aimed 
to study the resonances in the mass region from~1.3 to~2.3~GeV, where glueballs
are expected to be found. A tensor state $f_J(2220)$ was observed in the
radiative decays of $J/\Psi$ and was considered a good candidate for a
glueball. Since gluons do not couple to photons they cannot be
produced directly in $\GG$ collisions. The $\EPEM$ experiments did not
observe any resonance around the mass 2220~MeV, thus they support the
glueball interpretation of the $f_J(2220)$ state.

\paragraph*{\underline{$\GG\to c\bar{c}X$, $\GG\to b\bar{b}X$.}} Open
charm and beauty production was measured by the L3 collaboration
\cite{Acciarri:2000sc,Acciarri:2000kd}. The event selection was
performed in two steps. The first one selects hadronic states and the
second one identifies $c$ and $b$ quarks by their semileptonic decays.
The hadronic final states were selected by requiring at least 5 tracks
and a visible energy below $\sqrt{s}/3$. The visible mass of the events
was restricted to $W_{\rm vis}>3$~GeV. Then the unfolding procedure
based on either PYTHIA or PHOJET was applied to obtain the $\GG$ 
c.m. energy.

\paragraph*{\underline{$\GG\to D^{*\pm}X$.}} The inclusive production of
$D^*$-mesons was measured by L3 \cite{Acciarri:1999md} and OPAL
\cite{Csilling:2000xk}.  $D^{*\pm}$ mesons were detected via their
decay chains $D^{*\pm}\to D^0\pi^+_s$ and $D^0\to K^-\pi^+$, $D^0\to
K^-\pi^+\pi^0$.  The reconstruction of these decay chains requires a
sample of events containing hadronic final states. The hadronic events
were selected by cuts on the energy measured in the electromagnetic
and hadronic calorimeters and using tracing information.

\paragraph*{\underline{$\GG\to \eta_c, ~\chi_{c0}, ~\chi_{c2}$.}}
Charmomium production was studied by CLEO
\cite{Brandenburg:2000ry,Eisenstein:2001xe} and L3
\cite{Acciarri:1999rv}. The $\GG$ and total widths of these states
were measured. CLEO searched for $\eta_c$ meson in the $K^0_S K^\mp
\pi^\pm$ decay mode, and $\chi_{c0}$ and $\chi_{c2}$ via their decay
into $\pi^+\pi^-\pi^+\pi^-$. The L3 collaboration detected $\eta_c$ in
various decay modes with $\pi^\pm$, $K^\pm$, $\pi^0$, $K^0$, $\eta'$
and $\rho^\pm$ in the final state. Therefore the selection criteria were
the usual ones for exclusive events with such type of particles.

Summarizing the experience collected by $\EPEM$ experiments to study
$\GG$ interactions, one can see typical event selection criteria,
which can be adapted to heavy ion collisions. The $\GG$ events are
characterized by the absence of the scattered beam particles, a low (or
fixed in the case of exclusive reactions) multiplicity in the
final state, low transverse momentum of the final state, zero net
charge of the tracks, low (compared to the annihilation processes) energy
deposited in the detectors. Depending on the reaction to be studied,
different topology cuts can be applied. In the case of inclusive
reactions a Monte Carlo based unfolding procedure is used to obtain
the $\GG$ mass from the visible event mass.

\subsubsection{Experiments at Heavy Ion Colliders}
\label{ssec:experimentsHI}

The STAR collaboration is the first one to study $\GG$-processes
in heavy ion collisions in the AuAu collisions at 100--250~A~GeV at RHIC
\cite{KleinS95a,KleinS97b,NystrandK97,STAR347}. The scope of their
interest is lepton pair production, meson pair production, meson
spectroscopy, baryon pair production, and possibly charmonium
production. They studied background processes, trigger and
experimental techniques to select $\GG$-processes.

\paragraph*{\underline{\rm Background.}}
STAR studied several major sources of background for the two-photon
processes: peripheral hadronic nuclear collisions, beam-gas
interactions, cosmic rays, photon-nucleus collisions and Pomeron
processes. Cosmic rays and beam gas reactions are a problem mainly at
the trigger stage, because the more precise data analysis at the later stage
rejects such kind of events via track and vertex information. Detailed 
Monte Carlo
simulation has been performed to separate signal from background
\cite{NystrandK97}. Peripheral hadronic nuclear collisions were
simulated with two standard nuclear event generators, FRITIOF
\cite{Fritiof} and VENUS \cite{Venus} with impact parameters from 12
to 20~fm. Beam-gas interactions were simulated by the same event
generators as Au$p$ and AuN collisions where hydrogen $p$ and
nitrogen $N$ represent the beam-gas content. Photonuclear reactions were
studied by DTUNUC \cite{Dtunuc}. Only $\gamma$Au collisions with photon 
energies larger than 10~GeV were
simulated. Cosmic rays were simulated with a flux of $1.8\times
10^2~\mbox{m}^{-2}\mbox{s}^{-1}$ and the trigger rates in the TPC were
estimated. The most important cuts, which discriminate $\GG$ events
from background, were found to be:
\begin{itemize}
\item Multiplicity: many two-photon reactions, which can be detected at
  RHIC, have two or four charged particles in the final state;
\item the sum of the transverse momentum of the final state particles should 
  be small, of the order $\sqrt{2}\hbar c/R$; in the analysis a cut
  $|\Sigma \vec{p_T}| \leq 40$~MeV was used;
\item the c.m. rapidity distribution of the $\GG$ system is
  centered around zero with a narrow width; the cut $|y_{\GG}|\leq 1.0$
  was applied; this cut is realized automatically by the
  detector acceptance and reduces also the beam-gas and photonuclear
  interactions since these are characterized by an asymmetric particle
  emission.
\end{itemize}
Monte Carlo simulation performed by the STAR collaboration
\cite{NystrandK97} showed that these cuts are able to separate the signals
from background.

\paragraph*{\underline{\rm First experimental results from STAR.}}
RHIC had a first AuAu run in 2000, and the first reports on
observation of coherent interactions in nuclear collisions appeared at
the beginning of 2001. In \cite{Klein01b} the first observation
of exclusive $\rho^0$ production is presented. The $\rho^0$ are
produced with small perpendicular momentum, as expected, if they couple
coherently to both nuclei. Here we pay special attention to the event
selection used.

\begin{figure}[tbh]
\parbox[b]{0.44\hsize}{
\resizebox{\hsize}{!}{\includegraphics{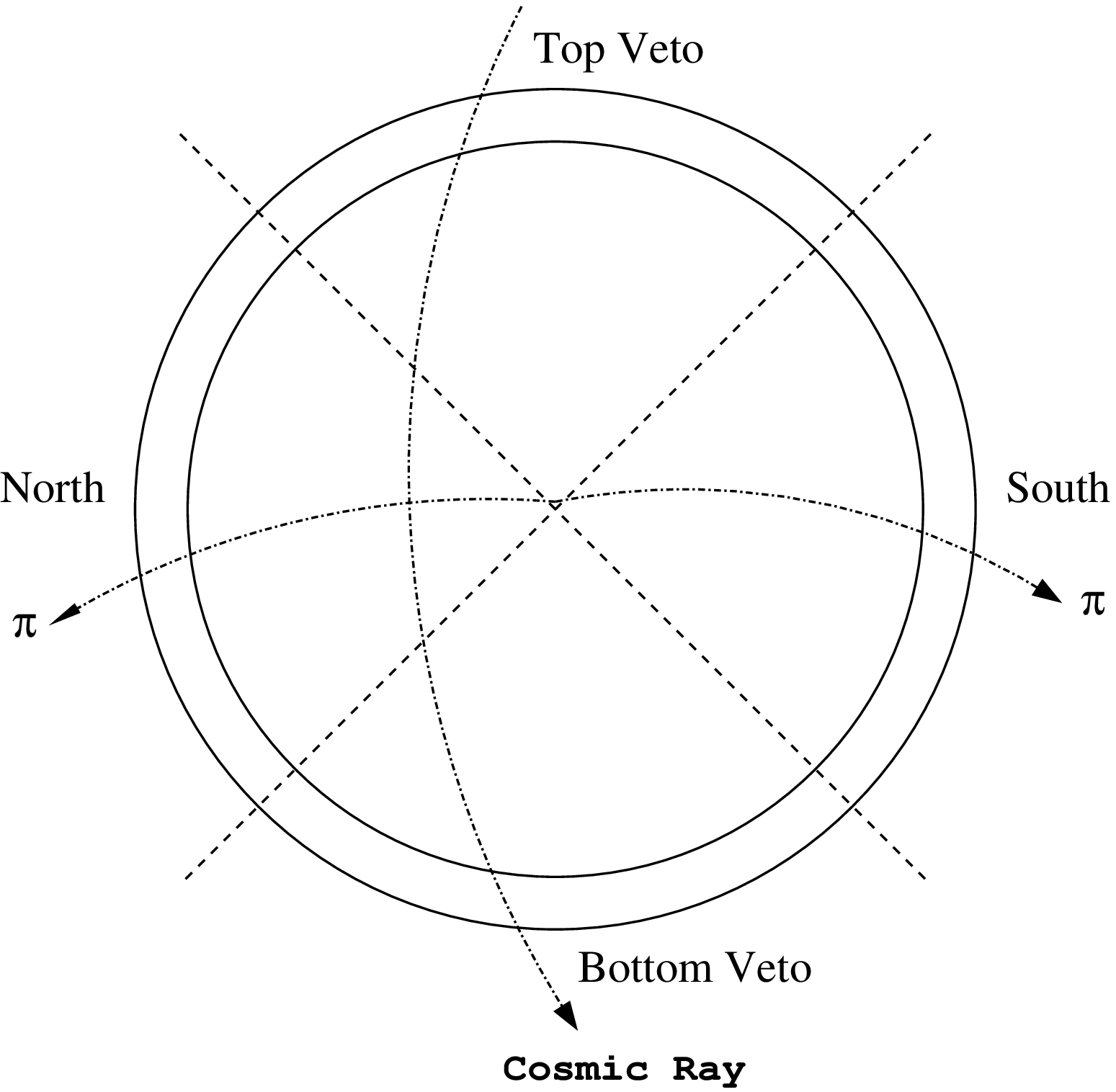}}
\caption{Topology trigger on coherent interactions at STAR. Figure
taken from \cite{Klein01b}.}
\label{STAR:topology}
}
\hfill
\parbox[b]{0.53\hsize}{
\resizebox{\hsize}{!}{\includegraphics[bb=37 19 521 380]{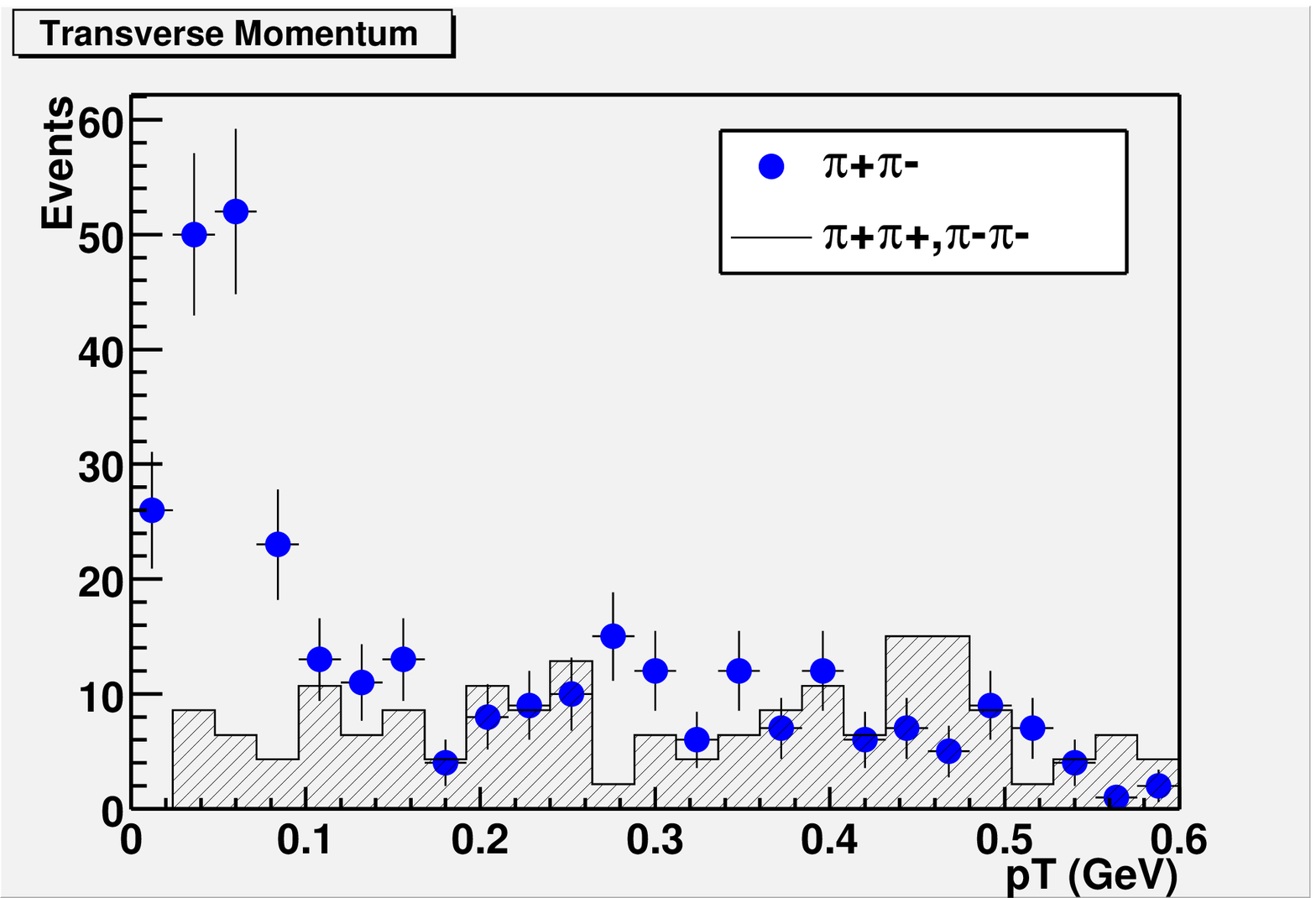}}
\caption{The transvers momentum $P_\perp$ of the two pions 
triggered with the topological trigger. The enhancement at small transverse
momentum, coming from the very peripheral collisions, is clearly seen. 
Figure taken from \cite{Klein01b}.}
\label{STAR:pt}
}
\end{figure}

The production of $\rho^0$s were studied in the decay channel
$\rho^0\to\pi^+\pi^-$. The data on AuAu collisions at
$\sqrt{s_{NN}}=130$~GeV was collected during the Summer 2000 run. The
$\rho$ production was studied with two separate triggers. The topology
trigger was designed to trigger $\rho^0$ decays detected in the
central trigger barrel (CTB). The topology trigger divided the CTB
into 4 azimuthal quadrants as shown in Fig.~\ref{STAR:topology}.
This trigger selects events with at least one hit in the North and the
South sectors. The top and bottom quadrants were used to reject most
cosmic rays. The selected events were then passed through a level-3
trigger, which rejects events with more that 15 tracks or a vertex far
outside the interaction point. These cuts rejected 95\% of the events and
rejected events from central collisions, beam-gas interactions and cosmic
rays.

A minimum bias trigger used both zero degree calorimeters (ZDC) to select
events, where both nuclei dissociated. This trigger required a
coincidence between the two ZDCs with a single (or more) neutron depositing
their energy
in each ZDC.

In the analysis STAR selected events with exactly two tracks with a
vertex within 2~cm of the center of the TPC in the transverse plane, and 
within 2~m along the beams. A slightly acoplanarity of the track pairs was
required to reject the remaining cosmic rays. The sum of the transverse 
momentum of the track pairs has a peak at low $p_T$, which shows that they 
are coming from coherent collisions, see Fig.~\ref{STAR:pt}. 
The invariant mass spectra of the $\pi^+\pi^-$-pairs in events selected by 
both kind of triggers shows a peak around the $\rho^0$ mass.

The STAR collaboration was the first one who studied experimentally
the coherent interaction of heavy ions. In 2001, STAR has
upgraded the detectors and the trigger, which will allow to increase the
data taken by several orders of magnitude. Also it allows to study other
reactions in coherent heavy ion interactions.

\subsection{Perspectives of ion experiments}

As was discussed before, $\GG$ processes have been studied 
thoroughly at $\EPEM$-experiments up to the $\GG$-invariant masses 
of about 3~GeV. Recently the $B$-factory at KEK, the $\EPEM$-collider 
KEKB started operating, and a few papers have  already appeared concerning 
the detection of $\GG$-processes in the experiment BELLE at KEKB. These 
recent reports have attracted some attention of the physics community 
towards KEKB as a new facility to study two-photon physics.
Actually KEKB provides an ``{\it unprecedented luminosity}'' 
\cite{CERN-Courier-41} of
$4.49\times 10^{33}~\mbox{cm}^{-2}\mbox{s}^{-1}$.
Thus the $\GG$-luminosity achieved at KEKB is several orders of magnitude 
higher than that at RHIC in the whole range of $\GG$ invariant masses possible 
in both colliders. So far the BELLE collaboration has reported 
only on a few measurements of $\GG$ processes \cite{Huang2001,Hou2001}, based 
on the first run stored in 2000, but one expects that a lot of
high-statistics results will appear from BELLE soon and the low-mass
$\GG$-physics will be exhausted there on the level of femtobarn cross 
sections. 
However the physics domain of photon-Pomeron and Pomeron-Pomeron processes 
at RHIC is out of the competition with any lepton collider.

For the experiments at LHC, the possibilities to study very peripheral ion 
collisions are currently discussed by the CMS and ALICE
collaborations.
The effective $\GG$-luminosity at LHC is large enough due to the high 
collision energy. In addition the $\GG$ mass range, that can be achieved 
in ion collisions there, allows to study $\GG$ processes well beyond 
the mass range studied at LEP, see Fig.~\ref{fig:gglum}.

CMS considers an extensive heavy ion program, which includes also 
the physics of very peripheral ion collisions
\cite{BaurHTS98,CMS-2000-060}.  This program is quite attractive, especially
as CMS provides a detector with a rather wide aperture and allowing thus to 
study various $\GG$-reactions.
The physics of very peripheral ion collisions at ALICE has been discussed in 
a number of papers, see \cite{ALICE95,Sadovsky93,Sadovsky95}.
Although ALICE is a quite promising setup for studies of this kind of physics
a trigger configuration for very peripheral events has not been developed
up to now. Nevertheless some possible options were proposed, see, i.e., 
\cite{Kolosov95}, and problem is still under discussion.

\section{Conclusion}
\label{sec_conclusion}

In this report we describe the  basic properties
of very peripheral ion-ion collisions. Due to the very strong 
electromagnetic fields of short duration, new possibilities
for interesting physics arise. The study of these electromagnetic
processes, that is, photon-photon and photon-ion collisions is an
interesting option, complementing the program for central collisions.
It is the study of events characterized by relatively small multiplicities
and a small background (especially when compared with the central collisions).
These are good conditions to search for new physics.
The method of equivalent photons is a well established tool to
describe these kinds of reactions. Reliable results of quasi-real
photon fluxes and $\GG$-luminosities are available. Unlike electrons
and positrons heavy ions and protons are particles with an internal
structure. Effects arising from this structure are well under
control and minor uncertainties coming from the exclusion of central
collisions and triggering can be eliminated by using a luminosity
monitor from $\mu$-- or $e$--pairs. 

The high photon fluxes open up possibilities for photon-photon as well
as photon-nucleus interaction studies up to energies hitherto
unexplored at the forthcoming colliders RHIC and LHC.
Heavy ion colliders are a unique tool to study 
(quasi-real) photon-nucleus collisions in a hitherto inaccessible 
energy range. First experimental results at RHIC on coherent $\rho$
production on Au nuclei are forthcoming, as well as, pion- and electron-pair 
production in very peripheral collisions.
A wealth of new data is to be expected.
Interesting physics can be explored at the high invariant
$\GG$-masses, where detecting new particles could be within
range. Also very interesting studies within the standard model, i.e.,
mainly QCD studies will be possible. This ranges from the study of the
total $\GG$-cross section into hadronic final states up to invariant
masses of about 100~GeV to the spectroscopy of light and heavy
mesons. The production via photon-photon fusion complements the
production from single timelike photons in
$e^+$--$e^-$ collider and also in hadronic collisions via other
partonic processes.

A good trigger for very peripheral collisions is essential in order to select 
photon-photon events. As it was shown in this report, such a trigger will be 
possible based on the 
survival of the nuclei after the collision and the use of the small 
transverse momenta of the produced system. A problem, which is difficult to 
judge quantitatively at the moment, is the influence of strong interactions 
in grazing collisions, i.e., effects arising from the nuclear stratosphere and
Pomeron interactions. 
With the pioneering experiments of the (Ultra)Peripheral Collisions Group
at STAR at RHIC this field has definitely left the area of theoretical
speculations and entered the area to be experimental feasible.

Very peripheral collisions with  Photon-Pomeron and Pomeron-Pomeron
interactions, that is, diffractive processes are an additional
interesting option. They use essentially the same triggering conditions and
therefore one should be able to record them at the same time as
photon-photon events.

This review is written in a time of rapid progress, both from an experimental 
and theoretical point of view. A few theorists are studying various aspects 
of very peripheral collisions, many study the field
of $\gamma$-hadron processes with, e.g., the interactions of vector mesons
in the nuclear medium and the vast field of $\GG$ collisions ranging from 
QCD studies to possible new physics. On the experimental side, experiments 
on very peripheral collisions are presently performed and analyzed at RHIC.
With these experiments the field has left the area of 
purely theoretical ideas and entered into the reality of experiments.
The first heavy ion beams are expected at LHC in 2007 and experiments
are in a construction stage. Certainly, the experience at RHIC will 
help to plan (even) better for LHC. The future is coming and one thing is 
clear: with all the new forthcoming experimental results the next review of 
the field of very peripheral collisions will be very different from this one.  

\subsection{Acknowledgments}

We are grateful to very many people for their collaboration, their
discussions, encouragements and suggestions at various stages on the
present subject. We thank them all. We especially wish to thank
N.N.~Achasov,A.~Alscher, A.~Aste, M.~Bedjidian, C.A.~Bertulani,
D.~Brand, D.~Denegri, O.~Drapier, V.V.~Ezhela I.F.~Gin\-zburg, S.~Datz,
K.~Eggert, R.~Engel, M.~Greiner, Z.~Halabuka, D.~Haas, B. ~Jeanneret,
S.R.~Klein, H.~Meier, N.N.~Nikolaev, J.~Norbury, J. ~Nystrand,
A.P.Samokhin, L.I.~Sarycheva, V.~Serbo, W.~Scheid,
Ch.~Schei\-den\-ber\-ger,
R.~Schuch, G.N.~Shestakov, G.~Soff, S.~White, U.~Wiedemann, and P.~Yepes.


\begin{thebibliography}{100}

\bibitem{QM2001}
Proceedings of the Quark Matter QM2001, Brookhaven, January 2001, available
  from {\tt http://www.rhic.bnl.gov/qm2001/}.

\bibitem{Angelis97}
A. Angelis {\it et~al.}, J. Phys. G {\bf 23},  2069  (1997).

\bibitem{Anselm1989}
A.~A. Anselm, Phys. Lett.~B {\bf 217},  169  (1989).

\bibitem{Bjorken1997}
J.~D. Bjorken, Acta Phys. Polon. {\bf B28},  2773  (1997).

\bibitem{Fermi24}
E. Fermi, Z. Phys. {\bf 29},  315  (1924).*

\bibitem{BertulaniB94}
C.~A. Bertulani and G. Baur, Physics Today, March,  22  (1994).

\bibitem{JacksonED}
J.~D. Jackson, {\em Classical Electrodynamics} (John Wiley, New York, 1975).*

\bibitem{Krasnitz:2001ph}
A. Krasnitz and R. Venugopalan, Small x physics and the initial conditions in
  heavy ion collisions, e-print hep-ph/0104168, 2001.

\bibitem{KleinS97a}
S. Klein and E. Scannapieco,  in {\em Photon '97, Egmond aan Zee}, edited by
  (World Scientific, Singapore, 1997), p.\ 369, e-print hep-ph/9706358.

\bibitem{BertulaniB88}
C.~A. Bertulani and G. Baur, Phys. Rep. {\bf 163},  299  (1988).***

\bibitem{BaurR94}
G. Baur and H. Rebel, Topical Review, J. Phys. G {\bf 20},  1  (1994).

\bibitem{BaurR96}
G. Baur and H. Rebel, Annu. Rev. Nucl. Part. Sci. {\bf 46},  321  (1996).

\bibitem{Primakoff51}
H. Primakoff, Phys. Rev. {\bf 31},  899  (1951).*

\bibitem{Moshammer97}
R. Moshammer {\it et~al.}, Phys. Rev. Lett. {\bf 79},  3621  (1997).

\bibitem{BaurHT98}
G. Baur, K. Hencken, and D. Trautmann, Topical Review, J. Phys. G {\bf 24},
  1657  (1998).***

\bibitem{KraussGS97}
F. Krauss, M. Greiner, and G. Soff, Prog. Part. Nucl. Phys. {\bf 39},  503
  (1997).***

\bibitem{BaurHT98b}
G. Baur, K. Hencken, and D. Trautmann,  in {\em Proceedings of the Lund
  workshop on photon interactions and the photon structure, Lund, September
  10-13, 1998}, edited by G. Jarlskog and T. Sjostrand (Dep. of Physics and
  Theoret. Physics, Lund University, Lund, Sweden, 1998).

\bibitem{HenckenSTB99}
K. Hencken, P. Stagnoli, D. Trautmann, and G. Baur, Nucl. Phys.~B {\bf 82},
  409  (2000).

\bibitem{KleinN99}
S. Klein and J. Nystrand, Phys. Rev.~C {\bf 60},  014903  (1999).***

\bibitem{Klein00b}
S.~R. Klein, Nonlinear QED Effects in Heavy Ion Collisions, LBNL-47144, e-print
  physics/0012021, Talk presented at the 18th Advanced ICFA Beam Dynamics
  Workshop on Quantum Aspects of Beam Physics, October 15-20, 2000, Capri,
  Italy, 2000.*

\bibitem{EngelRR97}
R. Engel {\it et~al.}, Z. Phys. C {\bf 74},  687  (1997).*

\bibitem{BaurHTS98}
G. Baur {\it et~al.}, Photon-Photon Physics with heavy ions at CMS, CMS Note
  1998/009, available from the CMS information server at
  http://cmsserver.cern.ch, 1998.

\bibitem{BaurHTS99}
G. Baur {\it et~al.}, Coherent Interactions with heavy ions at CMS, e-print
  hep-ph/9904361, 1999.*

\bibitem{Felix97}
K. Eggert {\it et~al.}, FELIX Letter of Intent, CERN/LHCC 97--45, LHCC/I10, to
  appear in J. Phys. G, 1997.***

\bibitem{LandauL34}
L.~D. Landau and E.~M. Lifshitz, Phys. Z. Sowjet. {\bf 6},  244  (1934).*

\bibitem{Racah37}
G. Racah, Nuovo Cim. {\bf 14},  93  (1937).*

\bibitem{BaurB88}
G. Baur and C.~A. Bertulani, Z. Phys. A {\bf 330},  77  (1988).**

\bibitem{GrabiakMG89}
M. Grabiak, B. M{\"u}ller, W. Greiner, and P. Koch, J. Phys.~G {\bf 15},  L25
  (1989).**

\bibitem{Papageorgiu89}
E. Papageorgiu, Phys. Rev.~D {\bf 40},  92  (1989).**

\bibitem{Baur90d}
G. Baur,  in {\em CBPF Int. Workshop on relativistic aspects of nuclear
  physics, Rio de Janeiro, Brazil 1989}, edited by T. Kodama {\it et~al.}
  (World Scientific, Singapore, 1990), p.\ 127.

\bibitem{BaurF90}
G. Baur and L.~G. {Ferreira Filho}, Nucl. Phys.~A {\bf 518},  786  (1990).**

\bibitem{CahnJ90}
N. Cahn and J.~D. Jackson, Phys. Rev.~D {\bf 42},  3690  (1990).**

\bibitem{Vannucci80}
F. Vannucci,  in {\em $\gamma\gamma$ Collisions, Proceedings, Amiens 1980},
  Vol.~134 of {\em Lecture Notes in Physics}, edited by G. Cochard
  (Springer-Verlag, Heidelberg, Berlin, New York, 1980), p.\ 238.

\bibitem{GinzburgS98}
I.~F. Ginzburg and A. Schiller, Phys. Rev.~D {\bf 57},  R6599  (1998).

\bibitem{ShamovT98}
{A. G. Shamov and V. I. Telnov, talk by A. Maslennikov},  in {\em Proceedings
  of the Lund workshop on photon interactions and the photon structure, Lund,
  September 10-13, 1998}, edited by G. Jarlskog and T. Sjostrand (Dep. of
  Physics and Theoret. Physics, Lund University, Lund, Sweden, 1998).

\bibitem{Martin01}
A.~D. Martin, First Workshop on Forward Physics and Luminosity Determination at
  the LHC, IPPP/01/13, e-print hep-ph/0103296, 2001.

\bibitem{AlderBHM56}
K. Alder {\it et~al.}, Rev. Mod. Phys. {\bf 28},  432  (1956).

\bibitem{AlderW66}
{\em Coulomb Excitation}, {\em Perspectives in Physics}, edited by K. Alder and
  A. Winther (Academic Press, New York, London, 1966).

\bibitem{AlderW75}
K. Alder and A. Winther, {\em Electromagnetic excitation} (North-Holland,
  Amsterdam, 1975).

\bibitem{WintherA79}
A. Winther and K. Alder, Nucl. Phys.~A {\bf 319},  518  (1979).

\bibitem{BaurB89}
G. Baur and C.~A. Bertulani, Nucl. Phys.~A {\bf 505},  835  (1989).**

\bibitem{HofmannSSG91}
C. Hofmann, G. Soff, A. Sch{\"a}fer, and W. Greiner, Phys. Lett.~B {\bf 262},
  210  (1991).**

\bibitem{BaronB93}
N. Baron and G. Baur, Phys. Rev.~C {\bf 48},  1999  (1993).

\bibitem{GreinerVHS95}
M. Greiner {\it et~al.}, Phys. Rev.~C {\bf 51},  911  (1995).

\bibitem{KleinNV00}
S.~R. Klein, J. Nystrand, and R. Vogt, Eur. Phys. J. C {\bf 21},  583  (2001).*

\bibitem{KleinS97b}
S. Klein and E. Scannapieco, Coherent Photons and Pomerons in Heavy Ion
  Collisions, presented at 6th Conference on the Intersections of Particle and
  Nuclear Physics, May 1997, Big Sky, Montana, STAR Note 298, LBNL-40495,
  e-print nucl-th/9707008, 1997.

\bibitem{KleinS95a}
S. Klein and E. Scannapieco, STAR Note 243, 1995.

\bibitem{KleinS95b}
S. Klein,  in {\em Photon '95, Sheffield}, edited by D.~J. Miller, S.~L.
  Cartwright, and V. Khoze (World Scientific, Singapore, 1995), p.\ 417.

\bibitem{NystrandK98}
J. Nystrand and S. Klein, Talk presented at Workshop on Photon Interactions and
  the Photon Structure, Lund, Sweden, 10-13 Sep 1998. In Lund 1998, Photon
  interactions and the photon structure, 263-277; LBNL-42524 1998, e-print
  nucl-th/9811007, 1998.**

\bibitem{Klein01b}
S. Klein and {the STAR collaboration}, Observation of $Au + Au \to Au + Au +
  \rho^0$ and $Au + Au \to Au^* + Au^* + \rho^0$ with STAR, LBNL-47723, e-print
  nucl-ex/0104016, 2001.***

\bibitem{Meissner01}
F. Meissner, Contribution to the Schlechting Conference, 2001.

\bibitem{Klein99}
S.~R. Klein, Coherent Photonuclear Interactions at RHIC: Theory and Experiment,
  Talk at the RHIC Winter Workshop, 7-9 January 1999, LBNL, Berkeley, USA,
  1999.

\bibitem{Brodsky99}
S.~J. Brodsky, Novel Peripheral Processes at RHIC, talk at the RHIC Winter
  Workshop, 7-9 January 1999, LBNL, Berkeley, USA, 1999.

\bibitem{HenckenKKS96}
K. Hencken {\it et~al.}, TPHIC, event generator of two photon interactions in
  heavy ion collisions, IHEP-96-38, 1996.

\bibitem{Sadovsky93}
S. Sadovsky, CERN-Note ALICE/92-07, 1993.*

\bibitem{Bjorken99}
J.~D. Bjorken, Nucl. Phys. B (Proc. Suppl.) {\bf 71},  484  (1999).

\bibitem{Freiburg99}
 in {\em Proceedings of the Photon'99 conference, Freiburg, May 23--27, 1999},
  edited by S. S{\"o}ldner-Rembold (Nucl. Phys. (Proc. Suppl.) 82B,
  North-Holland, 2000).

\bibitem{Ambleside00}
Proceedings of the Photon 2000 conference, Ambleside, England August 26--31,
  2000, to be published in the AIP Conference Proceeding Series.

\bibitem{BudnevGM75}
V.~M. Budnev, I.~F. Ginzburg, G.~V. Meledin, and V.~G. Serbo, Phys. Rep. {\bf
  15},  181  (1975).***

\bibitem{HenckenTB95}
K. Hencken, D. Trautmann, and G. Baur, Z. Phys. C {\bf 68},  473  (1995).*

\bibitem{HalzenM84}
F. Halzen and A.~D. Martin, {\em Quarks \& Leptons} (John Wiley \& Sons, New
  York, 1984).

\bibitem{LedererS78}
{\em Tables of Isotopes, 7th edition}, edited by C.~M. Lederer and V.~S.
  Shirley (Wiley, New York, 1978).

\bibitem{Bertulani93}
C. Bertulani, Phys. Lett.~B {\bf 319},  421  (1993).

\bibitem{Kniehl91}
B. Kniehl, Phys. Lett.~B {\bf 254},  267  (1991).

\bibitem{BlattW79}
J.~M. Blatt and V.~F. Weisskopf, {\em Theoretical Nuclear Physics}
  (Springer-Verlag, New York, 1979).

\bibitem{Walecka83}
J.~D. Walecka, ANL-83-50, 1983.

\bibitem{deForestW66}
T. deForest and J.~D. Walecka, Adv. Phys. {\bf 15},  1  (1966).

\bibitem{Chanfray93}
G. Chanfray, J. Delorme, M. Ericson, and A. Molinari, Nucl. Phys.~A {\bf 556},
  439  (1993).

\bibitem{Tsai74}
Y.-S. Tsai, Rev. Mod. Phys. {\bf 46},  815  (1974).

\bibitem{Moniz71}
E.~J. Moniz {\it et~al.}, Phys. Rev. Lett. {\bf 26},  445  (1971).

\bibitem{DreesGN94}
M. Drees, R.~M. Godbole, N. Nowakowski, and S.~D. Rindami, Phys. Rev.~D {\bf
  50},  2335  (1994).*

\bibitem{OhnemusWZ94}
J. Ohnemus, T.~F. Walsh, and P.~M. Zerwas, Phys. Lett.~B {\bf 328},  369
  (1994).*

\bibitem{DreesEZ89}
M. Drees, J. Ellis, and D. Zeppenfeld, Phys. Lett.~B {\bf 223},  454  (1989).*

\bibitem{DreesZ89}
M. Drees and D. Zeppenfeld, Phys. Rev.~D {\bf 39},  2536  (1989).

\bibitem{Levtchenko96}
B.~B. Levtchenko, Higgs Production and other Two-Photon Processes in eA
  Collisions at HERA Energies, SINP-96-71, e-print hep-ph/9608295.

\bibitem{Piotrzkowski00}
K. Piotrzkowski, Phys. Rev.~D {\bf 63},  071502  (2000).

\bibitem{BaurF91}
G. Baur and L.~G. {Ferreira Filho}, Phys. Lett.~B {\bf 254},  30  (1991).

\bibitem{Baron94}
N. Baron, Ph.D. thesis, Forschungszentrum J{\"u}lich, Institut f{\"u}r
  Kernphysik J{\"u}l-2846, 1994.

\bibitem{HenckenTB94}
K. Hencken, D. Trautmann, and G. Baur, Phys. Rev.~A {\bf 49},  1584  (1994).

\bibitem{HenckenTB00}
K. Hencken, D. Trautmann, and G. Baur, Phys. Rev.~C {\bf 61},  027901  (2000).

\bibitem{Lenkeit98}
{B. Lenkeit for the CERES Collaboration}, New results on low-mass lepton pair
  production in Pb-Au collisions at 158 GeV per nucleon, Paris conference?,
  1998.

\bibitem{FaeldtG90}
G. F{\"a}ldt and R. Glauber, Phys. Rev. {\bf 42},  395  (1990).

\bibitem{BeneshHF96}
C.~J. Benesh, A.~C. Hayes, and J.~L. Friar, Phys. Rev.~C {\bf 54},  1404
  (1996).

\bibitem{BaurB97}
G. Baur and C. Bertulani, Phys. Rev.~C {\bf 56},  581  (1997).

\bibitem{BeneshHF97}
C.~J. Benesh, A.~C. Hayes, and J.~L. Friar, Phys. Rev.~C {\bf 56},  583
  (1997).

\bibitem{RubehnMT97}
W.~F. J.~M. {Th}~Rubehn and W. Trautmann, Phys. Rev.~C {\bf 56},  1165  (1997).

\bibitem{Baur92}
G. Baur, Z. Phys. C {\bf 54},  419  (1992).

\bibitem{BaurB93}
G. Baur and N. Baron, Nucl. Phys.~A {\bf 561},  628  (1993).

\bibitem{VidovicGB93}
M. Vidovi{\'c}, M. Greiner, C. Best, and G. Soff, Phys. Rev.~C {\bf 47},  2308
  (1993).

\bibitem{BrandtEM94}
D. Brandt, K. Eggert, and A. Morsch, CERN AT/94-05(DI), 1994.

\bibitem{RHIC:CDR}
Conceptual design of the Relativistic Heavy Ion Collider, BNL-52195, 1989.

\bibitem{Brandt00}
D. Brandt, Review of the LHC Ion Programme, LHC Project Report 450, 2000.

\bibitem{Morsch2001}
A. Morsch, ALICE Internal Note ALICE-INT-2001-10, 2001.

\bibitem{Evans1993}
L.~R. Evans,  in {\em Particle Accelerator Conference, Washington, DC, USA, 17
  - 20 May 1993} (IEEE, New York, 1993), p.\ 1983.

\bibitem{KolanoskiZ88}
H. Kolanoski and P. Zerwas,  in {\em High Energy Electron-Positron Physics},
  edited by A. Ali and P. S{\"o}ding (World Scientific, Singapore, 1988).

\bibitem{BergerW87}
{Ch. Berger and W. Wagner}, Phys. Rep. {\bf 176C},  1  (1987).

\bibitem{Amiens80}
{\em {$\gamma\gamma$} Collisions, Proceedings, Amiens 1980}, Vol.~134 of {\em
  Lecture Notes in Physics}, edited by G. Cochard and P. Kessler (Springer,
  Berlin, 1980).

\bibitem{SanDiego92}
D.~O. Caldwell and H.~P. Paar, {\em Proc. 9th International Workshop on
  Photon-Photon Collisions, San Diego (1992)} (World Scientific, Singapore,
  1992).

\bibitem{Sheffield95}
{\em Photon'95, Xth International Workshop on Gamma-Gamma Collisions and
  related Processes}, edited by D.~J. Miller, S.~L. Cartwright, and V. Khoze
  (World Scientific, Singapore, 1995).

\bibitem{Egmond97}
{\em Photon'97, XIth International Workshop on Gamma-Gamma Collisions and
  related Processes, Egmond aan Zee}, edited by A. Buijs (World Scientific,
  Singapore, 1997).

\bibitem{L3:01}
{L3 Collaboration}, Total Cross Section in gamma gamma Collisions at LEP,
  CERN-EP/2001-012, L3 preprint 235, to appear in Phys. Lett. B (2001), 2001.

\bibitem{SchulerS97}
G. Schuler and T. Sj{\"o}strand, Z. Phys. C {\bf 73},  677  (1997).

\bibitem{DreesG85}
M. Drees and K. Grassie, Z. Phys. C {\bf 28},  451  (1985).

\bibitem{SchulerS95}
G. Schuler, Z. Phys. C {\bf 68},  607  (1995).

\bibitem{Gluck92}
M. Gl{\"u}ck, E. Reya, and A. Vogt, Z. Phys. C {\bf 53},  127  (1992).

\bibitem{PDG00}
{Particle Data Group}, Eur. Phys. J. C {\bf 15},  1  (2000).

\bibitem{Peskin00}
M.~E. Peskin, Theoretical Summary Lecture for EPS HEP99, SLAC-PUB-8351, e-print
  hep-ph/0002041, 2000.

\bibitem{L3:97}
{L3 collaboration}, Phys. Lett.~B {\bf 408},  450  (1997).*

\bibitem{Barklow90}
T. Barklow, SLAC-PUB-5364, 1990.

\bibitem{BrodskyKT70} 
S.J~Brodsky, T.~Kinoshita, and H.~Terazawa, 
  Phys. Rev. Lett. {\bf 25},  972  (1970).

\bibitem{BudnevGMS72}
V.~M. Budnev, I.~F. Ginzburg, G.~V. Meledin, and V.~G. Serbo, Phys. Lett.~B
  {\bf 39},  526  (1972).

\bibitem{BudnevGMS73}
V.~M. Budnev, I.~F. Ginzburg, G.~V. Meledin, and V.~G. Serbo, Nucl. Phys.~B
  {\bf 63},  519  (1973).*

\bibitem{Katuya83}
M. Katuya, Phys. Lett.~B {\bf 124},  421  (1983).

\bibitem{TupperS81}
G. Tupper and M.~A. Samuel, Phys. Rev.~D {\bf 23},  1933  (1981).

\bibitem{GribovP62}
V.~N. Gribov and I.~Y. Pomeranchuk, Phys. Rev. Lett. {\bf 8},  343  (1962).

\bibitem{GribovP62b}
V.~N. Gribov and I.~Y. Pomeranchuk, Phys. Rev. Lett. {\bf 8},  412  (1962).

\bibitem{ARGUS1994}
H. Albrecht {\it et~al.}, Phys. Lett.~B {\bf 332},  451  (1994).

\bibitem{Soldner1998}
S. {S\"oldner-Rembold},  in {\em 29th International Conference on High-Energy
  Physics, Vancouver, Canada, 23--29 July 1998}, edited by A. Astbury, D.~A.
  Axen, and J. Robinson (World Scientific, Singapore, 1999), p.\ 914.

\bibitem{Vogt1999} 
H. Vogt, Nucl. Phys. B (Proc. Suppl.) {\bf 79}, 283 (1999).

\bibitem{Achasov1995}
N.~N. Achasov and G.~N. Shestakov, Phys. Rev.~D {\bf 52},  6291  (1995).

\bibitem{Achasov1999}
N.~N. Achasov and G.~N. Shestakov, Phys. Rev.~D {\bf 60},  117503  (1999).

\bibitem{Yang48}
C.~N. Yang, Phys. Rev. {\bf 77},  242  (1948).*

\bibitem{Kwong88}
W. Kwong {\it et~al.}, Phys. Rev.~D {\bf 37},  3210  (1988).

\bibitem{Bodwin94}
Bodwin, Nucl. Phys. (Proc. Suppl.) {\bf 42},  306  (1995).

\bibitem{Cartwright98}
S. Cartwright {\it et~al.}, J. Phys.~G {\bf 24},  457  (1998).

\bibitem{Godang97}
R. Godang {\it et~al.}, Phys. Rev. Lett. {\bf 79},  3829  (1997).

\bibitem{Achasov1991}
N.N. Achasov and G.N. Shestakov, Sov. Phys. Usp. {\bf 34}, (No.6) 471
  (1991).

\bibitem{ARGUS1991}
H. Albrecht {\it et~al.}, Z. Phys., C {\bf 50} 1 (1991).

\bibitem{Wilczek00}
F. Wilczek, Nature {\bf 404},  452  (2000).

\bibitem{Savinov01}
V. Savinov and {CLEO Collaboration}, QCD Studies in Two-Photon Collisions at
  CLEO, Contribution W03 to the SLAC Workshop on $e^+e^-$ Physics at
  Intermediate Energies, eConf C010430 (2001) W03, 
  e-print hep-ex/0106013, 2001.

\bibitem{Telnov95}
V. Telnov,  in {\em Photon '95, Sheffield}, edited by D.~J. Miller, S.~L.
  Cartwright, and V. Khoze (World Scientific, Singapore, 1995), p.\ 369.

\bibitem{ginzburg95}
I.~F. Ginzburg,  in {\em Photon '95, Sheffield}, edited by D.~J. Miller, S.~L.
  Cartwright, and V. Khoze (World Scientific, Singapore, 1995), p.\ 399.

\bibitem{Ginzburg00}
I. Ginzburg, Nucl. Phys. (Proc. Suppl.) {\bf 82B},  367  (2000), contribution
  to DIFFRACTION 2000.

\bibitem{TESLATDR01}
{\em Tesla Technical Design Report}, edited by F. Richard {\it et~al.} (DESY,
  Hamburg, 2001), No.~DESY 2001-011, ECFA 2001-209, TESLA Report 2001-23,
  TESLA-FEL 2001-05, ISBN 3-935702-05-1.

\bibitem{Krawczyk96}
M. Krawczyk,  in {\em Future Physics at HERA}, edited by G. Ingelman, A. {De
  Roeck}, and R. Klanner (DESY, Hamburg, 1996), iFT 21/96, hep-ph/9609477.

\bibitem{herafuture96}
  {\em Future Physics at HERA}, edited by G. Ingelman, A. {De Roeck}, and R.
  Klanner (DESY, Hamburg, 1996).

\bibitem{HaberK85}
H.~E. Haber and G.~L. Kane, Phys. Rep. {\bf 117},  75  (1985).

\bibitem{FengS95}
J. Feng and M. Strassler, Phys. Rev.~D {\bf 51},  4661  (1995).

\bibitem{AchieserB62}
A.~I. Achieser and W.~B. Berestezki, {\em Quanten-Elektrodynamik}
  (Teubner-Verlag, Leipzig, 1962).

\bibitem{MuellerS90}
B. {M\"uller} and A.~J. Schramm, Phys. Rev.~D {\bf 42},  3699  (1990).

\bibitem{Papageorgiu95}
E. Papageorgiu, Phys. Lett.~B {\bf 352},  394  (1995).

\bibitem{Norbury90}
J. Norbury, Phys. Rev.~D {\bf 42},  3696  (1990).

\bibitem{LiettiNRR01}
S.~M. Lietti, A.~A. Natale, C.~G. Rold{\~a}o, and R. Rosenfeld, Phys. Lett.~B
  {\bf 497},  243  (2001).*

\bibitem{ChoudhuryK97}
D. Choudhury and M. Krawczyk, Phys. Rev.~D {\bf 55},  2774  (1997).

\bibitem{GreinerVS93}
G.~S. M.~Greiner, M.~Vidovic, Phys. Rev.~C {\bf 47},  2288  (1993).

\bibitem{Abrue2001}
P. Abrue {\it et~al.}, Phys. Lett.~B {\bf 499},  23  (2001).

\bibitem{Achard2001}
P. Achard {\it et~al.}, Preprint CERN-EP/2001-049.

\bibitem{Abbiendi2001}
G. Abbiendi {\it et~al.}, Phys. Lett.~B {\bf 499},  38  (2001).

\bibitem{LEPWGH2001}
{LEP working group for Higgs boson search}, LHWG Note/2001-03,
e-print hep-ex/0107029.

\bibitem{Cabibbo1979}
N. Cabibbo {\it et~al.}, Nucl. Phys.~B {\bf 158},  295  (1979).

\bibitem{Pomeran1970}
I.~Y. Pomeranchuk, Yad.Fiz. {\bf 11},  852  (1970).

\bibitem{Logunov1978}
A.~A. Logunov {\it et~al.}, Theor. and Math. Phys. {\bf 36},  147  (1978).

\bibitem{Linder1989}
M. Linder {\it et~al.}, Phys. Lett.~B {\bf 228},  139  (1989).

\bibitem{Sher1993}
M. Sher, Phys. Lett.~B {\bf 317},  159  (1993).

\bibitem{Sher1994}
M. Sher, Phys. Lett.~B {\bf 331},  448  (1994).

\bibitem{Altarelli1994}
G. Altarelli {\it et~al.}, Phys. Lett.~B {\bf 337},  141  (1994).

\bibitem{Casas1995}
J.~A. Casas {\it et~al.}, Phys. Lett.~B {\bf 342},  89  (1995).

\bibitem{AhernNP00}
S.~C. Ahern, J.~W. Norbury, and W.~J. Poyser, Phys. Rev.~D {\bf 62},  116001
  (2000).

\bibitem{Abbott98}
B. Abbott, Phys. Rev. Lett. {\bf 81},  524  (1998).

\bibitem{Renard83}
F.~M. Renard, Phys. Lett.~B {\bf 126},  59  (1983).

\bibitem{BaurFF84}
U. Baur, H. Fritzsch, and H. Faissner, Phys. Lett.~B {\bf 135},  313  (1984).

\bibitem{Jain1996}
P. Jain, B. Pire, and J.~P. Ralston, Phys. Rev. {\bf 271},  67  (1996).

\bibitem{Emling95}
H. Emling, Prog. Part. Nucl. Phys. {\bf 33},  729  (1995).

\bibitem{AumannBE98}
T. Aumann, P.~F. Bortignon, and H. Emling, Annu. Rev. Nucl. Part. Sci. {\bf
  48},  351  (1998).

\bibitem{BertulaniP99}
C.~A. Bertulani and V.~Y. Ponomarev, Phys. Rep. {\bf 321},  139  (1999).

\bibitem{Abreu98}
M.~C. Abreu {\it et~al.}, Phys. Rev.~C {\bf 59},  876  (1998).

\bibitem{Pshenichnov01}
I.~A. Pshenichnov {\it et~al.}, Phys. Rev.~C {\bf 64},  024903  (2001).*

\bibitem{Datz97}
S. Datz {\it et~al.}, Phys. Rev. Lett. {\bf 79},  3355  (1997).

\bibitem{Klein01}
S.~R. Klein, Nucl. Instrum. Methods {\bf A459},  51  (2001).**

\bibitem{Baltz98}
A. Baltz, C. Chasman, and S.~N. White, Nucl. Instrum. Methods {\bf 417},  1
  (1998).*

\bibitem{VidovicGS93}
M. Vidovi{\'c}, M. Greiner, and G. Soff, Phys. Rev.~C {\bf 48},  2011  (1993).

\bibitem{BaltzRW96}
A.~J. Baltz, M.~J. Rhoades-Brown, and J. Weneser, Phys. Rev.~E {\bf 54},  4233
  (1996).

\bibitem{BaltzS98}
A.~J. Baltz and M. Strikman, Phys. Rev.~D {\bf 57},  548  (1998).

\bibitem{Pshenichnov98}
I.~A. Pshenichnov {\it et~al.}, Phys. Rev.~C {\bf 57},  1920  (1998).

\bibitem{Gallio00}
M. Gallio,  in {\em CMS Heavy Ion Meeting in St. Petersburg}, edited by M.
  Bedjidian (PUBLISHER, ADDRESS, 2000), p.\ 329.

\bibitem{White01}
M. Chiu {\it et~al.}, Measurement of Mutual Coulomb Dissociation in
  {$\sqrt{s_NN}=130$} GeV Au+Au collisions at RHIC, e-print nucl-ex/0109018,
  2001.

\bibitem{AdlerDCM00}
C. Adler {\it et~al.}, The RHIC Zero Degree Calorimeters,
Nucl. Instrum. Methods {\bf A461}, 337 (2001).

\bibitem{AliceZDC_TDR}
G. Dellacasa {\it et~al.}, ALICE TRD 3, Zero Degree Calorimeter CERN/LHCC 99-5,
  1999.

\bibitem{PriceGW88}
P.~B. Price, R. Guaxiao, and W.~T. Williams, Phys. Rev. Lett. {\bf 61},  2193
  (1988).

\bibitem{KruscheABM01}
B. Krusche {\it et~al.}, Phys. Rev. Lett. {\bf 86},  4764  (2001).

\bibitem{Drell60}
S.~D. Drell, Phys. Rev. Lett. {\bf 5},  278  (1960).

\bibitem{ChikinKKS00}
K.~A. Chikin {\it et~al.}, Inclusive meson production in peripheral collisions
  of ultrarelativistic heavy ions, nucl-th/0002028, 2000.

\bibitem{KoelbigM68}
K.~S. Koelbig and B. Margolis, Nucl. Phys.~B {\bf 6},  85  (1968).

\bibitem{Crittenden97}
J.~A. Crittenden, {\em Exclusive production of neutral vector mesons at the
  electron-proton collider HERA}, Vol.~140 of {\em Springer tracts in modern
  physics} (Springer, Heidelberg, 1997).

\bibitem{BauerSYP78}
T.~H. Bauer, R.~D. Spital, D.~R. Yennie, and F.~M. Pipkin, Rev. Mod. Phys. {\bf
  50},  261  (1978).**

\bibitem{PautzS98}
A. Pautz and G. Shaw, Phys. Rev.~C {\bf 57},  2647  (1998).

\bibitem{DrellT66}
S.~D. Drell and J.~S. Refil, Phys. Rev. Lett. {\bf 16},  552  (1966).

\bibitem{KleinN00}
S. Klein and J. Nystrand, Phys. Rev. Lett. {\bf 84},  2330  (2000).*

\bibitem{DaphneHandbook}
{\em The second {$DA\Phi NE$} Physics Handbook Vol.~1}, edited by L. Maiani, G.
  Pancheri, and N. Paver (SIS Ufficio Pubblicazioni, P.O. Box 13, I-00044
  Frascati (Roma) Italy, 19XX).

\bibitem{Donnachie01}
A. Donnachie, Presented at e+ e- Physics at Intermediate Energies, SLAC,
  Stanford, California, 30 Apr--2 May 2001; hep-ph/0106197, 2001.

\bibitem{HKN1996}
J.Huefner, B.Kopeliovich, and J.Nemchik, Phys. Lett.~B {\bf 383},  362  (1996).*

\bibitem{HuefnerN81}
J. H{\"u}fner and M.~C. Nemes, Phys. Rev.~C {\bf 23},  2538  (1981).

\bibitem{Ackerstaff99}
K. Ackerstaff and {HERMES Collaboration}, Phys. Rev. Lett. {\bf 82},  3025
  (1999).

\bibitem{RenkPW00}
T. Renk, G. Piller, and W. Weise, Nucl. Phys.~A {\bf 689},  869  (2001).*

\bibitem{Frixione1993}
S.Frixione, M.L.Manango, and P. Nason, Phys. Lett.~B {\bf 308},  137  (1993).

\bibitem{Daum99}
K. Daum, Nucl. Phys. (Proc. Suppl.) {\bf 82B},  212  (2000), contribution to
  Photon 1999.

\bibitem{Hochman2000}
D. Hochman, Talk given at PHOTON 2000: International Workshop on Structure and
  Interactions of the Photon, Ambleside, Lake District, England, 26-031 Aug
  2000; hep-ex/0010050, 2000.

\bibitem{Frixione2001}
S. Frixione, M. Caccaiari, and P. Nason, hep-ph/0107063, 2001.

\bibitem{SchneiderGS92}
S.~M. Schneider, W. Greiner, and G. Soff, Phys. Rev.~D {\bf 46},  2930  (1992).*

\bibitem{Breitweg00}
J. Breitweg {\it et~al.}, Phys. Lett.~B {\bf 481},  213  (2000).

\bibitem{Arneodo94}
M. Arneodo, Phys. Rep. {\bf 240},  301  (1994).

\bibitem{EskolaKR98}
K.~J. Eskola, V.~J. Kolhinen, and P.~V. Ruuskanen, Nucl. Phys.~B {\bf 535},
  351  (1998).

\bibitem{EskolaKR99}
K.~J. Eskola, V.~J. Kolhinen, and P.~V. Ruuskanen, Eur. Phys. J. C {\bf 9},  61
   (1999).

\bibitem{GelisP01}
F.Gelis and A.Peshier, Probing colored glass via $q\bar{q}$ photoproduction.
  hep-ph/0107142 v2, 2001.

\bibitem{Abramowicz01}
H. Abramowicz, Nucl. Phys. (Proc. Suppl.) {\bf 99A},  79  (2001), contribution
  to DIFFRACTION 2000.

\bibitem{Cartiglia97}
N. Cartiglia, Diffraction at HERA, e-print hep-ph/9703245, 1997.

\bibitem{Levy97}
A. Levy, Low-x Physics at HERA, Bantz Lectures, DESY Preprint 97-013, 1993.

\bibitem{pp2pp98}
PP2PP Collaboration, Roman Pot Detectors for Small Angle {$pp$} Scattering,
  Technical Design Report, available from http://www.rhic.bnl.gov/pp2pp/, 1998.

\bibitem{GoulianosCDF01}
{K. Goulianos for the CDF Collaboration}, Diffraction in CDF: Run I results and
  plans for Run II, e-print hep-ex/0107069, 2001, contribution to DIS2001.

\bibitem{AffolderCDF01}
{T. Affolder for the CDF Collaboration}, Double Diffraction Dissociation at the
  Fermilab Tevatron Collider, 
Phys. Rev. Lett. {\bf 87}, 141802 (2001).

\bibitem{Santoro01}
A. Santoro, Nucl. Phys. (Proc. Suppl.) {\bf 99A},  289  (2001), contribution to
  DIFFRACTION 2000, also available hep-ex/0011023.

\bibitem{COMPASS}
G. Baum {\it et~al.}, COMPASS : a proposal for a common muon and proton
  apparatus for structure and spectroscopy. CERN-SPSLC-96-14; SPSLC-P-297,
  1996.

\bibitem{TOTEM99}
W. Kienzle {\it et~al.}, Total Cross Section, Elastic Scatterinf and
  Diffraction Dissociation at the LHC. TOTEM Technical Proposal, CERN/LHCC
  99-7, LHCC/P5, 1999.

\bibitem{PP2PP98}
{PP2PP Collaboration}, Roman Pot Detectors for Small Angle {$pp$} Scattering,
  Technical Design Report, http://www.rhic.bnl.gov/pp2pp/, 1998.

\bibitem{CavasinniPMS85}
V. Cavasinni {\it et~al.}, Z. Phys. C {\bf 28},  487  (1985).

\bibitem{Bjorken92}
J.~D. Bjorken, Int. J. Mod. Physics {\bf A7},  4189  (1992).

\bibitem{Bjorken93}
J.~D. Bjorken, Phys. Rev.~D {\bf 47},  101  (1993).

\bibitem{DonnachieL88}
A. Donnachie and P.~V. Landshoff, Phys. Lett.~B {\bf 207},  319  (1988).

\bibitem{DonnachieL88b}
A. Donnachie and P.~V. Landshoff, Phys. Lett.~B {\bf 303},  634  (1988).

\bibitem{DonnachieL88c}
A. Donnachie and P.~V. Landshoff, Nucl. Phys.~B {\bf 311},  509  (1988/89).

\bibitem{Landshoff01}
P.~V. Landshoff, Pomerons, e-print hep-ph/0108156, 2001.

\bibitem{MuellerS91}
B. {M\"uller} and A.~J. Schramm, Nucl. Phys.~A {\bf 523},  677  (1991).

\bibitem{SchaeferNS90}
A. Sch{\"a}fer, O. Nachtmann, and R. Sch{\"o}pf, Phys. Lett.~B {\bf 249},  331
  (1990).

\bibitem{SchrammR97}
A.~J. Schramm and D.~H. Reeves, Phys. Rev.~D {\bf 55},  7312  (1997).

\bibitem{CapellaSTT94}
A. Capella, U. Sukhatme, C. Tan, and J. {Tran Thanh Van}, Phys. Rep. {\bf 236},
   225  (1994).

\bibitem{Engel98}
R. Engel, private communication.

\bibitem{RoldaoN00}
C.~G. Rold{\~a}o and A.~A. Natale, Phys. Rev.~C {\bf 61},  064907  (2000).*

\bibitem{Chung91}
S.~U. Chung, D.~P. Weygand, and H.~J. Willutzki, Preprint BNL-46863, 1991.

\bibitem{ChatrchyanJZ97}
S.~A. Chatrchyan and P.~I. Zarubin,  in {\em Second CMS Heavy Ion Workshop,
  Dubna, Russia} (PUBLISHER, ADDRESS, 1997), Vol.~CMS Document 97-011, p.\ 257.

\bibitem{AlberiG81}
G. Alberi and G. Goggi, Phys. Rep. {\bf 74},  1  (1981).

\bibitem{Bottcher89}
C. Bottcher and M.~R. Strayer, Phys. Rev.~D {\bf 39},  1330  (1989).

\bibitem{MombergerGS91}
K. Momberger, N. Gr{\"u}n, and W. Scheid, Z. Phys. D {\bf 18},  133  (1991).

\bibitem{RumrichSG93}
K. Rumrich, G. Soff, and W. Greiner, Phys. Rev.~A {\bf 47},  215  (1993).

\bibitem{AlscherHT97}
A. Alscher, K. Hencken, D. Trautmann, and G. Baur, Phys. Rev.~A {\bf 55},  396
  (1997).

\bibitem{Eichler90}
J. Eichler, Phys. Rep. {\bf 193},  165  (1990).

\bibitem{EichlerM95}
J. Eichler and W.~E. Meyerhof, {\em Relativistic Atomic Collisions} (Academic
  Press, San Diego, 1995).**

\bibitem{LeeM00}
R.~N. Lee and A.~I. Milstein, Phys. Rev.~A {\bf 61},  032103  (2000).

\bibitem{LeeMS01}
R.~N. Lee, A.~I. Milstein, and V.~G. Serbo, Structure of the Coulomb and
  unitarity corrections to the cross section of {$e^{+}e^{-}$} pair production
  in ultra-relativistic nuclear collisions, e-print hep-ph/0108014, 2001.*

\bibitem{HenckenTB95b}
K. Hencken, D. Trautmann, and G. Baur, Phys. Rev.~A {\bf 51},  1874  (1995).

\bibitem{Guclu95}
M.~C. G{\"u\c c}l{\"u} {\it et~al.}, Phys. Rev.~A {\bf 51},  1836  (1995).

\bibitem{Heitler34}
W. Heitler, {\em The Quantum Theory of Radiation} (Oxford University Press,
  London, 1954).

\bibitem{Baur90}
G. Baur, Phys. Rev.~A {\bf 42},  5736  (1990).**

\bibitem{Baur90c}
G. Baur, Phys. Rev.~D {\bf 41},  3535  (1990).

\bibitem{RhoadesBrownW91}
M.~J. Rhoades-Brown and J. Weneser, Phys. Rev.~A {\bf 44},  330  (1991).

\bibitem{BestGS92}
C. Best, W. Greiner, and G. Soff, Phys. Rev.~A {\bf 46},  261  (1992).

\bibitem{Serbo70}
V.~G. Serbo, JETP Lett. {\bf 12},  39  (1970).

\bibitem{LipatovF71}
L.~N. Lipatov and G.~V. Drolov, Sov. J. Nucl. Phys. {\bf 13},  333  (1971).

\bibitem{HenckenTB95a}
K. Hencken, D. Trautmann, and G. Baur, Phys. Rev.~A {\bf 51},  998  (1995).**

\bibitem{BaltzGMP01}
A. Baltz, F. Gelis, L. McLerran, and A. Peshier, 
Nucl. Phys.~A {\bf 695}, 395 (2001).

\bibitem{Hencken94}
K. Hencken, Electromagnetic Production of Electron Positron Pairs in
  Relativistic Heavy Ion Collisions, Ph. D. thesis, U Basel, Institut f{\"u}r
  Physik, available from
  http://quasar.physik.unibas.ch/{$\sim$}hencken/Preprints/, 1994.

\bibitem{VaneDDD97}
C.~R. Vane {\it et~al.}, Phys. Rev.~A {\bf 56},  3682  (1997).

\bibitem{MombergerBS95}
K. Momberger, A. Belkacem, and A.~H. S{\o}rensen, Europhys. Lett. {\bf 32},
  401  (1995).

\bibitem{ThielHGS95}
J. Thiel, J. Hoffstadt, N. Gr{\"u}n, and W. Scheid, Z. Phys. D {\bf 34},  21
  (1995).

\bibitem{IvanovSS99}
D.~Y. Ivanov, A. Schiller, and V.~G. Servo, Phys. Lett.~B {\bf 454},  155
  (1999).

\bibitem{IvanovM97}
D. Ivanov and K. Melnikov, Phys. Rev.~D {\bf 57},  4025  (1998).

\bibitem{BetheM54}
H.~A. Bethe and L.~C. Maximon, Phys. Rev. {\bf 93},  768  (1954).

\bibitem{DaviesBM54}
H. Davies, H.~A. Bethe, and L.~C. Maximon, Phys. Rev. {\bf 93},  788  (1954).

\bibitem{LandauLQED}
L.~D. Landau and E.~M. Lifschitz, {\em Quantenelektrodynamik}, No.~IV in {\em
  Lehrbuch der theoretischen Physik} (Akademie Verlag, Berlin, 1986).

\bibitem{Baltz97}
A. Baltz, Phys. Rev. Lett. {\bf 78},  1231  (1997).

\bibitem{SegevW98}
B. Segev and J.~C. Wells, Phys. Rev.~A {\bf 57},  1849  (1998).

\bibitem{BaltzM98}
A.~J. Baltz and L. McLerran, Phys. Rev.~C {\bf 58},  1679  (1998).

\bibitem{SegevW98b}
B. Segev and J.~C. Wells, Phys. Rev.~C {\bf 59},  2753  (1999).

\bibitem{EichmannRSG99}
U. Eichmann, J. Reinhardt, S. Schramm, and W. Greiner, Phys. Rev.~A {\bf 59},
  1223  (1999).

\bibitem{HenckenTB99}
K. Hencken, D. Trautmann, and G. Baur, Phys. Rev.~C {\bf 59},  841  (1999).

\bibitem{Guclu00}
M.~C. G{\"u}{\c c}l{\"u}, Phys. Rev.~C {\bf 63},  049802  (2000).

\bibitem{EichmannRG00}
U. Eichmann, J. Reinhardt, and W. Greiner, Phys. Rev.~A {\bf 61},  062710
  (2000).

\bibitem{BlankenbeclerD87}
 R.~Blankenbecler and S.D.~Drell, Phys. Rev.~D {\bf 36}, 2846 (1987).

\bibitem{LeeM01}
R.~N. Lee and A.~I. Milstein, Coulomb corrections and multiple e+e- pair
  production in ultra-relativistic nuclear collisions, e-print hep-ph/0103212,
  Phys. Rev. A in press, 2001.

\bibitem{Feynman49}
R.~P. Feynman, Phys. Rev. {\bf 76},  749  (1949).

\bibitem{Schwinger54}
J. Schwinger, Phys. Rev. {\bf 93},  615  (1954).

\bibitem{ItzyksonZ80}
C. Itzykson and J.-B. Zuber, {\em Quantum Field Theory} (McGraw-Hill,
  Singapore, 1980).

\bibitem{BialynickiB75}
I. Bia{\-l}ynicki-Birula and Z. Bia{\-l}ynicki-Birula, {\em Quantum
  Electrodynamics} (Pergamon, ADDRESS, 1975).

\bibitem{Spencer01NEW}
S. Klein, Ultra-peripheral collisions of relativistic heavy ions, Presented at
  INPC, July 30--August 3, 2001, Berkeley. e-print nucl-ex/0108018., 2001.

\bibitem{BaurB93b}
G. Baur and N. Baron, Z. Phys. C {\bf 60},  95  (1993).

\bibitem{Baur92b}
G. Baur, Nucl. Phys.~A {\bf 538},  187c  (1992).

\bibitem{BaurCERES94}
R. Baur {\it et~al.}, Phys. Lett.~B {\bf 332},  471  (1994).

\bibitem{ChenZ75}
M. Chen and P. Zerwas, Phys. Rev.~D {\bf 12},  187  (1975).*

\bibitem{BaierFK73}
V.~N. Baier, V.~S. Fadin, and V.~H. Khoze, Nucl. Phys.~B {\bf 65},  381
  (1973).

\bibitem{DrellW64}
S.~D. Drell and J.~D. Walecka, Ann. Phys. {\bf 28},  18  (1964).

\bibitem{PDG96}
R.~M. Barnett {\it et~al.}, Phys. Rev.~D {\bf 54},  1  (1996).

\bibitem{KorotkikhC01}
V.~L. Korotkikh and K.~A. Chikin, {$\gamma$}-radiation of excited nuclear
  discrete levels in peripheral heavy ion collisions, INPH MSU 2001-1/641,
  e-print nucl-th/0103018, 2001.

\bibitem{MeierHTB98}
H. Meier {\it et~al.}, Eur. Phys. J. C {\bf 2},  741  (1998).

\bibitem{FadinK73}
V.~S. Fadin and V.~A. Khoze, Sov. Phys.-JETP {\bf 17},  313  (1973).

\bibitem{Weinberg97}
S. Weinberg, {\em The Quantum Theory of Fields} (Cambridge University Press,
  Cambridge, 1997), Vol.~1.

\bibitem{HenckenTB99b}
K. Hencken, D. Trautmann, and G. Baur, Phys. Rev.~C {\bf 60},  34901  (1999).

\bibitem{CMS-ECAL-TDR}
{CMS Collaboration}, CMS, the Electromagnetic Calorimeter Project: technical
  design report, CERN-LHCC-97-033, 1997.

\bibitem{Jeanneret00}
J.~B. Jeanneret, Electron capture in Pb-Pb collisions and quench limit, Beam
  Physics Note 41, 2000.*

\bibitem{MungerBS94}
C.~T. Munger, S.~J. Brodsky, and I. Schmidt, Phys. Rev.~D {\bf 49},  3228
  (1994).

\bibitem{Nemenov85}
L.~L. Nemenov, Sov. J. Nucl. Phys. {\bf 41},  629  (1985).

\bibitem{MeierHHT01}
H. Meier {\it et~al.}, Phys. Rev.~A {\bf 63},  032713  (2001).*

\bibitem{MeierHHT98}
H. Meier {\it et~al.}, Eur. Phys. J. C {\bf 5},  287  (1998).

\bibitem{BetheS57}
H.~A. Bethe and E.~E. Salpeter, {\em Quantum Mechanics of One- and Two-Electron
  Atoms} (Springer Verlag, Berlin, 1957).

\bibitem{BaltzRW91}
A.~J. Baltz, M.~J. Rhoades-Brown, and J. Weneser, Phys. Rev.~A {\bf 44},  5569
  (1991).

\bibitem{BaltzRW93}
A.~J. Baltz, M.~J. Rhoades-Brown, and J. Weneser, Phys. Rev.~A {\bf 48},  2002
  (1993).

\bibitem{BeckerGS87}
U. Becker, N. Gr{\"u}n, and W. Scheid, J. Phys.~B {\bf 20},  2075  (1987).

\bibitem{AsteHT94}
A. Aste, K. Hencken, D. Trautmann, and G. Baur, Phys. Rev.~A {\bf 50},  3980
  (1994).

\bibitem{AggerS97}
C.~K. Agger and A.~H. S{\o}rensen, Phys. Rev.~A {\bf 55},  402  (1997).

\bibitem{RhoadesBrownBS89}
M.~J. Rhoades-Brown, C. Bottcher, and M.~R. Strayer, Phys. Rev.~A {\bf 40},
  2831  (1989).

\bibitem{BertulaniD01}
C. Bertulani and D. Dolci, Nucl. Phys.~A {\bf 683},  635  (2001).

\bibitem{Belkacem93}
A. Belkacem {\it et~al.}, Phys. Rev. Lett. {\bf 71},  1514  (1993).

\bibitem{Belkacem94}
A. Belkacem {\it et~al.}, Phys. Rev. Lett. {\bf 73},  2432  (1994).

\bibitem{BelkacemGF97}
A. Belkacem {\it et~al.}, Phys. Rev.~A {\bf 56},  2806  (1997).

\bibitem{ClaytorBD97}
N. Claytor {\it et~al.}, Phys. Rev.~A {\bf 55},  R842  (1997).

\bibitem{BelkacemCD98}
A. Belkacem {\it et~al.}, Phys. Rev.~A {\bf 58},  1253  (1998).

\bibitem{VoitkivGS00}
A.~B. Voitkiv, N. Gr{\"u}n, and W. Scheid, Phys. Lett.~A {\bf 269},  325
  (2000).

\bibitem{Sorensen98}
A.~H. S{\o}rensen, Phys. Rev.~A {\bf 58},  2895  (1998).

\bibitem{AnholtB87}
R. Anholt and U. Becker, Phys. Rev.~A {\bf 36},  4628  (1987).

\bibitem{Krause98}
H.~F. Krause {\it et~al.}, Phys. Rev. Lett. {\bf 80},  1190  (1998).

\bibitem{Krause01}
H.~F. Krause {\it et~al.}, Phys. Rev.~A {\bf 63},  032711  (2001).

\bibitem{BaurO96}
G. Baur {\it et~al.}, Phys. Lett.~B {\bf 368},  251  (1996).

\bibitem{Blanford98}
G. Blanford {\it et~al.}, Phys. Rev. Lett. {\bf 80},  3037  (1998).

\bibitem{BertulaniB98}
C.~A. Bertulani and G. Baur, Phys. Rev.~D {\bf 58},  034005  (1998).

\bibitem{Baur90b}
G. Baur,  in {\em Perspectives on Photon Interactions with Hadrons and Nuclei},
  {\em Lecture Notes in Physics}, edited by M. Schumacher and G. Tamas
  (Springer Verlag, Berlin, Heidelberg, New York, 1990), p.\ 111.

\bibitem{Ginzburg97}
I.~F. Ginzburg, private communication.

\bibitem{KotkinKSS99}
G.~L. Kotkin, E.~A. Kuraev, A. Schiller, and V.~G. Serbo, Phys. Rev.~C {\bf
  59},  2734  (1999).*

\bibitem{BertulaniN01}
C. Bertulani and F. Navarra, 3photon production of Vector Mesons, e-print
  nucl-th/0107035, 2001.

\bibitem{Ginzburg98}
I.~F. Ginzburg, Phys. Rev.~C {\bf 58},  3565  (1998).

\bibitem{Gevorkyan98}
S.~R. Gevorkyan {\it et~al.}, Phys. Rev.~A {\bf 58},  4556  (1998).

\bibitem{Karsh1}
A. Karshenboim {\it et~al.}, Phys. Lett.~B {\bf 424},  397  (1998).

\bibitem{Jents}
U. Jentschura {\it et~al.}, Phys. Rev.~A {\bf 56},  4483  (1997).

\bibitem{Karsh2}
A. Karshenboim {\it et~al.}, Sov. Phys.-JETP {\bf 86},  226  (1998).

\bibitem{Nemenov}
L. Nemenov {\it et~al.}, Sov. J. Nucl. Phys. {\bf 15},  582  (1972).

\bibitem{Malenfant}
J. Malenfant, Phys. Rev.~D {\bf 36},  863  (1987).

\bibitem{Bilen}
S. Bilen'kii {\it et~al.}, Sov. J. Nucl. Phys. {\bf 10},  469  (1969).

\bibitem{Kozlov}
G.~A. Kozlov, Phys.At Nucl. {\bf 48},  167  (1988).

\bibitem{mohr}
P.~J. Mohr,  in {\em Atomic, Molecular and Optical Physics Handbook.}, edited
  by G.~W.~F. Drake (AIP, Woodbury, NY, 1996).

\bibitem{jungmann}
M.~G. Boshier {\it et~al.}, Comm. At. Mol. Phys. {\bf 33},  17  (1996).

\bibitem{Ginsburg}
I.~F. Ginzburg {\it et~al.}, Phys. Rev.~C {\bf 58},  3565  (1998).

\bibitem{Pythia}
T. {Sj\"ostrand}, Computer Phys. Commun. {\bf 82},  74  (1994).

\bibitem{Hijing}
X.~N. Wang and M. Gyulassy, HIJING 1.0: a Monte Carlo program for parton and
  particle production in high energy hadronic and nuclear collisions, Preprint
  LBL 34246; e-print nucl-th/9502021, 1995.

\bibitem{STAR347}
S. Klein and J. Nystrand, STAR Note 347, 1998.**

\bibitem{Acciarri:2000ev}
M. Acciarri {\it et~al.}, Phys. Lett. {\bf B501},  1  (2001).

\bibitem{Acciarri:2000ex}
M. Acciarri {\it et~al.}, Phys. Lett. {\bf B501},  173  (2001).

\bibitem{Huang2001}
H.C.Huang, e-print hep-ex/0104024, To appear in the proceedings of 4th
  International Conference on B Physics and CP Violation (BCP 4), Ago Town, Mie
  Prefecture, Japan, 19-23 Feb 2001, 2001.

\bibitem{Godang:1997hj}
R. Godang {\it et~al.}, Phys. Rev. Lett. {\bf 79},  3829  (1997).

\bibitem{Acciarri:2000sc}
M. Acciarri {\it et~al.}, Phys. Lett. {\bf B514},  19  (2001).

\bibitem{Acciarri:2000kd}
M. Acciarri {\it et~al.}, Phys. Lett. {\bf B503},  10  (2001).

\bibitem{Acciarri:1999md}
M. Acciarri {\it et~al.}, Phys. Lett. {\bf B467},  137  (1999).

\bibitem{Csilling:2000xk}
A. Csilling, Charm and bottom production in two-photon collisions with OPAL,
  Talk given at PHOTON 2000: Ambleside, Lake District, England, 26-31 Aug
  2000., 2000.

\bibitem{Brandenburg:2000ry}
G. Brandenburg {\it et~al.}, Phys. Rev. Lett. {\bf 85},  3095  (2000).

\bibitem{Eisenstein:2001xe}
B.~I. Eisenstein, Experimental investigation of the two-photon widths of the
  $\chi_{c0}$ and the $\chi_{c2}$ mesons, CLNS 01-1732, 2001.

\bibitem{Acciarri:1999rv}
M. Acciarri {\it et~al.}, Phys. Lett. {\bf B461},  155  (1999).

\bibitem{NystrandK97}
J. Nystrand and S. Klein, Two Photons Physics at RHIC: Separating Signals from
  Backgrounds, talk presented at ``Hadron'97'', Brookhaven National Laboratory,
  August 1997, STAR Note 315, LBNL-41111 Nov.97, e-print hep-e/9711021, 1997.

\bibitem{Fritiof}
H. Pi, Comp.Phys.Comm. {\bf 173},  71  (1992).

\bibitem{Venus}
K. Werner, Phys. Rev. {\bf 87},  232  (1993).

\bibitem{Dtunuc}
R. Engel, J. Ranft, and S. Roesler, Phys. Rev.~D {\bf 55},  6957  (1997).

\bibitem{CERN-Courier-41}
CERN Courier, vol.41, No.7, p.6, 2001.

\bibitem{Hou2001}
S. Hou, Plenary talk at the conference HADRON2001, 25 August -- 1 September
  2001, Protvino, Russia., 2001.

\bibitem{CMS-2000-060}
G. Baur {\it et~al.}, CMS Note 2000-060, 2000.

\bibitem{ALICE95}
N. Ahmad {\it et~al.}, Alice Technical Proposal, CERN/LHCC 95-71, 1995.

\bibitem{Sadovsky95}
S.~A. Sadovsky,  in {\em Proceedings of the 6th Intern. Conf. on Hadron
  Spectroscopy, Manchester, UK, 10-14 August 1995}, edited by M.~C. Birse,
  G.~D. Lafferty, and J.~A. McGovern (World Scientific, Singapore, 1996), p.\
  289.

\bibitem{Kolosov95}
V. Kolosov {\it et~al.}, ALICE Internal Note 95-45, 1995.

\end{thebibliography}
\end{document}